\documentclass[a4paper]{elsarticle}

\usepackage[colorlinks]{hyperref}

\usepackage[latin1]{inputenc}
\usepackage[T1]{fontenc}
\usepackage[english]{babel}
\addto\captionsenglish{}

\usepackage[export]{adjustbox}
\usepackage{booktabs}
\usepackage{multicol}

\usepackage{enumitem}

\usepackage{amssymb}
\usepackage{amsmath}
\usepackage{amsthm}
\usepackage{bm}
\usepackage[squaren,Gray]{SIunits}

\usepackage{xcolor}
\usepackage{subfig}
\graphicspath{{./pictures/}}

\usepackage{tikz}
\usepackage{pgf}
\usepackage{pgfplots}
\pgfplotsset{compat=1.14}
\usetikzlibrary{plotmarks}
\usetikzlibrary{shapes,arrows,backgrounds,patterns,calc}
\makeatletter
\def\input@path{{./tikz/}}
\makeatother

\numberwithin{equation}{section}

\journal{arXiv}

\begin{document}
	
\begin{frontmatter}
			
\title{Comparison of multiphase SPH and LBM approaches for the simulation of intermittent flows}
				
\author[1,2]{Thomas Douillet-Grellier} \corref{cor1}
\author[3,4]{S\'ebastien Leclaire}
\author[4]{David Vidal}
\author[4]{François Bertrand}
\author[5]{Florian De Vuyst} %\corref{cor2}

\address[1]{CMLA, CNRS, ENS Paris-Saclay, Universit\'e Paris-Saclay, Cachan, France.}
\address[2]{Total S.A., Tour Coupole, Paris La D\'efense, France}
\address[3]{Department of Mechanical Engineering, Polytechnique Montr\'eal, Qu\'ebec, Canada}
\address[4]{URPEI, Department of Chemical Engineering, Polytechnique Montr\'eal, Qu\'ebec, Canada}
\address[5]{LMAC, Universit\'e de Technologie de Compi\`egne, Compi\`egne, France}
		
\cortext[cor1]{thomas.douillet-grellier@ens-paris-saclay.fr}
		
\begin{abstract}
	
	Smoothed Particle Hydrodynamics (SPH) and Lattice Boltzmann Method (LBM) are increasingly popular and attractive methods that propose efficient multiphase formulations, each one with its own strengths and weaknesses. In this context, when it comes to study a given multi-fluid problem, it is helpful to rely on a quantitative comparison to decide which approach should be used and in which context. In particular, the simulation of intermittent two-phase flows in pipes such as slug flows is a complex problem involving moving and intersecting interfaces for which both SPH and LBM could be considered. It is a problem of interest in petroleum applications since the formation of slug flows that can occur in submarine pipelines connecting the wells to the production facility can cause undesired behaviors with hazardous consequences. In this work, we compare SPH and LBM multiphase formulations where surface tension effects are modeled respectively using the continuum surface force and the color gradient approaches on a collection of standard test cases, and on the simulation of intermittent flows in 2D. This paper aims to highlight the contributions and limitations of SPH and LBM when applied to these problems. First, we compare our implementations on static bubble problems with different density and viscosity ratios. Then, we focus on gravity driven simulations of slug flows in pipes for several Reynolds numbers. Finally, we conclude with simulations of slug flows with inlet/outlet boundary conditions. According to the results presented in this study, we confirm that the SPH approach is more robust and versatile whereas the LBM formulation is more accurate and faster.
	
\end{abstract}
		
		
\end{frontmatter}

\section{Introduction}
Two-phase flows in pipes are encountered in various industrial applications and research areas. For instance, gas-liquid flows are present in chemical and heat transfer systems such as evaporators or refrigerators, but they can also be found in the transportation of oil and gas through pipelines in the petroleum industry. In particular, intermittent flow patterns (also known as slug or plug flow~\cite{Fabre1992}) are undesirable in pipeline networks. Those slug patterns, that can measure up to tens of meters, are known to damage facilities (separator flooding, compressor starving, water hammer phenomenon, vibrations and fatigue) and to reduce flow efficiency. Therefore, it is crucial to be able to predict whether or not a slug flow regime will occur. If it does, the quantities of interest are the size of slugs, their transit time and their frequency.

In this context, numerical simulations have emerged as a tool of choice~\cite{lu2015experimental,pedersen2017} to understand the formation of intermittent flow patterns and to predict the appearance of a slug flow regime. At the industrial level, the first simulations~\cite{taitel1980,viggiani1988,sarica1991} were done in the mid 80's. Nowadays, in the oil and gas industry, two commercial software products are competing for slugging simulation : OLGA developed by SPT group~\cite{bendiksen1991} and LedaFlow, proposed by Kongsberg~\cite{kongsberg2014}. A detailed comparison of both codes~\cite{belt2011} concludes that although performing equally well on simple cases, they have trouble to simulate complex cases with a dominant gas phase. From an academic perspective, different models and methods have been used for slug flow modeling, for example volume-of-fluid~\cite{Taha2004,al2016numerical}, level-set~\cite{Fukagata2007,lizarraga2016study}, lattice Boltzmann method (LBM)~\cite{Yu2007}, smoothed particle hydrodynamics (SPH)~\cite{minier2016,DouilletGrellier2018} or phase field~\cite{He2008,xie2017}, but these are applied on microfluidic problems for the most part. In this work, we will focus on two of them~: SPH and LBM, because while they are very different numerical methods, both in their origin and in their nature, they have shown a strong potential to model multiphase flows~\cite{Huang2015,Wang2016,Li2016}. 

SPH is a Lagrangian meshfree method introduced in the late 70's for astrophysics applications~\cite{lucy1977numerical,gingold1977smoothed} and later expanded to standard fluid mechanics. SPH can be seen from two different perspectives. On one hand, it is a way to discretize partial differential equations such as the Navier-Stokes equations. On the other hand, it is a discrete Hamiltonian system composed of material points of constant mass that are tracked in time. Its pure meshless nature allows to get rid of many issues associated with meshing. However, it comes at some expenses too. Since there is no connectivity between particles, a neighbor searching procedure has to be carried for every particle at every time step, which is a serious bottleneck for code efficiency. Among the numerous applications of SPH, we can mention astrophysics~\cite{Springel2010}, hydrodynamics~\cite{violeau2016smoothed}, geophysics~\cite{Libersky1991,bui2008,spe,DouilletGrellier201673} and computer graphics~\cite{ihmsen2014sph}. Some extensive reviews have been published~\cite{monaghan2012review,price2012magneto,shadloo2016smoothed}.

LBM originates from two distinct approaches: the kinetic gas theory with discrete velocities and the lattice gas cellular automata method (LGCA or LGA~\cite{lgca}). In the late 80's, probability density functions of particles were introduced within LGCA, giving birth to LBM. It consists in solving the Boltzmann equation in a discrete velocity space, which has been proven to be equivalent to solving the Navier-Stokes equations in the limit of low Mach and Knudsen numbers (as can be shown by a multiscale Chapman-Enskog expansion~\cite{chapmanenskog}). In practice, LBM distinguishes itself from other methods for several reasons. First, the LBM operates on an uniform lattice (mostly square or hexagonal lattices). Then, LBM offers a local algorithm involving simple arithmetic operations with no differential terms, which makes it short, fast and particularly well suited for parallel execution~\cite{lbmparallelcomp}. Finally, traditional CFD methods are based on the calculation of macroscopic variables (velocity, pressure, density) whereas LBM tracks the evolution of probability distribution functions of particles~\cite{lbmtheory}. LBM has been used for decades for flow simulations in complex geometries, especially in porous media~\cite{lbmporousmedia,lbmporousmedia2,lbmflowporousmedia3,lbmflowporousmedia4}.

Our goal here is to propose a comparison of the multiphase SPH formulation presented in~\cite{Hu2006} and the color-gradient multiphase LBM formulation introduced in~\cite{Reis2007}, on a collection of standard 2D test cases and on the simulation of slug flow regimes with periodic and inlet/outlet boundary conditions. To the best of our knowledge, this the first time such a comparison is presented.

We first detail the multiphase LBM and SPH formulations used in this work including surface tension models and boundary conditions. Then, we compare both approaches on static bubble tests with different density and viscosity ratios. Finally, we extend the comparison to two cases of slug flows, one induced by gravity with periodic boundary conditions and the other one based on inlet/outlet boundary conditions.

In addition, we provide in Appendix~\ref{appendix} a comparison of SPH and LBM on two test cases where one is a single-phase flow and the other does not involve surface tension : the lid-driven cavity flow and the Rayleigh-Taylor instability problems.

\section{LBM immiscible multiphase model}

Four main multiphase formulations are available for LBM : the pseudo-potential model~\cite{Shan1993}, the free energy model~\cite{Swift1995}, the mean field model~\cite{He1999} and the color gradient model~\cite{Gunstensen1991}. We recommend the reading of~\cite{Liu2015,Leclaire2017} for those looking for a detailed comparison of these techniques. In this work, we choose to work with the color gradient model because among the diffuse interface approaches, it is the one with a thin interface compared to the pseudo potential approach for example. In addition, the pseudo-potential model is cumbersome to use and to parametrize because there is coupling between the basic parameters~\cite{Leclaire2017}. The free energy model requires solving a Poisson equation at every time step which is time consuming and the mean field approach is limited to small density ratios~\cite{CompMethForMultiphase}. Moreover, the color gradient model benefits from the large body of verification and validation cases available in the literature~\cite{Leclaire2011,Leclaire2012,Leclaire2013,Leclaire2014,Leclaire2015,Leclaire2016,Leclaire2017}.

The present LBM approach is the two-phase model introduced in~\cite{Reis2007} and completed with the improvements proposed in~\cite{Leclaire2012,Leclaire2011} for the recoloring operator and the color gradient. In addition, the contact angle ajustment is based on~\cite{Leclaire2016,Xu2017} and the corrective procedure to properly recover Navier-Stokes equations is borrowed from~\cite{Yan2016}. We work with $2$ sets of distribution functions, one for each fluid, moving on a D2Q9 lattice. The associated velocity vectors are $\bm{c}_{i}$ with $i \in \left[0, 8\right]$. As traditionally done in LBM, we set the lattice time step $\Delta t$ and the lattice space step $\Delta x$ to $1$. We can then define the 2D velocity vectors as follows :
\begin{align}
\bm{c}_{i}=\left\{
\begin{array}{ll}
(0,0), & i=0 \\
\left[\sin(\theta_{i}),\cos(\theta_{i})\right], & i=1,3,5,7 \\
\left[\sin(\theta_{i}),\cos(\theta_{i})\right]\sqrt{2}, & i=2,4,6,8
\end{array},\text{ with }\theta_{i}=\frac{\pi}{4}(4-i).\right.
\end{align}

The distribution functions for a fluid of color~$k$ (e.g. $k=r$ for red and $k=b$ for blue) are denoted $N^{k}_{i}(\bm{x},t)$, while $N_{i}(\bm{x},t)= N^{r}_{i}(\bm{x},t)+N^{b}_{i}(\bm{x},t)$ is used for the color-blind distribution function. In the rest of this section, the integer subscript~$i$ is varying between $\left[0, 8\right]$ while $k$ is referring to the color blue $b$ or red $r$ of the fluid. The lattice Boltzmann equation that describes the evolution of the system is then :
\begin{equation}
N^{k}_{i}(\bm{x}+\bm{c}_{i},t+1)= N^{k}_{i}(\bm{x},t) + \Omega^{k}_{i}\big(N^{k}_{i}(\bm{x},t)\big), \label{eq:evolution}
\end{equation}
\noindent where the collision operator $\Omega^{k}_{i}$ is the result of the combination of three sub operators (as previously done in~\cite{Tolke2002}) :
\begin{equation}
\Omega^{k}_{i}=(\Omega^{k}_{i})^{(3)}\left[(\Omega^{k}_{i})^{(1)}+(\Omega^{k}_{i})^{(2)}\right].
\end{equation}

\noindent The Eq.~\eqref{eq:evolution} is solved using four consecutive steps which make use of the following operators :
\begin{enumerate}[wide, labelwidth=!, labelindent=0pt]
	
	\item Single phase collision step (see Sec.~\ref{singlephase}) :
	\begin{equation}
	N^{k}_{i}(\bm{x},t_{*})=(\Omega^{k}_{i})^{(1)}\big(N^{k}_{i}(\bm{x},t)\big).\label{eq:subop1}
	\end{equation}
	\item Perturbation step (multiphase collision 1/2) (see Sec.~\ref{sec:lbm_st}) :
	\begin{equation}
	N^{k}_{i}(\bm{x},t_{**})=(\Omega^{k}_{i})^{(2)}\big(N^{k}_{i}(\bm{x},t_{*})\big).\label{eq:subop2}
	\end{equation}
	\item Recoloring step (multiphase collision 2/2) (see Sec.~\ref{recoloring}) :
	\begin{equation}
	N^{k}_{i}(\bm{x},t_{***})=(\Omega^{k}_{i})^{(3)}\big(N^{k}_{i}(\bm{x},t_{**})\big).\label{eq:subop3}
	\end{equation}
	\item Streaming step :
	\begin{equation}
	N^{k}_{i}(\bm{x}+\bm{c}_{i},t+1)=N^{k}_{i}(\bm{x},t_{***}).\label{eq:subop4}
	\end{equation}
	
\end{enumerate}

We will now examine in detail these steps as well as the specific treatments for the imposition of static contact angles and boundary conditions.

\subsection{Single phase collision operators}\label{singlephase}

\subsubsection{BGK operator}

The first sub operator, $(\Omega^{k}_{i})^{(1)}$ of Eq.~\eqref{eq:subop1}, is the usual BGK operator for single phase LBM. The distribution functions are relaxed towards a local equilibrium as follows :
\begin{equation}
(\Omega^{k}_{i})^{(1)}(N^{k}_{i})=N^{k}_{i}-\omega_{\text{eff}}\left(N^{k}_{i}-N^{k(e)}_{i}\right), \label{eq:singleCollision}
\end{equation}
\noindent where $\omega_{\text{eff}}$ is the effective relaxation rate. In practice, we first calculate the fluid densities based on the $0^{th}$ moment of the distribution functions :
\begin{equation}
\rho_{k}=\sum_{i}N^{k}_{i}=\sum_{i}N^{k(e)}_{i},
\end{equation}
\noindent where $N^{k(e)}_{i}$ are the equilibrium distribution functions. The total fluid density is given by $\rho=\rho_{r}+\rho_{b}$, while the total momentum is computed as the $1^{st}$ moment of the distribution functions :
\begin{equation}
\rho\bm{u}=\sum_{i}\sum_{k}N^{k}_{i}\bm{c}_{i}=\sum_{i}\sum_{k}N^{k(e)}_{i}\bm{c}_{i},
\end{equation}
\noindent where $\bm{u}$ is the density weighted arithmetic average velocity of the fluid. The equilibrium distribution functions $N^{k(e)}_{i}$ are defined in~\cite{Reis2007} by :
\begin{equation}
N^{k(e)}_{i}(\rho_k,\bm{u},\alpha_k)= \rho_{k}\left(\phi^{k}_{i}+W_{i}\left[ 3\bm{c}_{i}\cdot\bm{u}+\frac{9}{2}(\bm{c}_{i}\cdot\bm{u})^{2}-\frac{3}{2}(\bm{u})^{2}\right]\right). \label{eq:equilibrium}
\end{equation}

These equilibrium distribution functions $N^{k(e)}_{i}$ are chosen to satisfy mass and momentum conservation and are based on the Maxwell-Boltzmann distribution. The weights $W_{i}$ are those of a standard D2Q9 lattice while the values $\phi^{k}_{i}$ depend on the density ratio. They are expressed as follows :\newline
\begin{minipage}{.5\linewidth}
	\begin{align*}
	W_{i}=\left\{ 
	\begin{array}{ll}
	4/9, & i=0 \\
	1/9, & i=1,3,5,7 \\
	1/36,& i=2,4,6,8
	\end{array}\right.
	\end{align*}
\end{minipage}%
\begin{minipage}{.5\linewidth}
	\begin{align*}
	\phi^{k}_{i}=\left\{ 
	\begin{array}{ll}
	\alpha_{k}, & i=0 \\
	(1-\alpha_{k})/5, & i=1,3,5,7 \\
	(1-\alpha_{k})/20,& i=2,4,6,8
	\end{array}.\right.
	\end{align*}
\end{minipage}\newline

As introduced in~\cite{Grunau1993} for two-phase flows, the density ratio between red and blue fluids is $\gamma$, and must be computed as follows to obtain a stable interface :
\begin{align}
\gamma=\frac{\rho^0_{r}}{\rho^0_{b}}=\frac{1-\alpha_{r}}{1-\alpha_{b}},\label{eq:gammaAlphaBeta}
\end{align}
\noindent where the superscript $0$ refers to the initial value of the density. Besides, the pressure of the fluid of color~$k$ is :
\begin{equation}
p^{k}=\frac{3\rho_{k}(1-\alpha_{k})}{5}=\rho_{k}(c_{s}^{k})^2.\label{eq:pressureRelation}
\end{equation}

In Eq.~\eqref{eq:gammaAlphaBeta}, one of the $\alpha_{k}$ is actually a free parameter. In practice, if we assume that the least dense fluid is the blue one, we set a positive value for $\alpha_{b}>0$ ($0.2$ in this paper) and so we are certain that $0<\alpha_{b}\leq\alpha_{r}<1$ using Eq.~\eqref{eq:gammaAlphaBeta}. These parameters define the value of the sound speed~$c_{s}^{k}$ in each fluid of color~$k$~\cite{Reis2007}.

The effective relaxation parameter $\omega_{\text{eff}}$ is chosen so that the evolution Eq.~\eqref{eq:evolution} allows to recover the macroscopic Navier-Stokes equations for a single-phase flow in the single-phase areas. This parameter depends on the fluid kinematic viscosity $\nu_k$ through the following formula: $\omega_{k} = \left. 1 \middle/ \left( 3\nu_k + \frac{1}{2}\right) \right.$. However, when the viscosities of the fluids are different, the relaxation parameters are also different and an interpolation procedure is needed to define an effective relaxation parameter~$\omega_{\text{eff}}$ at the interface. It is common to use a quadratic interpolation~\cite{Reis2007,Grunau1993}. In order to detect in which area we are located (pure red fluid, pure blue fluid or interface), it is necessary to introduce a color field :
\begin{equation}
\psi = \left. \left(\frac{\rho_{r}}{\rho^0_{r}} - \frac{\rho_{b}}{\rho^0_{b}} \right)\middle/ \left(\frac{\rho_{r}}{\rho^0_{r}} + \frac{\rho_{b}}{\rho^0_{b}} \right).\right. \label{eq:color_field}
\end{equation}

The color field~$\psi$ lies between $-1$ and $+1$. Within an interface, the color field is strictly between $-1$ and $+1$ whereas it is equal to $-1$ or $+1$ when located in an area that contains respectively only red fluid or blue fluid. The relaxation factor $\omega_{\text{eff}}$ in Eq.~\eqref{eq:evolution} is then evaluated as follows :
\begin{align}
\omega_{\text{eff}}=\left\{ 
\begin{array}{ll}
\omega_{r}, & \psi > \delta \\
f_{r}(\psi), & \delta \geq \psi > 0 \\
f_{b}(\psi), & 0 \geq \psi \geq -\delta \\
\omega_{b},& \psi < -\delta
\end{array},\right.
\end{align}
in which $\delta$ is a free parameter and\newline
%\begin{minipage}{.5\linewidth}
\begin{align}
\begin{array}{ll}
f_{r}(\psi) = \chi + \eta \psi + \kappa \psi^2, \\
f_{b}(\psi) = \chi + \lambda \psi + \nu \psi^2,\\
\end{array}
\end{align}
%\end{minipage}
%\begin{minipage}{.5\linewidth}
with :
\begin{align}
\begin{array}{ll}
\chi  = \left. \left( 2\omega_{r}\omega_{b} \right) \middle/ \left(\omega_{r}+\omega_{b} \right),\right. \\
\eta  = \left. \left(2(\omega_{r} - \chi) \right)\middle/ \delta, \right.\\
\kappa = \left. -\eta \middle/ \left(2\delta \right),\right.\\
\lambda = \left.\left(2(\chi - \omega_{b}) \right)\middle/ \delta, \right.\\
\nu   = \left.\lambda \middle/ \left(2\delta\right). \right.
\end{array}
\end{align} 
%\end{minipage}\newline	
The parameter $\delta$ influences the thickness of the interface when the fluid viscosities are different. The larger $\delta$, the thicker the fluid interface. There is a trade off to find between robustness and interface sharpness. It is set to $0.1$ for all simulations in this paper. If the fluid viscosities are identical, $\delta$ is ineffective, because $\omega_{\text{eff}} = \omega_{r} = \omega_{b}$.

\subsubsection{MRT operator}

Alternatively, for the first sub operator, $(\Omega^{k}_{i})^{(1)}$, one can use the Multiple Relaxation Time (MRT) operator instead of the BGK operator. The MRT approach is more stable than its BGK counterpart~\cite{Lallemand2000,Yan2016}. It reads as follows :
\begin{equation}
(\Omega^{k}_{i})^{(1)}(N^{k}_{i})= N^{k}_{i}- \bm{M}^{-1} \bm{S} \bm{M} \left(N^{k}_{i}-N^{k(e)}_{i}\right), \label{eq:singleCollisionMRT}
\end{equation}
\noindent where $\bm{S}$ is a diagonal matrix given by :
\begin{equation}\label{relax_times}
\bm{S} = \text{diag}\left(s_0, s_1, s_2, s_3, s_4, s_5, s_6, s_7, s_8\right).
\end{equation}
The elements $s_i, i \in \left[0 \ldots 8 \right]$ are the relaxation parameters. Following~\cite{Leclaire2014}, $\forall i \in \left[0 \ldots 6 \right], s_i = \lambda \omega_{\text{eff}}$ while $s_7=s_8=\omega_{\text{eff}}$. As in~\cite{Leclaire2014}, we choose $\lambda=4/5$. Moreover, $\bm{M}$ is a linear orthogonal transformation matrix that projects the distribution functions into the moment space. The discrete moment matrices $\bm{M}$ and $\bm{M}^{-1}$ are explicitly given in Appendix~\ref{mrt_mat} for a D2Q9 lattice.

\subsubsection{Proper recovery of Navier-Stokes equations}

It has been emphasized in several papers~\cite{Liu2012,Huang2013,Leclaire2013} that the color gradient model does not fully recover the Navier-Stokes equations. An unwanted error term arises in the momentum equations when the density ratio is not one. Different techniques have been proposed to attenuate this issue~\cite{Leclaire2013,Holdych1998,Huang2013,Yan2016}. In the present work, we use the correction introduced in~\cite{Yan2016} for the MRT approach. It consists in two additions. First, a modified equilibrium distribution functions based on the $3^{rd}$ order Hermite expansion of the Maxwellian distribution~\cite{Shan2006,Li2012b} is used instead of Eq.~\eqref{eq:equilibrium}. It is expressed as follows :
\begin{align}
&N^{k(e)}_{i}(\rho_k,\bm{u},\alpha_k)=\rho_{k} \phi^{k}_{i} +&\notag\\
&\rho_{k}W_{i}\left(\left[ 3\bm{c}_{i}\cdot\bm{u} \left[ 1 + \frac{1}{2} \left( 3 (c_s^k)^2 -1 \right) \left( 3 \bm{c}_{i} -4 \right)\right]+\frac{9}{2}(\bm{c}_{i}\cdot\bm{u})^{2}-\frac{3}{2}(\bm{u})^{2}\right]\right).& \label{eq:equilibrium_hermite}
\end{align}
Second, a source term $\bm{U}^k$ is added in Eq.~\eqref{eq:singleCollisionMRT}. It reads :
\begin{equation}
\bm{U}^k = \bm{M}^{-1} \bm{C}^k,
\end{equation}
\noindent where $\bm{C}^k=\left[0,C^k_1,0,0,0,0,0,C^k_7,0 \right]^{T}$. The components $C^k_1$ and $C^k_7$ are computed as follows :\newline
\begin{minipage}{.5\linewidth}
	\begin{align}\nonumber
	\begin{array}{ll}
	C^k_1  = 3(1-s_1/2)(\partial_x Q^k_x + \partial_y Q^k_y), \\
	C^k_7  = 3(1-\omega_{\text{eff}}/2)(\partial_x Q^k_x + \partial_y Q^k_y),
	\end{array}\text{ with}
	\end{align} 
\end{minipage}
\begin{minipage}{.5\linewidth}
	\begin{align}
	\begin{array}{ll}
	Q^k_x  = (1.8 \alpha^k - 0.8) \rho^k u_x, \\
	Q^k_y  = (1.8 \alpha^k - 0.8) \rho^k u_y.
	\end{array}
	\end{align} 
\end{minipage}\newline\newline	
%In particular, the derivatives are evaluated using a $9$-point isotropic finite difference approximation. For example, for a given variable $\phi$, we can write
%\begin{equation}
%\partial_{\alpha} \phi(\bm{x}) = 3 \sum_i W_i \phi(\bm{x} + \bm{c}_{i}) c_{i \alpha}.
%\end{equation}
In particular, the derivatives are evaluated using a $9$-point isotropic finite difference approximation described shortly afterwards in Eq.~\eqref{eq:colorGradient}.

\subsection{Perturbation operator}\label{sec:lbm_st}
In the color gradient model, surface tension forces are introduced by means of a perturbation operator shown in Eq.~\eqref{eq:subop2}~\cite{halliday1998,Reis2007,Gunstensen1991}. To begin, it is needed to introduce the color gradient $\bm{F}$ of the color field $\psi$ (see Eq.~\eqref{eq:color_field}) that approximates the fluid-fluid interface normal. It is written as :
\begin{align}
\bm{F}= \bm{\nabla}\psi.\label{eq:FSpencer}
\end{align}
In this work, a bi-dimensional S$2$I$4$ (spatial order $S=2$, isotropic order $I=4$) discrete gradient operator is used~\cite{Leclaire2013b}. As shown in~\cite{Leclaire2011}, this kind of gradient operator enhances the accuracy of the color-gradient model significantly while having the advantage to only rest on the nearest neighboring nodes. It takes the following form : 
\begin{align}
\bm{\nabla}f(\bm{x})\approx \sum_{i} \xi_i \bm{c}_{i} f(\bm{x}+\bm{c}_{i}),\text{ with } \xi_i=\left\{ 
\begin{array}{ll}
0, & i=0 \\
1/3, & i=1,3,5,7 \\
1/12,& i=2,4,6,8
\end{array}.\right. \label{eq:colorGradient}
\end{align}
%\noindent with $\bm{\xi}=[0,4,4,4,4,1,1,1,1]/12$. 
The perturbation operator, for the fluid $k$, is defined by :
\begin{equation}
(\Omega_{i}^{k})^{(2)}(N^{k}_{i})= N^{k}_{i}+\sum_{\substack{l \\ l\neq k}} \frac{A}{2}|\bm{F}_{kl}|\left[W_{i}\frac{(\bm{F}\cdot\bm{c}_{i})^{2}}{|\bm{F}|^{2}}-B_{i}\right], \label{eq:perturbation}
\end{equation}
\noindent where :
\begin{align}
B_{i}=\left\{ 
\begin{array}{ll}
-4/27, & i=0 \\
2/27, & i=1,3,5,7 \\
5/108,& i=2,4,6,8
\end{array}.\right.
\end{align}

The parameter~$A$ which can vary in space and time handles the coupling between both fluids and is therefore linked with the surface tension coefficient $\sigma$. It is possible to predict the surface tension $\sigma$ between the two fluids using only the basic parameters of the model. As explained in~\cite{Leclaire2012}, knowing the form of the expression describing the surface tension and performing simulations on planar interfaces, one can derive an expression that links the surface tension across an interface to the model parameters. For isotropic color gradients defined as in Eq.~\eqref{eq:colorGradient}, the surface tension is set as follows :
\begin{equation}
\sigma = \frac{4}{9}\frac{A}{\omega_{\text{eff}}}. \label{eq:sigmaIsoAllen}
\end{equation} 
\noindent If $\sigma$ and $\omega_{\text{eff}}$ are fixed, one can obtain the value of $A$. Note that Eq.~\eqref{eq:sigmaIsoAllen} is not universal and is susceptible to change if one uses a different color gradient or a different gradient operator. It has been shown in~\cite{Reis2007} that the perturbation operator allows to recover, within the macroscopic limit, the capillary stress tensor responsible for the surface tension forces in the macroscopic two-phase flow equations.

\subsection{Recoloring operator}\label{recoloring}

The recoloring operator $(\Omega_{i}^{k})^{(3)}$ of Eq.~\eqref{eq:subop3} is used to maximize the amount of fluid $k$ at the interface that is sent to the $k$ fluid region. It guarantees the fluid's immiscibility. It respects the principles of mass and momentum conservation. 
The form of recoloring used in this paper is a combination of ideas taken from~\cite{Latva-Kokko2005} and~\cite{Halliday2007} and fully developed in~\cite{Leclaire2012}. It reads :
\begin{align}
\left. 
\begin{array}{ll}
(\Omega_{i}^{r})^{(3)}(N^{r}_{i})= \frac{\rho_{r}}{\rho}N_{i}+ \beta\frac{\rho_{r}\rho_{b}}{\rho^{2}}\cos(\varphi_i) \sum_k N^{k(e)}_{i}(\rho_k,\bm{0},\alpha_k),\\
(\Omega_{i}^{b})^{(3)}(N^{b}_{i})= \frac{\rho_{b}}{\rho}N_{i}- \beta\frac{\rho_{r}\rho_{b}}{\rho^{2}}\cos(\varphi_i) \sum_k N^{k(e)}_{i}(\rho_k,\bm{0},\alpha_k),
\end{array}\right.\label{eq:recoloringOperator}
\end{align} 
\noindent where $\beta \in [0\ldots 1]$ is a parameter controlling the interface thickness~\cite{Latva-Kokko2005} that will be set to $0.99$ unless otherwise stated. $\cos(\varphi_i) = \left. \left(\bm{c}_{i} \cdot \bm{F} \right)\middle/ \left(\lvert \bm{c}_{i} \rvert \lvert \bm{F} \rvert \right)\right.$ is the cosine of the angle between the color gradient $\bm{F}$ and the lattice direction vector $\bm{c}_{i}$. Note that for $i=0$ or $\lvert \bm{F} \rvert=0$, there is a division by~$0$. In such a case, we set the whole term equal to~0 to respect mass conservation.

\subsection{Adjustment of the color gradient for static contact angles\\}\label{sec:angleAdjust}
Based on~\cite{Leclaire2016}, the imposition of a contact angle is performed using the method described in~\cite{Xu2017}. In order to properly describe the wetting boundary conditions, we divide the lattice nodes in two categories $C_\text{f}$ and $C_\text{s}$, each category being also subdivided in two subcategories $C_{\text{fs}}$, $C_{\text{ff}}$, $C_{\text{sf}}$ and $C_{\text{ss}}$.
\begin{itemize}[leftmargin=*]
	\item[-] $C_\text{f}$ : the set of fluid lattice nodes
	\begin{itemize}
		\item $C_{\text{fs}}$ : fluid lattice nodes in contact with at least one solid lattice node
		\item $C_{\text{ff}}$ : fluid lattice nodes not in contact with any solid lattice node
	\end{itemize}
	\item[-] $C_\text{s}$ : the set of solid lattice nodes
	\begin{itemize}
		\item $C_{\text{sf}}$ : solid lattice nodes in contact with at least one fluid lattice node
		\item $C_{\text{ss}}$ : solid lattice nodes not in contact with any fluid lattice node
	\end{itemize}
\end{itemize}

When computing the color gradient in the fluid (i.e. for lattice nodes $\in C_\text{f}$), the knowledge of the color field at the boundary is required (i.e. for lattice nodes $\in C_{\text{sf}}$) because Eq.~\eqref{eq:colorGradient} involves the neighboring lattice nodes. Therefore, we need to extrapolate the color field to the boundary nodes, we do so using the following expression :

\begin{equation}
\forall \bm{x} \in C_{\text{sf}},\quad	\phi(\bm{x}) = \left. \sum\limits_{\substack{i \\ \bm{x} + \bm{c}_{i} \in C_{\text{fs}}}} W_i \phi(\bm{x} + \bm{c}_{i}) c_{i \alpha} \middle/ \sum\limits_{\substack{i \\ \bm{x} + \bm{c}_{i} \in C_{\text{fs}}}} W_i .\right.
\end{equation}
It is now possible to evaluate the orientation of the color gradient in the fluid using the expression hereafter :
\begin{equation}
\bm{n}^{*} = \frac{\nabla \phi(\bm{x})}{\lvert \nabla \phi(\bm{x}) \rvert}.
\end{equation}

In~\cite{Xu2017}, they use an $8^{th}$ order isotropic stencil to compute the surface normal $\bm{n}^s$. In the subsequent simulations, boundary surfaces are flat and normals are known so we directly input the exact value. The correct color gradient orientation $\bm{n}$ depends on the prescribed contact angle $\theta$ and is evaluated as follows :

\begin{align}
\bm{n}=\left\{ 
\begin{array}{ll}
\bm{n}^1, & \text{if }D_1 < D_2 \\
\bm{n}^2, & \text{if }D_1 > D_2 \\
\bm{n}^s,& \text{if }D_1 = D_2,
\end{array},\right.
\end{align}

\noindent where $D_1$ and $D_2$ are the Euclidean distances between the current unit normal vector $\bm{n}^{*}$ and the two possible theoretical unit normal vectors $\bm{n}^1$ and $\bm{n}^2$ of the interface at the contact line and are given by :
\begin{align}
\begin{array}{ll}
D_1 = \lvert \bm{n}^{*} - \bm{n}^1 \rvert, \\
D_2 = \lvert \bm{n}^{*} - \bm{n}^2 \rvert, 
\end{array}
\end{align}
\noindent with :
\begin{align}
\begin{array}{ll}
\bm{n}^1 = \left( \bm{n}^s_x \cos{\theta} - \bm{n}^s_y \sin{\theta} , \bm{n}^s_y  \cos{\theta} + \bm{n}^s_x \sin{\theta} \right), \\
\bm{n}^2 = \left( \bm{n}^s_x \cos{\theta} + \bm{n}^s_y \sin{\theta} , \bm{n}^s_y  \cos{\theta} - \bm{n}^s_x \sin{\theta} \right).
\end{array}
\end{align}

Finally, once $\bm{n}$ is known, the corrected color gradient value is computed by taking :
\begin{equation}
\nabla \phi(\bm{x}) = \lvert \nabla \phi(\bm{x}) \rvert \bm{n}.
\end{equation}

\subsection{Boundary conditions\\}\label{sec:lbm_io_bc}

An overview of the available techniques to impose boundary conditions for single phase LBM can be found in~\cite{Guo2013}. For multiphase LBM, the literature is less abundant. In particular, the case of inlet/outlet boundary conditions is particularly difficult because specific treatments are needed to handle the interface when the fluids are entering and/or leaving the domain. Being able to have efficient inlet/outlet boundary conditions is attractive because it extends the range of two-phase flow simulations that could be explored~\cite{Lou2013,Li2017,Huang2017,Hou2018,AziziTarksalooyeh2018} and is mandatory for open channels. Three kinds of boundary conditions are used in this work :
\begin{enumerate}[wide, labelwidth=!, labelindent=0pt]
	\item \underline{No slip boundary conditions} are imposed using full bounceback~\cite{He1997}. Then, free slip boundary conditions are also imposed using full bounceback except that the diagonally traveling distribution functions are sent forward along the wall rather than reflected back the way they came. Finally, velocity and/or pressure boundary conditions are imposed following~\cite{bc_zouhe}.
	
	\item \underline{Periodic boundary conditions} is a very useful tool in computational simulations because it allows to reduce the size of the simulated domain. The implementation of these boundary conditions consists in sending the distribution functions that are leaving the domain on one end to the other end of the domain as if the two sides of the domain were connected.
	
	\item \underline{Inlet/outlet boundary conditions} are achieved using a modified version of Zou-He boundary conditions~\cite{bc_zouhe}. The approach detailed in this paper is very similar to what is described for two-phase pressure boundary conditions in~\cite{Huang2017}. First, we will describe how we inject two phase flows with two different velocities $\bm{u}^{b}_{\text{inlet}}$ and $\bm{u}^{r}_{\text{inlet}}$ from the north wall. The prescribed velocity fields $\bm{v}^{b,\text{in}}$ and $\bm{v}^{r,\text{in}}$ are designed so that $\bm{v}^{b,\text{in}}= \bm{u}^{b}_{\text{inlet}}$ on blue boundary lattice nodes and $\bm{v}^{b,\text{in}}=0$ on red boundary lattice nodes. Similarly, $\bm{v}^{r,\text{in}}= \bm{u}^{r}_{\text{inlet}}$ on red boundary lattice nodes and $\bm{v}^{r,\text{in}}=0$ on blue boundary lattice nodes. It is then possible to generate a color-blind prescribed velocity field $\bm{v}^{\text{in}} = \bm{v}^{b,\text{in}} + \bm{v}^{r,\text{in}}$. Following the classic Zou-He procedure described in~\cite{bc_zouhe}, we can then compute the modified density and distribution functions. It reads :
	\begin{align}
	\begin{array}{ll}
	\rho^{\text{in}} = \frac{1}{1+ \bm{v}^{\text{in}}_{y}} \left( N_0 + N_1 + N_3 + 2 \left( N_2 + N_5 + N_6 \right) \right), \\
	N_4 = N_2 - \frac{2}{3} \rho^{\text{in}} \bm{v}^{\text{in}}_{y} + H^{\text{in}}, \\
	N_7 = N_5 + \frac{1}{2} \left( N_1 - N_3 \right) - \frac{1}{2} \rho^{\text{in}} \bm{v}^{\text{in}}_{x} - \frac{1}{6} \rho^{\text{in}} \bm{v}^{\text{in}}_{y} - \frac{1}{2} H^{\text{in}}, \\
	N_8 = N_6 + \frac{1}{2} \left( N_3 - N_1 \right) + \frac{1}{2} \rho^{\text{in}} \bm{v}^{\text{in}}_{x} - \frac{1}{6} \rho^{\text{in}} \bm{v}^{\text{in}}_{y} - \frac{1}{2}H^{\text{in}}, 
	\end{array}
	\end{align}
	\noindent where $H^{\text{in}}$ is a corrective term that depends if we use Eq.~\eqref{eq:equilibrium} or Eq.~\eqref{eq:equilibrium_hermite} for the equilibrium. In practice, to derive Zou-He boundary conditions, one has to solve a linear system where one term comes from the equilibrium distribution function. If we use Eq.~\eqref{eq:equilibrium_hermite}, we obtain extra terms compared to the classical Zou-He approach due to the Hermite expansion. It is computed as follows :
	\begin{align}
	H^{\text{in}} =
	\begin{cases}
	0\quad\text{if we use Eq.~\eqref{eq:equilibrium}},\\[1.3ex]
	\left[\frac{\phi + 1}{2}3\rho^{\text{in}} (c_s^b)^2 + \left( 1 - \frac{\phi + 1}{2} \right) 3\rho^{\text{in}} (c_s^r)^2 -  \rho^{\text{in}}\right] \frac{1}{3}\bm{v}^{\text{in}}_{y} \quad\text{if we use  Eq.~\eqref{eq:equilibrium_hermite}}.
	\end{cases}
	\end{align}
	
	It is then needed to redistribute these quantities in function of the color field value :
	\begin{align}
	\begin{array}{ll}
	\rho^{b} = \frac{\phi + 1}{2} \rho^{\text{in}},& \rho^{r} = \left( 1 - \frac{\phi + 1}{2} \right) \rho^{\text{in}}, \label{eq:redistribution} \\
	N^{b}_4 = \frac{\phi + 1}{2} N_4,& N^{r}_4=\left( 1 - \frac{\phi + 1}{2} \right) N_4, \\
	N^{b}_7 = \frac{\phi + 1}{2} N_7,& N^{r}_7=\left( 1 - \frac{\phi + 1}{2} \right) N_7, \\
	N^{b}_8 = \frac{\phi + 1}{2} N_8,& N^{r}_8=\left( 1 - \frac{\phi + 1}{2} \right) N_8. 
	\end{array}
	\end{align}
	Second, we will describe how we impose a constant pressure $p^{\text{out}}$ at the outlet located on the south wall. The corresponding prescribed densities are $\rho^{b,out} = \left. p^{\text{out}} \middle/ (c_{s}^{b})^2 \right.$ and $\rho^{r,out} = \left. p^{\text{out}} \middle/ (c_{s}^{r})^2 \right. $. The color-blind prescribed density is then $\rho^{\text{out}} = \rho^{b,out} + \rho^{r,out}$. In addition, we also have :
	\begin{align}
	\begin{array}{ll}
	v^{\text{out}}_x = 0 \\
	v^{\text{out}}_y = \frac{1}{\rho^{\text{out}}} \left( N_0 + N_1 + N_3 + 2 \left( N_4 + N_7 + N_8 \right) \right), \\
	N_2 = N_4 + \frac{2}{3}\rho^{\text{out}} v^{\text{out}}_y + H^{\text{out}}, \\
	N_5 = N_7 - \frac{1}{2} \left( N_1- N_3 \right) - \frac{1}{2} \rho^{\text{out}} v^{\text{out}}_x + \frac{1}{6} \rho^{\text{out}} v^{\text{out}}_y - \frac{1}{2}H^{\text{out}}, \\
	N_6 = N_8 - \frac{1}{2} \left( N_3- N_1 \right) + \frac{1}{2} \rho^{\text{out}} v^{\text{out}}_x + \frac{1}{6} \rho^{\text{out}} v^{\text{out}}_y - \frac{1}{2}H^{\text{out}},
	\end{array}
	\end{align}
	\noindent where $H^{\text{out}}$ is evaluated as follows :
	\begin{align}
	H^{\text{out}} =
	\begin{cases}
	0\quad\text{if we use Eq.~\eqref{eq:equilibrium}},\\[1.3ex]
	-\left[\frac{\phi + 1}{2}3\rho^{\text{out}} (c_s^b)^2 + \left( 1 - \frac{\phi + 1}{2} \right) 3\rho^{\text{out}} (c_s^r)^2 -  \rho^{\text{out}}\right] \frac{1}{3}\bm{v}^{\text{out}}_{y} \quad\text{if we use  Eq.~\eqref{eq:equilibrium_hermite}}.
	\end{cases}
	\end{align}
	We can then redistribute these quantities similarly with what was done in Eq.~\eqref{eq:redistribution}~:
	\begin{align}
	\begin{array}{ll}
	\rho^{b} = \frac{\phi + 1}{2} \rho^{\text{out}},& \rho^{r} = \left( 1 - \frac{\phi + 1}{2} \right) \rho^{\text{out}}, \\
	N^{b}_2 = \frac{\phi + 1}{2} N_2,& N^{r}_2=\left( 1 - \frac{\phi + 1}{2} \right) N_2, \\
	N^{b}_5 = \frac{\phi + 1}{2} N_5,& N^{r}_5=\left( 1 - \frac{\phi + 1}{2} \right) N_5, \\
	N^{b}_6 = \frac{\phi + 1}{2} N_6,& N^{r}_6=\left( 1 - \frac{\phi + 1}{2} \right) N_6. 
	\end{array}
	\end{align}
	
	A test case has been setup to check the performance of these boundary conditions. Initial and final configurations can be found in Fig.~\ref{init_geo}. In Fig.~\ref{IOS_BC_lbm_it} is shown the evolution of the inlet velocities and outlet pressure with the number of iterations. Note that quantities have been averaged along the height of the pipe. We can see that that we are recovering the prescribed values at the inlet and at the outlet after a transient period with a maximum error $\leq 2\%$. In Fig.~\ref{IOS_BC_lbm_height}, we see the distribution of the color field, the inlet velocity and the outlet pressure along the height of the pipe. It is possible to observe a velocity peak and a pressure peak located at the interface. This is likely due to the fact that fluids are mixed at the interface resulting in governing equations not being properly solved at this location. Moreover, slight discrepancies can be seen at the walls due to boundary conditions. Overall, the previously described boundary conditions are giving satisfactory results and will be used later in the paper.

	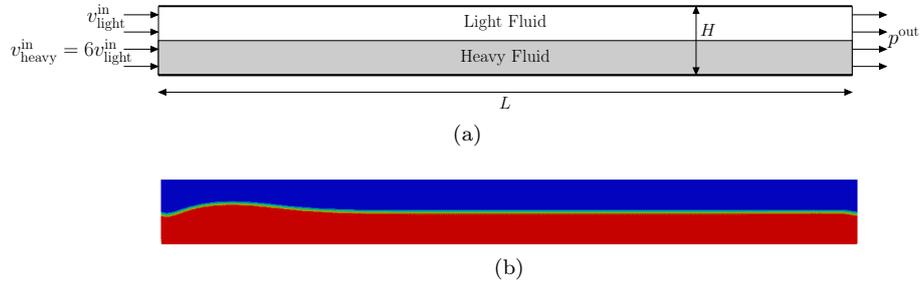
\begin{figure}[bthp]
		\begin{center}
			\subfloat[] {\resizebox{1.0\textwidth}{!}{\begin{tikzpicture}

\node[circle,fill=black!0](bl) at (0,0) {};

\node[circle,fill=black!0](ml) at (0,1) {};

\node[circle,fill=black!0](tl) at (0,2) {};

\node[circle,fill=black!0](br) at (20,0) {};

\node[circle,fill=black!0](mr) at (20,1) {};

\node[circle,fill=black!0](tr) at (20,2) {};

%\node[circle,fill=black!0,label={$x$}](x) at (12,0) {};
%\node[circle,fill=black!0,label={$y$}](y) at (0,12) {};

\draw[ultra thick] (bl.center) -- (br.center);
\draw (br.center) -- (tr.center);
\draw[ultra thick] (tr.center) -- (tl.center);
\draw (tl.center) -- (bl.center);

%\draw[dashed]  (ml.center) -- (mr.center);

\fill[fill=black!20, draw=black] (bl.center) rectangle (mr.center);

\draw[triangle 45-triangle 45] (0.0,-0.5) -- node[below] {\Large $L$} (20.0,-0.5);
\draw[triangle 45-triangle 45] (15.5,0) -- node[above right] {\Large $H$} (15.5,2.0);

%\draw[triangle 45-triangle 45] (16.5,0) -- node[right] {\Large $\alpha_l$} (16.5,1.0);
%\draw[triangle 45-triangle 45] (16.5,1.0) -- node[right] {\Large $\alpha_g$} (16.5,2.0);

\draw[-triangle 45] (-1.0,1.75) --  (0.0,1.75);
\draw[-triangle 45] (-1.0,1.25) --  (0.0,1.25);
\draw[-triangle 45] (-1.0,0.75) --  (0.0,0.75);
\draw[-triangle 45] (-1.0,0.25) --  (0.0,0.25);

\node[circle,fill=black!0,label={\LARGE $v^{\text{in}}_{\text{light}}$}](ug) at (-1.5,1.0) {};
\node[circle,fill=black!0,label={\LARGE $v^{\text{in}}_{\text{heavy}}=6 v^{\text{in}}_{\text{light}}$}](ul) at (-2.5,0.0) {};

\draw[-triangle 45] (20.0,1.75) --  (21.0,1.75);
\draw[-triangle 45] (20.0,1.25) --  (21.0,1.25);
\draw[-triangle 45] (20.0,0.75) --  (21.0,0.75);
\draw[-triangle 45] (20.0,0.25) --  (21.0,0.25);

\node[circle,fill=black!0,label={\LARGE $p^{\text{out}}$}](ug) at (21.5,0.5) {};

\node[circle,fill=none,label={\Large Light Fluid}](tr) at (10,1.0) {};
\node[circle,fill=none,label={\Large Heavy Fluid}](tr) at (10,0.0) {};

\draw[ultra thick] (bl.center) -- (br.center);

\end{tikzpicture}}}\\
			\hspace{1.1cm}\subfloat[] {\includegraphics[width=0.76\textwidth]{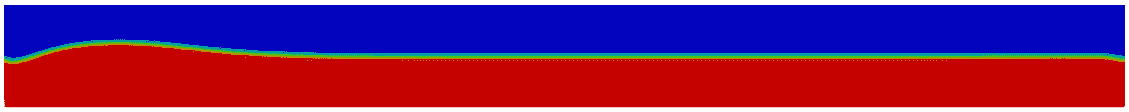}}
			\caption {(a) Initial configuration sketch. (b) Phases distribution after $50000$ iterations.}
			\label{init_geo}
		\end{center}
	\end{figure}

	\begin{figure}[bthp]
		\begin{center}
			\makebox[\textwidth][c]{
				\subfloat[Inlet light phase velocity] {\resizebox{0.333\textwidth}{!}{%\pgfplotsset{label style={font=\tiny},
%	tick label style={font=\tiny},scaled y ticks=false, scaled x ticks=false }
%
\begin{tikzpicture}

\begin{axis}[%
width=1.5in,
height=1.5in,
at={(0in,0in)},
scale only axis,
tick label style={/pgf/number format/fixed},
xmajorgrids=false,
ymajorgrids=true,
grid style={dotted,gray},
xmin=0,
xmax=50000,
xlabel={$N_{\text{iter}}$},
ymin=0.97,
ymax=1.03,
ylabel={$v_{\text{light}} / v^{\text{in}}_{\text{light}}$},
ylabel near ticks,
xlabel near ticks,
xtick pos=left,
xtick={0,10000,20000,30000,40000},
xticklabels={0k,10k,20k,30k,40k},
scaled x ticks = false,
ytick pos=left,
ytick={0.985,1.0,1.015},
yticklabels={$0.985$,$1$,$1.015$},
axis background/.style={fill=white}
]
\addplot [color=black,solid,forget plot]
  table[row sep=crcr]{%
0	1.0000000000125\\
1000	1.00051139962143\\
2000	1.0087950995875\\
3000	0.993256989042857\\
4000	0.987817834046428\\
5000	1.02171976556429\\
6000	1.02667744948393\\
7000	1.02774361450893\\
8000	1.02411605802321\\
9000	1.01884985257321\\
10000	1.01286924004643\\
11000	1.01495888559107\\
12000	1.01820143241071\\
13000	1.01905670170179\\
14000	1.01779480341607\\
15000	1.01764349364107\\
16000	1.01843611047143\\
17000	1.01872389711429\\
18000	1.0180171199375\\
19000	1.017552526225\\
20000	1.01801915330893\\
21000	1.01843508085\\
22000	1.01804173018214\\
23000	1.01766286613393\\
24000	1.01797859652321\\
25000	1.01828643812143\\
26000	1.01804372416786\\
27000	1.0178021093875\\
28000	1.01798339140536\\
29000	1.01815772357321\\
30000	1.01801438374286\\
31000	1.01787843661964\\
32000	1.01797785929643\\
33000	1.01806738009643\\
34000	1.01798400272679\\
35000	1.01791390760893\\
36000	1.01797291275536\\
37000	1.01801885760536\\
38000	1.01796871710536\\
39000	1.01793162800179\\
40000	1.01796627108929\\
41000	1.0179878809625\\
42000	1.01795461271071\\
43000	1.01793239826964\\
44000	1.01795257997857\\
45000	1.01796353313214\\
46000	1.01794261463214\\
47000	1.01792990220893\\
48000	1.01794227670179\\
49000	1.01794813868571\\
50000	1.01793469650893\\
};
%\addplot [color=red,solid,forget plot]
%  table[row sep=crcr]{%
%0	1.0000000000125\\
%500	1.00025569981696\\
%1000	1.00310216640714\\
%1500	1.00064087206607\\
%2000	0.998076264462143\\
%2500	1.00201684797917\\
%3000	1.00553979105128\\
%3500	1.00831526898348\\
%4000	1.01007091221012\\
%4500	1.01094880624643\\
%5000	1.01112339113734\\
%6000	1.0124832898263\\
%7000	1.01409147462532\\
%8000	1.01502434754481\\
%9000	1.01725505794237\\
%10000	1.01996648154188\\
%11000	1.01966796744253\\
%12000	1.01894491722711\\
%13000	1.01806069044789\\
%14000	1.01746400573896\\
%15000	1.01738848762403\\
%16000	1.01789447315162\\
%17000	1.01817473175081\\
%18000	1.01812577118019\\
%19000	1.01802776161851\\
%20000	1.01807245568263\\
%21000	1.01810884027597\\
%22000	1.0180512038138\\
%23000	1.01798388511299\\
%24000	1.01799666726169\\
%25000	1.01803865430877\\
%26000	1.01802586188247\\
%27000	1.01798429628669\\
%28000	1.01798662809708\\
%29000	1.01801582233279\\
%30000	1.0180099415224\\
%31000	1.0179814392164\\
%32000	1.01797917861981\\
%33000	1.01799432477597\\
%34000	1.01798961901201\\
%35000	1.01797221424075\\
%36000	1.01796980489708\\
%37000	1.01797672999627\\
%38000	1.01797259717565\\
%39000	1.0179621608013\\
%40000	1.01796029992906\\
%41000	1.01796290965844\\
%42000	1.01795899960877\\
%43000	1.01795203770844\\
%44000	1.01795016694302\\
%45000	1.01795044589821\\
%};
\end{axis}
\end{tikzpicture}%
}}
				\subfloat[Inlet heavy phase velocity] {\resizebox{0.333\textwidth}{!}{%\pgfplotsset{label style={font=\tiny},
%	tick label style={font=\tiny},scaled y ticks=false, scaled x ticks=false }
%
\begin{tikzpicture}

\begin{axis}[%
width=1.5in,
height=1.5in,
at={(0in,0in)},
scale only axis,
tick label style={/pgf/number format/fixed},
xmajorgrids=false,
ymajorgrids=true,
grid style={dotted,gray},
xmin=0,
xmax=50000,
xlabel={$N_{\text{iter}}$},
ymin=0.99,
ymax=1.01,
ylabel={$v_{\text{heavy}} / v^{\text{in}}_{\text{heavy}}$},
ylabel near ticks,
xlabel near ticks,
xtick pos=left,
xtick={0,10000,20000,30000,40000},
xticklabels={0k,10k,20k,30k,40k},
scaled x ticks = false,
ytick pos=left,
ytick={0.995,1.0,1.005},
yticklabels={$0.995$,$1$,$1.005$},
axis background/.style={fill=white}
]
\addplot [color=black,solid,forget plot]
  table[row sep=crcr]{%
0	1\\
1000	1.000112322375\\
2000	1.00541871905921\\
3000	1.00138695356908\\
4000	1.0042426832796\\
5000	1.00216479900658\\
6000	0.998498840326389\\
7000	0.997053211260417\\
8000	0.998958433770833\\
9000	0.999733152149306\\
10000	1.00009510007639\\
11000	1.00003353220833\\
12000	0.999952316510417\\
13000	0.999708876618056\\
14000	0.999512920756944\\
15000	0.999510338524306\\
16000	0.999658753871528\\
17000	0.99971954784375\\
18000	0.999638983579861\\
19000	0.999586496458333\\
20000	0.999662194878472\\
21000	0.999730233600695\\
22000	0.999675987173611\\
23000	0.999617918555555\\
24000	0.99965987403125\\
25000	0.999708030725694\\
26000	0.999676010770833\\
27000	0.999639365034722\\
28000	0.999664463298611\\
29000	0.999693695246528\\
30000	0.999676069204861\\
31000	0.999656298333333\\
32000	0.999671966708333\\
33000	0.999688675520833\\
34000	0.999678737170139\\
35000	0.999668341552083\\
36000	0.999677635336805\\
37000	0.999686663111111\\
38000	0.999680842600694\\
39000	0.9996754703125\\
40000	0.999681396447917\\
41000	0.999686840065972\\
42000	0.999683828642361\\
43000	0.999681564864583\\
44000	0.999685766065972\\
45000	0.999689165552083\\
46000	0.999687590548611\\
47000	0.999686666465278\\
48000	0.999689524489583\\
49000	0.999691760149306\\
50000	0.999690961135417\\
};
%\addplot [color=red,solid,forget plot]
%  table[row sep=crcr]{%
%0	1\\
%500	1.0000561611875\\
%1000	1.00184368047807\\
%1500	1.00172949875082\\
%2000	1.00223213565658\\
%2500	1.00222091288158\\
%3000	1.00168918823084\\
%3500	1.00110969110953\\
%4000	1.00087066251635\\
%4500	1.00075691147964\\
%5000	1.00069674680662\\
%6000	1.00069979518919\\
%7000	1.00068524920151\\
%8000	1.00016617261595\\
%9000	0.99999580599666\\
%10000	0.999565592837088\\
%11000	0.999337770552083\\
%12000	0.999448743962753\\
%13000	0.999683814173611\\
%14000	0.999740910781566\\
%15000	0.999734460120581\\
%16000	0.999701290440972\\
%17000	0.999668786346906\\
%18000	0.999638386532828\\
%19000	0.999633931752209\\
%20000	0.999651669022096\\
%21000	0.999666730135417\\
%22000	0.999664967513889\\
%23000	0.999659959827967\\
%24000	0.999664933615846\\
%25000	0.999673076592803\\
%26000	0.999672540543245\\
%27000	0.99966724355303\\
%28000	0.999668397039141\\
%29000	0.999673926004103\\
%30000	0.999674695778725\\
%31000	0.999671932561553\\
%32000	0.999672900956124\\
%33000	0.999676671643939\\
%34000	0.999677672281565\\
%35000	0.999676554208964\\
%36000	0.999677533378157\\
%37000	0.999680036133523\\
%38000	0.999680908693182\\
%39000	0.999680644197285\\
%40000	0.999681592232008\\
%41000	0.999683342140783\\
%42000	0.999684163152462\\
%43000	0.999684423277777\\
%44000	0.999685415782197\\
%45000	0.999686824038826\\
%};
\end{axis}
\end{tikzpicture}%
}}
				\subfloat[Outlet pressure] {\resizebox{0.333\textwidth}{!}{%\pgfplotsset{label style={font=\tiny},
%	tick label style={font=\tiny},scaled y ticks=false, scaled x ticks=false,yticklabel style={
%		/pgf/number format/fixed,
%		/pgf/number format/precision=5 }}
%%
\begin{tikzpicture}

\begin{axis}[%
width=1.5in,
height=1.5in,
at={(0in,0in)},
scale only axis,
tick label style={/pgf/number format/fixed},
xmajorgrids=false,
ymajorgrids=true,
grid style={dotted,gray},
xmin=0,
xmax=50000,
xlabel={$N_{\text{iter}}$},
ymin=0.998,
ymax=1.002,
ylabel={$p / p^{\text{out}}$},
ylabel near ticks,
xlabel near ticks,
xtick pos=left,
scaled x ticks = false,
xtick={0,10000,20000,30000,40000},
xticklabels={0k,10k,20k,30k,40k},
ytick pos=left,
ytick={0.999,1.0,1.001},
yticklabels={$0.999$,$1$,$1.001$},
axis background/.style={fill=white}
]
\addplot [color=black,solid,forget plot]
  table[row sep=crcr]{%
0	1\\
1000	1.00003543925181\\
2000	1.00000468509058\\
3000	0.999999473112772\\
4000	1.00001119890263\\
5000	1.00001648952264\\
6000	1.00001047885371\\
7000	1.00000745051087\\
8000	1.00001011013315\\
9000	1.00001331004303\\
10000	1.00001185326721\\
11000	1.00000992273777\\
12000	1.0000113088587\\
13000	1.0000125632346\\
14000	1.00001147213134\\
15000	1.00001057127989\\
16000	1.00001150045245\\
17000	1.00001218357926\\
18000	1.00001137738632\\
19000	1.00001089950861\\
20000	1.00001162376902\\
21000	1.00001197666893\\
22000	1.00001136195924\\
23000	1.00001110477672\\
24000	1.00001161459511\\
25000	1.00001179999185\\
26000	1.00001138219022\\
27000	1.00001124225861\\
28000	1.00001157560009\\
29000	1.00001168142482\\
30000	1.00001141907518\\
31000	1.00001133962908\\
32000	1.00001154616893\\
33000	1.00001160894746\\
34000	1.00001145124728\\
35000	1.00001140622011\\
36000	1.00001153101359\\
37000	1.00001156823596\\
38000	1.00001147507926\\
39000	1.00001145005208\\
40000	1.00001152476766\\
41000	1.00001154647328\\
42000	1.00001149160236\\
43000	1.00001147803216\\
44000	1.00001152278804\\
45000	1.00001153503668\\
46000	1.00001150261322\\
47000	1.00001149545154\\
48000	1.00001152227808\\
49000	1.000011528976\\
50000	1.00001150953623\\
};
%\addplot [color=red,solid,forget plot]
%  table[row sep=crcr]{%
%0	1\\
%500	1.00001771962591\\
%1000	1.0000133747808\\
%1500	1.00000989936379\\
%2000	1.00001015927156\\
%2500	1.00001121431341\\
%3000	1.00001110924774\\
%3500	1.00001065190563\\
%4000	1.00001059170869\\
%4500	1.00001086354212\\
%5000	1.00001095351713\\
%6000	1.0000118555842\\
%7000	1.0000096619121\\
%8000	1.00001037810701\\
%9000	1.00001146892688\\
%10000	1.00001141187027\\
%11000	1.00001095831843\\
%12000	1.00001111329348\\
%13000	1.00001147028216\\
%14000	1.00001154204356\\
%15000	1.00001138874592\\
%16000	1.00001139996426\\
%17000	1.00001153080258\\
%18000	1.00001151224967\\
%19000	1.00001142600972\\
%20000	1.00001145581522\\
%21000	1.00001152953434\\
%22000	1.00001150606217\\
%23000	1.00001145079134\\
%24000	1.0000114784312\\
%25000	1.00001152566453\\
%26000	1.00001149983362\\
%27000	1.00001146069726\\
%28000	1.00001148315073\\
%29000	1.00001151464806\\
%30000	1.00001149570487\\
%31000	1.00001147125231\\
%32000	1.00001148816556\\
%33000	1.00001150933107\\
%34000	1.00001149791761\\
%35000	1.00001148367605\\
%36000	1.0000114952577\\
%37000	1.00001150907345\\
%38000	1.0000115028792\\
%39000	1.00001149504652\\
%40000	1.00001150266374\\
%41000	1.00001151142675\\
%42000	1.00001150819384\\
%43000	1.00001150401585\\
%44000	1.00001150891556\\
%45000	1.00001151432321\\
%};
\end{axis}
\end{tikzpicture}%
}}}		
			\caption {Evolution of selected quantities with the number of iterations}
			\label{IOS_BC_lbm_it}
		\end{center}
	\end{figure}
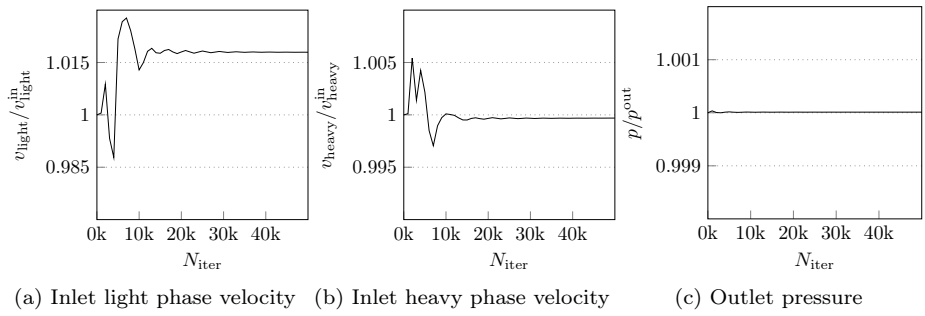
	
	\begin{figure}[bthp]
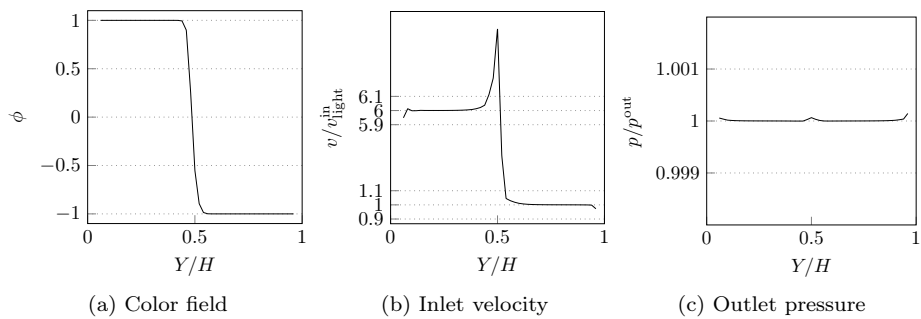

		\begin{center}
			\makebox[\textwidth][c]{
				\subfloat[Color field] {\resizebox{0.338\textwidth}{!}{\input{LBM_Color_Field}}}
				\subfloat[Inlet velocity] {\resizebox{0.328\textwidth}{!}{\input{LBM_InletVelocityX}}}
				\subfloat[Outlet pressure] {\resizebox{0.338\textwidth}{!}{\input{LBM_OutletPressureX}}}}		
			\caption {Variation of selected quantities along the height of the pipe at steady state}
			\label{IOS_BC_lbm_height}
		\end{center}
	\end{figure}
\end{enumerate}

\section{SPH immiscible multiphase model}\label{governing}

In this section, we will describe in details the SPH immisible multiphase model used in this work. This formulation and the associated open boundary conditions are taken from~\cite{Hu2006,DouilletGrellier2018}.

\subsection{Governing equations}\label{sec:sph_st}

For an incompressible fluid with a constant viscosity, the mass and momentum equation (completed with the equation of state) in a Lagrangian system are given as :
\begin{align}
\frac{D \rho}{D t} &= - \rho \nabla \cdot \bm{u}, \label{continuity_equation}\\ 
\frac{D \bm{u}}{D t} &= -\frac{\nabla p}{\rho} + \nu \nabla^2 \bm{u} +\frac{\bm{F}^{st}}{\rho}+ \bm{g},\label{momentum_equation}\\
p &= \frac{c^2 \rho_0}{\gamma} \left[ \left( \frac{\rho}{\rho_0} \right)^{\gamma} -1 \right] + p_0,\label{tait_eos}
\end{align}
\noindent with $\bm{u}$ fluid velocity, $\rho$ fluid density, $\mu$ the fluid viscosity, $\nu=\frac{\mu}{\rho}$ the kinematic viscosity, $p$ fluid pressure, $\mu$ fluid dynamic viscosity, $\bm{g}$ gravity, $c$ fluid speed of sound (here constant), $\gamma$ fluid adiabatic index, $\rho_0$ fluid initial density, $p_0$ background pressure, $\bm{F}^{st}$ is the surface tension force and $D/Dt$ denotes the material derivative following the motion.

The Tait's equation of state~\eqref{tait_eos} is added to Eqs.~\eqref{continuity_equation}-\eqref{momentum_equation} to close the system. In this work, $\gamma$ will be set to $7.0$ for all fluids considered in all subsequent simulations. This is the so-called Weakly Compressible SPH formulation (WCSPH). Just like LBM, it is not a truly incompressible approach since the density is allowed to vary. This artificial compressibility has to be as weak as possible and is controlled by the speed of sound $c$. In this paper, given a reference length $L_{\text{ref}}$ and a reference speed $U_{\text{ref}}$, we used the following formulas~\cite{morris1999surfacetension} to set a value for $c$ and $p_0$ :

\begin{equation}\label{sspb}
\left. 
\begin{array}{l l}
c_\alpha &= \max\left( \frac{U_{\text{ref}}}{\sqrt{\Delta \rho}},\sqrt{\frac{\lvert \bm{g} \rvert L_{\text{ref}}}{\Delta \rho}}, \sqrt{\frac{\sigma^{\alpha\beta}}{\rho_{0\alpha} L_{\text{ref}}}}, \sqrt{\frac{\mu_\alpha U_{\text{ref}}}{\rho_{0\alpha} L_{\text{ref}} \Delta \rho}} \right),\quad \forall \alpha \in \{1,\ldots,N_{\text{phases}}\},\\
p_0 &= \max_{\alpha \in \{1,\ldots,N_{\text{phases}}\}} \frac{c^2_\alpha \rho_{0\alpha}}{\gamma_\alpha},
\end{array} \right.
\end{equation}
\noindent with $\Delta \rho = 0.01$ to enforce (not strictly) a maximum variation of $1\%$ of the density field and $\sigma^{\alpha \beta}$ the surface tension coefficient between phases $\alpha$ and $\beta$. The integer $N_{\text{phases}}$ is the number of different phases.

In order to model surface tension between fluids, an interaction force $\bm{F}^{st}$ is added to the momentum Eq.~\eqref{momentum_equation}. Following~\cite{lafaurie1994}, the continuum surface stress method introduces a surface tension force per unit volume that is expressed as the divergence of the capillary pressure tensor :

\begin{equation}
\bm{F}^{st} = -\nabla \cdot \Pi,
\end{equation}

\noindent with $\Pi$ the capillary pressure tensor defined by :

\begin{equation}\label{st_pi}
\Pi = \sum_{\alpha,\beta \mid \alpha < \beta} \Pi^{\alpha \beta},
\end{equation}

\noindent where $\alpha, \beta \in \{1,\ldots,N_{\text{phases}}\}$ and $\Pi^{\alpha \beta}$ is expressed as :

\begin{equation}\label{st_piab}
\Pi^{\alpha \beta} = -\sigma^{\alpha \beta} \left( \bm{I} - \tilde{\bm{n}}^{\alpha \beta} \otimes \tilde{\bm{n}}^{\alpha \beta} \right) \delta^{\alpha \beta},
\end{equation}

\noindent with $\tilde{\bm{n}}^{\alpha \beta}$ the unit normal vector from phase $\alpha$ to phase $\beta$, $\sigma^{\alpha \beta}$ the surface tension coefficient between phase $\alpha$ and phase $\beta$, $\delta^{\alpha \beta}$ a well-chosen surface delta function and $\bm{I}$ the identity matrix. For example, in the case of a three-phase system with a wetting phase $s$, a non-wetting phase $n$ and a solid phase $s$, the stress tensor reads $\Pi = \Pi^{ns} + \Pi^{ws} + \Pi^{nw}$.

\subsection{SPH formulation}\label{sphform}

The SPH formulation adopted for this paper is identical to the one used in~\cite{DouilletGrellier2018}. We will not here recall in detail the SPH discretization process and will assume that the reader already has experience with SPH. For those who are looking for an exhaustive description of this method, we recommend for example the reading of~\cite{violeau2012fluid}.

An SPH discretization consists of a set of points with fixed mass, which possess material properties and interact with its neighboring particles through a weighting function (or smoothing kernel)~\cite{gingold1977smoothed}. A particle's support domain, $\Lambda$, is defined by its smoothing length, $h$, which is the radius of the smoothing kernel $W$. In all simulations presented in this paper, $h=2\Delta r$ where $\Delta r$ is the initial particle spacing. To obtain the value of a function at a given particle location, values of that function are found by taking a weighted (by the smoothing function) interpolation from all particles within the given particle's support domain. An analytical differentiation of the smoothing kernel is used to find gradients of this function. 

The interpolated value of a function $\bm{A}$ at the position $\bm{x}_a$ of particle $a$ can be expressed using SPH smoothing as :

\begin{equation}\label{eq:4.2}
\bm{A}(\bm{x}_a)=\sum_{b\in\Lambda_{a}}\bm{A}_{b}\frac{m_b}{\rho_{b}}W(\bm{x}_a-\bm{x}_{b},h)=\sum_{b\in\Lambda_{a}}\bm{A}_{b}\frac{m_b}{\rho_{b}}W_{ab},
\end{equation}

\noindent in which $\bm{A}_{b}=\bm{A}(\bm{x}_b)$, $m_b$ and $\rho_{b}$ are the mass and the density of neighboring particle $b$. The set of particles $\Lambda_{a}=\{ b\in \mathbb{N}\mid \lvert\bm{x}_{a}-\bm{x}_{b}\rvert \leq \kappa h \}$ contains neighbors of particle $a$ that lie within its defined support domain. The coefficient $\kappa$ depends on the choice of the kernel, it is equal to $2$ for the $5^{th}$ order $\mathcal{C}^2$ Wendland kernel function~\cite{Wendland1995,Dehnen11092012}) used in this paper. For the sake of clarity, $W(\bm{x}_a-\bm{x}_{b},h)$ has been denoted $W_{ab}$. In 2D, this kernel is expressed as follows :

\begin{equation}
W(q,h) = \frac{7}{4\pi h^2} \left(1-\frac{q}{2}\right)_+^4 \left(1 +2q\right), \text{ with } q=\frac{\lvert \bm{x}_a-\bm{x}_{b} \rvert}{h}.
\end{equation}
%\noindent with $q=\frac{\lvert \bm{x}_a-\bm{x}_{b} \rvert}{h}$, $q \leq 2$.

Several multiphase formulations~\cite{colagrossi2003,grenier2009,tofighi2013} have been proposed for SPH throughout the years. In this work, the formalism presented in~\cite{Hu2006} has been adopted. The density is directly evaluated through a kernel summation which gives an exact solution to the continuity equation. It reads :

\begin{equation}
\rho_{a} = m_{a}\sum_{b\in\Lambda_{a}} W_{ab}.
\end{equation}

\noindent Discrete gradient and divergence operators in this formalism are given by :

\begin{equation}
\left.
\begin{array}{ll}
\nabla \bm{A}(\bm{x}_a) = \sum_{b\in\Lambda_{a}} \left(\frac{\bm{A}_{a}}{\Theta_a^2} + \frac{\bm{A}_{b}}{\Theta_b^2}\right) \Theta_a \nabla_a W_{ab},\\
\nabla \cdot \bm{A}(\bm{x}_a) = \sum_{b\in\Lambda_{a}} \left(\frac{\bm{A}_{a}}{\Theta_a^2} + \frac{\bm{A}_{b}}{\Theta_b^2}\right) \Theta_a \nabla_a \cdot W_{ab},
\end{array} \right.
\end{equation}

\noindent where $\Theta_a=\left. \rho_{a} \middle/ m_{a} \right.$. It follows that the full multiphase SPH formulation for a particle $a$ is :

\begin{align}\label{fullformulation}
\begin{array}{l l}
\rho_{a} &= m_{a}\sum_{b\in\Lambda_{a}} W_{ab}, \\
\frac{D \bm{u}}{D t} &= -\frac{1}{m_a} \sum_{b\in\Lambda_{a}} \left(\frac{p_{a}\bm{I}+\Pi_{a}}{\Theta_a^2} + \frac{p_{b}\bm{I}+\Pi_{b}}{\Theta_b^2}\right) \nabla_a W_{ab}\\
&\quad+ \frac{1}{m_a} \sum_{b\in\Lambda_{a}} \frac{2 \mu_a \mu_b}{\mu_a + \mu_b} \left(\frac{1}{\Theta_a^2} + \frac{1}{\Theta_b^2}\right) \frac{\bm{x}_{ab} \cdot \nabla_a W_{ab}}{\lvert \bm{x}_{ab} \rvert^2 + \eta^2} \bm{u}_{ab}\\
&\quad+ \bm{g},\\
p_a &= \frac{c^2_a \rho_{0a}}{\gamma_a} \left[ \left( \frac{\rho_a}{\rho_{0a}} \right)^{\gamma_a} -1 \right] + p_0,\\
\frac{D \bm{x}_a}{Dt} &= \bm{u}_a, \end{array}
\end{align}

\noindent where the viscous term $\nu \nabla^2 \bm{u}$ has been discretized using the inter-particle averaged shear stress~\cite{flekkoy2000} and $\eta = 0.01 h$ is a safety factor to avoid a division by zero. Moreover, the evaluation of normal vectors is performed through the computation of the gradient of a color function $\chi$ defined for a given particle $a$ and a given phase $\alpha$ as :
\begin{align}
\chi^{\alpha}_a =
\begin{cases}
1\quad\text{if }a\in\text{phase }\alpha,\\
0\quad\text{else}.
\end{cases}
\end{align}
\noindent The normal vector $\bm{n}^{\alpha \beta}_a$ of particle $a$ belonging to phase $\alpha$ to the interface $\alpha\beta$ is then evaluated by :

\begin{equation}
\bm{n}^{\alpha \beta}_a = \nabla \chi^{\alpha \beta}_a = \sum_{b\in\Lambda_{a}} \left(\frac{\bm{\chi}^{\beta}_{a}}{\Theta_a^2} + \frac{\bm{\chi}^{\beta}_{b}}{\Theta_b^2}\right) \Theta_a \nabla_a W_{ab}.
\end{equation}

\noindent The surface delta function $\delta^{\alpha\beta}_a$ is chosen to be equal to $\lvert \bm{n}^{\alpha \beta}_a \rvert$ and $\tilde{\bm{n}}^{\alpha \beta} = \left. \bm{n}^{\alpha \beta}_a \middle/ \lvert \bm{n}^{\alpha \beta}_a \rvert \right.$ for use in Eq.~\eqref{st_piab}.

\subsection{Corrective terms}
Corrective terms are commonly used in SPH to remediate intrinsic issues of this formulation such as particles disorder or micro-mixing at the interface. Three different SPH corrections have been used in this work :

\begin{enumerate}[wide, labelwidth=!, labelindent=0pt]
	\item As suggested in~\cite{bonet1999}, the kernel gradient is enhanced in order to restore consistency. For a given particle $a$, it reads :
	\begin{align}
	\widetilde{\nabla W}_{ab} &= \bm{L}_a \nabla W_{ab},
	\end{align}
	\noindent where $\bm{L}_a = \left( \sum_{b\in\Lambda_{a}}\frac{m_a}{\rho_{a}}\nabla W_{ab} \otimes (\bm{x}_b - \bm{x}_a) \right)^{-1}$. Note that the tilde notation will be dropped in the rest of paper although the kernel gradient correction will be always used.
	
	\item In order to maintain a good spatial distribution of particles and ensure a better accuracy, a shifting technique for multiphase flows has been used~\cite{Mokos2016}. At the end of every timestep, all particles are shifted by a distance $\delta \bm{r}^s$ from their original position. This shifting distance of a particle $a$ is implemented through : 
	\begin{equation}
	\delta \bm{r}^s_a =
	\left\{ 
	\begin{array}{l l}
	- D_a \nabla C_a,\quad\text{if }a\in\text{light phase}\\
	- D_a \left(\frac{\partial C_a}{\partial s}\bm{s} + \alpha_n \left(\frac{\partial C_a}{\partial n}\bm{n} - \beta_n \right) \right),\quad\text{if }a\in\text{heavy phase}
	\end{array} \right.
	\end{equation}
	\noindent where $C_a = \sum_{b\in\Lambda_{a}}\frac{m_a}{\rho_{a}}W_{ab}$ is the particle concentration, $\nabla C_a=\sum_{b\in\Lambda_{a}}\frac{m_a}{\rho_{a}} (C_b-C_a) \nabla W_{ab}$ is the particle concentration gradient, $D_a$ is the diffusion coefficient, $\bm{s}$ and $\bm{n}$ are respectively the tangent and normal vectors to the interface light/heavy phase (with $\bm{n}$ oriented towards the light phase), $\beta_n$ is a reference concentration gradient (taken equal to its initial value) and $\alpha_n$ is the normal diffusion parameter set equal to $0.1$.
	The diffusion coefficient $D_a$ is computed as follows :
	\begin{equation}
	D_a = A_s \lvert \bm{u}_a \rvert \Delta t,
	\end{equation}
	\noindent where $A_s$ is a parameter set to $2$, $\bm{u}_a$ is the velocity of particle $a$, and $\Delta t$ is the time step.
	
	\item Multiphase SPH can suffer from sub-kernel micro-mixing phenomena as highlighted in numerous publications~\cite{colagrossi2003,grenier2009,szewc2013,ghai2017}. Around the interface, within a distance corresponding to the range of the smoothing length, particles tend to mix. It is because there is no mechanism that guarantees phases immiscibility in the surface tension's continuum surface stress model. As suggested by the previously mentioned authors, we introduce a small repulsive force between phases as follows :
	
	\begin{equation}
	\bm{F}^{corr}_a = \varepsilon \sum_{b\in\Lambda_{a},b\notin\Omega_{a}} \left(\frac{1}{\Theta_a^2} + \frac{1}{\Theta_b^2}\right) \nabla_a W_{ab},
	\end{equation}
	\noindent where $\varepsilon=\left. L_{\text{ref}} \middle/ h \right.$ for all simulations as suggested in~\cite{Szewc2016} and where $L_{\text{ref}}$ is a reference length, typically the diameter of the pipe. The impact of this corrective force on the simulation of intermittent flows is evaluated in~\cite{douillet2018}.
	
\end{enumerate}

\subsection{Time discretization and integration}

Concerning time integration in SPH, different schemes are eligible. Among them, one can mention the Runge-Kutta, the Verlet or the Leapfrog schemes. In this work, the Predictor-Corrector Leapfrog scheme was adopted. It is a symplectic integrator which is recommended for SPH because of its conservative nature~\cite{violeau2012fluid}. Indeed, SPH generally requires very small time steps resulting in a high number of iterations. The algorithm proceeds through the following steps. For every particle $a$,

\begin{enumerate}[wide, labelwidth=!, labelindent=0pt]
	\item Predictor step :
	\begin{equation*}
	\bm{u}^{n} = 
	\left\{ 
	\begin{array}{l l}
	\bm{u}^{0},\quad\text{if }t=0,\\
	\bm{u}^{n-\frac{1}{2}} + \frac{\Delta t}{2} {\frac{D \bm{u}}{D t}}^{n-1},\quad\text{if }t>0.
	\end{array} \right.
	\end{equation*}
	\item Compute $\rho^{n}_a$ and $p^{n}_a$ using the corresponding expressions in Eq.~\eqref{fullformulation}.
	\item Evaluate $\frac{D \bm{u}}{D t}^{n}$ using the momentum equation in Eq.~\eqref{fullformulation}.
	\item Corrector step :
	\begin{equation*}
	\bm{u}^{n+\frac{1}{2}} =
	\left\{ 
	\begin{array}{l l}
	\bm{u}^{n} + \frac{\Delta t}{2} {\frac{D \bm{u}}{D t}}^{n},\quad\text{if }t=0,\\
	\bm{u}^{n-\frac{1}{2}} + \Delta t {\frac{D \bm{u}}{D t}}^{n},\quad\text{if }t>0,
	\end{array} \right.
	\end{equation*}
	\begin{equation*}
	%\bm{u}^{n+\frac{1}{2}} = \bm{u}^{n-\frac{1}{2}} + \Delta t {\frac{D \bm{u}}{D t}}^{n}
	\bm{x}^{n+1} = \bm{x}^{n} + \Delta t \bm{u}^{n+\frac{1}{2}}.
	\end{equation*}
\end{enumerate}

The constant time step $\Delta t$ has to respect the Courant-Friedrichs-Lewy (CFL) criteria to ensure a stable evolution of the system, e.g.

\begin{equation}
\Delta t = \min \left( \Delta t_{\text{visc}}, \Delta t_{\text{grav}}, \Delta t_{\text{speed}}, \Delta t_{\text{st}} \right),
\end{equation}

\noindent where, following~\cite{morris1999surfacetension}, we have : 

\begin{equation}
\left\{ 
\begin{array}{l l}
\Delta t_{\text{visc}} &= 0.125\, \min_{\alpha \in \{1,\ldots,N_{\text{phases}}\}} \frac{ h^2 \rho^0_{\alpha}}{\mu_{\alpha}},\\
\Delta t_{\text{grav}} &= 0.25\, \sqrt{\frac{h}{\lvert\bm{g} \rvert}},\\
\Delta t_{\text{speed}} &= 0.25\, \min_{\alpha \in \{1,\ldots,N_{\text{phases}}\}} \frac{h}{c_{\alpha}},\\
\Delta t_{\text{st}} &= 0.25\, \min_{\alpha,\beta \in \{1,\ldots,N_{\text{phases}}\}} \sqrt{ \frac{h^3 \rho^0_{\alpha} } {2 \pi \sigma^{\alpha \beta}} }.
\end{array} \right.
\end{equation}

A recent article~\cite{violeau2014timestep} proposes a detailed investigation on maximum admissible time steps for WCSPH. 

\subsection{Boundary conditions}\label{sec:bc}

\begin{figure}[bthp]
	\begin{center}
		\resizebox{0.65\textwidth}{!}{\begin{tikzpicture}[yscale=0.5]

% 1
\draw  (-2.0,1.0) ellipse (0.1 and 0.2);
\draw  (-1.0,1.0) ellipse (0.1 and 0.2);
\draw[fill=gray!80]  (-0.0,1.0) ellipse (0.1 and 0.2);
\draw[fill=gray!80]  (1.0,1.0) ellipse (0.1 and 0.2);
\draw[fill=gray!80]  (2.0,1.0) ellipse (0.1 and 0.2);
% 2
\draw  (-2.5,0.5) ellipse (0.1 and 0.2);
\draw  (-1.5,0.5) ellipse (0.1 and 0.2);
\draw  (-0.5,0.5) ellipse (0.1 and 0.2);
\draw[fill=gray!80]  (0.5,0.5) ellipse (0.1 and 0.2);
\draw[fill=gray!80]  (1.5,0.5) ellipse (0.1 and 0.2);
\draw[fill=gray!80]  (2.5,0.5) ellipse (0.1 and 0.2);
% 3
\draw  (-2.0,0.0) ellipse (0.1 and 0.2);
\draw  (-1.0,0.0) ellipse (0.1 and 0.2);
\draw[fill=gray!80]  (-0.0,0.0) ellipse (0.1 and 0.2);
\draw[fill=gray!80]  (1.0,0.0) ellipse (0.1 and 0.2);
\draw[fill=gray!80]  (2.0,0.0) ellipse (0.1 and 0.2);
% 4
\draw  (-2.5,-0.5) ellipse (0.1 and 0.2);
\draw  (-1.5,-0.5) ellipse (0.1 and 0.2);
\draw[fill=gray!80]  (-0.5,-0.5) ellipse (0.1 and 0.2);
\draw[fill=gray!80]  (0.5,-0.5) ellipse (0.1 and 0.2);
\draw[fill=gray!80]  (1.5,-0.5) ellipse (0.1 and 0.2);
\draw[fill=gray!80]  (2.5,-0.5) ellipse (0.1 and 0.2);
% 5
\draw  (-2.0,-1.0) ellipse (0.1 and 0.2);
\draw  (-1.0,-1.0) ellipse (0.1 and 0.2);
\draw[fill=gray!80]  (-0.0,-1.0) ellipse (0.1 and 0.2);
\draw[fill=gray!80]  (1.0,-1.0) ellipse (0.1 and 0.2);
\draw[fill=gray!80]  (2.0,-1.0) ellipse (0.1 and 0.2);
% 6
\draw  (-2.5,-1.5) ellipse (0.1 and 0.2);
\draw  (-1.5,-1.5) ellipse (0.1 and 0.2);
\draw  (-0.5,-1.5) ellipse (0.1 and 0.2);
\draw[fill=gray!80]  (0.5,-1.5) ellipse (0.1 and 0.2);
\draw[fill=gray!80]  (1.5,-1.5) ellipse (0.1 and 0.2);
\draw[fill=gray!80]  (2.5,-1.5) ellipse (0.1 and 0.2);
% 7
\draw  (-2.0,-2.0) ellipse (0.1 and 0.2);
\draw  (-1.0,-2.0) ellipse (0.1 and 0.2);
\draw[fill=black]  (-0.0,-2.0) ellipse (0.1 and 0.2);
\draw[fill=gray!80]  (1.0,-2.0) ellipse (0.1 and 0.2);
\draw  (2.0,-2.0) ellipse (0.1 and 0.2);
% 8
\draw  (-2.5,-2.5) ellipse (0.1 and 0.2);
\draw  (-1.5,-2.5) ellipse (0.1 and 0.2);
\draw  (-0.5,-2.5) ellipse (0.1 and 0.2);
\draw  (0.5,-2.5) ellipse (0.1 and 0.2);
\draw  (1.5,-2.5) ellipse (0.1 and 0.2);
\draw  (2.5,-2.5) ellipse (0.1 and 0.2);

% Interface
\draw  plot[smooth, tension=.7] coordinates {(0,1.5) (-0.5,-1) (1,-2) (3,-1.5)};

%\fill[gray!20,pattern=north east lines] (0,-2.0) circle (1cm);
\fill[gray!20,pattern=north east lines] (0,-2.0) circle ellipse (0.75 and 1.5);

% Labels
\node at (1.5,1.25) {$\Omega_f$};
\node at (-2.0,-3.0) {$\Omega_g$};
\node at (0.0,-2.75) {$\bm{\Lambda_g}$};
\node at (3.25,-1.5) {$\Gamma_{fg}$};
\node (v1) at (-0.15,-1.55) {};
\node (v2) at (-0.7,-2.2) {};
%\path[->, thick]  (v1.center) edge (v2.center);
%\node at (-0.9,-2.4) {$\mathbf{n}_i$};
\end{tikzpicture}}
		\caption{Schematic of a ghost particle $g$ (in black) and its associated support domain $\Lambda_g$ (hatched area) intersecting with the fluid domain $\Omega_f$ (gray particles) and the ghost domain $\Omega_g$ (white particles) separated by $\Gamma_{fg}$.}
		\label{bc_dummy}
	\end{center}
\end{figure}
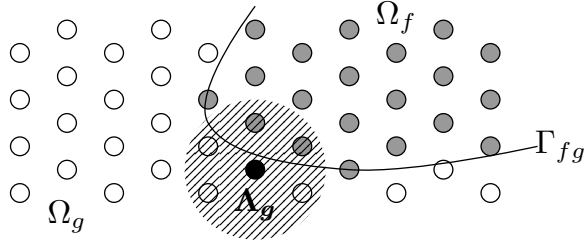

In the subsequent simulations, we used three types of~boundary conditions (BC): no-slip wall (superscript $w$), periodic and inlet/outlet (superscripts $in$ and $out$). 

For no-slip wall BC, the ghost particle method has been used along with the following prescribed values for the pressure $p^w$, density $\rho^w$ and velocity $\bm{v}^w$ for a given ghost particle $g$ :

\begin{align}
{p^w_{g}}&=\frac{1}{V_{ga}}\sum_{a\in\Omega_{f}\cap\Lambda_{g}}p_{a}\frac{m_a}{\rho_{a}}W_{ga},\label{bc_wall1}\\
{\rho^w_{g}}&=\frac{1}{V_{ga}}\sum_{a\in\Omega_{f}\cap\Lambda_{g}}\rho_{a}\frac{m_a}{\rho_{a}}W_{ga},\label{bc_wall2}\\
{\bm{v}^w_{g}}&=\frac{-1}{V_{ga}}\sum_{a\in\Omega_{f}\cap\Lambda_{g}}\bm{v}_{a}\frac{m_a}{\rho_{a}}W_{ga},\label{bc_wall3}
\end{align}

\noindent with $V_{ga}=\sum_{a\in\Omega_{f}\cap\Lambda_{g}}\frac{m_a}{\rho_{a}}W_{ga}$, $\Omega_{f}$ the set of fluid particles and $\Lambda_{g}$ the set of neighboring particles of ghost particle $g$. A schematic drawn in Fig.~\ref{bc_dummy} helps to visualize what is the intersection $\Omega_{f}\cap\Lambda_{g}$. Note that, for free-slip wall BC, one can use the same equations and change the sign of Eq.~\eqref{bc_wall3} for the direction where the free slip is allowed. Additionally, for velocity wall BC, the term $2\bm{v}_0$ (where $\bm{v}_0$ is the prescribed velocity) can be added to Eq.~\eqref{bc_wall3}.

For periodic BC, different variants are available in the literature. In essence, particles close to a periodic boundary are allowed to interact with particles near an associated boundary. Quantities are exchanged both ways and if a particle leaves the domain through one side, it reenters the domain from the other side.

For inlet/outlet BC, the method extensively described in~\cite{DouilletGrellier2018} (combining ideas from~\cite{tafuni2018,alvaro2017}) has been used. The inlet and outlet boundaries are extended with a buffer layer of size $\kappa h$ to ensure a full kernel support. At the inlet, the goal is to inject particles with a prescribed velocity profile. At the outlet, particles need to leave the domain smoothly while imposing a prescribed pressure profile (or density since they are connected through Eq.~\eqref{tait_eos}). 

On one hand, a particle $i$ in the inlet buffer is moving with a prescribed velocity profile $\bm{v}^p$ and it carries the following values of pressure $p^{\text{in}}$, density $\rho^{\text{in}}$ and velocity~$\bm{v}^{\text{in}}$ :

\begin{align}
{p^{\text{in}}_{i}}&=\frac{1}{V_{ia}}\sum_{a\in\Omega_{f}\cap\Lambda_{i}}p_{a}\frac{m_a}{\rho_{a}}W_{ia},\\
{\rho^{\text{in}}_{i}}&=\frac{1}{V_{ia}}\sum_{a\in\Omega_{f}\cap\Lambda_{i}}\rho_{a}\frac{m_a}{\rho_{a}}W_{ia},\\
{\bm{v}^{\text{in}}_{i}}&= 2\bm{v}^p - \frac{1}{V_{ia}}\sum_{a\in\Omega_{f}\cap\Lambda_{i}}\bm{v}_{a}\frac{m_a}{\rho_{a}}W_{ia}.
\end{align}

\noindent with $V_{ia}=\sum_{a\in\Omega_{f}\cap\Lambda_{i}}\frac{m_a}{\rho_{a}}W_{ia}$ and $\Lambda_{i}$ the set of neighboring particles of inlet particle $i$.

On the other hand, at the outlet, a particle $o$ in the buffer is moved according to a smoothed convective velocity $\bm{v}^{conv}$. For example, if the outlet boundary is vertical and the flow leaves along the $x$ direction, it reads :

\begin{equation}\label{conv}
{\bm{v}^{out,conv}_{o}}=\frac{1}{V'_{oa}}\sum_{a\in\Lambda_{o}}\bm{v}_{a}\frac{m_a}{\rho_{a}}W_{oa},
\end{equation}

\noindent with $V'_{oa}=\sum_{a\in\Lambda_{o}}\frac{m_a}{\rho_{a}}W_{oa}$ the set of neighboring particles of outlet particle $o$. Note that in Eq.~\eqref{conv}, the summation is over the full kernel support $\Lambda_{o}$ including fluid and outlet particles and not only over the intersection $\Omega_{f}\cap\Lambda_{o}$.
Besides, particle~$o$ also carries the following values of pressure $p^{\text{out}}$, density $\rho^{\text{out}}$ and velocity $\bm{v}^{\text{out}}$ :

\begin{align}
{p^{\text{out}}_{o}}&= 2p^p - \frac{1}{V_{oa}}\sum_{a\in\Omega_{f}\cap\Lambda_{o}}p_{a}\frac{m_a}{\rho_{a}}W_{oa},\\
{\rho^{\text{out}}_{o}}&= 2\rho^p - \frac{1}{V_{oa}}\sum_{a\in\Omega_{f}\cap\Lambda_{o}}\rho_{a}\frac{m_a}{\rho_{a}}W_{oa},\\
{{v}^{\text{out}}_{o,x}}&=\frac{1}{V_{oa}}\sum_{a\in\Omega_{f}\cap\Lambda_{o}}{v}_{a,x}\frac{m_a}{\rho_{a}}W_{oa},\\
{{v}^{\text{out}}_{o,y}}&=\frac{-1}{V_{oa}}\sum_{a\in\Omega_{f}\cap\Lambda_{o}}{v}_{a,y}\frac{m_a}{\rho_{a}}W_{oa},
\end{align}

\noindent with $V_{oa}=\sum_{a\in\Omega_{f}\cap\Lambda_{o}}\frac{m_a}{\rho_{a}}W_{oa}$, $p^p$ and $\rho^p$ the prescribed pressure and density. Concerning the velocity, null cross velocities (here $v_y$) are enforced to ensure a divergence free velocity field at the outlet. An evaluation of these boundary conditions on a collection of test cases can be found in~\cite{DouilletGrellier2018}.

%\section{Results}
%\subsection{Static Bubble}\label{sec:staticbubble}

\section{Static bubble tests}\label{sec:staticbubble}

In this section, the goal is to validate and compare the implementation of LBM and SPH surface tension models respectively described in Secs.~\ref{sec:lbm_st} and~\ref{sec:sph_st}. 

\paragraph{Square-to-droplet case.}
The standard square-to-droplet test case is simulated and when a steady state is reached, the pressure difference between the exterior and the interior of the bubble is measured and compared to Laplace's formula :
\begin{equation}
\Delta P = \frac{\sigma}{R} = \frac{\sigma \sqrt{\pi}}{a},\label{laplaceslaw}
\end{equation}
\noindent with $\Delta P$ the pressure difference, $\sigma$ the surface tension coefficient, $R$ the bubble's radius and $a$ the lateral dimension of the initial square bubble. Simulations are performed for three different resolutions : $60\times 60$, $100\times 100$ and $200\times 200$ nodes/particles. Four different combinations of density and viscosity ratios were tested : $(\frac{\rho_{\text{heavy}}}{\rho_{\text{light}}},\frac{\mu_{\text{heavy}}}{\mu_{\text{light}}})=(1,1)$, $(5,2)$, $(1000,1000)$ and $(1,100)$. The surface tension coefficient is $\sigma = 1.88~\newton\per\meter$ for SPH and $\sigma = 0.01~l.u.$ for LBM (where $l.u.$ stands for lattice units). The whole domain is $1~\meter \times 1~\meter$ and the lateral size of the initial square droplet is $a=0.33~\meter$. The time is normalized by $t_\sigma=\sqrt{\rho a^3 /\sigma}$. Note that, following~\cite{Leclaire2015}, parameter $\beta$ in Eq.\ref{eq:recoloringOperator} is adjusted when the resolution is increased taking the lowest resolution as a reference i.e. :
\begin{equation}
\beta = \beta_{60\times 60} \left( \frac{\Delta x}{\Delta x_{60\times 60}} \right)^{5/8},
\end{equation}
\noindent with $\beta_{60\times 60}=0.99$.

\begin{figure}[bthp]
	\centering
	\subfloat[{$t / t_{\sigma} = 0.2$}] {\includegraphics[width=0.25\textwidth]{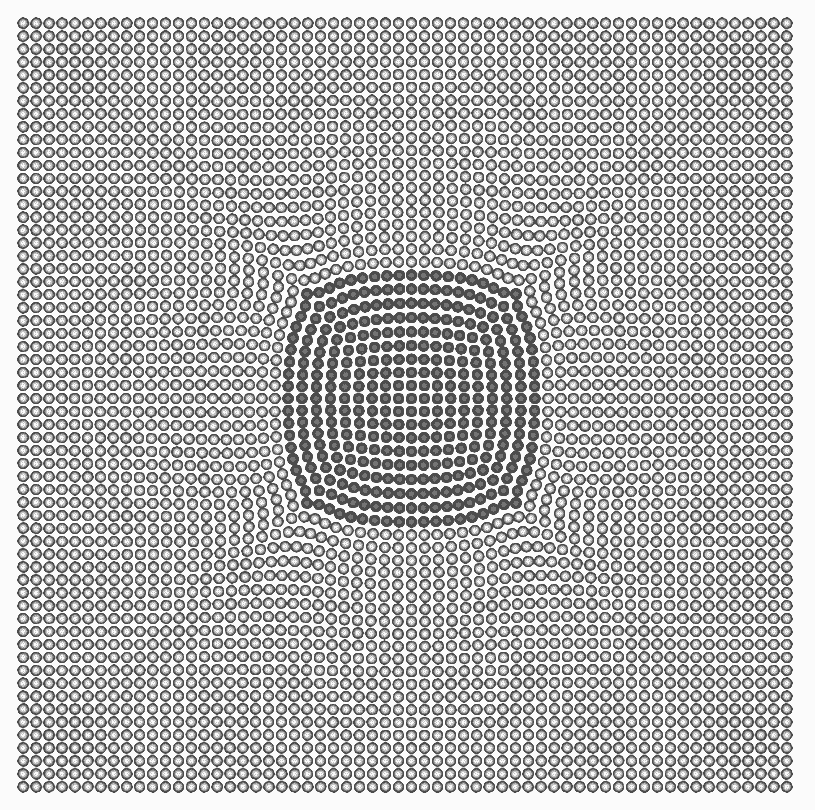}}\hfill
	\subfloat[{$t / t_{\sigma} = 0.6$}] {\includegraphics[width=0.25\textwidth]{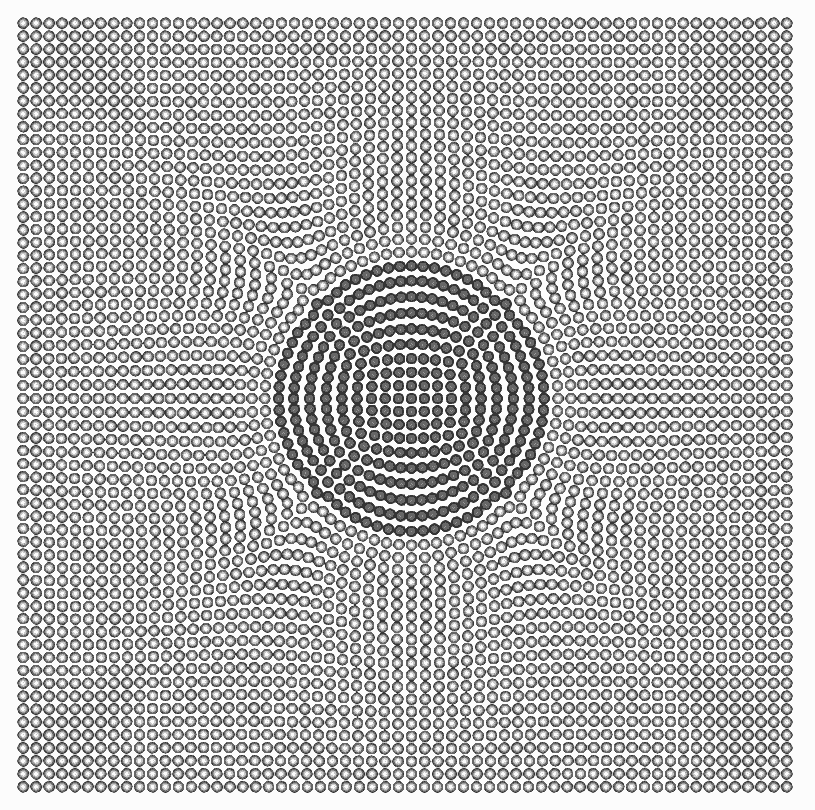}}\hfill  
	\subfloat[{$t / t_{\sigma} = 1.5$}] {\includegraphics[width=0.25\textwidth]{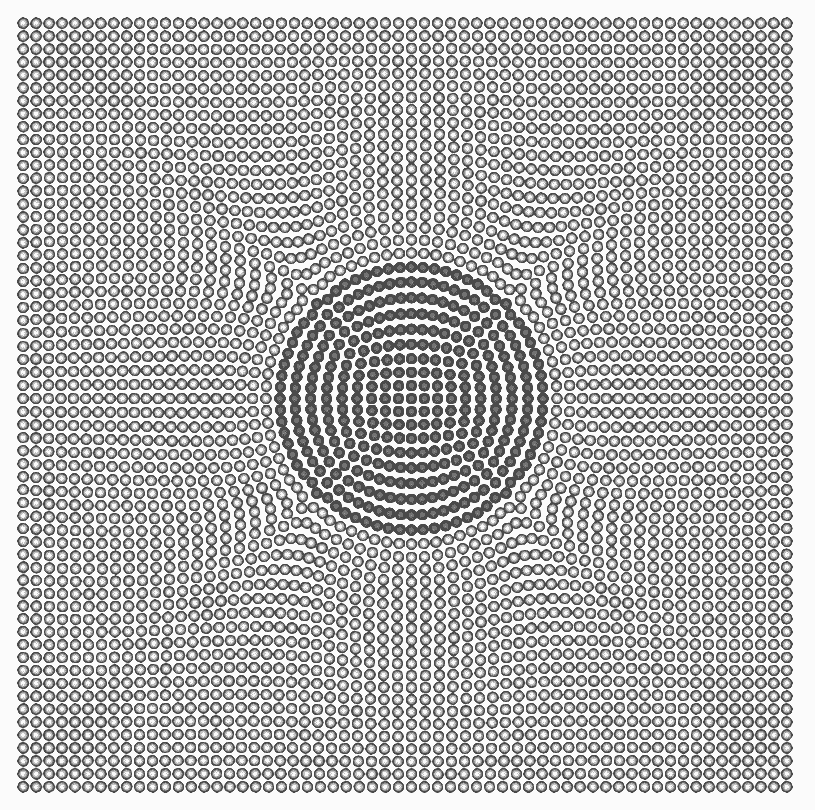}}\\
	\subfloat[{$t / t_{\sigma} = 0.2$}] {\includegraphics[width=0.249\textwidth]{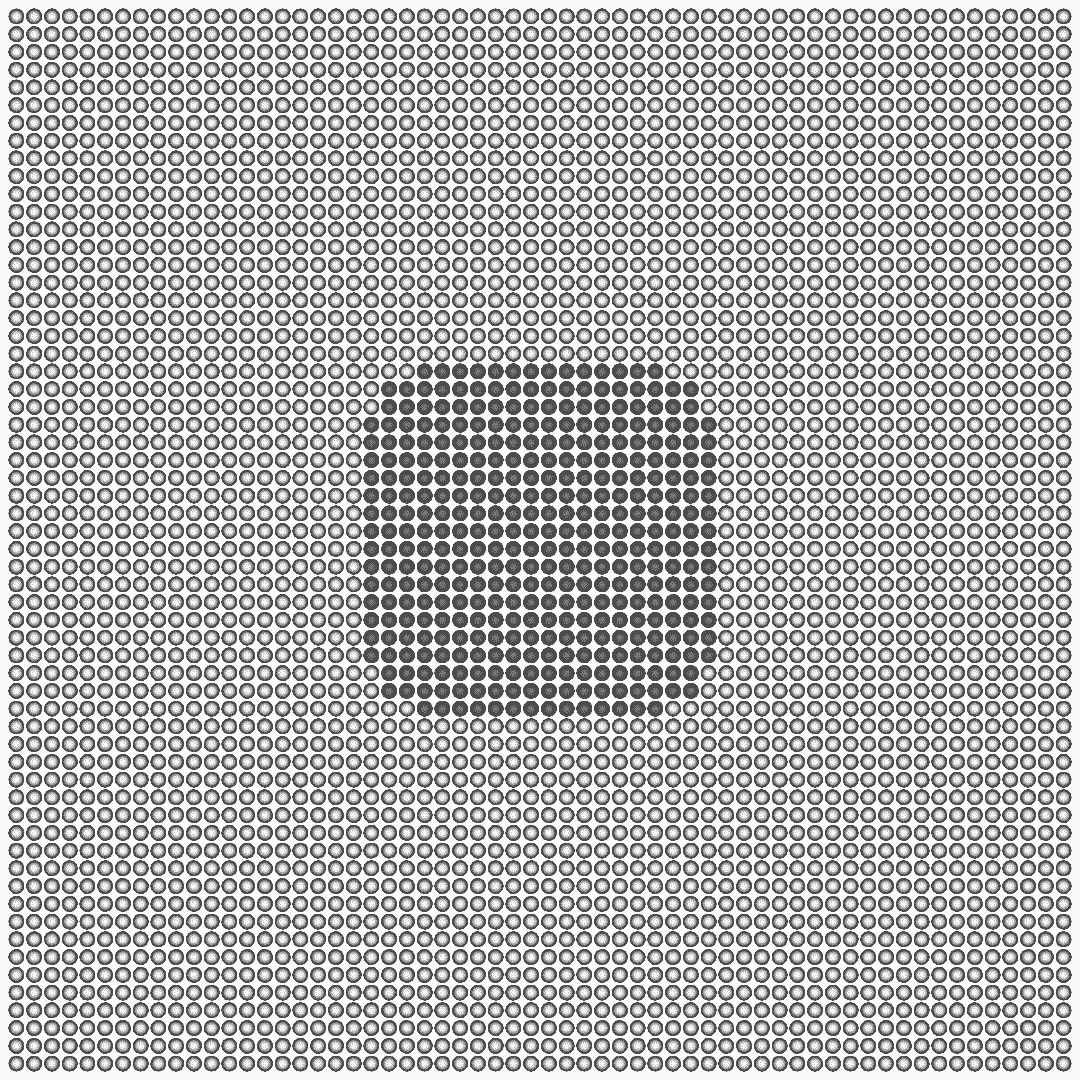}}\hfill
	\subfloat[{$t / t_{\sigma} = 0.6$}] {\includegraphics[width=0.249\textwidth]{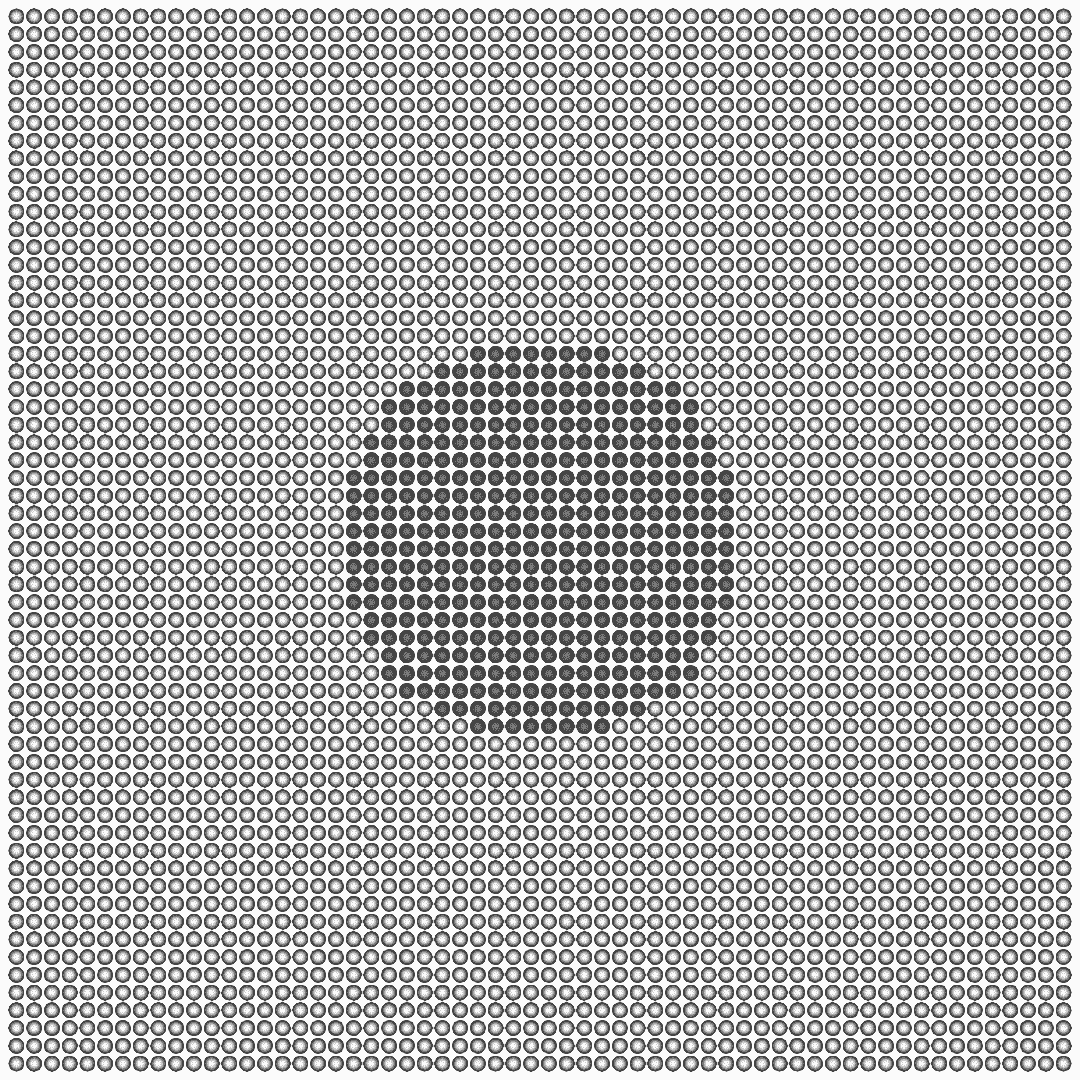}}\hfill  
	\subfloat[{$t / t_{\sigma} = 1.5$}] {\includegraphics[width=0.249\textwidth]{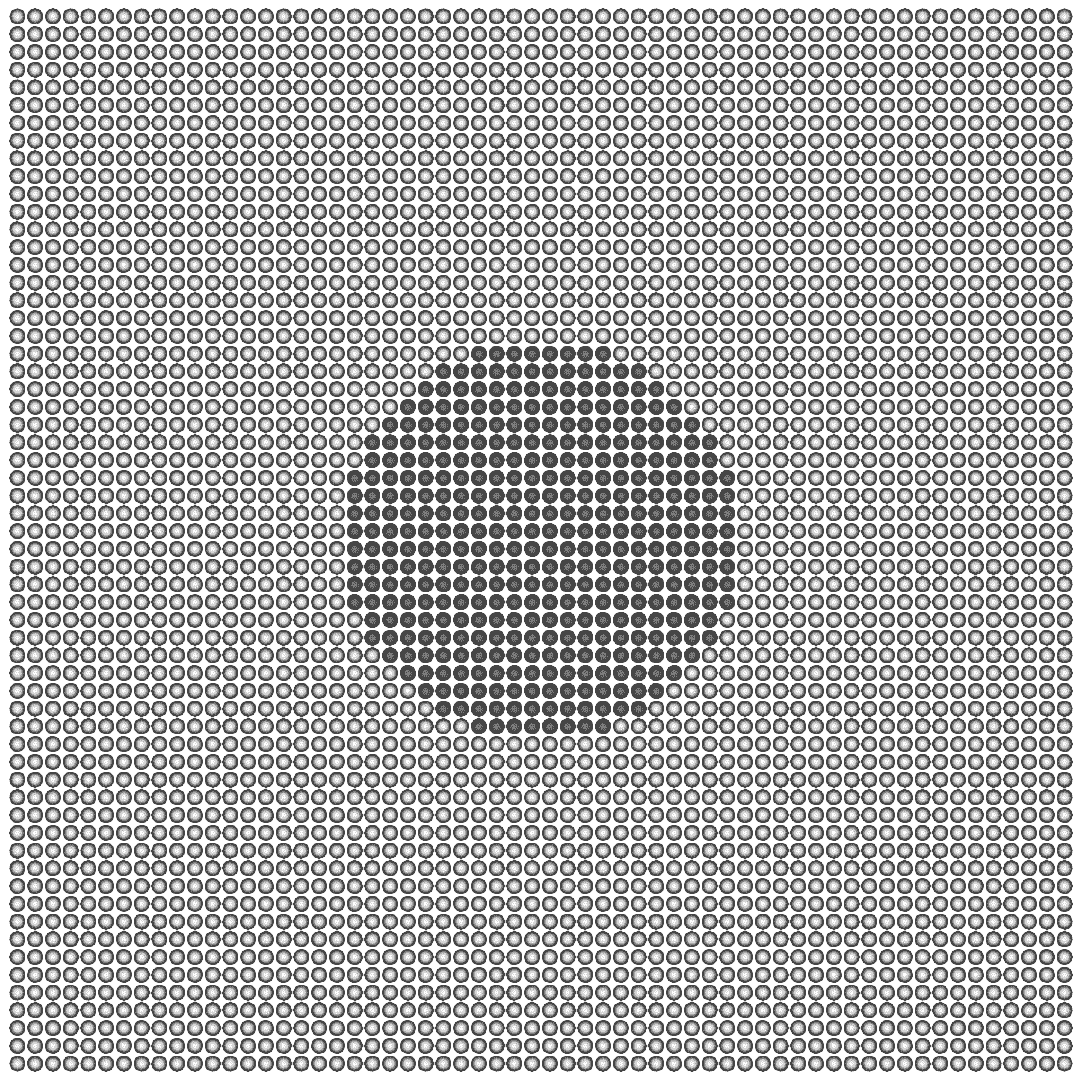}}
	\caption{Evolution of the initial square bubble at selected timesteps. (a,b,c) SPH. (d,e,f) LBM.}
	\label{sqtodrop}
\end{figure}

\begin{figure}[bthp]
	\centering	
	\includegraphics[width=0.65\textwidth]{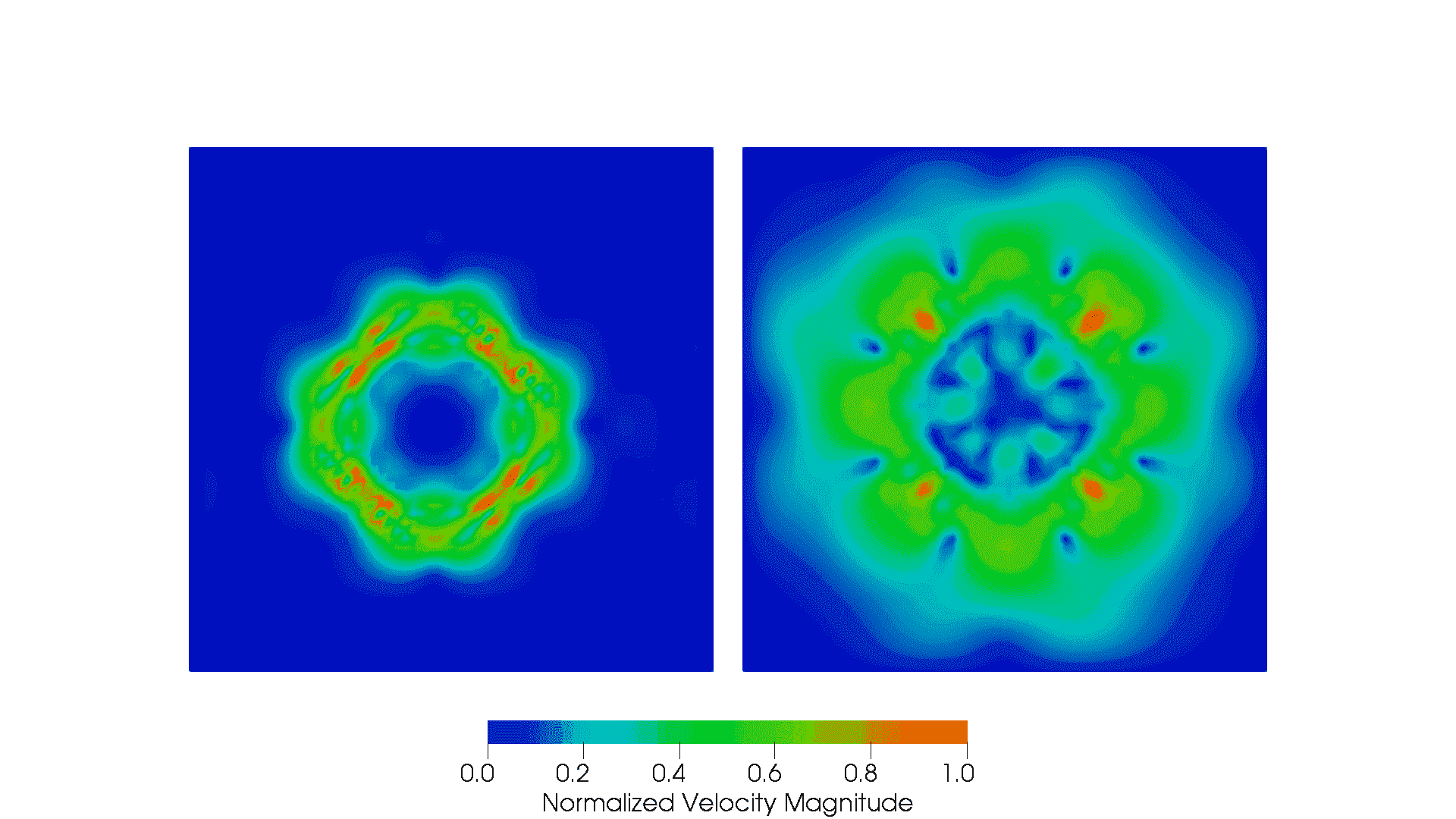}
	\caption{Normalized velocity field $\frac{\lvert\bm{v}\rvert}{\lvert\bm{v}_{max}\rvert}$ for LBM (left) and SPH (right) at $t / t_{\sigma} = 1.5$.}
	\label{bubble_spurious_vels}
\end{figure}

Initially, when the density and viscosity ratios are set to one, one can observe in Fig.~\ref{sqtodrop} the deformation of an initial square bubble towards a circular bubble under the influence of the surface tension. The Lagrangian/Eulerian difference between SPH and LBM is magnified in Fig.~\ref{sqtodrop}. We clearly see that, in SPH, particles of each phase move to form a circular bubble over time whereas in LBM, nodes are fixed and they switch phase to form the expect circular bubble. Besides, when the circular bubble is stabilized, both methods present residual velocities around the interface as shown in Fig.~\ref{bubble_spurious_vels}. However, those spurious currents are much more spread into the domain in SPH compared to LBM where they are localized around the interface. Note that, in the LBM color gradient framework, it is possible to significantly reduce the amplitude of spurious currents by choosing a more isotropic gradient operator~\cite{Leclaire2011} but, as it involves second range neighbors, it is more computationally expensive. it leads to. For SPH, Hu and Adams' formulation~\cite{Hu2006} used in this paper has been reported to generate stronger spurious currents than other formulations~\cite{kunz2015}.

\begin{figure}[bthp]
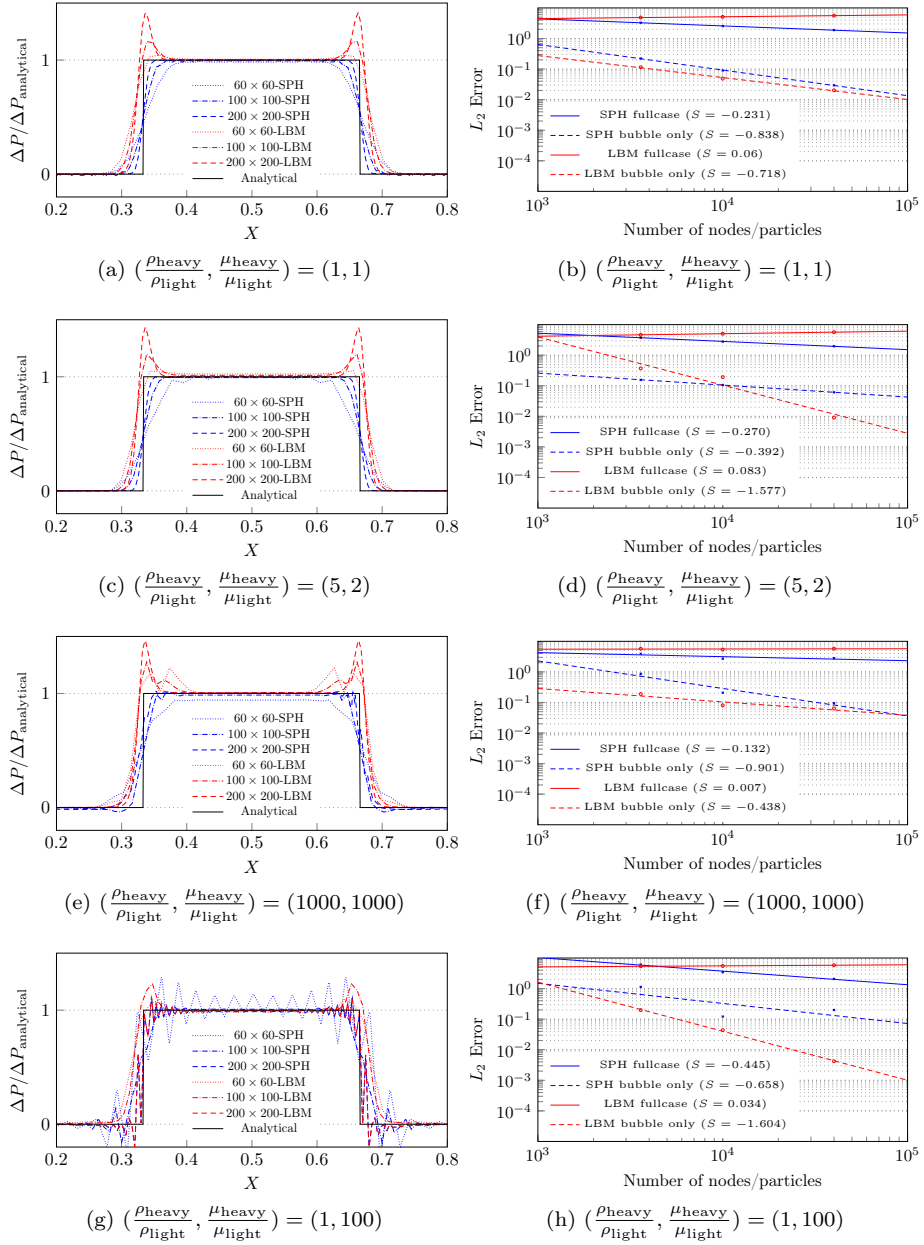

	\centering
	\makebox[\textwidth][c]{	
		\subfloat[$(\frac{\rho_{\text{heavy}}}{\rho_{\text{light}}},\frac{\mu_{\text{heavy}}}{\mu_{\text{light}}})=(1,1)$\label{laplace1}]{\resizebox{0.5\textwidth}{!}{\input{Laplace_test_100x100}}}
		\subfloat[$(\frac{\rho_{\text{heavy}}}{\rho_{\text{light}}},\frac{\mu_{\text{heavy}}}{\mu_{\text{light}}})=(1,1)$\label{laplace2}]{\resizebox{0.5\textwidth}{!}{%\pgfplotsset{label style={font=\tiny},
%	tick label style={font=\tiny} }
%
\begin{tikzpicture}

\begin{axis}[%
scaled ticks=false, 
tick label style={/pgf/number format/fixed},
xmajorgrids=false,
ymajorgrids=true,
grid style={dotted,gray},
width=3.0in,
height=1.5in,
at={(0in,0in)},
scale only axis,
xmode=log,
xmin=1000,
xmax=100000,
xminorticks=true,
yminorgrids=true,
ymode=log,
ymin=0.00001,
ymax=10,
yminorticks=true,
ylabel near ticks,
xlabel near ticks,
xtick pos=left,
ytick pos=left,
ytick={0.0001,0.001,0.01,0.1,1},
xlabel={Number of nodes/particles},
ylabel={$L_2$ Error},
axis background/.style={fill=white},
legend style={legend style={nodes={scale=1.25, transform shape}},fill=white,align=left,draw=none,at={(0.7,0.485)}}
]
\addplot[color=blue,solid, domain=1000:100000, samples=100, smooth] 
plot (\x, { (\x)^(-0.230894957261117) *exp(3.074799034603247) } );
\addlegendentry{\tiny SPH fullcase ($S=-0.231$)};
\addplot[color=blue,densely dashed, domain=1000:100000, samples=100, smooth] 
plot (\x, { (\x)^(-0.837862573969365) *exp(5.339971560823567) } );
\addlegendentry{\tiny SPH bubble only ($S=-0.838$)};
\addplot [color=blue,only marks,mark=x,mark size=0.85, mark repeat=2, forget plot]
table[row sep=crcr]{%
3600	3.292456044444545\\
};
\addplot [color=blue,only marks,mark=x,mark size=0.85, mark repeat=10, forget plot]
table[row sep=crcr]{%
10000	2.547323780072372\\
};
\addplot [color=blue,only marks,mark=x,mark size=0.85, mark repeat=20, forget plot]
table[row sep=crcr]{%
40000	1.884499547575960\\
};
\addplot [color=blue,only marks,mark=x,mark size=0.85, mark repeat=2, forget plot]
table[row sep=crcr]{%
	3600    0.222356802544865\\
};
\addplot [color=blue,only marks,mark=x,mark size=0.85, mark repeat=10, forget plot]
table[row sep=crcr]{%
	10000	0.090038652956643\\
};
\addplot [color=blue,only marks,mark=x,mark size=0.85, mark repeat=20, forget plot]
table[row sep=crcr]{%
	40000	0.029433658378439\\
};
\addplot[color=red,solid, domain=1000:100000, samples=100, smooth] 
plot (\x, { (\x)^(0.060961118484167) *exp(1.076734073622228) } );
\addlegendentry{\tiny LBM fullcase ($S=0.06$)};
\addplot[color=red,densely dashed, domain=1000:100000, samples=100, smooth] 
plot (\x, { (\x)^(-0.718754228416176) *exp(3.679187124533889) } );
\addlegendentry{\tiny LBM bubble only ($S=-0.718$)};
\addplot [color=red,only marks,mark=o,mark options={solid},mark size=0.85, mark repeat=5, forget plot]
table[row sep=crcr]{%
3600	4.865243062599003\\
};
\addplot [color=red,only marks,mark=o,mark options={solid},mark size=0.85, mark repeat=10, forget plot]
table[row sep=crcr]{%
	10000	5.090892961728061\\
};
\addplot [color=red,only marks,mark=o,mark options={solid,rotate=90},mark size=0.85, mark repeat=20, forget plot]
table[row sep=crcr]{%
	40000	5.625328235893352\\
};
\addplot [color=red,only marks,mark=o,mark options={solid},mark size=0.85, mark repeat=5, forget plot]
table[row sep=crcr]{%
	3600	0.115247621302047\\
};
\addplot [color=red,only marks,mark=o,mark options={solid},mark size=0.85, mark repeat=10, forget plot]
table[row sep=crcr]{%
	10000	0.048785150960651\\	
};
\addplot [color=red,only marks,mark=o,mark options={solid,rotate=90},mark size=0.85, mark repeat=20, forget plot]
table[row sep=crcr]{%
	40000	0.020172416331880\\	
};
\end{axis}
\end{tikzpicture}%
}}}\\
	\makebox[\textwidth][c]{
		\subfloat[$(\frac{\rho_{\text{heavy}}}{\rho_{\text{light}}},\frac{\mu_{\text{heavy}}}{\mu_{\text{light}}})=(5,2)$\label{laplace3}]{\resizebox{0.5\textwidth}{!}{\input{Laplace_test_RhoRatio5_MuRatio2}}}
		\subfloat[$(\frac{\rho_{\text{heavy}}}{\rho_{\text{light}}},\frac{\mu_{\text{heavy}}}{\mu_{\text{light}}})=(5,2)$\label{laplace4}]{\resizebox{0.5\textwidth}{!}{%\pgfplotsset{label style={font=\tiny},
%	tick label style={font=\tiny} }
%
\begin{tikzpicture}

\begin{axis}[%
scaled ticks=false, 
tick label style={/pgf/number format/fixed},
xmajorgrids=false,
ymajorgrids=true,
grid style={dotted,gray},
width=3.0in,
height=1.5in,
at={(0in,0in)},
scale only axis,
xmode=log,
xmin=1000,
xmax=100000,
xminorticks=true,
yminorgrids=true,
ymode=log,
ymin=0.00001,
ymax=10,
yminorticks=true,
ylabel near ticks,
xlabel near ticks,
xtick pos=left,
ytick pos=left,
ytick={0.0001,0.001,0.01,0.1,1},
xlabel={Number of nodes/particles},
ylabel={$L_2$ Error},
axis background/.style={fill=white},
legend style={legend style={nodes={scale=1.25, transform shape}},fill=white,align=left,draw=none,at={(0.7,0.485)}}
]
\addplot[color=blue,solid, domain=1000:100000, samples=100, smooth] 
plot (\x, { (\x)^(-0.269664296543075) *exp(3.523786955193346) } );
\addlegendentry{\tiny SPH fullcase ($S=-0.270$)};
\addplot[color=blue,densely dashed, domain=1000:100000, samples=100, smooth] 
plot (\x, { (\x)^(-0.391632112375951 ) *exp(1.357508044913314) } );
\addlegendentry{\tiny SPH bubble only ($S=-0.392$)};
\addplot [color=black,only marks,mark=x,mark size=0.85, mark repeat=2, forget plot]
table[row sep=crcr]{%
3600	3.764211835632049\\
};
\addplot [color=blue,only marks,mark=x,mark size=0.85, mark repeat=10, forget plot]
table[row sep=crcr]{%
10000	2.780988828657281\\
};
\addplot [color=blue,only marks,mark=x,mark size=0.85, mark repeat=20, forget plot]
table[row sep=crcr]{%
40000	1.961255260186354\\
};
\addplot [color=blue,only marks,mark=x,mark size=0.85, mark repeat=2, forget plot]
table[row sep=crcr]{%
	3600    0.156421764020914\\
};
\addplot [color=blue,only marks,mark=x,mark size=0.85, mark repeat=10, forget plot]
table[row sep=crcr]{%
	10000	0.106505039069393\\
};
\addplot [color=blue,only marks,mark=x,mark size=0.85, mark repeat=20, forget plot]
table[row sep=crcr]{%
	40000	0.061010170409688\\
};
\addplot[color=red,solid, domain=1000:100000, samples=100, smooth] 
plot (\x, { (\x)^(0.082652895935550) *exp(0.859283235318851) } );
\addlegendentry{\tiny LBM fullcase ($S=0.083$)};
\addplot[color=red,densely dashed, domain=1000:100000, samples=100, smooth] 
plot (\x, { (\x)^(-1.576642526645714 ) *exp(12.267664201124079) } );
\addlegendentry{\tiny LBM bubble only ($S=-1.577$)};
\addplot [color=red,only marks,mark=o,mark options={solid},mark size=0.85, mark repeat=5, forget plot]
table[row sep=crcr]{%
3600	4.664981320931299\\
};
\addplot [color=red,only marks,mark=o,mark options={solid},mark size=0.85, mark repeat=10, forget plot]
table[row sep=crcr]{%
	10000	5.020996502979505\\
};
\addplot [color=red,only marks,mark=o,mark options={solid,rotate=90},mark size=0.85, mark repeat=20, forget plot]
table[row sep=crcr]{%
	40000	5.686302299446294\\
};
\addplot [color=red,only marks,mark=o,mark options={solid},mark size=0.85, mark repeat=5, forget plot]
table[row sep=crcr]{%
	3600	0.370917075865403\\
};
\addplot [color=red,only marks,mark=o,mark options={solid},mark size=0.85, mark repeat=10, forget plot]
table[row sep=crcr]{%
	10000	0.192456128633359\\	
};
\addplot [color=red,only marks,mark=o,mark options={solid,rotate=90},mark size=0.85, mark repeat=20, forget plot]
table[row sep=crcr]{%
	40000	0.009127181207859\\	
};
\end{axis}
\end{tikzpicture}%
}}}\\
	\makebox[\textwidth][c]{	
		\subfloat[$(\frac{\rho_{\text{heavy}}}{\rho_{\text{light}}},\frac{\mu_{\text{heavy}}}{\mu_{\text{light}}})=(1000,1000)$\label{laplacerho1000}]{\resizebox{0.5\textwidth}{!}{\input{Laplace_test_RhoRatio1000_MuRatio1000}}}
		\subfloat[$(\frac{\rho_{\text{heavy}}}{\rho_{\text{light}}},\frac{\mu_{\text{heavy}}}{\mu_{\text{light}}})=(1000,1000)$\label{laplace5}]{\resizebox{0.5\textwidth}{!}{%\pgfplotsset{label style={font=\tiny},
%	tick label style={font=\tiny} }
%
\begin{tikzpicture}

\begin{axis}[%
scaled ticks=false, 
tick label style={/pgf/number format/fixed},
xmajorgrids=false,
ymajorgrids=true,
grid style={dotted,gray},
width=3.0in,
height=1.5in,
at={(0in,0in)},
scale only axis,
xmode=log,
xmin=1000,
xmax=100000,
xminorticks=true,
yminorgrids=true,
ymode=log,
ymin=0.00001,
ymax=10,
yminorticks=true,
ylabel near ticks,
xlabel near ticks,
xtick pos=left,
ytick pos=left,
ytick={0.0001,0.001,0.01,0.1,1},
xlabel={Number of nodes/particles},
ylabel={$L_2$ Error},
axis background/.style={fill=white},
legend style={legend style={nodes={scale=1.25, transform shape}},fill=white,align=left,draw=none,at={(0.7,0.485)}}
]
\addplot[color=blue,solid, domain=1000:100000, samples=100, smooth] 
plot (\x, { (\x)^(-0.132484056024296) *exp(2.368225020581479) } );
\addlegendentry{\tiny SPH fullcase ($S=-0.132$)};
\addplot[color=blue,densely dashed, domain=1000:100000, samples=100, smooth] 
plot (\x, { (\x)^(-0.901265400114207) *exp(7.054195963058999) } );
\addlegendentry{\tiny SPH bubble only ($S=-0.901$)};
\addplot [color=blue,only marks,mark=x,mark size=0.85, mark repeat=2, forget plot]
table[row sep=crcr]{%
3600	3.941745819501293\\
};
\addplot [color=blue,only marks,mark=x,mark size=0.85, mark repeat=10, forget plot]
table[row sep=crcr]{%
10000	2.703861036102414\\
};
\addplot [color=blue,only marks,mark=x,mark size=0.85, mark repeat=20, forget plot]
table[row sep=crcr]{%
40000	2.799361686463250\\
};
\addplot [color=blue,only marks,mark=x,mark size=0.85, mark repeat=2, forget plot]
table[row sep=crcr]{%
	3600    0.867041253286235\\
};
\addplot [color=blue,only marks,mark=x,mark size=0.85, mark repeat=10, forget plot]
table[row sep=crcr]{%
	10000	0.209050104812117\\
};
\addplot [color=blue,only marks,mark=x,mark size=0.85, mark repeat=20, forget plot]
table[row sep=crcr]{%
	40000	0.094317776917633\\
};
\addplot[color=red,solid, domain=1000:100000, samples=100, smooth] 
plot (\x, { (\x)^(0.006978682038192 ) *exp(1.661081919294864) } );
\addlegendentry{\tiny LBM fullcase ($S=0.007$)};
\addplot[color=red,densely dashed, domain=1000:100000, samples=100, smooth] 
plot (\x, { (\x)^(-0.437850980482187 ) *exp(1.773218252149892) } );
\addlegendentry{\tiny LBM bubble only ($S=-0.438$)};
\addplot [color=red,only marks,mark=o,mark options={solid},mark size=0.85, mark repeat=5, forget plot]
table[row sep=crcr]{%
3600	5.713741775408960\\
};
\addplot [color=red,only marks,mark=o,mark options={solid},mark size=0.85, mark repeat=10, forget plot]
table[row sep=crcr]{%
	10000	5.379249231767769\\
};
\addplot [color=red,only marks,mark=o,mark options={solid,rotate=90},mark size=0.85, mark repeat=20, forget plot]
table[row sep=crcr]{%
	40000	5.773021898472355\\
};
\addplot [color=red,only marks,mark=o,mark options={solid},mark size=0.85, mark repeat=5, forget plot]
table[row sep=crcr]{%
	3600	0.189538773921467\\
};
\addplot [color=red,only marks,mark=o,mark options={solid},mark size=0.85, mark repeat=10, forget plot]
table[row sep=crcr]{%
	10000	0.080586392627978\\	
};
\addplot [color=red,only marks,mark=o,mark options={solid,rotate=90},mark size=0.85, mark repeat=20, forget plot]
table[row sep=crcr]{%
	40000	0.063501792676923\\	
};
\end{axis}
\end{tikzpicture}%
}}}\\
	\makebox[\textwidth][c]{
		\subfloat[$(\frac{\rho_{\text{heavy}}}{\rho_{\text{light}}},\frac{\mu_{\text{heavy}}}{\mu_{\text{light}}})=(1,100)$\label{laplacenu100}]{\resizebox{0.5\textwidth}{!}{\input{Laplace_test_RhoRatio1_MuRatio100}}}
		\subfloat[$(\frac{\rho_{\text{heavy}}}{\rho_{\text{light}}},\frac{\mu_{\text{heavy}}}{\mu_{\text{light}}})=(1,100)$\label{laplace6}]{\resizebox{0.5\textwidth}{!}{%\pgfplotsset{label style={font=\tiny},
%	tick label style={font=\tiny} }
%
\begin{tikzpicture}

\begin{axis}[%
scaled ticks=false, 
tick label style={/pgf/number format/fixed},
xmajorgrids=false,
ymajorgrids=true,
grid style={dotted,gray},
width=3.0in,
height=1.5in,
at={(0in,0in)},
scale only axis,
xmode=log,
xmin=1000,
xmax=100000,
xminorticks=true,
yminorgrids=true,
ymode=log,
ymin=0.00001,
ymax=10,
yminorticks=true,
ylabel near ticks,
xlabel near ticks,
xtick pos=left,
ytick pos=left,
ytick={0.0001,0.001,0.01,0.1,1},
xlabel={Number of nodes/particles},
ylabel={$L_2$ Error},
axis background/.style={fill=white},
legend style={legend style={nodes={scale=1.25, transform shape}},fill=white,align=left,draw=none,at={(0.7,0.485)}}
]
\addplot[color=blue,solid, domain=1000:100000, samples=100, smooth] 
plot (\x, { (\x)^(-0.444643352558098 ) *exp(5.409397790737189) } );
\addlegendentry{\tiny SPH fullcase ($S=-0.445$)};
\addplot[color=blue,densely dashed, domain=1000:100000, samples=100, smooth] 
plot (\x, { (\x)^(-0.657625598289187  ) *exp(4.941180065083363) } );
\addlegendentry{\tiny SPH bubble only ($S=-0.658$)};
\addplot [color=blue,only marks,mark=x,mark size=0.85, mark repeat=2, forget plot]
table[row sep=crcr]{%
3600	6.132876214977943\\
};
\addplot [color=blue,only marks,mark=x,mark size=0.85, mark repeat=10, forget plot]
table[row sep=crcr]{%
10000	3.439669433343475\\
};
\addplot [color=blue,only marks,mark=x,mark size=0.85, mark repeat=20, forget plot]
table[row sep=crcr]{%
40000	2.077298304379131\\
};
\addplot [color=blue,only marks,mark=x,mark size=0.85, mark repeat=2, forget plot]
table[row sep=crcr]{%
	3600    1.134783777720900\\
};
\addplot [color=blue,only marks,mark=x,mark size=0.85, mark repeat=10, forget plot]
table[row sep=crcr]{%
	10000	0.121667885556911\\
};
\addplot [color=blue,only marks,mark=x,mark size=0.85, mark repeat=20, forget plot]
table[row sep=crcr]{%
	40000	0.200466690958104\\
};
\addplot[color=red,solid, domain=1000:100000, samples=100, smooth] 
plot (\x, { (\x)^(0.033748561836304 ) *exp(1.402255083950357) } );
\addlegendentry{\tiny LBM fullcase ($S=0.034$)};
\addplot[color=red,densely dashed, domain=1000:100000, samples=100, smooth] 
plot (\x, { (\x)^(-1.604049836884864 ) *exp(11.562584546127246) } );
\addlegendentry{\tiny LBM bubble only ($S=-1.604$)};
\addplot [color=red,only marks,mark=o,mark options={solid},mark size=0.85, mark repeat=5, forget plot]
table[row sep=crcr]{%
3600	5.384714835600942\\
};
\addplot [color=red,only marks,mark=o,mark options={solid},mark size=0.85, mark repeat=10, forget plot]
table[row sep=crcr]{%
	10000	5.498569689855182\\
};
\addplot [color=red,only marks,mark=o,mark options={solid,rotate=90},mark size=0.85, mark repeat=20, forget plot]
table[row sep=crcr]{%
	40000	5.832969980915278\\
};
\addplot [color=red,only marks,mark=o,mark options={solid},mark size=0.85, mark repeat=5, forget plot]
table[row sep=crcr]{%
	3600	0.197652143360860\\
};
\addplot [color=red,only marks,mark=o,mark options={solid},mark size=0.85, mark repeat=10, forget plot]
table[row sep=crcr]{%
	10000	0.043868620120114\\	
};
\addplot [color=red,only marks,mark=o,mark options={solid,rotate=90},mark size=0.85, mark repeat=20, forget plot]
table[row sep=crcr]{%
	40000	0.004207509646349\\	
};
\end{axis}
\end{tikzpicture}%
}}}	
	\caption{ (a,c,e,g) Pressure profiles at steady state for different resolutions and different density and viscosity ratios. (b,d,f,h) Log-log $L_2$ error plots as function of resolution (superposed with linear regressions of slope $S$). }
	\label{laplace}
\end{figure}

In Fig.~\ref{laplace}, the pressure profiles at steady state for the different resolutions and the different density and viscosity ratios considered are shown along with the corresponding $L_2$ error plots. First, one can clearly see that LBM is returning incorrect pressure values at the interface. Indeed, LBM presents non-physical pressure peaks at the bubble's interface that tend to grow with the number of nodes. In fact, in the LBM color gradient method, the pressure is not well-defined at the interface. The pressure formula of Eq.~\eqref{eq:pressureRelation} does not make sense at the interface where fluids are mixed and there is no mixture pressure defined in the considered framework. Hence, summing the fluid pressures is just an analytical construction that depends on the density profiles and the interface width. Thus, if we look at the LBM $L_2$ error along the whole horizontal centerline, we do not have mesh convergence since the error is growing at the interface. However, when we restrict the calculation of the $L_2$ error inside the bubble (i.e. when $0.4~\meter < X < 0.6~\meter$), we do obtain a negative slope indicating mesh convergence. 

Next, analyzing the impact of the density and viscosity ratios, for the case where $(\frac{\rho_{\text{heavy}}}{\rho_{\text{light}}},\frac{\mu_{\text{heavy}}}{\mu_{\text{light}}})=(1,1)$ in Figs.~\ref{laplace1} and \ref{laplace2}, we get approximately the same order of convergence for both methods ($0.718$ and $0.838$ for LBM and SPH respectively) and the same error levels ($\leq 1\%$ at the bubble's center) even though SPH always has a slightly higher error level. Next, when the density and viscosity ratios remains moderate (i.e. respectively up to $5$ and $2$ in Figs.~\ref{laplace3} and \ref{laplace4}), both methods are under $2.5 \%$ error compared to the reference solution. Additionally, we see that LBM offers a better order of convergence than SPH. In fact, LBM sees its order of convergence increased (from $0.718$ to $1.577$) compared to the previous case unlike SPH where it decreases (from $0.838$ to $0.392$). Moreover, LBM is less accurate than SPH for the two lowest resolutions $60 \times 60$ and $100 \times 100$ but performs better for the $200\times 200$ case thanks to its higher order of convergence. Overall, although both methods are returning satisfactory results for this case, we begin to observe a fall in performance whether it is for the order of convergence (for SPH) or for the error levels (for LBM and SPH) because gradients at stake are steeper. Then, for the high density ratio case where $(\frac{\rho_{\text{heavy}}}{\rho_{\text{light}}},\frac{\mu_{\text{heavy}}}{\mu_{\text{light}}})=(1000,1000)$ shown in Figs.~\ref{laplacerho1000} and \ref{laplace5}, we can see that the order of convergence of SPH remains roughly the same compared to the first case ($0.838$ vs $0.901$). The maximum error level is $\leq 6\%$ i.e. higher than the first case. This tends to indicate that when the density ratio increases, SPH is a quite robust and offers a reasonable accuracy for the same order of convergence. On the other hand, LBM sees its order of convergence strongly affected by the presence of this density ratio ($0.438$ vs $0.718$) while maintaining approximately the same error level. Finally, we looked at one last case, shown in Figs.~\ref{laplacenu100} and \ref{laplace6}, where the density ratio is equal to $1$ and the viscosity ratio increased up to $100$. One can immediately note that, for both methods, the pressure profiles are heavily impacted at the interface (oscillations) in particular for the $60 \times 60$ case. However, when looking at the pressure jump at the center of the bubble, LBM appears very robust to the presence of such a strong viscosity ratio. Indeed, we see that the error levels are of the same order than those of the first case and that the order of convergence is even higher ($0.718$ vs $1.604$). It shows that refining the lattice strongly helps to stabilize the pressure field. On the contrary, SPH appears more affected. The error levels are the highest of all four cases considered and the order of convergence is inferior to the first case ($0.838$ vs $0.658$). Moreover, the error does not seem to decrease anymore exponentially with the number of particles although more simulations would be needed to further check that statement. 

To sum up, for limited density ratios and viscosity ratios, both methods are able to reproduce the pressure jump predicted by Eq.~\eqref{laplaceslaw} with a good accuracy and with steep and clean pressure/density profiles. When the the density ratio increases up to $1000$, SPH seems to be more resilient than LBM in the sense that its order of convergence is not impacted by the presence of such a strong density ratio. LBM seems less robust in the same situation. On the contrary, when the viscosity ratio goes up to $100$, both methods render perturbed pressure profiles. However, it is SPH that appears to have more problem to handle a strong viscosity ratio whereas LBM maintains its performance level.

\begin{figure}[bthp]
	\centering
	%	\subfloat[LBM] {
	%	\resizebox{0.4\textwidth}{!}{\input{contact_angle_lbm2}}}
	%	\subfloat[SPH] {
	\resizebox{0.65\textwidth}{!}{\input{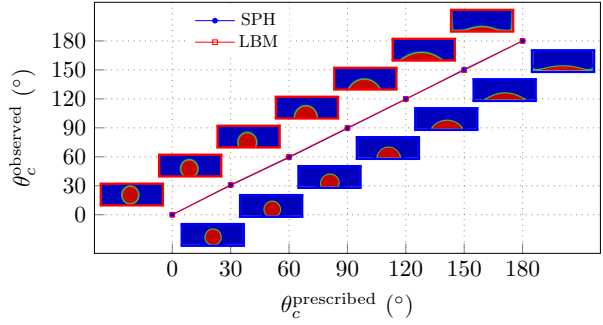}}
	\caption{Comparison between prescribed and observed contact angles for both SPH and LBM (density and viscosity ratios are both equal to one).}
	\label{cangle}
\end{figure}

\paragraph{Contact angle case.}
In addition, we compared the ability of the previously described implementations of SPH and LBM to prescribe a contact angle between a wetting phase, a non-wetting phase and a solid phase. Simulations are done with $100 \times 100$ nodes/particles. The density and viscosity ratios are both equal to one. The whole domain is $1~\meter \times 1~\meter$ and the lateral dimensions of the initial rectangular droplet is $0.33~\meter \times 0.165~\meter$. For SPH, the surface tension coefficient between the wetting and non-wetting phase is $\sigma^{nw} = 1.88~\newton\per\meter$ whereas the one between the wetting and the solid phase is set to $\sigma^{sw}=0~\newton\per\meter$ and the one between the non-wetting and the solid phase $\sigma^{sn}$ is adjusted to prescribed the desired contact angle $\theta^{\text{prescribed}}_c$ using the Young-Laplace equation $\theta_c = \frac{\sigma^{sw} - \sigma^{sn}}{\sigma^{nw}}$. For LBM, we follow the procedure described in Sec.~\ref{sec:angleAdjust}. The surface tension coefficient is set to $\sigma = 0.01~l.u.$. At steady state, the observed contact angle $\theta^{\text{observed}}_c$ is measured and reported in Fig.~\ref{cangle}. The coefficient of determination is $\geq 0.99$ for both methods confirming that they can accurately reproduce a prescribed contact angle. Finally, one can observe the normalized velocity field at steady state for $\theta^{\text{prescribed}}_c=150^{\circ}$. The same comments made before are still valid, LBM spurious currents are less spread throughout the domain than in SPH. This is likely due to the Lagrangian nature of SPH where particles have to rearrange to match the simulated physics.

%\subsection{RhoRatio1 MuRatio1}

%\begin{table}[bthp]
%	\centering
%	\subfloat[SPH] {\resizebox{0.49\textwidth}{!}{
%			\begin{tabular}{lll}
%				\toprule
%		Case       & $E(\%)$           & $\lvert \bm{u} \lvert_{max}$ \\
%				\toprule 
%		$0^{\circ}$   & $\color{black}\bm{0.00\%}$  & $2.12 \times 10^{-3}$ \\ 	 
%		$30^{\circ}$   & $\color{black}\bm{1.53\%}$  & $2.93 \times 10^{-3}$ \\ 
%		$60^{\circ}$   & $\color{black}\bm{1.26\%}$  & $4.14 \times 10^{-3}$ \\
%		$90^{\circ}$   & $\color{black}\bm{0.72\%}$  & $5.56 \times 10^{-3}$ \\
%		$120^{\circ}$  & $\color{black}\bm{0.28\%}$  & $4.47 \times 10^{-3}$ \\ 
%		$150^{\circ}$  & $\color{black}\bm{0.30\%}$  & $2.14 \times 10^{-3}$ \\ 
%		$180^{\circ}$  & $\color{black}\bm{0.00\%}$  & $2.93 \times 10^{-3}$ \\ 
%				\bottomrule
%	\end{tabular}}}\hfill
%	\subfloat[LBM] {\resizebox{0.49\textwidth}{!}{
%			\begin{tabular}{lll}
%				\toprule
%		Case       & $E(\%)$           & $\lvert \bm{u} \lvert_{max}$ \\
%				\toprule 
%		$0^{\circ}$   & $\color{black}\bm{0.00\%}$  & $3.62 \times 10^{-4}$ \\ 	 
%		$30^{\circ}$   & $\color{black}\bm{3.30\%}$  & $1.60 \times 10^{-3}$ \\ 
%		$60^{\circ}$   & $\color{black}\bm{0.12\%}$  & $2.81 \times 10^{-4}$ \\
%		$90^{\circ}$   & $\color{black}\bm{0.05\%}$  & $4.37 \times 10^{-4}$ \\
%		$120^{\circ}$  & $\color{black}\bm{0.06\%}$  & $1.73 \times 10^{-4}$ \\ 
%		$150^{\circ}$  & $\color{black}\bm{0.23\%}$  & $1.24 \times 10^{-3}$ \\ 
%		$180^{\circ}$  & $\color{black}\bm{0.00\%}$  & $6.15 \times 10^{-3}$ \\ 
%				\bottomrule
%	\end{tabular}}}  
%	\caption{Contact angle verification : errors and spurious currents}
%	\label{table_contactangle}
%\end{table}

\begin{figure}[bthp]
	\centering	
	\includegraphics[width=0.65\textwidth]{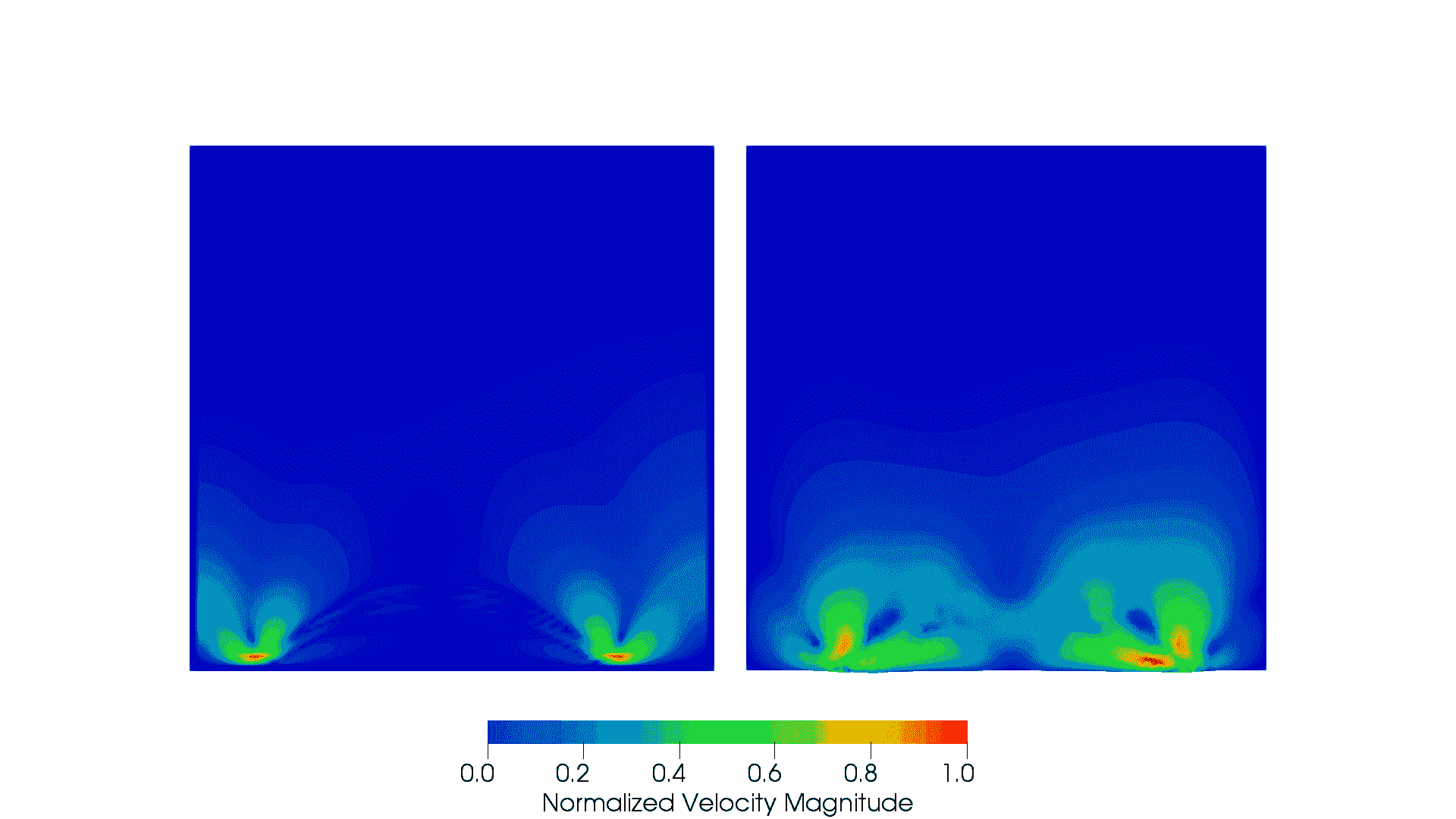}
	\caption{Normalized velocity field for LBM (left) and SPH (right) for $\theta^{\text{prescribed}}_c=150^{\circ}$.}
	\label{laplace_angle}
\end{figure}

\section{Intermittent two-phase flows in pipes}

In the following section, we extend our comparative study to two cases of intermittent two-phase flows in pipes for different Reynolds numbers. The first one is periodic and gravity-driven while the second one is generated by a velocity inlet and a pressure boundary condition respectively at the inlet and outlet of the pipe.

\subsection{Periodic case}

\begin{figure}[bthp]
	%\begin{center}
	%\hspace{-4cm}
	\centering
	\resizebox{1.0\textwidth}{!}{\begin{tikzpicture}

\path[use as bounding box] (-1.5, -1) rectangle (21.5, 4.0);

\node[circle,fill=black!0](bl) at (0,0) {};

\node[circle,fill=black!0](ml) at (0,1) {};

\node[circle,fill=black!0](tl) at (0,2) {};

\node[circle,fill=black!0](br) at (20,0) {};

\node[circle,fill=black!0](mr) at (20,1) {};

\node[circle,fill=black!0](tr) at (20,2) {};

%\node[circle,fill=black!0,label={$x$}](x) at (12,0) {};
%\node[circle,fill=black!0,label={$y$}](y) at (0,12) {};

\draw[ultra thick] (bl.center) -- (br.center);
\draw (br.center) -- (tr.center);
\draw[ultra thick] (tr.center) -- (tl.center);
\draw (tl.center) -- (bl.center);

%\draw[dashed]  (ml.center) -- (mr.center);

\fill[fill=black!20, draw=black] (bl.center) rectangle (tr.center);

%\draw[triangle 45-triangle 45] (16.5,0) -- node[right] {\Large $\alpha_l$} (16.5,1.0);
%\draw[triangle 45-triangle 45] (16.5,1.0) -- node[right] {\Large $\alpha_g$} (16.5,2.0);

%\draw[-triangle 45] (-1.0,1.75) --  (0.0,1.75);
%\draw[-triangle 45] (-1.0,1.25) --  (0.0,1.25);
%\draw[-triangle 45] (-1.0,0.75) --  (0.0,0.75);
%\draw[-triangle 45] (-1.0,0.25) --  (0.0,0.25);

%\node[circle,fill=black!0,label={\Large $v^{\text{in}}_{\text{light}}$}](ug) at (-1.5,1.0) {};
%\node[circle,fill=black!0,label={\Large $v^{\text{in}}_{\text{heavy}}=6 v^{\text{in}}_{\text{light}}$}](ul) at (-2.5,0.0) {};

%\draw[-triangle 45] (20.0,1.75) --  (21.0,1.75);
%\draw[-triangle 45] (20.0,1.25) --  (21.0,1.25);
%\draw[-triangle 45] (20.0,0.75) --  (21.0,0.75);
%\draw[-triangle 45] (20.0,0.25) --  (21.0,0.25);

%\node[circle,fill=black!0,label={\Large $p^{\text{out}}$}](ug) at (21.5,0.5) {};

%\node[circle,fill=none,label={\Large Light Fluid}](tr) at (10,1.0) {};
%\node[circle,fill=none,label={\Large Heavy Fluid}](tr) at (10,0.0) {};

\draw[ultra thick] (bl.center) -- (br.center);

%%%%%%%%
\draw[fill=black!0] (0.83333333333,0.5) circle [radius=0.4];
%\node[color=black] at (0.83333333333,0.5) {$R=0.2 H$};

\draw[fill=black!0] (2.5,0.5) circle [radius=0.4];
%\node[color=black] {text};

\draw[fill=black!0] (4.16666666667,0.5) circle [radius=0.4];
%\node[color=black] {text};

\draw[fill=black!0] (5.83333333333,0.5) circle [radius=0.4];
%\node[color=black] {text};

\draw[fill=black!0] (7.5,0.5) circle [radius=0.4];
%\node[color=black] {text};

\draw[fill=black!0] (9.16666666667,0.5) circle [radius=0.4];
%\node[color=black] {text};

\draw[fill=black!0] (10.8333333333,0.5) circle [radius=0.4];
%\node[color=black] {text};

\draw[fill=black!0] (12.5,0.5) circle [radius=0.4];
%\node[color=black] {text};

\draw[fill=black!0] (14.1666666667,0.5) circle [radius=0.4];
%\node[color=black] {text};

\draw[fill=black!0] (15.8333333333,0.5) circle [radius=0.4];
%\node[color=black] {text};

\draw[fill=black!0] (17.5,0.5) circle [radius=0.4];
%\node[color=black] {text};

\draw[fill=black!0] (19.1666666667,0.5) circle [radius=0.4];
%\node[color=black] {text};
%%%%%%%
\draw[fill=black!0] (0.83333333333,1.5) circle [radius=0.4];
%\node[color=black] {text};

\draw[fill=black!0] (2.5,1.5) circle [radius=0.4];
%\node[color=black] {text};

\draw[fill=black!0] (4.16666666667,1.5) circle [radius=0.4];
%\node[color=black] {text};

\draw[fill=black!0] (5.83333333333,1.5) circle [radius=0.4];
%\node[color=black] {text};

\draw[fill=black!0] (7.5,1.5) circle [radius=0.4];
%\node[color=black] {text};

\draw[fill=black!0] (9.16666666667,1.5) circle [radius=0.4];
%\node[color=black] {text};

\draw[fill=black!0] (10.8333333333,1.5) circle [radius=0.4];
%\node[color=black] {text};

\draw[fill=black!0] (12.5,1.5) circle [radius=0.4];
%\node[color=black] {text};

\draw[fill=black!0] (14.1666666667,1.5) circle [radius=0.4];
%\node[color=black] {text};

\draw[fill=black!0] (15.8333333333,1.5) circle [radius=0.4];
%\node[color=black] {text};

\draw[fill=black!0] (17.5,1.5) circle [radius=0.4];
%\node[color=black] {text};

\draw[fill=black!0] (19.1666666667,1.5) circle [radius=0.4];
%\node[color=black] {CHOCO};
%%%%%%
\draw[fill=black!0] (1.66666666667,1.0) circle [radius=0.13333333333];
%\node[color=black] at (1.66666666667,1.0)   {$R_2=R/3$};
\draw[fill=black!0] (3.33333333333,1.0) circle [radius=0.13333333333];
%\node[color=black] {text};
\draw[fill=black!0] (5,1.0) circle [radius=0.13333333333];
%\node[color=black] {text};
\draw[fill=black!0] (8.33333333333,1.0) circle [radius=0.13333333333];
%\node[color=black] {text};
\draw[fill=black!0] (10,1.0) circle [radius=0.13333333333];
%\node[color=black] {text};
\draw[fill=black!0] (11.6666666667,1.0) circle [radius=0.13333333333];
%\node[color=black] {text};
\draw[fill=black!0] (15,1.0) circle [radius=0.13333333333];
%\node[color=black] {text};
\draw[fill=black!0] (16.6666666667,1.0) circle [radius=0.13333333333];
%\node[color=black] {text};
\draw[fill=black!0] (18.3333333333,1.0) circle [radius=0.13333333333];
%\node[color=black] {CHOCO};

\draw [-latex,black,thick] (20,1) to [out=30,in=150] (0,1);

\draw[triangle 45-triangle 45] (0.0,-0.5) -- node[below] {\Large $L=6H$} (20.0,-0.5);
\draw[triangle 45-triangle 45] (6.3,0) -- node[above right] {\Large $H$} (6.3,2.0);
\draw[triangle 45-triangle 45] (12.5,1.5) -- node[below] { $L/12$} (14.1666666667,1.5);
\draw[triangle 45-triangle 45] (15,1.0) -- node[right] { $H/2$} (15,2.0);
\draw[triangle 45-triangle 45] (5,1.5) -- node[right] { $H/4$} (5,2.0);
\draw[triangle 45-triangle 45] (7.5,1.5) -- node[below] { $L/24$} (8.33333333333,1.5);

\node[circle,fill=none,label={\Large (periodic)}](tr) at (10,3.0) {};

\draw[triangle 45-] (9,2.5) -- node[above] {\Large $\bm{g}$} (11,2.5);

\end{tikzpicture}}
	\caption {Initial configuration sketch. The large bubbles radius is $R=0.2H$ and the small bubbles' radius is $R_2=R/3$. The pipe height is $H=0.1~\meter$.}
	\label{init_geo2}
	%\end{center}
\end{figure}
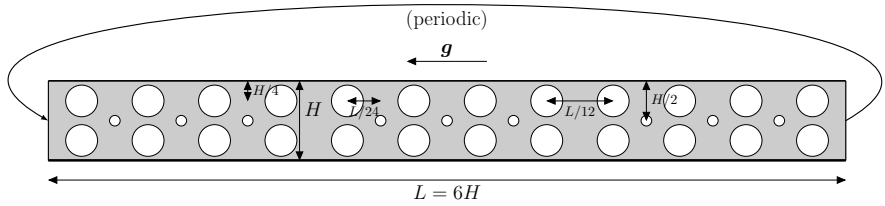

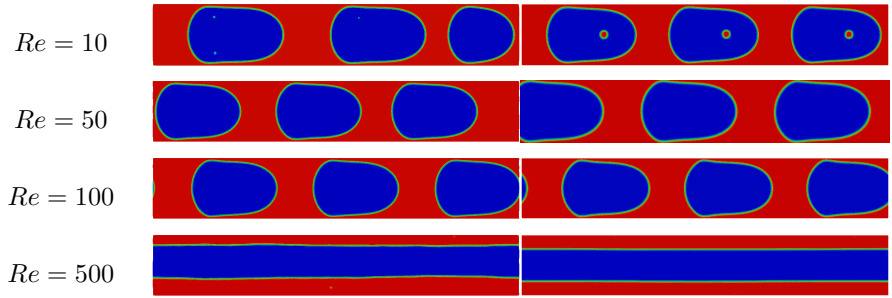
\begin{figure}[bthp]
	\begin{center}
		\captionsetup[subfigure]{labelformat=empty}
		\resizebox{0.2\textwidth}{!}{\begin{tikzpicture}
\path[use as bounding box] (-1.25, 0.1) rectangle (1.1, 1.1);
\node[circle,fill=none,label={$Re=10$}](tr) at (0,0) {};
\end{tikzpicture} }\hfill
		\includegraphics[width=0.4\textwidth]{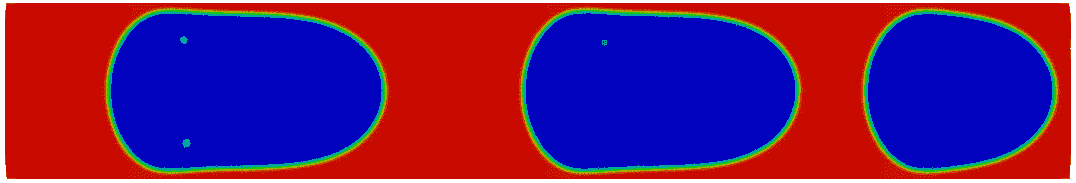}\hfill
		\includegraphics[width=0.4\textwidth]{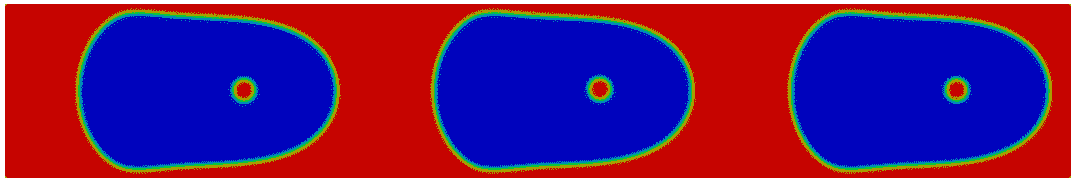}\\
		\resizebox{0.2\textwidth}{!}{\begin{tikzpicture}
\path[use as bounding box] (-1.25, 0.1) rectangle (1.1, 1.1);
\node[circle,fill=none,label={$Re=50$}](tr) at (0,0) {};
\end{tikzpicture} }\hfill
		\includegraphics[width=0.4\textwidth]{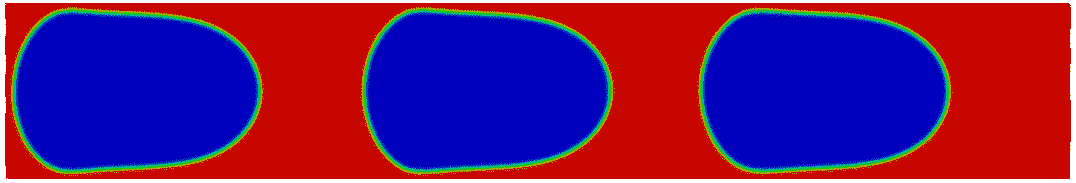}\hfill
		\includegraphics[width=0.4\textwidth]{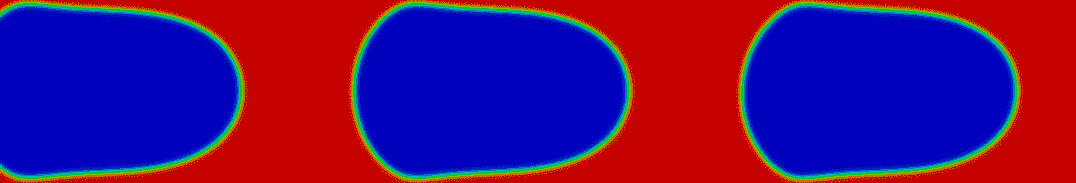}\\
		\resizebox{0.2\textwidth}{!}{\begin{tikzpicture}
\path[use as bounding box] (-1.25, 0.1) rectangle (1.1, 1.1);
\node[circle,fill=none,label={$Re=100$}](tr) at (0,0) {};
\end{tikzpicture} }\hfill
		\includegraphics[width=0.4\textwidth]{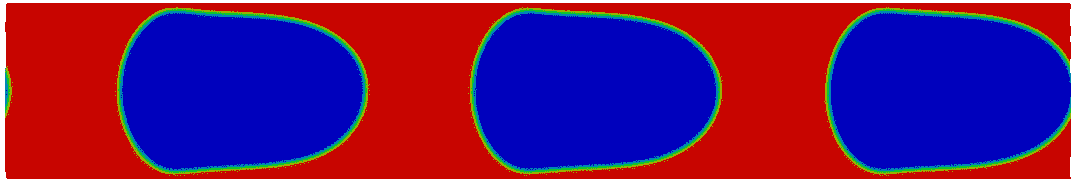}\hfill
		\includegraphics[width=0.4\textwidth]{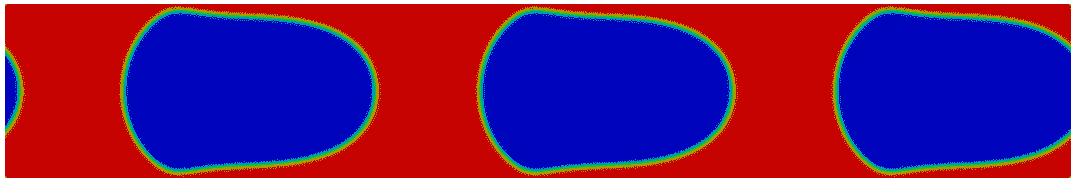}\\
		\resizebox{0.2\textwidth}{!}{\begin{tikzpicture}
\path[use as bounding box] (-1.25, 0.1) rectangle (1.1, 1.1);
\node[circle,fill=none,label={$Re=500$}](tr) at (0,0) {};
\end{tikzpicture} }\hfill
		\includegraphics[width=0.4\textwidth]{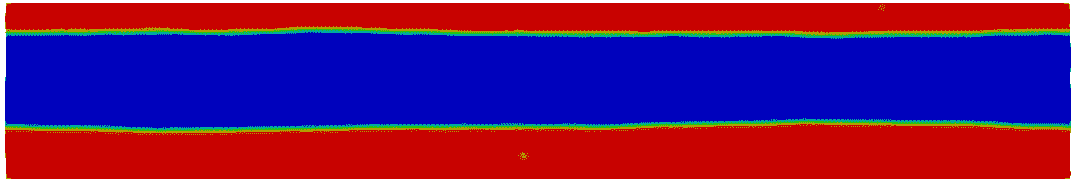}\hfill
		\includegraphics[width=0.4\textwidth]{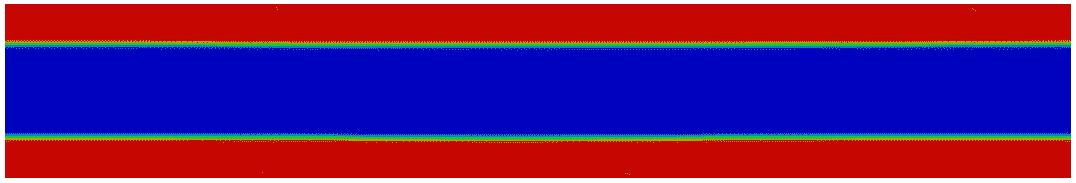}\\
		\caption {Steady state for different $Re$. SPH and LBM results are on the left and right columns respectively.}
		\label{periodic_slugs_re}
	\end{center}
\end{figure}

In this section, we study the establishment of different periodic two-phase flow patterns under the influence of gravity $\bm{g}$ starting from a given bubbly flow. Following~\cite{minier2016}, the initial configuration is composed of $27.25 \%$ of light phase and $72.75\%$ of heavy phase and is described in Fig.~\ref{init_geo2}. All physical properties and simulation properties are in Tab.~\ref{tab:physprop}. The heavy phase viscosity $\mu_l$ is adjusted as function of the Reynolds number $Re = \frac{g H^3}{8 \nu_l^2}$. The initial velocity field $\bm{V}$ is $\bm{V}(x,y) = \frac{g}{2 \nu_l} y (H-y)$. Viscosity values and dimensionless numbers for each case are reported in Tabs.~\ref{tab:phys_prop_cases2} and~\ref{tab:adim_cases2}. No-slip boundary conditions are applied to the walls. The simulations is done with 50000 nodes/particles for $t=30~\second$. Four different Reynolds numbers were tested : $Re=10$, $50$, $100$ and $500$. The phase distributions, pressure fields and velocity fields at final state are shown in Figs.~\ref{periodic_slugs_re}, \ref{periodic_slugs_re_press} and \ref{periodic_slugs_re_vel} respectively.

\begin{table}[bthp]
	\centering
	\subfloat[SPH]{\begin{adjustbox}{max width=0.62\textwidth}
			\begin{tabular}{llll}
				\toprule
				\textbf{Property}     & \textbf{Light Phase}  & \textbf{Heavy Phase} & \textbf{Units}\\
				\midrule
				\textbf{Density} ($\rho$) & $1.0$     & $5.0$     & $\kilogram\per\meter^{3}$\\			
				\textbf{Viscosity} ($\mu$)& $\mu_g = \mu_l/2$ & $\mu_l$ & $\pascal .\second$\\
				\textbf{Contact Angle} ($\theta_c$) & \multicolumn{2}{l}{$0$} & $\degree$\\
				\textbf{Surface Tension} ($\sigma^{nw}$) & \multicolumn{2}{l}{$5.0968\times 10^{-2}$} & $\newton\per\meter$\\
				\textbf{Gravity} ($g_x$) & \multicolumn{2}{l}{$1.0$} & $\meter\per\second^2$\\
				\textbf{Space step} ($\Delta x$) & \multicolumn{2}{l}{$1.09\times 10^{-3}$} & $\meter$\\
				\textbf{Domain size} ($L_x \times L_y$) & \multicolumn{2}{l}{$0.6\times 0.1$} & $\meter$\\							
				\bottomrule
			\end{tabular}
	\end{adjustbox}}
	\subfloat[LBM]{	\begin{adjustbox}{max width=0.38\textwidth}
			\begin{tabular}{lll}
				\toprule
				\textbf{Light Phase}  & \textbf{Heavy Phase} & \textbf{Units}\\
				\midrule
				$1.0$     & $5.0$     & $l.u.$\\	
				$\mu_g = \mu_l/2$     & $\mu_l$     & $l.u.$\\	
				\multicolumn{2}{l}{$0$} & $\degree$\\
				\multicolumn{2}{l}{$2.11\times 10^{-2}$} & $l.u.$\\	
				\multicolumn{2}{l}{$5.0\times 10^{-7}$} & $l.u.$\\
				\multicolumn{2}{l}{$1.09\times 10^{-3}$} & $l.u.$\\
				\multicolumn{2}{l}{$551\times 91$} & $l.u.$\\							
				\bottomrule
			\end{tabular}
	\end{adjustbox}}
	\caption{Simulation parameters.}
	\label{tab:physprop}
\end{table}

\begin{table}[bthp]
	\centering
	\subfloat[SPH]{	\begin{adjustbox}{max width=0.5\textwidth}
			\begin{tabular}{llll}
				\toprule
				\textbf{Case}     & \textbf{$\mu_g$ ($\pascal .\second$)}  & \textbf{$\mu_l$ ($\pascal .\second$)}\\
				\midrule
				1 ($Re=10$)     &$8.84\times 10^{-3}$ & $1.77\times 10^{-2}$\\
				2 ($Re=50$)     &$3.95\times 10^{-3}$ & $7.91\times 10^{-3}$\\								
				3 ($Re=100$)    &$2.79\times 10^{-3}$ & $5.59\times 10^{-3}$\\
				4 ($Re=500$)    &$1.25\times 10^{-2}$ & $2.50\times 10^{-3}$\\
				\bottomrule
			\end{tabular}
	\end{adjustbox}}
	\subfloat[LBM]{		\begin{adjustbox}{max width=0.5\textwidth}
			\begin{tabular}{llll}
				\toprule
				\textbf{$\mu_g$ ($l.u.$)}  & \textbf{$\mu_l$ ($l.u.$)} & \textbf{$\tau_g$} & \textbf{$\tau_l$}\\
				\midrule
				$6.86\times 10^{-2}$ & $3.43\times 10^{-1}$ & $1.014$ & $0.706$\\								
				$7.67\times 10^{-2}$ & $1.53\times 10^{-2}$ & $0.730$ & $0.592$\\					
				$5.42\times 10^{-2}$ & $1.08\times 10^{-2}$ & $0.663$ & $0.565$\\	
				$2.43\times 10^{-2}$ & $4.85\times 10^{-2}$ & $0.572$ & $0.529$\\	
				\bottomrule
			\end{tabular}
	\end{adjustbox}}
	\caption{Viscosity values for each case.}
	\label{tab:phys_prop_cases2}
\end{table}

\begin{table}[bthp]
	\centering
	\begin{adjustbox}{max width=0.5\textwidth}
		\begin{tabular}{llll}
			\toprule
			\textbf{Case}     &\textbf{$Re = \frac{g L_y^3}{8 \nu_l^2}$}     & \textbf{$La = \frac{\sigma \rho_l}{\mu_l^2}$}   & \textbf{$Bo = \frac{\Delta \rho g L_y^2}{\sigma}$} \\
			\midrule
			1      & $10$        &$82$   & $0.7848$\\
			2      & $50$        &$408$  & $0.7848$\\								
			3      & $100$       &$816$  & $0.7848$\\	
			4      & $500$       &$4077$ & $0.7848$\\	
			
			\bottomrule
		\end{tabular}
	\end{adjustbox}
	\caption{Reynolds, Laplace and Bond numbers for each case.}
	\label{tab:adim_cases2}
\end{table}

\begin{figure}[bthp]
	\begin{center}
		\captionsetup[subfigure]{labelformat=empty}
		\subfloat[$Re=10$]{\includegraphics[width=0.33\textwidth]{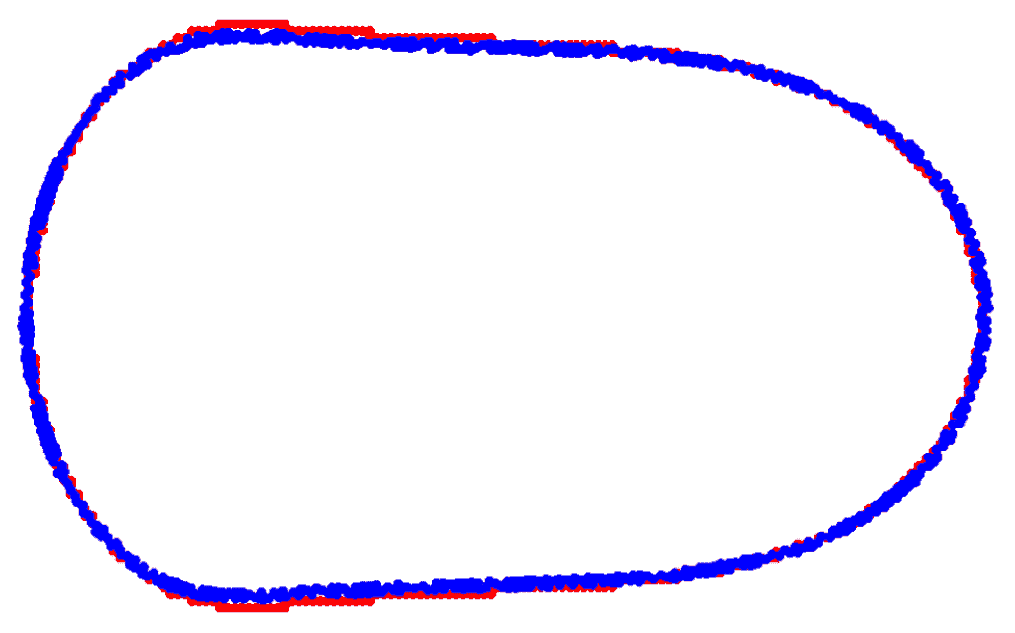}}\hfill
		\subfloat[$Re=50$]{\includegraphics[width=0.33\textwidth]{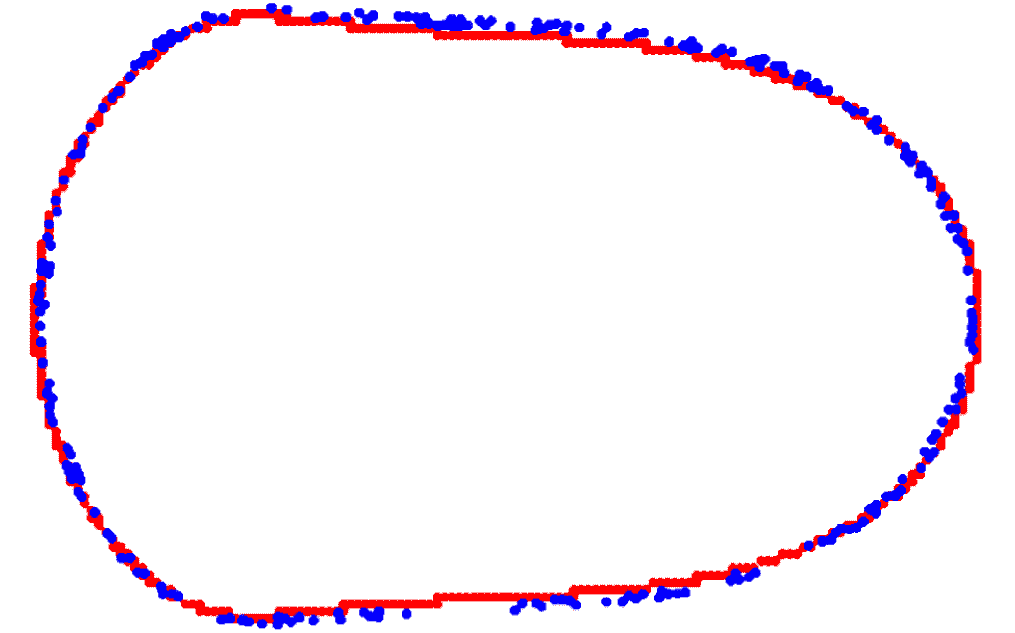}}\hfill
		\subfloat[$Re=100$]{\includegraphics[width=0.33\textwidth]{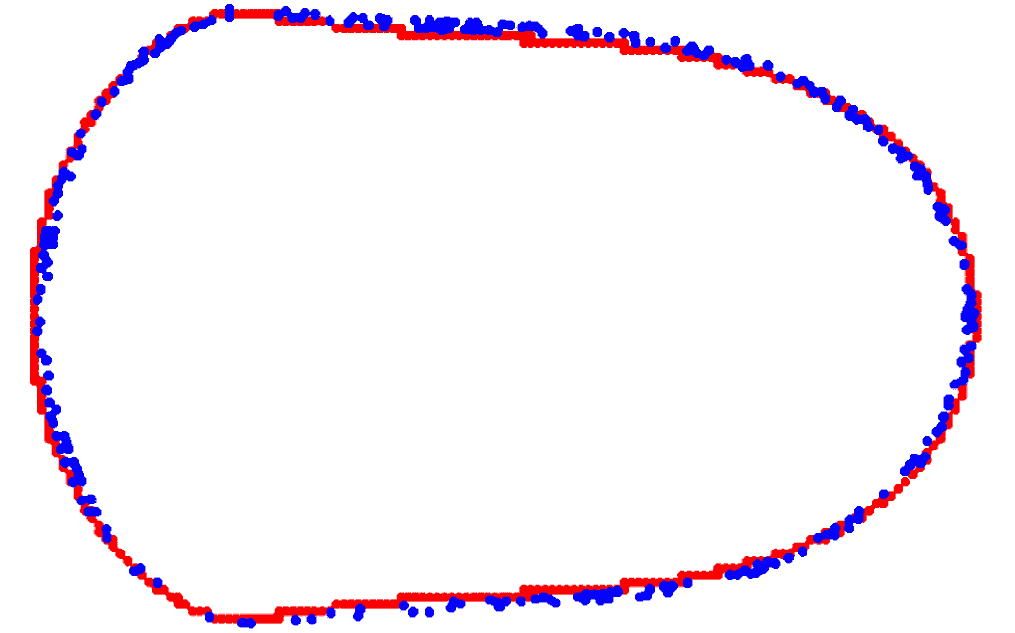}}\hfill
		\caption {Superposition of bubbles' shapes obtained with SPH (blue) and LBM (red) for different Reynolds numbers.}
		\label{periodic_slugs_bubbleshape_comp}
	\end{center}
\end{figure}

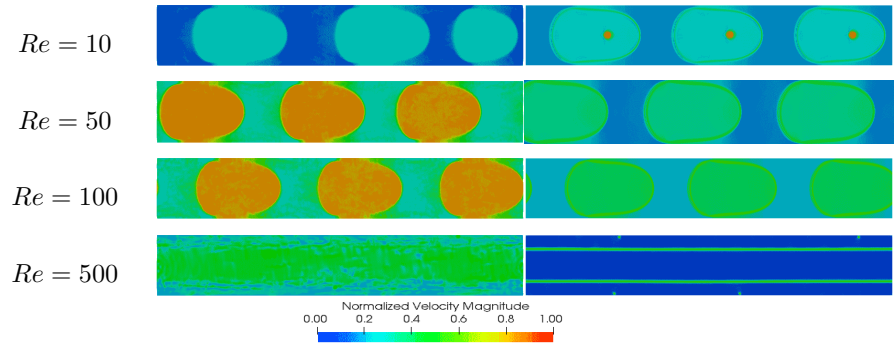
\begin{figure}[bthp]
	\begin{center}
		\captionsetup[subfigure]{labelformat=empty}
		\resizebox{0.2\textwidth}{!}{\begin{tikzpicture}
\path[use as bounding box] (-1.25, 0.1) rectangle (1.1, 1.1);
\node[circle,fill=none,label={$Re=10$}](tr) at (0,0) {};
\end{tikzpicture} }\hfill
		\includegraphics[width=0.4\textwidth]{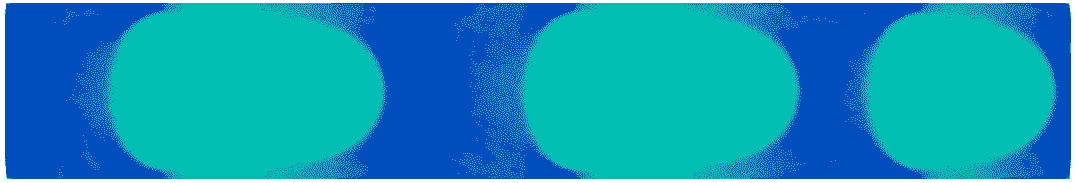}\hfill
		\includegraphics[width=0.4\textwidth]{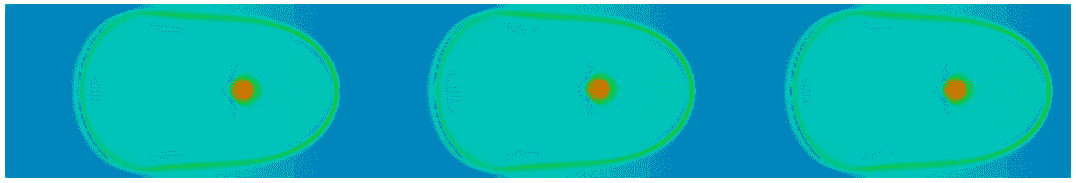}\\
		\resizebox{0.2\textwidth}{!}{\begin{tikzpicture}
\path[use as bounding box] (-1.25, 0.1) rectangle (1.1, 1.1);
\node[circle,fill=none,label={$Re=50$}](tr) at (0,0) {};
\end{tikzpicture} }\hfill
		\includegraphics[width=0.4\textwidth]{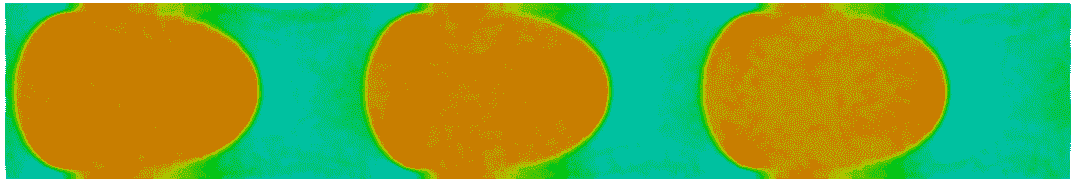}\hfill
		\includegraphics[width=0.4\textwidth]{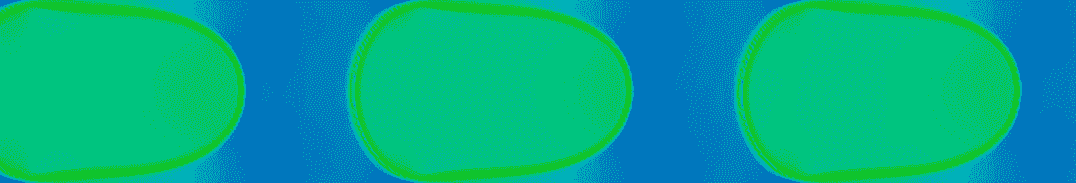}\\
		\resizebox{0.2\textwidth}{!}{\begin{tikzpicture}
\path[use as bounding box] (-1.25, 0.1) rectangle (1.1, 1.1);
\node[circle,fill=none,label={$Re=100$}](tr) at (0,0) {};
\end{tikzpicture} }\hfill
		\includegraphics[width=0.4\textwidth]{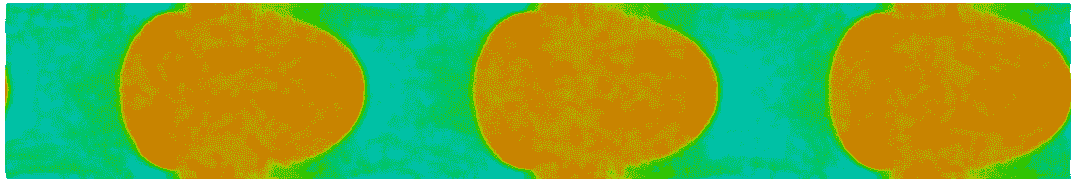}\hfill
		\includegraphics[width=0.4\textwidth]{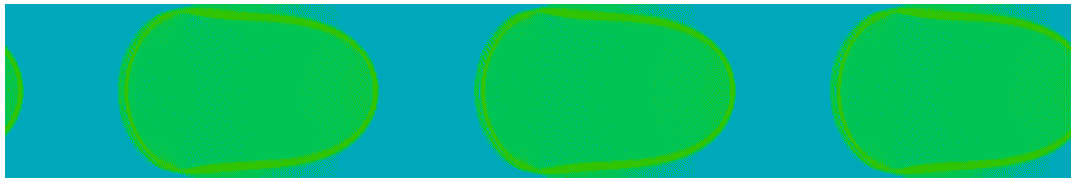}\\
		\resizebox{0.2\textwidth}{!}{\begin{tikzpicture}
\path[use as bounding box] (-1.25, 0.1) rectangle (1.1, 1.1);
\node[circle,fill=none,label={$Re=500$}](tr) at (0,0) {};
\end{tikzpicture} }\hfill
		\includegraphics[width=0.4\textwidth]{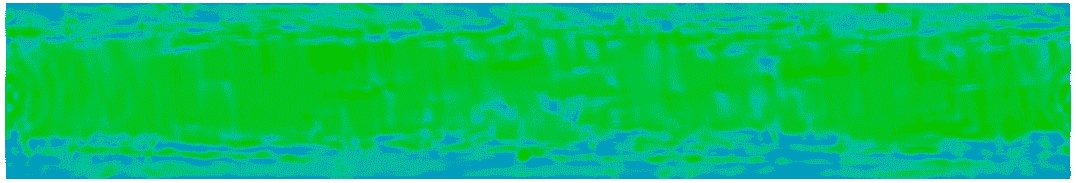}\hfill
		\includegraphics[width=0.4\textwidth]{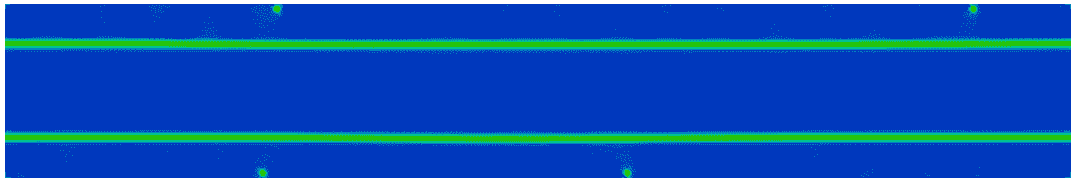}\\
		\includegraphics[width=0.3\textwidth,height=0.05\textwidth]{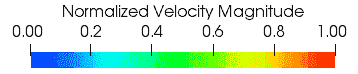}
		\caption {Normalized pressure fields at steady state for different $Re$. SPH and LBM results are on the left and right columns respectively. Pressure is normalized by the maximum pressure inside the bubbles.}
		\label{periodic_slugs_re_press}
	\end{center}
\end{figure}

\begin{figure}[bthp]
	\begin{center}
		\captionsetup[subfigure]{labelformat=empty}
		\resizebox{0.2\textwidth}{!}{\begin{tikzpicture}
\path[use as bounding box] (-1.25, 0.1) rectangle (1.1, 1.1);
\node[circle,fill=none,label={$Re=10$}](tr) at (0,0) {};
\end{tikzpicture} }\hfill
		\includegraphics[width=0.4\textwidth]{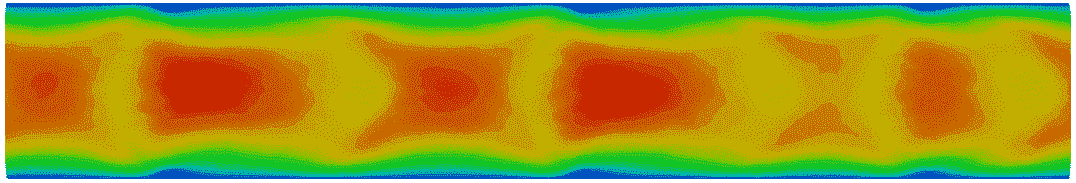}\hfill
		\includegraphics[width=0.4\textwidth]{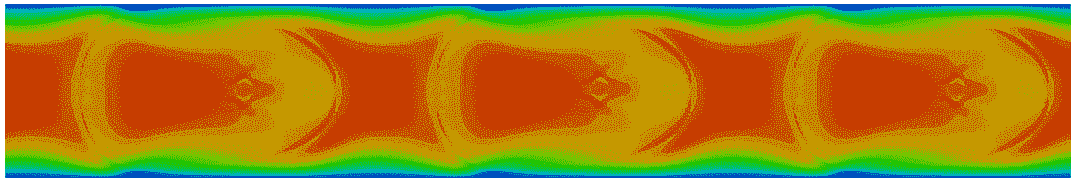}\\
		\resizebox{0.2\textwidth}{!}{\begin{tikzpicture}
\path[use as bounding box] (-1.25, 0.1) rectangle (1.1, 1.1);
\node[circle,fill=none,label={$Re=50$}](tr) at (0,0) {};
\end{tikzpicture} }\hfill
		\includegraphics[width=0.4\textwidth]{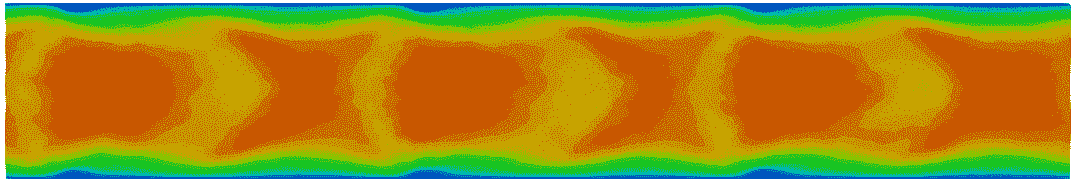}\hfill
		\includegraphics[width=0.4\textwidth]{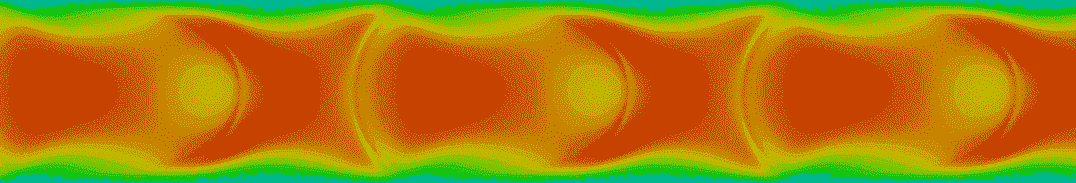}\\
		\resizebox{0.2\textwidth}{!}{\begin{tikzpicture}
\path[use as bounding box] (-1.25, 0.1) rectangle (1.1, 1.1);
\node[circle,fill=none,label={$Re=100$}](tr) at (0,0) {};
\end{tikzpicture} }\hfill
		\includegraphics[width=0.4\textwidth]{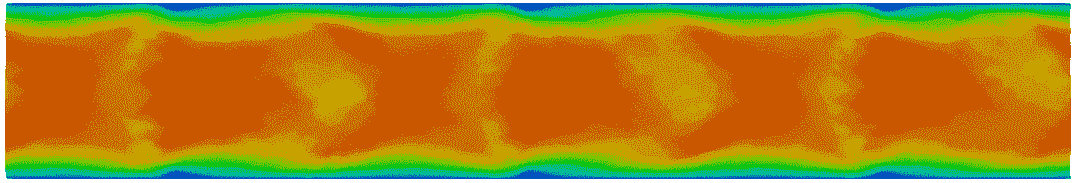}\hfill
		\includegraphics[width=0.4\textwidth]{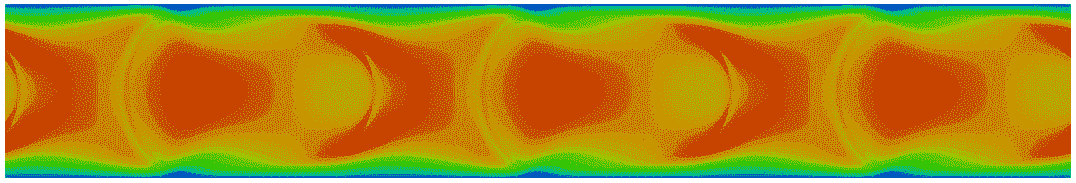}\\
		\resizebox{0.2\textwidth}{!}{\begin{tikzpicture}
\path[use as bounding box] (-1.25, 0.1) rectangle (1.1, 1.1);
\node[circle,fill=none,label={$Re=500$}](tr) at (0,0) {};
\end{tikzpicture} }\hfill
		\includegraphics[width=0.4\textwidth]{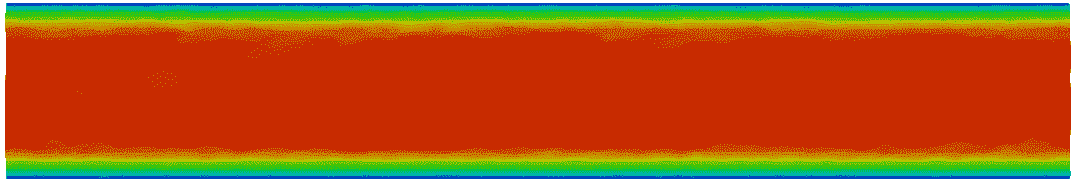}\hfill
		\includegraphics[width=0.4\textwidth]{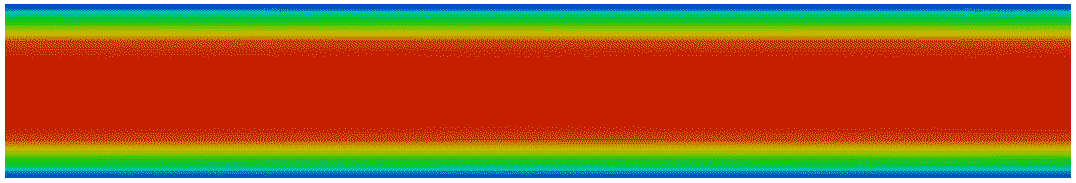}\\
		\includegraphics[width=0.3\textwidth,height=0.05\textwidth]{NormalizedVel.png}
		\caption {Normalized velocity fields at steady state for different $Re$. SPH and LBM results are on the left and right columns respectively.}
		\label{periodic_slugs_re_vel}
	\end{center}
\end{figure}

In Fig.~\ref{periodic_slugs_re}, it is possible to see that both methods reproduce the same flow pattern for all four $Re$ numbers considered. For $10\leq Re \leq 100$, we obtain a bubbly flow composed of three different Taylor bubbles whereas for $Re=500$, we have an annular flow where the heavy phase is in contact with the pipe and the light phase travels in the middle. For $Re=10$, we observe that SPH present a bubbly flow where one bubble is clearly smaller than the two others. It is not the case in LBM where all bubbles are identical within each case. Besides, for this case, we have heavy phase droplets than are captured inside light phase bubbles. Note that these small bubbles are to be absorbed by the main flow if the simulation lasted longer because they are moving slower than their environment.  For $Re=50$ and $Re=100$, we obtain in all cases the same pattern made of three identical Taylor bubbles. Moreover, as shown in Fig.~\ref{periodic_slugs_bubbleshape_comp}, the bubbles' shapes between SPH and LBM for $10\leq Re \leq 100$ are very similar. Finally, as $Re$ grows, the Taylor bubbles are getting slightly shorter and higher in size. For $Re=500$, we again see that LBM offers a perfectly symmetric annular pattern. On the contrary, the bottom heavy phase layer in SPH is thicker than the top one. In general, LBM provides more symmetric results than SPH because of its Eulerian nature.

In Fig.~\ref{periodic_slugs_re_press}, we can see that for $Re=10$, the pressure fields is dominated by the captured droplets of heavy phase inside the bubbles. For $Re=50$ and $100$, the pressure field reaches a maximum for SPH inside the bubbles whereas for LBM it is at the interface. Nevertheless, as predicted by Laplace's law, the pressure is higher inside the light phase's bubbles than in the heavy phase bulk. When looking at the velocity fields in Fig.~\ref{periodic_slugs_re_vel}, we see that they are also very similar. The same patterns surrounding the bubbles can be observed. For the annular case where $Re=500$, it is possible to see that the no-slip condition on the walls affects the flow more strongly in LBM than in SPH which results in a flatter velocity profile for the latter.

\begin{figure}[bthp]
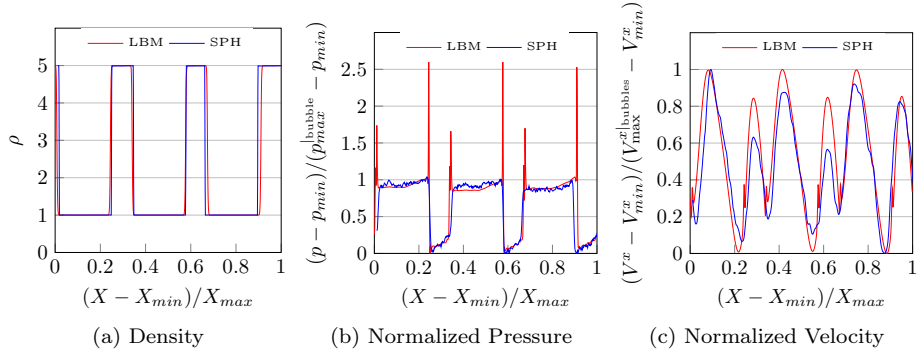

	\centering
	\makebox[\textwidth][c]{
		\subfloat[Density\label{periodic_slugs_re50_comp1}] {\resizebox{0.315\textwidth}{!}{\input{taylor_bubbles_dens_Re50}}}
		\subfloat[Normalized Pressure\label{periodic_slugs_re50_comp2}] {\resizebox{0.3425\textwidth}{!}{\input{taylor_bubbles_press_Re50}}}
		\subfloat[Normalized Velocity\label{periodic_slugs_re50_comp3}] {\resizebox{0.3425\textwidth}{!}{\input{taylor_bubbles_vel_Re50}}}}	
	\caption{Case $Re=50$. (a) Superposed densities. (b) Superposed normalized pressures. (c) Superposed normalized velocities.}
	\label{periodic_slugs_re50_comp}
\end{figure}

\begin{figure}[bthp]
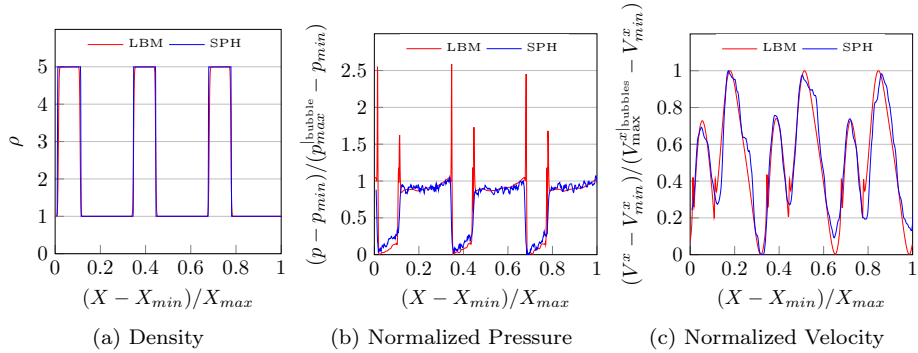

	\centering
	\makebox[\textwidth][c]{
		\subfloat[Density\label{periodic_slugs_re100_comp1}] {\resizebox{0.315\textwidth}{!}{\input{taylor_bubbles_dens}}}
		\subfloat[Normalized Pressure\label{periodic_slugs_re100_comp2}] {\resizebox{0.3425\textwidth}{!}{\input{taylor_bubbles_press}}}
		\subfloat[Normalized Velocity\label{periodic_slugs_re100_comp3}] {\resizebox{0.3425\textwidth}{!}{\input{taylor_bubbles_vel}}}}	
	\caption{Case $Re=100$. (a) Superposed densities. (b) Superposed normalized pressures. (c) Superposed normalized velocities.}
	\label{periodic_slugs_re100_comp}
\end{figure}

For $Re=50$ and $Re=100$, where the SPH and LBM patterns are the closest, we compared the density, velocity and pressure fields along the centerline on Figs.~\ref{periodic_slugs_re50_comp} and \ref{periodic_slugs_re100_comp}. Note that because the bubbles do not have the exact same position, we have shifted the LBM profiles from a fixed distance to be able to superpose the profiles. On the density plots of Figs.~\ref{periodic_slugs_re50_comp1} and \ref{periodic_slugs_re100_comp1}, the different density treatment in both methods clearly appears. In LBM, the density is smoothed at the interface whereas in SPH, thanks to its Lagrangian nature, there is no interface smoothing in the density field because a given particle belongs to one phase or not, there is no intermediate state. Concerning the pressure fields shown in Figs.~\ref{periodic_slugs_re50_comp2} and \ref{periodic_slugs_re100_comp2}, we observe that LBM suffers from the same overshoots at the interface that were described and explained in Sec.~\ref{sec:staticbubble}. On the other hand, the SPH pressure field is polluted with noise. Despite these discrepancies, both profiles are very close. Finally, in Figs.~\ref{periodic_slugs_re50_comp3} and \ref{periodic_slugs_re100_comp3}, we see the velocity profiles in both methods have the same shape. The bubbles are moving at a much higher speed than the surrounding fluid (about $30\%$ faster). At each interface, the velocity reaches a local minimum. The only differences between both profiles is that in certain areas, the SPH velocity peaks have a smaller amplitude than in LBM. For example, the bubble velocity is the same for all three bubbles in LBM for $Re=50$ whereas for SPH the last bubble travels about $10\%$ faster than the other ones. One last comment is that in LBM at the interface, the velocity field present non-physical oscillations due to the fluids mixing at the interface. It is not the case in SPH because the pressure field does not suffer from pressure overshoots and fluids are not mixed at the interface.

To conclude, we can add that SPH and LBM are both well capable of simulating the transition from a given bubbly flow to a slug flow composed of Taylor bubbles for $Re \leq 500$. To further assess their relative performance, an extended comparison with other methods or with experimental data would be of great interest. Note that we have limited our study to $Re \leq 500$ because, for higher velocities and/or smaller viscosities, we lie outside LBM stability region whether because the low Mach rule is violated or because the relaxation time is too close from $0.5$.

\subsection{Open channel case (inlet/outlet)}

%Note that unlike previous sections in which only the dimensionless numbers such as $Re$ were similar between SPH and LBM, in this section, we simulate the exact same test case in both methods. It means when converting LBM lattice units from Tab.~\ref{tab:phys_prop_lbm} to physical units following~\cite{latt2008}, one can re-obtain the values shown in Tab.~\ref{tab:phys_prop_sph}.

\begin{figure}[bthp]
	\centering
	\resizebox{1.0\textwidth}{!}{\begin{tikzpicture}

\node[circle,fill=black!0](bl) at (0,0) {};

\node[circle,fill=black!0](ml) at (0,1.6) {};

\node[circle,fill=black!0](tl) at (0,2) {};

\node[circle,fill=black!0](br) at (20,0) {};

\node[circle,fill=black!0](mr) at (20,1.6) {};

\node[circle,fill=black!0](tr) at (20,2) {};

%\node[circle,fill=black!0,label={$x$}](x) at (12,0) {};
%\node[circle,fill=black!0,label={$y$}](y) at (0,12) {};

\draw[ultra thick] (bl.center) -- (br.center);
\draw (br.center) -- (tr.center);
\draw[ultra thick] (tr.center) -- (tl.center);
\draw (tl.center) -- (bl.center);

%\draw[dashed]  (ml.center) -- (mr.center);

\fill[fill=black!20, draw=black] (bl.center) rectangle (mr.center);

\draw[triangle 45-triangle 45] (0.0,-0.5) -- node[below] {\Large $L=10\meter$} (20.0,-0.5);
\draw[triangle 45-triangle 45] (15.5,0) -- node[above right] {\Large $D=1\meter$} (15.5,2.0);

%\draw[triangle 45-triangle 45] (16.5,0) -- node[right] {\Large $\alpha_l$} (16.5,1.0);
%\draw[triangle 45-triangle 45] (16.5,1.0) -- node[right] {\Large $\alpha_g$} (16.5,2.0);

\draw[-triangle 45] (-1.0,1.75) --  (0.0,1.75);
\draw[-triangle 45] (-1.0,1.25) --  (0.0,1.25);
\draw[-triangle 45] (-1.0,0.75) --  (0.0,0.75);
\draw[-triangle 45] (-1.0,0.25) --  (0.0,0.25);

\node[circle,fill=black!0,label={\LARGE $u_g$}](ug) at (-1.5,1.32) {};
\node[circle,fill=black!0,label={\LARGE $u_l$}](ul) at (-1.5,0.32) {};

\draw[-triangle 45] (20.0,1.75) --  (21.0,1.75);
\draw[-triangle 45] (20.0,1.25) --  (21.0,1.25);
\draw[-triangle 45] (20.0,0.75) --  (21.0,0.75);
\draw[-triangle 45] (20.0,0.25) --  (21.0,0.25);

\node[circle,fill=black!0,label={\LARGE $p_0$}](ug) at (21.5,0.5) {};

\node[circle,fill=none,label={\large Light Fluid ($\alpha_g$)}](tr) at (10,1.26) {};
\node[circle,fill=none,label={\large Heavy Fluid ($\alpha_l$)}](tr) at (10,0.32) {};

\draw[ultra thick] (bl.center) -- (br.center);

\end{tikzpicture}}
	\caption {Initial configuration sketch.}
	\label{init_geo3}
\end{figure}
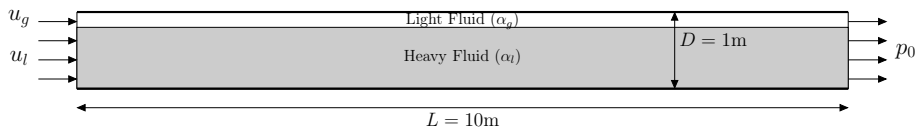

\begin{figure}[bthp]
	\begin{center}
		\resizebox{0.65\textwidth}{!}{\input{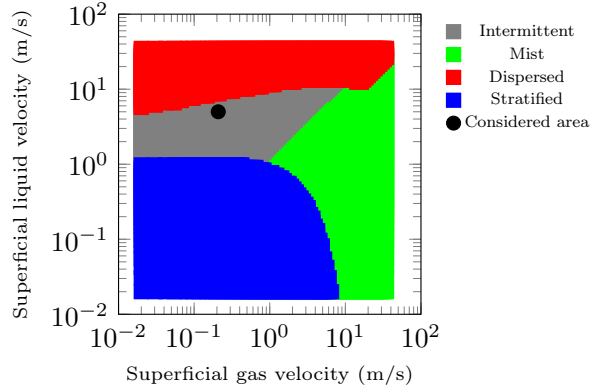}}
		\caption {Flow regime maps}
		\label{flow_map_res}
	\end{center}
\end{figure}

\begin{figure}[bthp]
	\begin{center}
		\subfloat[SPH - $Re=125$]{\includegraphics[width=0.495\textwidth,height=0.2\textwidth]{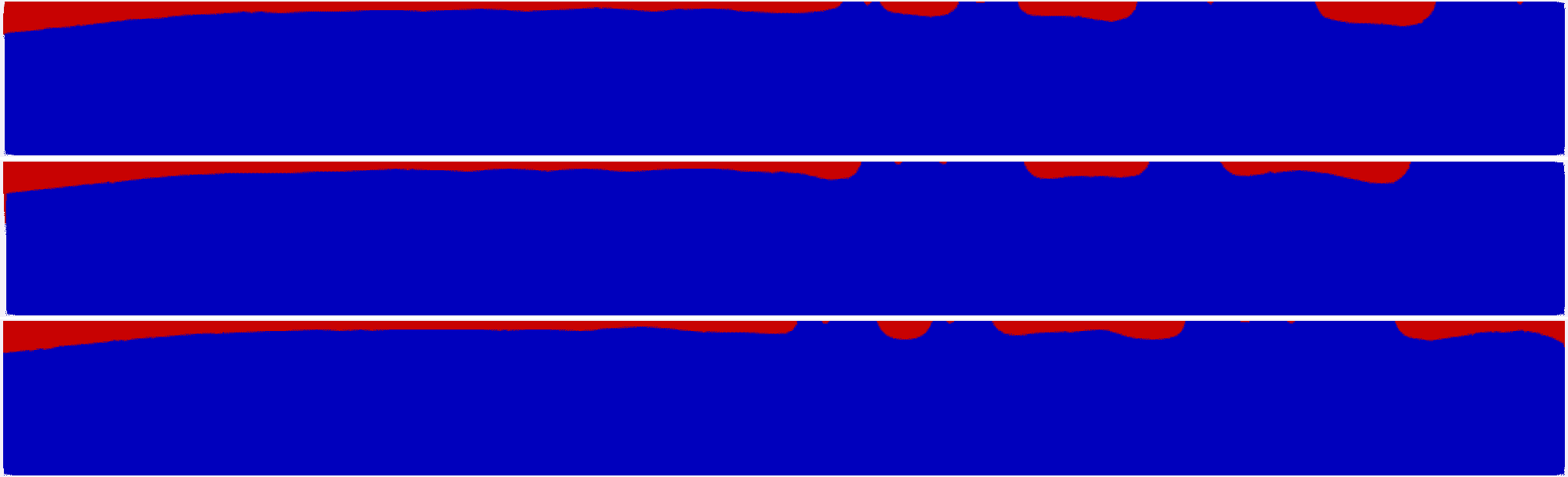}}\hfill
		\subfloat[LBM - $Re=125$]{\includegraphics[width=0.495\textwidth,height=0.2\textwidth]{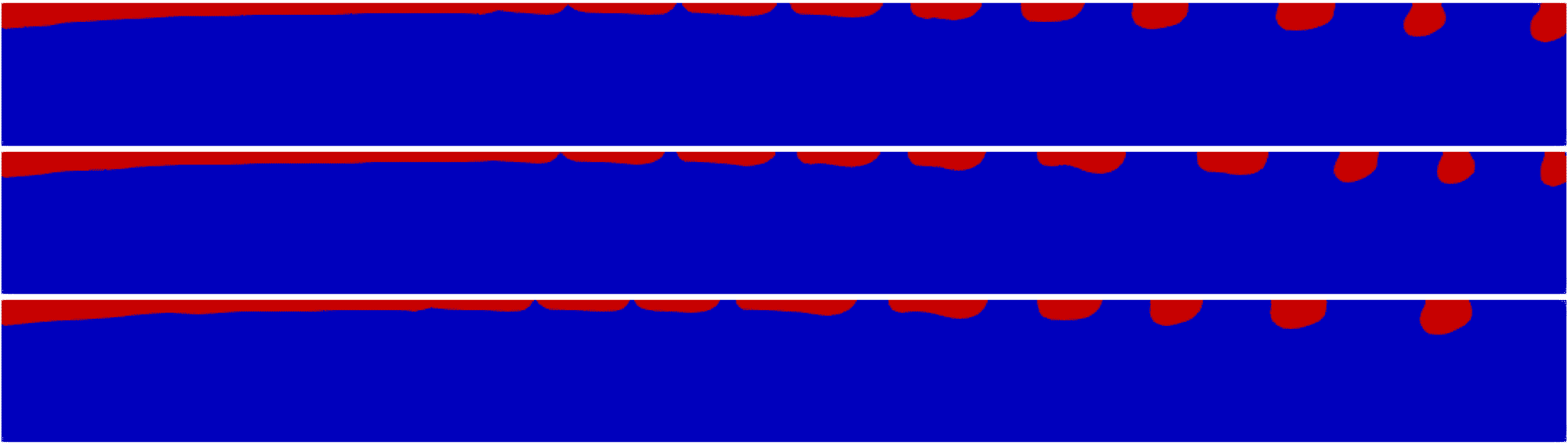}}\\
		\subfloat[SPH - $Re=312.5$]{\includegraphics[width=0.495\textwidth,height=0.2\textwidth]{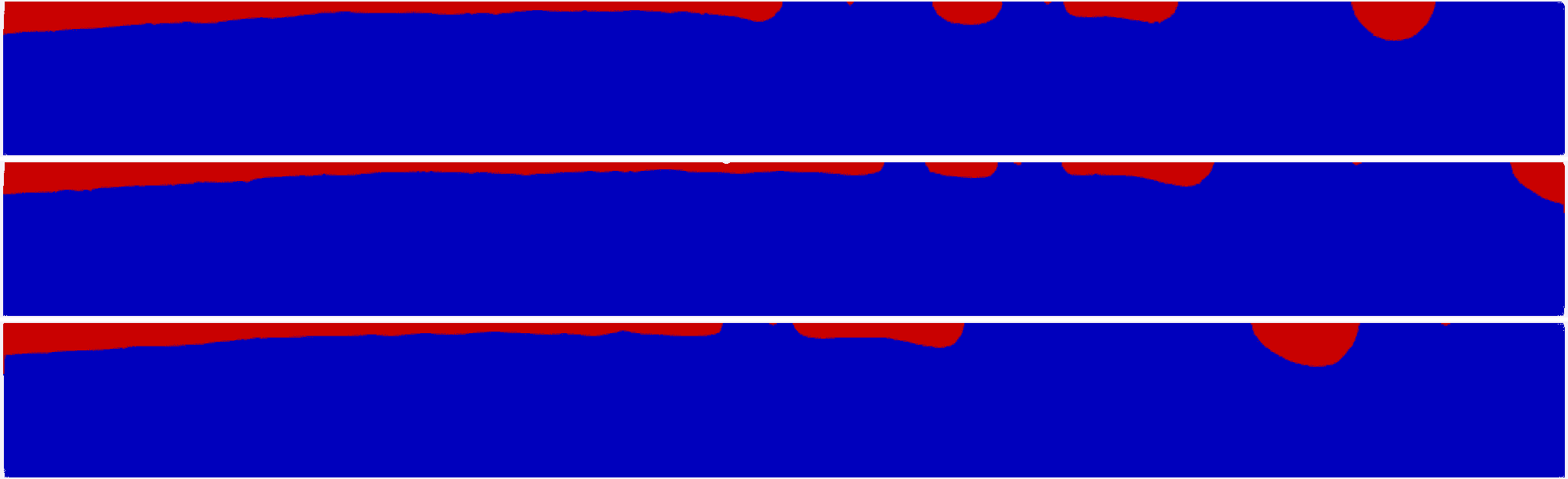}}\hfill
		\subfloat[LBM - $Re=312.5$]{\includegraphics[width=0.495\textwidth,height=0.2\textwidth]{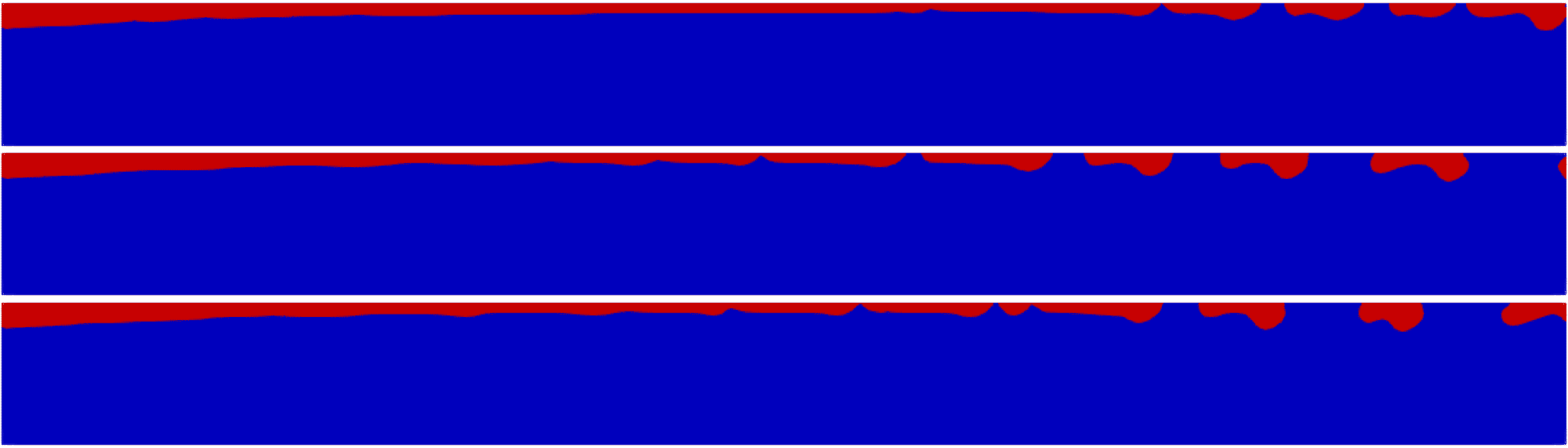}}
		\caption { (a, b) From top to bottom : snapshots at $t=4.7~\second, 13.5~\second, 25~\second$. (c, d) From top to bottom : snapshots at $t=4.2~\second, 16.4~\second, 25.6~\second$.}
		\label{ios_slugs_re125_comp}
	\end{center}
\end{figure}

\begin{figure}[bthp]
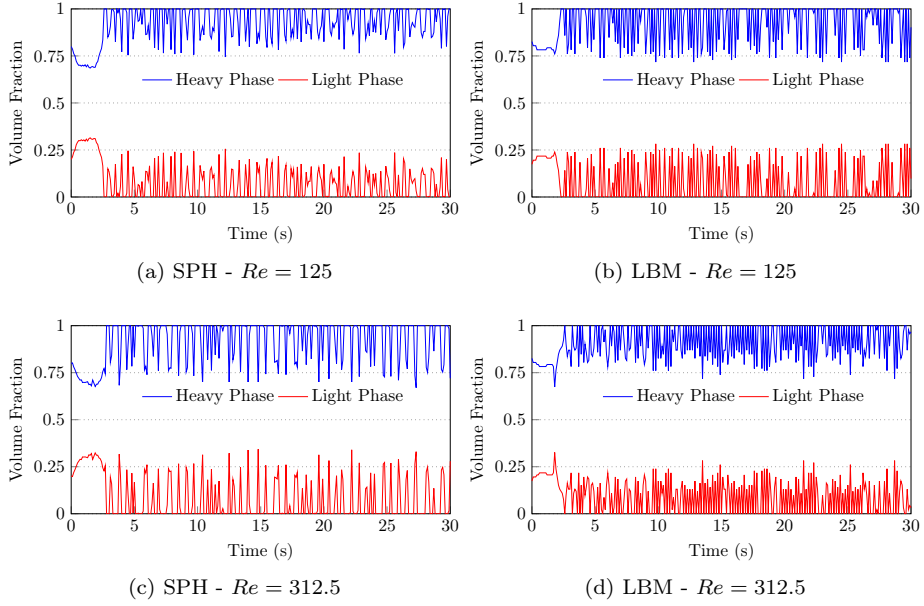

	\begin{center}
		\makebox[\textwidth][c]{
			\subfloat[SPH - $Re=125$] {\resizebox{0.5\textwidth}{!}{\input{SPH_Re125_fig1}}}
			\subfloat[LBM - $Re=125$] {\resizebox{0.5\textwidth}{!}{%\pgfplotsset{label style={font=\tiny},
%	tick label style={font=\tiny} }
%
\begin{tikzpicture}

\begin{axis}[%
width=3.0in,
height=1.5in,
at={(0in,0in)},
scale only axis,
xmin=0,
xmax=30,
xlabel={Time ($\second$)},
ymin=0,
ymax=1.0,
ylabel near ticks,
xlabel near ticks,
xmajorgrids=false,
ymajorgrids=true,
xtick pos=left,
ytick pos=left,
grid style={dotted,gray},
ytick={0,0.25,0.5,0.75,1},
ylabel={Volume Fraction},
axis background/.style={fill=white},
legend style={legend style={nodes={scale=1.0, transform shape}},legend columns=-1,fill=white,align=left,draw=none,at={(0.9,0.7)}}
]
\addplot [color=blue,solid]
  table[row sep=crcr]{%
0	0.826086956521739\\
0.1	0.804347826086957\\
0.2	0.804347826086957\\
0.3	0.804347826086957\\
0.4	0.782608695652174\\
0.5	0.782608695652174\\
0.6	0.782608695652174\\
0.7	0.782608695652174\\
0.8	0.782608695652174\\
0.9	0.782608695652174\\
1	0.782608695652174\\
1.1	0.782608695652174\\
1.2	0.793478260869565\\
1.3	0.793478260869565\\
1.4	0.793478260869565\\
1.5	0.793478260869565\\
1.6	0.782608695652174\\
1.7	0.782608695652174\\
1.8	0.760869565217391\\
1.9	0.782608695652174\\
2	0.826086956521739\\
2.1	0.858695652173913\\
2.2	0.934782608695652\\
2.3	1\\
2.4	1\\
2.5	1\\
2.6	0.858695652173913\\
2.7	1\\
2.8	0.760869565217391\\
2.9	0.815217391304348\\
3	1\\
3.1	0.782608695652174\\
3.2	1\\
3.3	1\\
3.4	0.760869565217391\\
3.5	1\\
3.6	1\\
3.7	1\\
3.8	0.815217391304348\\
3.9	1\\
4	1\\
4.1	1\\
4.2	1\\
4.3	0.956521739130435\\
4.4	0.793478260869565\\
4.5	1\\
4.6	1\\
4.7	0.771739130434783\\
4.8	1\\
4.9	0.83695652173913\\
5	1\\
5.1	1\\
5.2	1\\
5.3	1\\
5.4	0.739130434782609\\
5.5	1\\
5.6	1\\
5.7	0.739130434782609\\
5.8	1\\
5.9	1\\
6	1\\
6.1	0.847826086956522\\
6.2	1\\
6.3	1\\
6.4	0.923913043478261\\
6.5	1\\
6.6	0.826086956521739\\
6.7	1\\
6.8	0.815217391304348\\
6.9	1\\
7	0.739130434782609\\
7.1	1\\
7.2	1\\
7.3	0.793478260869565\\
7.4	1\\
7.5	1\\
7.6	0.760869565217391\\
7.7	1\\
7.8	1\\
7.9	0.75\\
8	1\\
8.1	0.771739130434783\\
8.2	1\\
8.3	1\\
8.4	0.815217391304348\\
8.5	1\\
8.6	0.934782608695652\\
8.7	1\\
8.8	1\\
8.9	1\\
9	0.760869565217391\\
9.1	0.869565217391304\\
9.2	1\\
9.3	1\\
9.4	0.760869565217391\\
9.5	1\\
9.6	0.771739130434783\\
9.7	1\\
9.8	0.717391304347826\\
9.9	1\\
10	1\\
10.1	0.75\\
10.2	1\\
10.3	1\\
10.4	1\\
10.5	0.739130434782609\\
10.6	1\\
10.7	1\\
10.8	1\\
10.9	0.782608695652174\\
11	1\\
11.1	0.739130434782609\\
11.2	1\\
11.3	0.83695652173913\\
11.4	1\\
11.5	1\\
11.6	0.739130434782609\\
11.7	1\\
11.8	1\\
11.9	0.760869565217391\\
12	0.956521739130435\\
12.1	1\\
12.2	1\\
12.3	1\\
12.4	0.782608695652174\\
12.5	1\\
12.6	0.793478260869565\\
12.7	1\\
12.8	0.717391304347826\\
12.9	1\\
13	1\\
13.1	0.739130434782609\\
13.2	1\\
13.3	1\\
13.4	0.760869565217391\\
13.5	0.793478260869565\\
13.6	1\\
13.7	1\\
13.8	1\\
13.9	0.771739130434783\\
14	1\\
14.1	0.75\\
14.2	1\\
14.3	0.728260869565217\\
14.4	1\\
14.5	1\\
14.6	0.771739130434783\\
14.7	1\\
14.8	0.83695652173913\\
14.9	1\\
15	1\\
15.1	1\\
15.2	0.782608695652174\\
15.3	0.869565217391304\\
15.4	0.989130434782609\\
15.5	1\\
15.6	0.760869565217391\\
15.7	1\\
15.8	0.728260869565217\\
15.9	1\\
16	1\\
16.1	0.739130434782609\\
16.2	1\\
16.3	0.83695652173913\\
16.4	1\\
16.5	1\\
16.6	1\\
16.7	1\\
16.8	1\\
16.9	1\\
17	1\\
17.1	0.826086956521739\\
17.2	1\\
17.3	0.782608695652174\\
17.4	1\\
17.5	1\\
17.6	0.804347826086957\\
17.7	1\\
17.8	0.760869565217391\\
17.9	1\\
18	0.771739130434783\\
18.1	0.91304347826087\\
18.2	1\\
18.3	1\\
18.4	1\\
18.5	1\\
18.6	0.902173913043478\\
18.7	1\\
18.8	0.739130434782609\\
18.9	1\\
19	1\\
19.1	0.826086956521739\\
19.2	1\\
19.3	0.739130434782609\\
19.4	1\\
19.5	0.83695652173913\\
19.6	1\\
19.7	1\\
19.8	0.782608695652174\\
19.9	1\\
20	1\\
20.1	1\\
20.2	1\\
20.3	0.869565217391304\\
20.4	1\\
20.5	1\\
20.6	1\\
20.7	1\\
20.8	0.956521739130435\\
20.9	1\\
21	0.760869565217391\\
21.1	1\\
21.2	1\\
21.3	0.782608695652174\\
21.4	1\\
21.5	1\\
21.6	1\\
21.7	1\\
21.8	0.739130434782609\\
21.9	1\\
22	1\\
22.1	1\\
22.2	1\\
22.3	0.978260869565217\\
22.4	1\\
22.5	0.760869565217391\\
22.6	1\\
22.7	1\\
22.8	0.739130434782609\\
22.9	1\\
23	0.717391304347826\\
23.1	1\\
23.2	1\\
23.3	0.760869565217391\\
23.4	1\\
23.5	1\\
23.6	1\\
23.7	1\\
23.8	1\\
23.9	1\\
24	0.858695652173913\\
24.1	1\\
24.2	1\\
24.3	0.739130434782609\\
24.4	1\\
24.5	0.717391304347826\\
24.6	1\\
24.7	1\\
24.8	0.760869565217391\\
24.9	1\\
25	1\\
25.1	1\\
25.2	1\\
25.3	1\\
25.4	1\\
25.5	0.782608695652174\\
25.6	1\\
25.7	1\\
25.8	0.793478260869565\\
25.9	1\\
26	0.75\\
26.1	1\\
26.2	0.782608695652174\\
26.3	0.782608695652174\\
26.4	1\\
26.5	0.956521739130435\\
26.6	1\\
26.7	0.978260869565217\\
26.8	1\\
26.9	1\\
27	0.91304347826087\\
27.1	1\\
27.2	1\\
27.3	0.91304347826087\\
27.4	1\\
27.5	0.815217391304348\\
27.6	1\\
27.7	0.739130434782609\\
27.8	0.847826086956522\\
27.9	1\\
28	0.717391304347826\\
28.1	1\\
28.2	0.717391304347826\\
28.3	1\\
28.4	1\\
28.5	0.815217391304348\\
28.6	1\\
28.7	1\\
28.8	1\\
28.9	1\\
29	0.978260869565217\\
29.1	1\\
29.2	0.739130434782609\\
29.3	0.869565217391304\\
29.4	1\\
29.5	0.717391304347826\\
29.6	1\\
29.7	0.717391304347826\\
29.8	1\\
29.9	0.739130434782609\\
30	0.902173913043478\\
};
\addlegendentry{Heavy Phase};

\addplot [color=red,solid]
  table[row sep=crcr]{%
0	0.173913043478261\\
0.1	0.195652173913043\\
0.2	0.195652173913043\\
0.3	0.195652173913043\\
0.4	0.217391304347826\\
0.5	0.217391304347826\\
0.6	0.217391304347826\\
0.7	0.217391304347826\\
0.8	0.217391304347826\\
0.9	0.217391304347826\\
1	0.217391304347826\\
1.1	0.217391304347826\\
1.2	0.206521739130435\\
1.3	0.206521739130435\\
1.4	0.206521739130435\\
1.5	0.206521739130435\\
1.6	0.217391304347826\\
1.7	0.217391304347826\\
1.8	0.239130434782609\\
1.9	0.217391304347826\\
2	0.173913043478261\\
2.1	0.141304347826087\\
2.2	0.0652173913043478\\
2.3	0\\
2.4	0\\
2.5	0\\
2.6	0.141304347826087\\
2.7	0\\
2.8	0.239130434782609\\
2.9	0.184782608695652\\
3	0\\
3.1	0.217391304347826\\
3.2	0\\
3.3	0\\
3.4	0.239130434782609\\
3.5	0\\
3.6	0\\
3.7	0\\
3.8	0.184782608695652\\
3.9	0\\
4	0\\
4.1	0\\
4.2	0\\
4.3	0.0434782608695652\\
4.4	0.206521739130435\\
4.5	0\\
4.6	0\\
4.7	0.228260869565217\\
4.8	0\\
4.9	0.16304347826087\\
5	0\\
5.1	0\\
5.2	0\\
5.3	0\\
5.4	0.260869565217391\\
5.5	0\\
5.6	0\\
5.7	0.260869565217391\\
5.8	0\\
5.9	0\\
6	0\\
6.1	0.152173913043478\\
6.2	0\\
6.3	0\\
6.4	0.0760869565217391\\
6.5	0\\
6.6	0.173913043478261\\
6.7	0\\
6.8	0.184782608695652\\
6.9	0\\
7	0.260869565217391\\
7.1	0\\
7.2	0\\
7.3	0.206521739130435\\
7.4	0\\
7.5	0\\
7.6	0.239130434782609\\
7.7	0\\
7.8	0\\
7.9	0.25\\
8	0\\
8.1	0.228260869565217\\
8.2	0\\
8.3	0\\
8.4	0.184782608695652\\
8.5	0\\
8.6	0.0652173913043478\\
8.7	0\\
8.8	0\\
8.9	0\\
9	0.239130434782609\\
9.1	0.130434782608696\\
9.2	0\\
9.3	0\\
9.4	0.239130434782609\\
9.5	0\\
9.6	0.228260869565217\\
9.7	0\\
9.8	0.282608695652174\\
9.9	0\\
10	0\\
10.1	0.25\\
10.2	0\\
10.3	0\\
10.4	0\\
10.5	0.260869565217391\\
10.6	0\\
10.7	0\\
10.8	0\\
10.9	0.217391304347826\\
11	0\\
11.1	0.260869565217391\\
11.2	0\\
11.3	0.16304347826087\\
11.4	0\\
11.5	0\\
11.6	0.260869565217391\\
11.7	0\\
11.8	0\\
11.9	0.239130434782609\\
12	0.0434782608695652\\
12.1	0\\
12.2	0\\
12.3	0\\
12.4	0.217391304347826\\
12.5	0\\
12.6	0.206521739130435\\
12.7	0\\
12.8	0.282608695652174\\
12.9	0\\
13	0\\
13.1	0.260869565217391\\
13.2	0\\
13.3	0\\
13.4	0.239130434782609\\
13.5	0.206521739130435\\
13.6	0\\
13.7	0\\
13.8	0\\
13.9	0.228260869565217\\
14	0\\
14.1	0.25\\
14.2	0\\
14.3	0.271739130434783\\
14.4	0\\
14.5	0\\
14.6	0.228260869565217\\
14.7	0\\
14.8	0.16304347826087\\
14.9	0\\
15	0\\
15.1	0\\
15.2	0.217391304347826\\
15.3	0.130434782608696\\
15.4	0.0108695652173913\\
15.5	0\\
15.6	0.239130434782609\\
15.7	0\\
15.8	0.271739130434783\\
15.9	0\\
16	0\\
16.1	0.260869565217391\\
16.2	0\\
16.3	0.16304347826087\\
16.4	0\\
16.5	0\\
16.6	0\\
16.7	0\\
16.8	0\\
16.9	0\\
17	0\\
17.1	0.173913043478261\\
17.2	0\\
17.3	0.217391304347826\\
17.4	0\\
17.5	0\\
17.6	0.195652173913043\\
17.7	0\\
17.8	0.239130434782609\\
17.9	0\\
18	0.228260869565217\\
18.1	0.0869565217391304\\
18.2	0\\
18.3	0\\
18.4	0\\
18.5	0\\
18.6	0.0978260869565217\\
18.7	0\\
18.8	0.260869565217391\\
18.9	0\\
19	0\\
19.1	0.173913043478261\\
19.2	0\\
19.3	0.260869565217391\\
19.4	0\\
19.5	0.16304347826087\\
19.6	0\\
19.7	0\\
19.8	0.217391304347826\\
19.9	0\\
20	0\\
20.1	0\\
20.2	0\\
20.3	0.130434782608696\\
20.4	0\\
20.5	0\\
20.6	0\\
20.7	0\\
20.8	0.0434782608695652\\
20.9	0\\
21	0.239130434782609\\
21.1	0\\
21.2	0\\
21.3	0.217391304347826\\
21.4	0\\
21.5	0\\
21.6	0\\
21.7	0\\
21.8	0.260869565217391\\
21.9	0\\
22	0\\
22.1	0\\
22.2	0\\
22.3	0.0217391304347826\\
22.4	0\\
22.5	0.239130434782609\\
22.6	0\\
22.7	0\\
22.8	0.260869565217391\\
22.9	0\\
23	0.282608695652174\\
23.1	0\\
23.2	0\\
23.3	0.239130434782609\\
23.4	0\\
23.5	0\\
23.6	0\\
23.7	0\\
23.8	0\\
23.9	0\\
24	0.141304347826087\\
24.1	0\\
24.2	0\\
24.3	0.260869565217391\\
24.4	0\\
24.5	0.282608695652174\\
24.6	0\\
24.7	0\\
24.8	0.239130434782609\\
24.9	0\\
25	0\\
25.1	0\\
25.2	0\\
25.3	0\\
25.4	0\\
25.5	0.217391304347826\\
25.6	0\\
25.7	0\\
25.8	0.206521739130435\\
25.9	0\\
26	0.25\\
26.1	0\\
26.2	0.217391304347826\\
26.3	0.217391304347826\\
26.4	0\\
26.5	0.0434782608695652\\
26.6	0\\
26.7	0.0217391304347826\\
26.8	0\\
26.9	0\\
27	0.0869565217391304\\
27.1	0\\
27.2	0\\
27.3	0.0869565217391304\\
27.4	0\\
27.5	0.184782608695652\\
27.6	0\\
27.7	0.260869565217391\\
27.8	0.152173913043478\\
27.9	0\\
28	0.282608695652174\\
28.1	0\\
28.2	0.282608695652174\\
28.3	0\\
28.4	0\\
28.5	0.184782608695652\\
28.6	0\\
28.7	0\\
28.8	0\\
28.9	0\\
29	0.0217391304347826\\
29.1	0\\
29.2	0.260869565217391\\
29.3	0.130434782608696\\
29.4	0\\
29.5	0.282608695652174\\
29.6	0\\
29.7	0.282608695652174\\
29.8	0\\
29.9	0.260869565217391\\
30	0.0978260869565217\\
};
\addlegendentry{Light Phase};

\end{axis}
\end{tikzpicture}%
}}}\\
		\makebox[\textwidth][c]{
			\subfloat[SPH - $Re=312.5$] {\resizebox{0.5\textwidth}{!}{%\pgfplotsset{label style={font=\tiny},
%	tick label style={font=\tiny} }
%
\begin{tikzpicture}

\begin{axis}[%
width=3.0in,
height=1.5in,
at={(0in,0in)},
scale only axis,
xmin=0,
xmax=30,
xlabel={Time ($\second$)},
ymin=0,
ymax=1.0,
ylabel near ticks,
xlabel near ticks,
xmajorgrids=false,
ymajorgrids=true,
xtick pos=left,
ytick pos=left,
grid style={dotted,gray},
ytick={0,0.25,0.5,0.75,1},
ylabel={Volume Fraction},
axis background/.style={fill=white},
legend style={legend style={nodes={scale=1.0, transform shape}},legend columns=-1,fill=white,align=left,draw=none,at={(0.9,0.7)}}
]
\addplot [color=blue,solid]
  table[row sep=crcr]{%
0	0.8\\
0.1	0.801980198019802\\
0.2	0.780405405405405\\
0.3	0.765676567656766\\
0.4	0.748344370860927\\
0.5	0.730263157894737\\
0.6	0.720394736842105\\
0.7	0.710526315789474\\
0.8	0.711920529801324\\
0.9	0.697674418604651\\
1	0.698675496688742\\
1.1	0.697674418604651\\
1.2	0.7\\
1.3	0.701986754966887\\
1.4	0.686666666666667\\
1.5	0.688963210702341\\
1.6	0.67986798679868\\
1.7	0.710963455149502\\
1.8	0.694078947368421\\
1.9	0.676567656765677\\
2	0.688741721854305\\
2.1	0.688963210702341\\
2.2	0.702970297029703\\
2.3	0.707641196013289\\
2.4	0.723333333333333\\
2.5	0.755852842809365\\
2.6	0.772277227722772\\
2.7	0.744262295081967\\
2.8	1\\
2.9	1\\
3	1\\
3.1	0.805280528052805\\
3.2	0.841059602649007\\
3.3	1\\
3.4	1\\
3.5	1\\
3.6	1\\
3.7	1\\
3.8	0.683006535947712\\
3.9	0.796666666666667\\
4	1\\
4.1	1\\
4.2	1\\
4.3	1\\
4.4	0.765886287625418\\
4.5	0.896666666666667\\
4.6	1\\
4.7	1\\
4.8	0.784053156146179\\
4.9	0.996666666666667\\
5	0.832236842105263\\
5.1	1\\
5.2	1\\
5.3	0.996688741721854\\
5.4	1\\
5.5	1\\
5.6	1\\
5.7	0.986622073578595\\
5.8	0.764119601328904\\
5.9	0.783783783783784\\
6	0.739273927392739\\
6.1	1\\
6.2	1\\
6.3	1\\
6.4	0.943708609271523\\
6.5	0.836666666666667\\
6.6	1\\
6.7	1\\
6.8	0.996655518394649\\
6.9	0.760797342192691\\
7	0.977049180327869\\
7.1	1\\
7.2	1\\
7.3	1\\
7.4	1\\
7.5	1\\
7.6	0.953947368421053\\
7.7	0.760655737704918\\
7.8	0.775919732441472\\
7.9	0.773333333333333\\
8	1\\
8.1	1\\
8.2	1\\
8.3	1\\
8.4	1\\
8.5	0.73421926910299\\
8.6	0.966555183946488\\
8.7	1\\
8.8	1\\
8.9	1\\
9	1\\
9.1	0.753333333333333\\
9.2	0.776666666666667\\
9.3	1\\
9.4	1\\
9.5	1\\
9.6	0.957236842105263\\
9.7	0.733333333333333\\
9.8	1\\
9.9	1\\
10	1\\
10.1	1\\
10.2	1\\
10.3	0.854237288135593\\
10.4	0.686468646864686\\
10.5	1\\
10.6	1\\
10.7	1\\
10.8	0.819397993311037\\
10.9	1\\
11	1\\
11.1	0.762376237623762\\
11.2	1\\
11.3	0.996666666666667\\
11.4	0.844884488448845\\
11.5	1\\
11.6	0.95959595959596\\
11.7	0.826086956521739\\
11.8	1\\
11.9	0.996655518394649\\
12	0.973244147157191\\
12.1	1\\
12.2	1\\
12.3	1\\
12.4	1\\
12.5	0.677852348993289\\
12.6	1\\
12.7	1\\
12.8	0.817880794701987\\
12.9	0.853820598006645\\
13	1\\
13.1	1\\
13.2	1\\
13.3	1\\
13.4	0.753333333333333\\
13.5	1\\
13.6	1\\
13.7	1\\
13.8	0.996677740863787\\
13.9	1\\
14	0.661073825503356\\
14.1	0.95049504950495\\
14.2	1\\
14.3	1\\
14.4	1\\
14.5	1\\
14.6	1\\
14.7	1\\
14.8	0.656565656565657\\
14.9	1\\
15	1\\
15.1	1\\
15.2	1\\
15.3	0.986754966887417\\
15.4	0.742574257425743\\
15.5	1\\
15.6	1\\
15.7	1\\
15.8	1\\
15.9	1\\
16	1\\
16.1	0.805369127516778\\
16.2	0.808724832214765\\
16.3	0.814814814814815\\
16.4	0.743333333333333\\
16.5	1\\
16.6	1\\
16.7	0.99672131147541\\
16.8	1\\
16.9	1\\
17	0.739273927392739\\
17.1	1\\
17.2	0.996666666666667\\
17.3	0.957792207792208\\
17.4	1\\
17.5	1\\
17.6	1\\
17.7	0.796052631578947\\
17.8	0.806666666666667\\
17.9	0.778877887788779\\
18	1\\
18.1	1\\
18.2	0.996699669966997\\
18.3	0.864686468646865\\
18.4	1\\
18.5	1\\
18.6	0.792642140468227\\
18.7	0.95\\
18.8	1\\
18.9	1\\
19	0.823333333333333\\
19.1	1\\
19.2	1\\
19.3	0.996677740863787\\
19.4	1\\
19.5	0.983443708609272\\
19.6	0.686666666666667\\
19.7	1\\
19.8	1\\
19.9	0.814569536423841\\
20	1\\
20.1	1\\
20.2	0.996666666666667\\
20.3	1\\
20.4	0.717171717171717\\
20.5	1\\
20.6	1\\
20.7	1\\
20.8	0.996655518394649\\
20.9	0.894039735099338\\
21	0.797979797979798\\
21.1	1\\
21.2	1\\
21.3	1\\
21.4	0.766101694915254\\
21.5	0.816993464052288\\
21.6	1\\
21.7	1\\
21.8	1\\
21.9	1\\
22	1\\
22.1	0.993399339933993\\
22.2	0.66\\
22.3	0.986666666666667\\
22.4	1\\
22.5	1\\
22.6	1\\
22.7	0.983277591973244\\
22.8	0.729372937293729\\
22.9	1\\
23	1\\
23.1	1\\
23.2	1\\
23.3	0.866666666666667\\
23.4	0.858552631578947\\
23.5	1\\
23.6	1\\
23.7	0.782608695652174\\
23.8	1\\
23.9	1\\
24	0.787878787878788\\
24.1	1\\
24.2	1\\
24.3	1\\
24.4	1\\
24.5	1\\
24.6	1\\
24.7	0.735973597359736\\
24.8	0.727574750830565\\
24.9	1\\
25	1\\
25.1	1\\
25.2	0.99009900990099\\
25.3	0.730263157894737\\
25.4	1\\
25.5	1\\
25.6	1\\
25.7	1\\
25.8	1\\
25.9	0.71\\
26	0.819397993311037\\
26.1	1\\
26.2	1\\
26.3	1\\
26.4	1\\
26.5	0.818481848184819\\
26.6	0.77\\
26.7	1\\
26.8	1\\
26.9	1\\
27	1\\
27.1	1\\
27.2	0.73489932885906\\
27.3	0.669934640522876\\
27.4	1\\
27.5	1\\
27.6	1\\
27.7	1\\
27.8	1\\
27.9	1\\
28	1\\
28.1	0.778145695364238\\
28.2	0.751655629139073\\
28.3	0.83\\
28.4	1\\
28.5	1\\
28.6	1\\
28.7	0.864686468646865\\
28.8	1\\
28.9	1\\
29	1\\
29.1	1\\
29.2	1\\
29.3	0.767441860465116\\
29.4	0.753289473684211\\
29.5	0.792642140468227\\
29.6	1\\
29.7	1\\
29.8	1\\
29.9	0.983552631578947\\
30	0.720394736842105\\
};
\addlegendentry{Heavy Phase};

\addplot [color=red,solid]
  table[row sep=crcr]{%
0	0.2\\
0.1	0.198019801980198\\
0.2	0.219594594594595\\
0.3	0.234323432343234\\
0.4	0.251655629139073\\
0.5	0.269736842105263\\
0.6	0.279605263157895\\
0.7	0.289473684210526\\
0.8	0.288079470198675\\
0.9	0.302325581395349\\
1	0.301324503311258\\
1.1	0.302325581395349\\
1.2	0.3\\
1.3	0.298013245033113\\
1.4	0.313333333333333\\
1.5	0.311036789297659\\
1.6	0.32013201320132\\
1.7	0.289036544850498\\
1.8	0.305921052631579\\
1.9	0.323432343234323\\
2	0.311258278145695\\
2.1	0.311036789297659\\
2.2	0.297029702970297\\
2.3	0.292358803986711\\
2.4	0.276666666666667\\
2.5	0.244147157190635\\
2.6	0.227722772277228\\
2.7	0.255737704918033\\
2.8	0\\
2.9	0\\
3	0\\
3.1	0.194719471947195\\
3.2	0.158940397350993\\
3.3	0\\
3.4	0\\
3.5	0\\
3.6	0\\
3.7	0\\
3.8	0.316993464052288\\
3.9	0.203333333333333\\
4	0\\
4.1	0\\
4.2	0\\
4.3	0\\
4.4	0.234113712374582\\
4.5	0.103333333333333\\
4.6	0\\
4.7	0\\
4.8	0.215946843853821\\
4.9	0.00333333333333333\\
5	0.167763157894737\\
5.1	0\\
5.2	0\\
5.3	0.0033112582781457\\
5.4	0\\
5.5	0\\
5.6	0\\
5.7	0.0133779264214047\\
5.8	0.235880398671096\\
5.9	0.216216216216216\\
6	0.260726072607261\\
6.1	0\\
6.2	0\\
6.3	0\\
6.4	0.0562913907284768\\
6.5	0.163333333333333\\
6.6	0\\
6.7	0\\
6.8	0.00334448160535117\\
6.9	0.239202657807309\\
7	0.0229508196721311\\
7.1	0\\
7.2	0\\
7.3	0\\
7.4	0\\
7.5	0\\
7.6	0.0460526315789474\\
7.7	0.239344262295082\\
7.8	0.224080267558528\\
7.9	0.226666666666667\\
8	0\\
8.1	0\\
8.2	0\\
8.3	0\\
8.4	0\\
8.5	0.26578073089701\\
8.6	0.0334448160535117\\
8.7	0\\
8.8	0\\
8.9	0\\
9	0\\
9.1	0.246666666666667\\
9.2	0.223333333333333\\
9.3	0\\
9.4	0\\
9.5	0\\
9.6	0.0427631578947368\\
9.7	0.266666666666667\\
9.8	0\\
9.9	0\\
10	0\\
10.1	0\\
10.2	0\\
10.3	0.145762711864407\\
10.4	0.313531353135314\\
10.5	0\\
10.6	0\\
10.7	0\\
10.8	0.180602006688963\\
10.9	0\\
11	0\\
11.1	0.237623762376238\\
11.2	0\\
11.3	0.00333333333333333\\
11.4	0.155115511551155\\
11.5	0\\
11.6	0.0404040404040404\\
11.7	0.173913043478261\\
11.8	0\\
11.9	0.00334448160535117\\
12	0.0267558528428094\\
12.1	0\\
12.2	0\\
12.3	0\\
12.4	0\\
12.5	0.322147651006711\\
12.6	0\\
12.7	0\\
12.8	0.182119205298013\\
12.9	0.146179401993355\\
13	0\\
13.1	0\\
13.2	0\\
13.3	0\\
13.4	0.246666666666667\\
13.5	0\\
13.6	0\\
13.7	0\\
13.8	0.00332225913621262\\
13.9	0\\
14	0.338926174496644\\
14.1	0.0495049504950495\\
14.2	0\\
14.3	0\\
14.4	0\\
14.5	0\\
14.6	0\\
14.7	0\\
14.8	0.343434343434343\\
14.9	0\\
15	0\\
15.1	0\\
15.2	0\\
15.3	0.0132450331125828\\
15.4	0.257425742574257\\
15.5	0\\
15.6	0\\
15.7	0\\
15.8	0\\
15.9	0\\
16	0\\
16.1	0.194630872483221\\
16.2	0.191275167785235\\
16.3	0.185185185185185\\
16.4	0.256666666666667\\
16.5	0\\
16.6	0\\
16.7	0.00327868852459016\\
16.8	0\\
16.9	0\\
17	0.260726072607261\\
17.1	0\\
17.2	0.00333333333333333\\
17.3	0.0422077922077922\\
17.4	0\\
17.5	0\\
17.6	0\\
17.7	0.203947368421053\\
17.8	0.193333333333333\\
17.9	0.221122112211221\\
18	0\\
18.1	0\\
18.2	0.0033003300330033\\
18.3	0.135313531353135\\
18.4	0\\
18.5	0\\
18.6	0.207357859531773\\
18.7	0.05\\
18.8	0\\
18.9	0\\
19	0.176666666666667\\
19.1	0\\
19.2	0\\
19.3	0.00332225913621262\\
19.4	0\\
19.5	0.0165562913907285\\
19.6	0.313333333333333\\
19.7	0\\
19.8	0\\
19.9	0.185430463576159\\
20	0\\
20.1	0\\
20.2	0.00333333333333333\\
20.3	0\\
20.4	0.282828282828283\\
20.5	0\\
20.6	0\\
20.7	0\\
20.8	0.00334448160535117\\
20.9	0.105960264900662\\
21	0.202020202020202\\
21.1	0\\
21.2	0\\
21.3	0\\
21.4	0.233898305084746\\
21.5	0.183006535947712\\
21.6	0\\
21.7	0\\
21.8	0\\
21.9	0\\
22	0\\
22.1	0.0066006600660066\\
22.2	0.34\\
22.3	0.0133333333333333\\
22.4	0\\
22.5	0\\
22.6	0\\
22.7	0.0167224080267559\\
22.8	0.270627062706271\\
22.9	0\\
23	0\\
23.1	0\\
23.2	0\\
23.3	0.133333333333333\\
23.4	0.141447368421053\\
23.5	0\\
23.6	0\\
23.7	0.217391304347826\\
23.8	0\\
23.9	0\\
24	0.212121212121212\\
24.1	0\\
24.2	0\\
24.3	0\\
24.4	0\\
24.5	0\\
24.6	0\\
24.7	0.264026402640264\\
24.8	0.272425249169435\\
24.9	0\\
25	0\\
25.1	0\\
25.2	0.0099009900990099\\
25.3	0.269736842105263\\
25.4	0\\
25.5	0\\
25.6	0\\
25.7	0\\
25.8	0\\
25.9	0.29\\
26	0.180602006688963\\
26.1	0\\
26.2	0\\
26.3	0\\
26.4	0\\
26.5	0.181518151815182\\
26.6	0.23\\
26.7	0\\
26.8	0\\
26.9	0\\
27	0\\
27.1	0\\
27.2	0.26510067114094\\
27.3	0.330065359477124\\
27.4	0\\
27.5	0\\
27.6	0\\
27.7	0\\
27.8	0\\
27.9	0\\
28	0\\
28.1	0.221854304635762\\
28.2	0.248344370860927\\
28.3	0.17\\
28.4	0\\
28.5	0\\
28.6	0\\
28.7	0.135313531353135\\
28.8	0\\
28.9	0\\
29	0\\
29.1	0\\
29.2	0\\
29.3	0.232558139534884\\
29.4	0.246710526315789\\
29.5	0.207357859531773\\
29.6	0\\
29.7	0\\
29.8	0\\
29.9	0.0164473684210526\\
30	0.279605263157895\\
};
\addlegendentry{Light Phase};

\end{axis}
\end{tikzpicture}%
}}
			\subfloat[LBM - $Re=312.5$] {\resizebox{0.5\textwidth}{!}{\input{LBM_Re312_fig1}}}}
		\caption {Evolution of the volume fractions at the outlet over time.}
		\label{fig1}
	\end{center}
\end{figure}

In this section, we study the ability of both methods to simulate a predicted intermittent flow regime. We consider an horizontal pipe of diameter $D=1~\meter$ and length $L=10D$. The light phase and heavy phase are denoted with a $g$ and $l$ subscript respectively. The flow enters from the inlet (left) and is assumed to be stratified with given volume fractions for each phase $\alpha_g=0.2$ and $\alpha_l=0.8$. All the physical properties are summarized in Tab.~\ref{tab:phys_prop}. Using these properties, it is possible to plot the flow regime map, see Fig.~\ref{flow_map_res}\footnote{In order to plot the map, one has to compute the Lockhart-Martelli~\cite{lockhart1949} parameter which depends on $n$, $m$, $C_g$ and $C_l$. In this study, we used $n=m=2$ and $C_g=C_l=0.042$}, and to pick an area to be investigated in the intermittent region. In this area, we adjust the viscosity to choose two cases that correspond to $Re=125$ and $Re=312.5$. Viscosity values and dimensionless numbers for each case are reported in Tabs.~\ref{tab:phys_prop_cases} and~\ref{tab:adim_cases}. Both cases were simulated in 2D with $25000$ nodes/particles. The simulation time was $30~\second$. At the inlet, each phase is injected with a constant velocity corresponding to its superficial velocity $u_{g,l}^s = \alpha_{g,l} u_{g,l}$. At the outlet, a constant pressure equal to the initial pressure $p_0$ is prescribed. Free slip boundary conditions are applied on the top and bottom walls (we were not able to generate a slug flow in LBM with no-slip boundary conditions under the same conditions). The initial setup is presented in Fig.~\ref{init_geo3}. The phases distributions for each case are shown in Fig.~\ref{ios_slugs_re125_comp} along with the associated plots showing the volume fraction evolution over time, the average pressure drop evolution over time, the heavy/light phase velocity evolution over time respectively in Fig.~\ref{fig1}, Fig.~\ref{fig2} and Fig.~\ref{fig4}.

\begin{table}[bthp]
	\centering
	\subfloat[SPH\label{tab:phys_prop_sph}]{	\begin{adjustbox}{max width=0.62\textwidth}
			\begin{tabular}{llll}
				\toprule
				\textbf{Property}     & \textbf{Light Phase}  & \textbf{Heavy Phase} & \textbf{Units}\\
				\midrule
				\textbf{Density} ($\rho$) & $1.0$     & $5.0$     & $\kilogram\per\meter^{3}$\\
				\textbf{Viscosity} ($\mu$)& $\mu_g = \mu_l/2$ & $\mu_l$ & $\pascal .\second$\\			
				\textbf{Sound speed} ($c_s$) & $153.73$ & $68.75$ & $\meter\per\second$\\
				\textbf{Surface Tension} ($\sigma^{nw}$) & \multicolumn{2}{l}{$0.01$} & $\newton\per\meter$\\
				\textbf{Contact Angle} ($\theta_c$) & \multicolumn{2}{l}{$90$} & $\degree$\\
				\textbf{Gravity} ($g_z$) & \multicolumn{2}{l}{$5.556$} & $\meter\per\second^2$\\
				\textbf{Space step} ($\Delta x$) & \multicolumn{2}{l}{$0.02$} & $\meter$\\
				\textbf{Time step} ($\Delta t$) & \multicolumn{2}{l}{$6.53\times 10^{-5}$} & $\second$\\
				\textbf{Domain size} ($L_x \times L_y$) & \multicolumn{2}{l}{$10\times 1$} & $\meter$\\	
				\textbf{Inlet velocity} ($u$)& $1.0416$ & $6.25$ & $\meter\per\second$\\							
				\bottomrule
			\end{tabular}
	\end{adjustbox}}
	\subfloat[LBM\label{tab:phys_prop_lbm}]{		\begin{adjustbox}{max width=0.38\textwidth}
			\begin{tabular}{llll}
				\toprule
				\textbf{Light Phase}  & \textbf{Heavy Phase} & \textbf{Units}\\
				\midrule
				$1.0$     & $5.0$     & $l.u.$\\
				$\mu_g = \mu_l/2$ & $\mu_l$ & $l.u.$\\			
				\multicolumn{2}{l}{$0.5773503$} & $l.u.$\\
				\multicolumn{2}{l}{$7.2\times 10^{-5}$} & $l.u.$\\
				\multicolumn{2}{l}{$90$} & $\degree$\\
				\multicolumn{2}{l}{$1.6\times 10^{-5}$} & $l.u.$\\
				\multicolumn{2}{l}{$1.0\times 10^{-2}$} & $l.u.$\\
				\multicolumn{2}{l}{$2.4\times 10^{-4}$} & $l.u.$\\
				\multicolumn{2}{l}{$500\times 50$} & $l.u.$\\
				$1.25\times 10^{-2}$ & $7.5\times 10^{-2}$ & $l.u.$\\							
				\bottomrule
			\end{tabular}
	\end{adjustbox}}
	\caption{Simulation parameters.}
	\label{tab:phys_prop}
\end{table}

\begin{table}[bthp]
	\centering
	\subfloat[SPH]{	\begin{adjustbox}{max width=0.5\textwidth}
			\begin{tabular}{llll}
				\toprule
				\textbf{Case}     & \textbf{$\mu_g$ ($\pascal .\second$)}  & \textbf{$\mu_l$ ($\pascal .\second$)}\\
				\midrule
				1 ($Re=125$)      &$1.25\times 10^{-1}$ & $2.5\times 10^{-1}$\\
				2 ($Re=312.5$)    &$5\times 10^{-2}$ & $1\times 10^{-1}$\\								
				\bottomrule
			\end{tabular}
	\end{adjustbox}}
	\subfloat[LBM]{		\begin{adjustbox}{max width=0.5\textwidth}
			\begin{tabular}{llll}
				\toprule
				\textbf{$\mu_g$ ($l.u.$)}  & \textbf{$\mu_l$ ($l.u.$)} & \textbf{$\tau_g$} & \textbf{$\tau_l$}\\
				\midrule
				$7.5\times 10^{-2}$ & $1.5\times 10^{-1}$ & $0.725$ & $0.59$\\								
				$3.0\times 10^{-2}$ & $6.0\times 10^{-2}$ & $0.59$ & $0.536$\\					
				\bottomrule
			\end{tabular}
	\end{adjustbox}}
	\caption{Viscosity values for each case.}
	\label{tab:phys_prop_cases}
\end{table}

\begin{table}[bthp]
	\centering
	\begin{adjustbox}{max width=0.5\textwidth}
		\begin{tabular}{llll}
			\toprule
			\textbf{Case}     &\textbf{$Re = \frac{L_y u_l}{\nu_l}$}     & \textbf{$La = \frac{\sigma \rho_l}{\mu_l^2}$}   & \textbf{$Bo = \frac{\Delta \rho g L_y^2}{\sigma}$} \\
			\midrule
			1      & $125$        &$0.8$ & $2222$\\
			2      & $312.5$       &$5$ & $2222$\\								
			\bottomrule
		\end{tabular}
	\end{adjustbox}
	\caption{Reynolds, Laplace and Bond numbers for each case.}
	\label{tab:adim_cases}
\end{table}

\begin{figure}[bthp]
	\begin{center}
		\makebox[\textwidth][c]{
			\subfloat[SPH - $Re=125$] {\resizebox{0.5\textwidth}{!}{%\pgfplotsset{label style={font=\footnotesize},
%	tick label style={font=\footnotesize} }
%
\begin{tikzpicture}

\begin{axis}[%
width=3.0in,
height=1.5in,
at={(0in,0in)},
scale only axis,
xmin=0,
xmax=5,
xlabel={Frequency $f$ ($\hertz$)},
ymin=0,
ymax=25,
ylabel near ticks,
xlabel near ticks,
xmajorgrids=false,
ymajorgrids=true,
xtick pos=left,
ytick pos=left,
grid style={dotted,gray},
xtick={0,1,2,3,4,5},
ylabel={$\lvert \mathcal{F}(f) \rvert^2$},
axis background/.style={fill=white}
]
\addplot [color=black,solid,forget plot]
  table[row sep=crcr]{%
0	3.47543765151596e-27\\
0.036231884057971	0.0570585239132698\\
0.072463768115942	0.055107043537487\\
0.108695652173913	0.13871174495742\\
0.144927536231884	0.00839837063350157\\
0.181159420289855	0.0102576005101518\\
0.217391304347826	0.247166356827483\\
0.253623188405797	0.299084257949098\\
0.289855072463768	0.150022016492709\\
0.326086956521739	0.188291963700356\\
0.36231884057971	0.0613972916778234\\
0.398550724637681	0.157051156731367\\
0.434782608695652	0.0728207944007692\\
0.471014492753623	0.457173278309738\\
0.507246376811594	0.00966689683652677\\
0.543478260869565	0.561371383591941\\
0.579710144927536	0.0918237666836142\\
0.615942028985507	0.542132604283378\\
0.652173913043478	0.27466363117405\\
0.688405797101449	0.130684338592724\\
0.72463768115942	0.477325373870416\\
0.760869565217391	0.368901717061404\\
0.797101449275362	0.716155755764127\\
0.833333333333333	0.241662205625913\\
0.869565217391304	0.651329934709248\\
0.905797101449275	0.440519432433894\\
0.942028985507246	0.705934290961149\\
0.978260869565217	0.631313649107282\\
1.01449275362319	2.78201609701108\\
1.05072463768116	0.846571669733872\\
1.08695652173913	2.54119641148846\\
1.1231884057971	0.412205526078879\\
1.15942028985507	2.60145715679763\\
1.19565217391304	1.06982496809884\\
1.23188405797101	0.63119618121886\\
1.26811594202899	0.184112160879024\\
1.30434782608696	2.87353902193794\\
1.34057971014493	0.268078561199226\\
1.3768115942029	13.5995550449632\\
1.41304347826087	0.719953740218654\\
1.44927536231884	3.17092128568652\\
1.48550724637681	0.86395310571603\\
1.52173913043478	0.187420774247039\\
1.55797101449275	0.662711305243512\\
1.59420289855072	1.23204858549997\\
1.6304347826087	4.13567676208031\\
1.66666666666667	1.31023953961029\\
1.70289855072464	5.9151582255535\\
1.73913043478261	1.81173715459559\\
1.77536231884058	1.93610931441135\\
1.81159420289855	0.116935329627821\\
1.84782608695652	6.72635398214907\\
1.88405797101449	2.42882727729718\\
1.92028985507246	13.4659734906662\\
1.95652173913043	7.17140009799675\\
1.99275362318841	0.0949957580472393\\
2.02898550724638	0.390057688236685\\
2.06521739130435	4.83429087417348\\
2.10144927536232	8.13093567677214\\
2.13768115942029	1.26070826102734\\
2.17391304347826	1.80918843094504\\
2.21014492753623	2.91941599548586\\
2.2463768115942	0.219480041030813\\
2.28260869565217	1.99789557605217\\
2.31884057971015	2.09889634401398\\
2.35507246376812	0.0748558954423911\\
2.39130434782609	1.15887251144296\\
2.42753623188406	0.531290883939201\\
2.46376811594203	7.10944033638949\\
2.5	0.938437729760144\\
2.53623188405797	0.0768331896893409\\
2.57246376811594	0.455229356671508\\
2.60869565217391	2.31798930165621\\
2.64492753623188	0.299494576767827\\
2.68115942028985	2.85938914932995\\
2.71739130434783	3.39374619476932\\
2.7536231884058	1.00285751956361\\
2.78985507246377	0.109639117519887\\
2.82608695652174	1.05431151040858\\
2.86231884057971	0.512792017771436\\
2.89855072463768	4.05564735340393\\
2.93478260869565	4.98511656958348\\
2.97101449275362	0.366793329439216\\
3.00724637681159	2.68082282141839\\
3.04347826086957	0.946081025877839\\
3.07971014492754	2.45516709399924\\
3.11594202898551	0.344601575260037\\
3.15217391304348	2.71550406231134\\
3.18840579710145	0.78223102939003\\
3.22463768115942	0.967325191148518\\
3.26086956521739	0.709674510185059\\
3.29710144927536	1.15166381444532\\
3.33333333333333	0.106131911270265\\
3.3695652173913	2.44901318136545\\
3.40579710144928	0.319460473077401\\
3.44202898550725	1.76008777087106\\
3.47826086956522	1.37220185984025\\
3.51449275362319	1.99349791832468\\
3.55072463768116	1.50327375274816\\
3.58695652173913	1.8405450083013\\
3.6231884057971	0.738422224194043\\
3.65942028985507	3.04365489623593\\
3.69565217391304	2.19435729360595\\
3.73188405797101	0.284292592183254\\
3.76811594202899	0.421091579283094\\
3.80434782608696	1.38951533205159\\
3.84057971014493	1.91782156018276\\
3.8768115942029	9.81581153832982\\
3.91304347826087	1.78218358209924\\
3.94927536231884	0.798879992614371\\
3.98550724637681	1.28727736938187\\
4.02173913043478	3.68123753258223\\
4.05797101449275	0.0822403640179857\\
4.09420289855072	0.071678402303594\\
4.1304347826087	0.324652945969716\\
4.16666666666667	1.21067450260562\\
4.20289855072464	2.09483284660447\\
4.23913043478261	2.87419444239559\\
4.27536231884058	0.919269625030302\\
4.31159420289855	3.70630321343424\\
4.34782608695652	0.0857653909781811\\
4.38405797101449	0.668840105677963\\
4.42028985507246	0.275716733251383\\
4.45652173913043	0.331828924698137\\
4.49275362318841	0.552157171209011\\
4.52898550724638	2.71177922038434\\
4.56521739130435	0.128716203566095\\
4.60144927536232	0.794780395879932\\
4.63768115942029	0.227486375677255\\
4.67391304347826	0.459866504309593\\
4.71014492753623	0.981136492218544\\
4.7463768115942	0.0912116800945127\\
4.78260869565217	1.13325339089669\\
4.81884057971014	1.64151747705832\\
4.85507246376812	0.980075636399123\\
4.89130434782609	0.124619117778357\\
4.92753623188406	1.41169583364441\\
4.96376811594203	3.67477640976674\\
5	0.21168225835905\\
};
\end{axis}
\end{tikzpicture}%
}}
			\subfloat[LBM - $Re=125$] {\resizebox{0.5\textwidth}{!}{%\pgfplotsset{label style={font=\footnotesize},
%	tick label style={font=\footnotesize} }
%
\begin{tikzpicture}

\begin{axis}[%
width=3.0in,
height=1.5in,
at={(0in,0in)},
scale only axis,
xmin=0,
xmax=5,
xlabel={Frequency $f$ ($\hertz$)},
ymin=0,
ymax=120,
ylabel near ticks,
xlabel near ticks,
xmajorgrids=false,
ymajorgrids=true,
xtick pos=left,
ytick pos=left,
grid style={dotted,gray},
xtick={0,1,2,3,4,5},
ylabel={$\lvert \mathcal{F}(f) \rvert^2$},
axis background/.style={fill=white}
]
\addplot [color=black,solid,forget plot]
  table[row sep=crcr]{%
0	1.10933564796705e-27\\
0.036231884057971	1.65916478838947\\
0.072463768115942	2.50246946835501\\
0.108695652173913	0.222145425032325\\
0.144927536231884	0.469920935173872\\
0.181159420289855	1.52662376107396\\
0.217391304347826	0.13213144293766\\
0.253623188405797	1.23349908711628\\
0.289855072463768	0.179161414470852\\
0.326086956521739	0.141887891290327\\
0.36231884057971	0.204383207773418\\
0.398550724637681	1.59288922081698\\
0.434782608695652	1.39746072346179\\
0.471014492753623	2.66405361686506\\
0.507246376811594	0.756672878269826\\
0.543478260869565	2.18208635467575\\
0.579710144927536	9.51320153990567\\
0.615942028985507	9.93653607704733\\
0.652173913043478	2.28217342154152\\
0.688405797101449	1.51277589548786\\
0.72463768115942	0.00598458394513843\\
0.760869565217391	1.29052105953133\\
0.797101449275362	2.22730163763141\\
0.833333333333333	0.721637610274129\\
0.869565217391304	0.0912091973154662\\
0.905797101449275	0.109134654671831\\
0.942028985507246	0.207286166256409\\
0.978260869565217	1.39328690561481\\
1.01449275362319	0.0413645322891429\\
1.05072463768116	0.845546443272379\\
1.08695652173913	0.326197521741908\\
1.1231884057971	1.59872894336897\\
1.15942028985507	0.95684086231056\\
1.19565217391304	0.913737029960095\\
1.23188405797101	0.0771229495988288\\
1.26811594202899	0.858148884765892\\
1.30434782608696	1.9668752773391\\
1.34057971014493	3.51986901916304\\
1.3768115942029	0.59679764341808\\
1.41304347826087	1.14045491961526\\
1.44927536231884	1.24850700307167\\
1.48550724637681	0.254474161320621\\
1.52173913043478	0.641786534273813\\
1.55797101449275	0.410447911572304\\
1.59420289855072	1.90930033318466\\
1.6304347826087	0.529519132666277\\
1.66666666666667	0.353379017013231\\
1.70289855072464	0.750144231162551\\
1.73913043478261	1.00329104900897\\
1.77536231884058	2.48214271002002\\
1.81159420289855	0.00281557159310013\\
1.84782608695652	0.503080062093241\\
1.88405797101449	1.36980640067727\\
1.92028985507246	3.40430351653114\\
1.95652173913043	7.67432131100523\\
1.99275362318841	0.665467793195074\\
2.02898550724638	1.11475774150555\\
2.06521739130435	0.189385333575614\\
2.10144927536232	6.20448942831914\\
2.13768115942029	1.25377140747399\\
2.17391304347826	2.15313438182165\\
2.21014492753623	1.37727956393993\\
2.2463768115942	0.0779038822668397\\
2.28260869565217	0.5043509079694\\
2.31884057971015	0.264800177226407\\
2.35507246376812	0.175592759171192\\
2.39130434782609	0.959589679000602\\
2.42753623188406	2.38082887064764\\
2.46376811594203	0.447410975795814\\
2.5	0.587074669187145\\
2.53623188405797	0.535264532841669\\
2.57246376811594	2.99111097182268\\
2.60869565217391	0.0877981654637403\\
2.64492753623188	4.71047678166128\\
2.68115942028985	17.9955571081938\\
2.71739130434783	5.89497708731498\\
2.7536231884058	3.60971877086387\\
2.78985507246377	5.83305339124643\\
2.82608695652174	0.0897160636708387\\
2.86231884057971	0.573041968432595\\
2.89855072463768	0.156385878354508\\
2.93478260869565	2.92118680818868\\
2.97101449275362	2.31236701901338\\
3.00724637681159	0.316183973404203\\
3.04347826086957	0.329903009864392\\
3.07971014492754	2.93797392916678\\
3.11594202898551	2.36901358889704\\
3.15217391304348	2.18304620345605\\
3.18840579710145	3.26532208111819\\
3.22463768115942	4.14276105855703\\
3.26086956521739	0.0678316846602745\\
3.29710144927536	7.786834693519\\
3.33333333333333	4.57455103969754\\
3.3695652173913	3.61824490007863\\
3.40579710144928	12.2660141612755\\
3.44202898550725	2.88117457679219\\
3.47826086956522	2.3844357021685\\
3.51449275362319	0.499889026346387\\
3.55072463768116	1.02778163972501\\
3.58695652173913	2.16475715158393\\
3.6231884057971	1.97117561382067\\
3.65942028985507	0.794754298092715\\
3.69565217391304	1.0373430117158\\
3.73188405797101	3.02878696789292\\
3.76811594202899	4.51727322073206\\
3.80434782608696	0.920604788933133\\
3.84057971014493	0.245536506027207\\
3.8768115942029	4.5779073330134\\
3.91304347826087	1.38958049101869\\
3.94927536231884	0.161815222313299\\
3.98550724637681	0.0281090916061745\\
4.02173913043478	106.96456141964\\
4.05797101449275	11.9091623396165\\
4.09420289855072	1.42009234546031\\
4.1304347826087	3.76162332839087\\
4.16666666666667	2.68124518745744\\
4.20289855072464	1.15591962211705\\
4.23913043478261	2.11106111974777\\
4.27536231884058	2.09903335934903\\
4.31159420289855	2.77284980685565\\
4.34782608695652	0.864872561480008\\
4.38405797101449	2.77833338964628\\
4.42028985507246	0.392531124705434\\
4.45652173913043	0.30395941391105\\
4.49275362318841	1.13269350103605\\
4.52898550724638	2.77883802125055\\
4.56521739130435	0.13988051901987\\
4.60144927536232	2.45382041886776\\
4.63768115942029	32.2122116422378\\
4.67391304347826	18.2424045153072\\
4.71014492753623	1.52609822187944\\
4.7463768115942	0.111240713872092\\
4.78260869565217	3.69994133141685\\
4.81884057971014	0.130750093236346\\
4.85507246376812	0.467615274251901\\
4.89130434782609	1.1510364658442\\
4.92753623188406	0.738116747370844\\
4.96376811594203	0.0409614254841841\\
5	0.179702268431002\\
};
\end{axis}
\end{tikzpicture}%
}}}\\
		\makebox[\textwidth][c]{
			\subfloat[SPH - $Re=312.5$] {\resizebox{0.5\textwidth}{!}{%\pgfplotsset{label style={font=\footnotesize},
%	tick label style={font=\footnotesize} }
%
\begin{tikzpicture}

\begin{axis}[%
width=3.0in,
height=1.5in,
at={(0in,0in)},
scale only axis,
xmin=0,
xmax=5,
xlabel={Frequency $f$ ($\hertz$)},
ymin=0,
ymax=25,
ylabel near ticks,
xlabel near ticks,
xmajorgrids=false,
ymajorgrids=true,
xtick pos=left,
ytick pos=left,
grid style={dotted,gray},
xtick={0,1,2,3,4,5},
ylabel={$\lvert \mathcal{F}(f) \rvert^2$},
axis background/.style={fill=white}
]
\addplot [color=black,solid,forget plot]
  table[row sep=crcr]{%
0	1.08912108727076e-28\\
0.036231884057971	1.61869948726818\\
0.072463768115942	0.750020408877863\\
0.108695652173913	0.375911318275645\\
0.144927536231884	0.763897144523243\\
0.181159420289855	0.0850581834313445\\
0.217391304347826	0.795954355399132\\
0.253623188405797	0.27889707154134\\
0.289855072463768	0.568571956106301\\
0.326086956521739	0.220908841705911\\
0.36231884057971	0.99137846175924\\
0.398550724637681	0.316788593406112\\
0.434782608695652	0.325289532604978\\
0.471014492753623	0.36551653759129\\
0.507246376811594	1.56330692667772\\
0.543478260869565	0.194456473526118\\
0.579710144927536	8.07275322100911\\
0.615942028985507	0.532539005668292\\
0.652173913043478	0.212898820620171\\
0.688405797101449	1.30097049348375\\
0.72463768115942	0.12892121795092\\
0.760869565217391	0.470802614934206\\
0.797101449275362	1.1432149170304\\
0.833333333333333	7.65438262873443\\
0.869565217391304	0.360624130963613\\
0.905797101449275	5.54002581079013\\
0.942028985507246	5.62651468801318\\
0.978260869565217	5.17333190668354\\
1.01449275362319	4.22555512449809\\
1.05072463768116	1.48945224331185\\
1.08695652173913	1.35781343351243\\
1.1231884057971	3.65508302892679\\
1.15942028985507	12.6192573922069\\
1.19565217391304	1.24590464386513\\
1.23188405797101	0.714198135023059\\
1.26811594202899	3.25963505022652\\
1.30434782608696	3.19937269208599\\
1.34057971014493	3.14518526898059\\
1.3768115942029	13.0117287781297\\
1.41304347826087	3.88262225492876\\
1.44927536231884	10.2590806374217\\
1.48550724637681	14.6883456025286\\
1.52173913043478	22.940742781396\\
1.55797101449275	0.107695395099383\\
1.59420289855072	17.3535402081516\\
1.6304347826087	12.4791488828998\\
1.66666666666667	8.82933974383661\\
1.70289855072464	0.787912392691594\\
1.73913043478261	0.561424548342349\\
1.77536231884058	2.91509437916249\\
1.81159420289855	0.628126826233103\\
1.84782608695652	9.16031041889189\\
1.88405797101449	2.08303435312456\\
1.92028985507246	2.71696720644135\\
1.95652173913043	1.05195284058674\\
1.99275362318841	3.79562271202159\\
2.02898550724638	3.27558409161292\\
2.06521739130435	2.79093225250225\\
2.10144927536232	4.9030689698595\\
2.13768115942029	5.36648115499928\\
2.17391304347826	1.21338276116916\\
2.21014492753623	2.03755958328475\\
2.2463768115942	1.58497257254688\\
2.28260869565217	2.57456219950414\\
2.31884057971015	3.3077835248841\\
2.35507246376812	4.49802365470828\\
2.39130434782609	1.97742214376303\\
2.42753623188406	4.93929449806258\\
2.46376811594203	0.96422073695015\\
2.5	0.522572980288576\\
2.53623188405797	0.664008857947002\\
2.57246376811594	3.01284446258857\\
2.60869565217391	0.217865145259737\\
2.64492753623188	0.328620756607984\\
2.68115942028985	4.92579411166643\\
2.71739130434783	5.83165706861589\\
2.7536231884058	0.752657780997871\\
2.78985507246377	5.09517268652074\\
2.82608695652174	0.262713407742885\\
2.86231884057971	0.887898289197033\\
2.89855072463768	5.8854186926739\\
2.93478260869565	0.324853895337884\\
2.97101449275362	1.4043649170748\\
3.00724637681159	5.31298732124753\\
3.04347826086957	6.24610180166962\\
3.07971014492754	1.53558788469044\\
3.11594202898551	4.51247227731291\\
3.15217391304348	13.5614848787753\\
3.18840579710145	0.0767317995389982\\
3.22463768115942	2.01583281444725\\
3.26086956521739	3.55605507293006\\
3.29710144927536	0.183229257792498\\
3.33333333333333	0.982759828344917\\
3.3695652173913	2.42869282836517\\
3.40579710144928	2.6874548192041\\
3.44202898550725	8.89163564769648\\
3.47826086956522	0.946806578276538\\
3.51449275362319	7.64634668116206\\
3.55072463768116	2.66531708469344\\
3.58695652173913	6.68521065303257\\
3.6231884057971	1.76010157558131\\
3.65942028985507	0.307787955753606\\
3.69565217391304	1.00347659759002\\
3.73188405797101	1.57772022701971\\
3.76811594202899	4.9560118266207\\
3.80434782608696	2.1622973870517\\
3.84057971014493	1.50123323302149\\
3.8768115942029	1.5156842978137\\
3.91304347826087	2.6002497250555\\
3.94927536231884	1.56478750341489\\
3.98550724637681	1.05049226226144\\
4.02173913043478	0.273985508659011\\
4.05797101449275	0.136864385828364\\
4.09420289855072	1.10548557213806\\
4.1304347826087	1.94469477727184\\
4.16666666666667	1.40233990737485\\
4.20289855072464	0.0446936414328027\\
4.23913043478261	0.700546962801193\\
4.27536231884058	1.69879915705555\\
4.31159420289855	2.13369117053277\\
4.34782608695652	0.975762065563834\\
4.38405797101449	1.0718085790153\\
4.42028985507246	0.687361171271876\\
4.45652173913043	3.96474318330267\\
4.49275362318841	1.7883134526939\\
4.52898550724638	0.225402487524349\\
4.56521739130435	0.346126654875468\\
4.60144927536232	0.623904849554562\\
4.63768115942029	2.13720143450771\\
4.67391304347826	1.35764367868084\\
4.71014492753623	0.106800929468225\\
4.7463768115942	0.540306100422451\\
4.78260869565217	2.49960644427606\\
4.81884057971014	4.28592957321611\\
4.85507246376812	0.161225490807328\\
4.89130434782609	5.85742212649336\\
4.92753623188406	1.83019557294604\\
4.96376811594203	3.40817677985651\\
5	1.87024473333351\\
};
\end{axis}
\end{tikzpicture}%
}}
			\subfloat[LBM - $Re=312.5$] {\resizebox{0.5\textwidth}{!}{%\pgfplotsset{label style={font=\footnotesize},
%	tick label style={font=\footnotesize} }
%
\begin{tikzpicture}

\begin{axis}[%
width=3.0in,
height=1.5in,
at={(0in,0in)},
scale only axis,
xmin=0,
xmax=5,
xlabel={Frequency $f$ ($\hertz$)},
ymin=0,
ymax=120,
ylabel near ticks,
xlabel near ticks,
xmajorgrids=false,
ymajorgrids=true,
xtick pos=left,
ytick pos=left,
grid style={dotted,gray},
xtick={0,1,2,3,4,5},
ylabel={$\lvert \mathcal{F}(f) \rvert^2$},
axis background/.style={fill=white}
]
\addplot [color=black,solid,forget plot]
  table[row sep=crcr]{%
0	1.09143959877501e-26\\
0.036231884057971	0.540383270239104\\
0.072463768115942	0.373848240063567\\
0.108695652173913	0.657047544413358\\
0.144927536231884	0.808263441689281\\
0.181159420289855	1.26494151839722\\
0.217391304347826	0.0456539530555818\\
0.253623188405797	1.18197915183934\\
0.289855072463768	1.61888524002687\\
0.326086956521739	0.551662979290768\\
0.36231884057971	0.436426109200391\\
0.398550724637681	0.684240035504645\\
0.434782608695652	0.901381272332246\\
0.471014492753623	0.338200993252702\\
0.507246376811594	0.76506681559612\\
0.543478260869565	0.188437149645895\\
0.579710144927536	0.913009970704448\\
0.615942028985507	0.790010388502791\\
0.652173913043478	1.33616151087513\\
0.688405797101449	3.37696242657263\\
0.72463768115942	1.6944320210403\\
0.760869565217391	0.624525503929752\\
0.797101449275362	1.0756927153281\\
0.833333333333333	0.112408879770167\\
0.869565217391304	0.296849006330974\\
0.905797101449275	0.538364731992033\\
0.942028985507246	0.0599340297844132\\
0.978260869565217	0.453723083894293\\
1.01449275362319	0.912484184734572\\
1.05072463768116	1.96980880939226\\
1.08695652173913	0.63932884953306\\
1.1231884057971	0.167454610665414\\
1.15942028985507	0.338139820987697\\
1.19565217391304	0.285131227281588\\
1.23188405797101	0.633164055561659\\
1.26811594202899	2.43281947397383\\
1.30434782608696	1.64297735011718\\
1.34057971014493	1.85455731458948\\
1.3768115942029	1.45827183626628\\
1.41304347826087	0.885936911688592\\
1.44927536231884	0.487610129921223\\
1.48550724637681	0.826893127069408\\
1.52173913043478	0.626293092072082\\
1.55797101449275	0.105764602583004\\
1.59420289855072	0.277430179499748\\
1.6304347826087	2.4398490469307\\
1.66666666666667	3.23263232514179\\
1.70289855072464	0.176179180710965\\
1.73913043478261	0.644339628670012\\
1.77536231884058	0.103325667931386\\
1.81159420289855	1.13798635161429\\
1.84782608695652	0.848247290062848\\
1.88405797101449	1.78568560392546\\
1.92028985507246	1.81611546507334\\
1.95652173913043	0.155604701685992\\
1.99275362318841	0.655633311236602\\
2.02898550724638	1.1586487237146\\
2.06521739130435	0.0876951177703066\\
2.10144927536232	1.24073804584098\\
2.13768115942029	0.459431622659509\\
2.17391304347826	1.38184007434971\\
2.21014492753623	2.13903622898985\\
2.2463768115942	0.859881709172969\\
2.28260869565217	0.916294469659153\\
2.31884057971015	0.187750209700803\\
2.35507246376812	0.19223304251287\\
2.39130434782609	0.0385139034207282\\
2.42753623188406	5.90974911772912\\
2.46376811594203	0.11334019555035\\
2.5	0.116375236294896\\
2.53623188405797	0.188049394342108\\
2.57246376811594	0.155326653027876\\
2.60869565217391	0.0550877029620787\\
2.64492753623188	1.20168645951008\\
2.68115942028985	6.67551967458714\\
2.71739130434783	2.53397269251405\\
2.7536231884058	2.78452130996749\\
2.78985507246377	2.14698502101253\\
2.82608695652174	0.199744848608983\\
2.86231884057971	0.664209779025377\\
2.89855072463768	0.93790902482886\\
2.93478260869565	0.00516390410788595\\
2.97101449275362	4.09622193668866\\
3.00724637681159	0.317914515795653\\
3.04347826086957	1.57404119408142\\
3.07971014492754	1.10514934192675\\
3.11594202898551	0.711373766561732\\
3.15217391304348	0.0528482233463939\\
3.18840579710145	0.505953488776268\\
3.22463768115942	0.13391616713492\\
3.26086956521739	0.256697835757859\\
3.29710144927536	0.254943455282161\\
3.33333333333333	0.117793005671077\\
3.3695652173913	1.6116501809602\\
3.40579710144928	0.160570823924085\\
3.44202898550725	1.88326775572683\\
3.47826086956522	2.12095244960731\\
3.51449275362319	0.372398132307512\\
3.55072463768116	0.768577831947211\\
3.58695652173913	0.401405946067219\\
3.6231884057971	1.14027268144433\\
3.65942028985507	0.143178539410004\\
3.69565217391304	1.41715924171477\\
3.73188405797101	0.0751936289466035\\
3.76811594202899	1.21940698319063\\
3.80434782608696	4.93637195570155\\
3.84057971014493	0.506321805921039\\
3.8768115942029	1.56287845069306\\
3.91304347826087	0.551987040950594\\
3.94927536231884	0.781904485585455\\
3.98550724637681	0.370777486045645\\
4.02173913043478	0.398357336443262\\
4.05797101449275	0.543906719710446\\
4.09420289855072	2.80078270925202\\
4.1304347826087	0.595183901734271\\
4.16666666666667	0.589623256335694\\
4.20289855072464	0.374231682947869\\
4.23913043478261	2.64240427865317\\
4.27536231884058	0.369008008692764\\
4.31159420289855	4.59842600614743\\
4.34782608695652	1.69628214217557\\
4.38405797101449	1.3753904423\\
4.42028985507246	1.52338585553128\\
4.45652173913043	0.478464296797393\\
4.49275362318841	1.1550217875975\\
4.52898550724638	0.286542808525341\\
4.56521739130435	4.29492753772299\\
4.60144927536232	2.89787567463575\\
4.63768115942029	2.66773753917527\\
4.67391304347826	0.476739644421739\\
4.71014492753623	0.586764653898919\\
4.7463768115942	1.62413345119054\\
4.78260869565217	3.83074134447204\\
4.81884057971014	2.24150080730506\\
4.85507246376812	72.3105473978542\\
4.89130434782609	58.7038916322989\\
4.92753623188406	9.53202257693401\\
4.96376811594203	1.31332296599652\\
5	0.357396030245743\\
};
\end{axis}
\end{tikzpicture}%
}}}
		\caption {Periodograms of the heavy phase volume fraction time series (in blue in Fig.~\ref{fig1}). The mean has been removed from the signal before performing the Fourier transform to remove the $0~\hertz$ frequency component.}
		\label{fft}
	\end{center}
\end{figure}
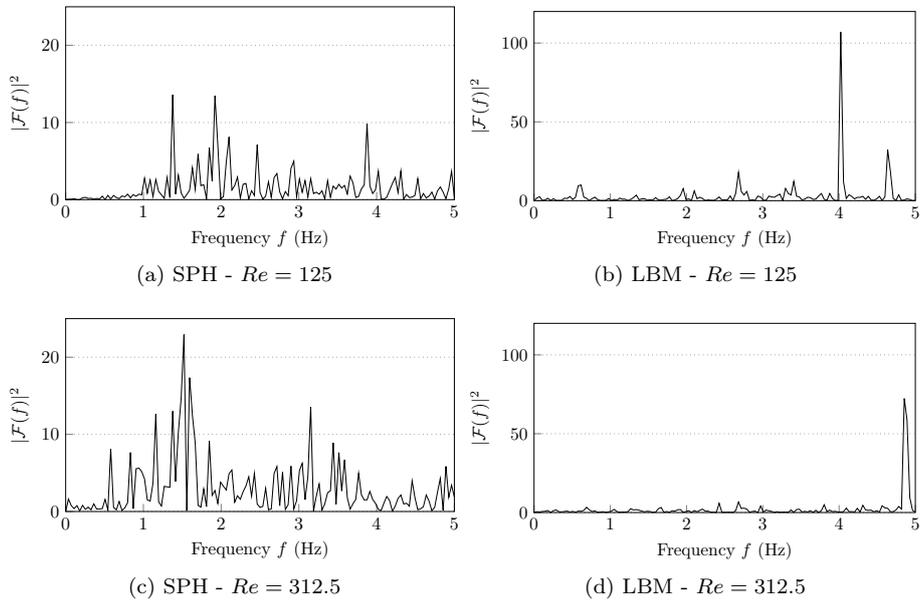

\begin{figure}[bthp]
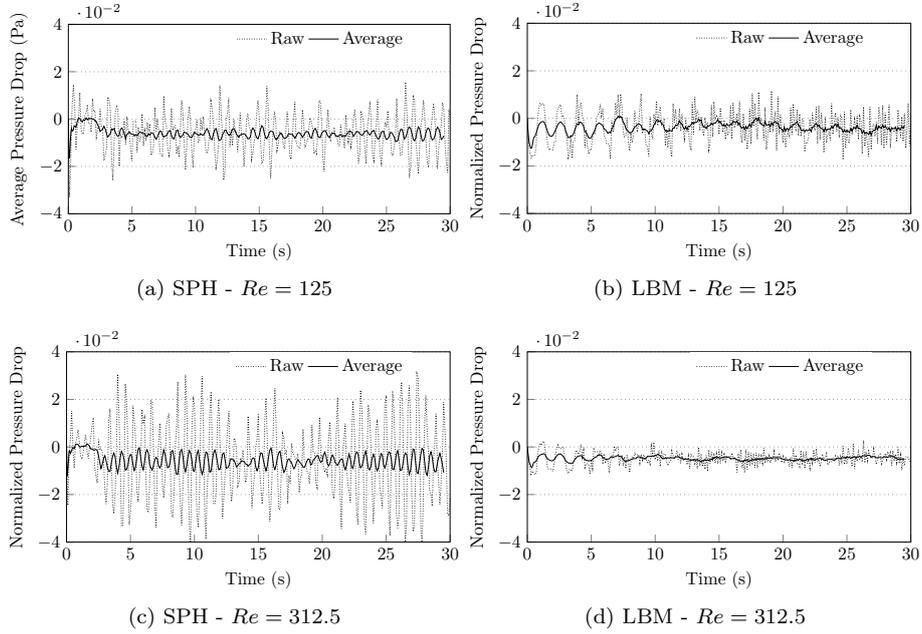

	\begin{center}
		\makebox[\textwidth][c]{
			\subfloat[SPH - $Re=125$] {\resizebox{0.5\textwidth}{!}{\input{SPH_Re125_fig2}}}
			\subfloat[LBM - $Re=125$] {\resizebox{0.5\textwidth}{!}{\input{LBM_Re125_fig2}}}}\\
		\makebox[\textwidth][c]{
			\subfloat[SPH - $Re=312.5$] {\resizebox{0.5\textwidth}{!}{\input{SPH_Re312_fig2}}}
			\subfloat[LBM - $Re=312.5$] {\resizebox{0.5\textwidth}{!}{\input{LBM_Re312_fig2}}}}
		\caption {Evolution of the normalized pressure drops over time.}
		\label{fig2}
	\end{center}
\end{figure}

\begin{figure}[bthp]
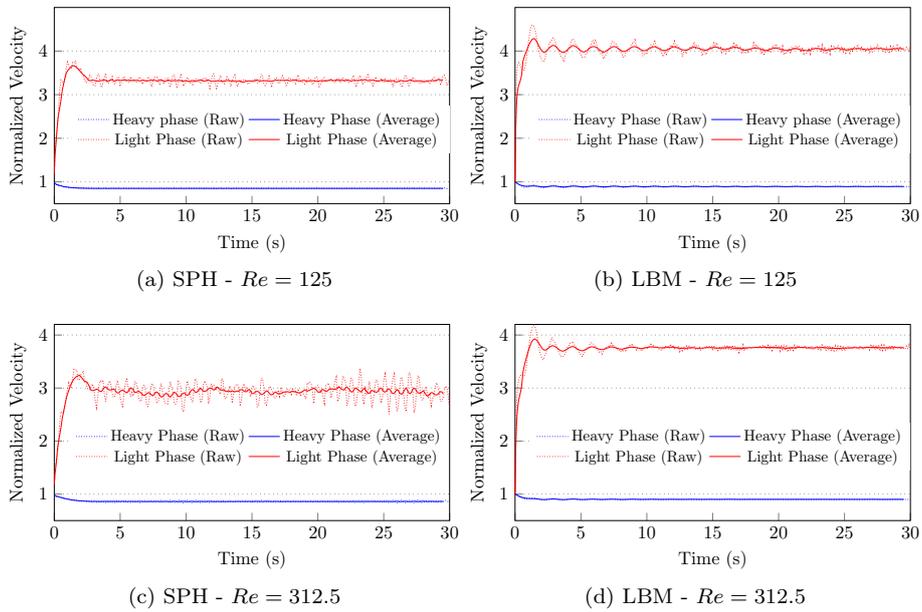

	\begin{center}
		\makebox[\textwidth][c]{
			\subfloat[SPH - $Re=125$] {\resizebox{0.5\textwidth}{!}{\input{SPH_Re125_fig4}}}
			\subfloat[LBM - $Re=125$] {\resizebox{0.5\textwidth}{!}{\input{LBM_Re125_fig4}}}}\\
		\makebox[\textwidth][c]{
			\subfloat[SPH - $Re=312.5$] {\resizebox{0.5\textwidth}{!}{\input{SPH_Re312_fig4}}}
			\subfloat[LBM - $Re=312.5$] {\resizebox{0.5\textwidth}{!}{\input{LBM_Re312_fig4}}}}
		\caption {Evolution of the normalized velocities over time.}
		\label{fig4}
	\end{center}
\end{figure}

In Fig.~\ref{ios_slugs_re125_comp}, one can see snapshots of the phases' distribution for both methods at selected timesteps. It is clear that both methods are able to generate a slug flow as predicted by the flow map of Fig.~\ref{flow_map_res}. However, LBM produces a much more regular intermittent flow pattern with a higher slug frequency than SPH. This can also be seen in Figs.~\ref{fig1} and~\ref{fft}. For example, for $Re=312.5$, the LBM slug frequency at the outlet is approximately $4.85~\hertz$ whereas for SPH it is close to $ 1.52~\hertz$. The slug frequency seems to remain roughly stable or to slightly increase (about $+1~\hertz$ for LBM and about $+0.16~\hertz$ for SPH for the highest peak) when $Re$ changes from $125$ to $312.5$ although it is less obvious in SPH periodograms due to the noise and composition of the signal. It is expected that the slug frequency increases when $Re$ increases but we could not raise $Re$ higher without making LBM simulations unstable ($\tau \to 0$ or $Ma \to 1$). In addition, SPH periodograms are noisier with $3$ to $4$ major frequency components unlike LBM where one frequency clearly emerges. It indicates that SPH volume fraction signals are less regular and are composed of signals with different frequencies. Moreover, we can observe in both methods that the point where the first slug appears is in general closer from the pipe entry when $Re$ is smaller. This is expected since when velocities are smaller at the entry, the first slug tends to form earlier in the pipe.

In Fig.~\ref{fig2}, the raw and average pressure drops evolution over time for all the different cases studied are shown. One can notice that in all cases, the pressure drops are $\leq 0$ on average (even though strong oscillations are observed). It means that the average pressure along the pipe's height is higher at the outlet than at the inlet. This phenomenon was already seen in~\cite{DouilletGrellier2018} and may be due to several factors : the quality of the boundary conditions, the recirculation areas at the entry that tend to lower the pressure and the averaging area chosen (i.e. $0.1~\meter$ (from the inlet) $\times D$). Nevertheless, we observed that both methods are returning the same average pressure drop level around $\approx -0.05 p_0$. SPH plots present much stronger oscillations than LBM ones (up to $7$ times the mean value for SPH against $3$ times the mean value for LBM). This is a known issue in weakly compressible SPH where the particles spatial distribution is directly linked to the density/pressure. 

%\begin{figure}[bthp]
%	\begin{center}
%		\makebox[\textwidth][c]{
%			\subfloat[SPH - $Re=125$] {\resizebox{0.5\textwidth}{!}{\input{SPH_Re125_fig3}}}
%			\subfloat[LBM - $Re=125$] {\resizebox{0.5\textwidth}{!}{\input{LBM_Re125_fig3}}}}\\
%		\makebox[\textwidth][c]{
%			\subfloat[SPH - $Re=312.5$] {\resizebox{0.5\textwidth}{!}{\input{SPH_Re312_fig3}}}
%			\subfloat[LBM - $Re=312.5$] {\resizebox{0.5\textwidth}{!}{\input{LBM_Re312_fig3}}}}
%		\caption {Evolution of the normalized heavy phase velocities over time.}
%		\label{fig3}
%	\end{center}
%\end{figure}

In Fig.~\ref{fig4}, the evolution over time of the heavy phase and light phase velocities within the whole pipe are shown for all cases considered. After a transient period of $\approx 4~\second$, the light and heavy phase velocities stabilize around a fixed value when a periodic state is reached. The main difference between SPH and LBM on this aspect is that the final velocity values are higher in LBM than they are in SPH. Overall, the average velocity field in LBM is $5\% v^{\text{in}}$ higher than in SPH. In our opinion, this is due to the wall boundary conditions that are handled in a very different way in both methods (full bounce back approach in LBM and interpolation-based approach in SPH) and to the wetting boundary conditions that is also handled differently. Those can strongly affect the flow, especially in dynamic cases like the ones considered in this section. From a general point of view, one can add that the oscillations observed in Figs.~\ref{fig2} and~\ref{fig4} are gaining amplitude when $Re$ increases in SPH whereas it is not the case in LBM for which they tend to reduce.

In a nutshell, as predicted by Taitel and Dukler's flow map, both methods are capable to generate a slug flow pattern starting from the exact same simulation setup. Nevertheless, unlike all previous test cases for which the results were globally similar, we obtain significantly different flow patterns. Indeed, SPH produces bigger bubbles and in a more irregular way compared to LBM. We tend to believe that it is due to boundary conditions. In addition, we could not extend our study to higher $Re$ number because LBM was not stable anymore. Although, we tend to think that LBM is entitled to propose the best solution, in particular because of its superior accuracy, its narrow stability range may be a serious drawback to simulate two-phase flows in pipes at realistic $Re$ numbers.

One last point that has not been addressed yet is the computational efficiency of both methods. On that aspect and despite very important progresses made in the SPH community, LBM remains superior in terms of speed, thanks to its local lattice-based algorithm that does not require a nearest neighbor search at every time step. For this paper, we have implemented both methods in the same framework in Fortran combined with an OpenMP library to handle multi-threading. For a simulation involving $40000$ nodes/particles, LBM was about $4$ times faster than SPH on a laptop equipped with a $2.9~\giga\hertz$ Intel Core i7 processor with $4$ cores and $16$~Gb of RAM. This number is given as an indicative number and should be taken with caution since the codes were not fully optimized.

\section{Conclusions}

In this paper, we propose a 2D multiphase comparison of two particle methods, SPH and LBM, that are very different by nature. Despite these differences, both methods have been extensively used to model multiphase flows with success. We have chosen a multiphase formulation for each method among those available in the literature : the continuum surface force technique for SPH and the color gradient method for LBM. Then, we have simulated a collection of static bubble tests with different density and viscosity ratios with both methods. Finally, we have prolonged the comparison to more realistic cases, i.e. to the generation of slug flows in pipes. To this end, we have extended LBM Zou-He boundary conditions to be able to handle velocity inlet and pressure outlet conditions in a multiphase context.

From a general point of view, we have confirmed that LBM offers a better order of convergence and a better accuracy than SPH although it suffers from a more narrow stability range than SPH. In many situations for which the Mach number is too high or the viscosity is too low, LBM will be unstable contrary to SPH which is only controlled by the CFL condition. Note that research on the extension of LBM to high Mach numbers is very active. In addition, SPH tends to generate pressure fields that are noisier than with LBM because of the Lagrangian behavior of particles whose position is directly linked to pressure through the density evaluation. This problem has been the subject of many investigations and several treatments are available. Moreover, LBM is more computationally efficient than SPH by construction. 

On the multiphase aspects, both methods are very capable to simulate a variety of dynamic incompressible multiphase flow problems with good precision in the case of moderate density and viscosity ratios. However, we have noted that, on one hand, SPH appears more robust for high density ratios than LBM and, on the other hand, SPH has more trouble to handle high viscosity ratios than LBM. Another difference is that, at the interface, the fluids are mixed resulting in a diffuse interface whereas with SPH, particles are affected to one phase or the other without any mixing. More specifically, both methods have been able to simulate slug flows where expected but, due to boundary conditions, there might be differences in slug frequency and/or slug sizes.

To conclude, according to the results presented in this paper, our recommendation would be to use LBM when stability is not an issue and SPH otherwise. In future works, we would like to extend our comparison to other methods, experimental measurements and commercial software. This would help to characterize more precisely which method is best adapted to a given situation.

\section*{Acknowledgements}
The authors would like to thank Total for its scientific and financial support.

\section*{References}
\bibliographystyle{unsrt}
\bibliography{references}

\appendix

\section{Additional test cases}\label{appendix}

For all the following additional test cases, no kernel gradient correction or shifting or interface correction were used.

\subsection{Lid-driven Cavity Flow}

\begin{figure}[bthp]
	\begin{center}
		\resizebox{0.5\textwidth}{!}{\begin{tikzpicture}

\node[circle,fill=black!0,label=left:{\Large $(0,0)$}](bl) at (0,0) {};

\node[circle,fill=black!0,label=left:{\Large $(0,1)$}](tl) at (0,10) {};

\node[circle,fill=black!0,label=below:{\Large $(1,0)$}](br) at (10,0) {};

\node[circle,fill=black!0,label={\Large $(1,1)$}](tr) at (10,10) {};

\node[circle,fill=black!0,label={\Large $x$}](x) at (12,0) {};
\node[circle,fill=black!0,label={\Large $y$}](y) at (0,12) {};

\draw (bl.center) -- node[below] {\huge $\mathbf{v_x,v_y=0}$} (br.center);
\draw (br.center) -- node[right] {\huge $\mathbf{v_x,v_y=0}$} (tr.center);
\draw (tr.center) -- node[above] {\huge $\mathbf{v_x=U_{lid},v_y=0}$}(tl.center);
\draw (tl.center) -- node[left] {\huge $\mathbf{v_x,v_y=0}$} (bl.center);

\draw[-triangle 45,dashed] (bl.center) -- (x.center);
\draw[-triangle 45,dashed] (bl.center) -- (y.center);

\draw [red,-<, very thick] (8,5) arc (0:264:30mm) (8,5);

\end{tikzpicture}}
		\caption {The 2D Lid-driven Cavity Flow problem}
		\label{cavity_flow}
	\end{center}
\end{figure}
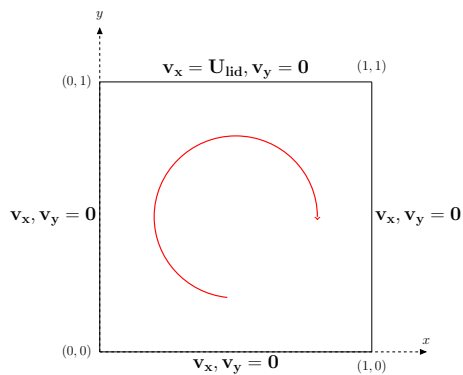

The goal of this section is to validate the implementation of SPH and LBM for the single phase Navier-Stokes case. The test case chosen for this purpose is the well-known 2D lid-driven cavity flow problem shown in Fig.~\ref{cavity_flow}. This is a common problem in the fluid mechanics community and numerous reference solutions performed with different numerical methods are available in the literature. In this case, we use Ghia et al. solution as a reference~\cite{ghia1982}. Note that Ghia's solution is also numerical. 

The Reynolds number for this problem is defined as follows $Re = \frac{U_{\text{lid}} L}{\nu}$ where $U_{\text{lid}}$ is the velocity of the imposed at the top boundary, $\nu$ is the kinematic viscosity and $L$ is the characteristic length of the problem. The simulations were performed for $Re=100$, $400$, $1000$ and $10000$ and for $50\times 50=2500$, $100\times 100=10000$ and $200\times 200=40000$ particles or nodes (respectively for SPH and LBM). The density is set to $1000~\kilogram/\meter^{3}$, the velocity of the lid is $U_{\text{lid}}=0.1~l.u.$ for LBM and $1~\meter/\second$ for SPH, the domain is $L_x \times L_y = 1~\meter\times 1~\meter$ and the viscosity $\nu$ is adjusted to reach the desired Reynolds number. 

For LBM, due to stability issues, the MRT collision operator was used. The standard set of relaxation times $\bm{S}$ defined in Eq.~\eqref{relax_times}. In order to have stable results for at least one lattice size for every Reynolds number, a specific setup was used where indicated (referred as $LBM*$). The relaxation times are the following :
\begin{equation}
\bm{S}^{*} = \text{diag}(1.0,1.0,1.0,1.0,1.0,1.0,1.0,\omega_{\text{eff}},\omega_{\text{eff}})
\end{equation}
\noindent and the lid velocity is increased $U^*_{\text{lid}} = 0.4 l.u.$.

The velocity boundary condition at the top boundary has been applied using the procedure described in Sec.~\ref{sec:bc} for SPH and in Sec.~\ref{sec:lbm_io_bc} for LBM. For the other boundaries, a no-slip boundary conditions has been applied. The simulations are terminated when a steady state is reached (i.e. $\sqrt{\sum_a\frac{\lvert\rho_a^{n+1}-\rho_a^{n}\rvert}{\rho_a^{n}}}< 1e^{-2}$ or after $60~\second$ of real simulated time).

\subsubsection{$Re=100$}

\begin{figure}[bthp]
	\begin{minipage}{0.45\textwidth}
		\captionsetup[subfigure]{labelformat=empty}
		\centering
		\subfloat[{\tiny $50\times 50$ - Velocity (left) - Streamlines (right)}] {\includegraphics[width=0.12\textwidth]{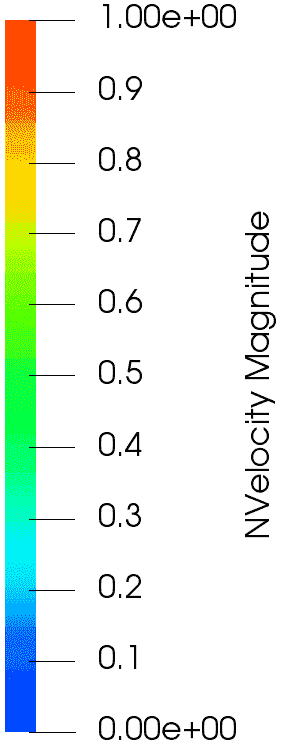}\includegraphics[width=0.4\textwidth]{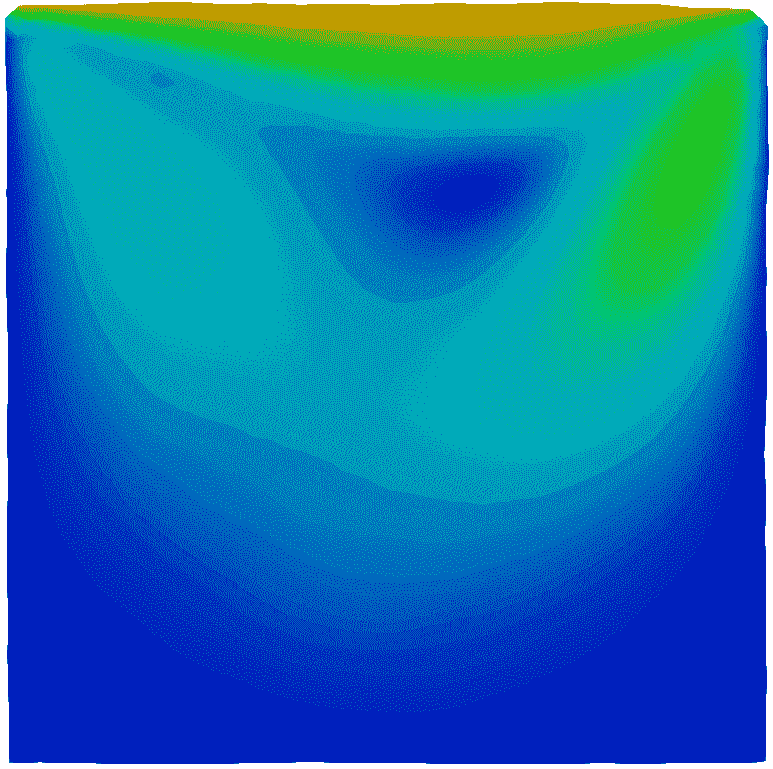}\includegraphics[width=0.4\textwidth]{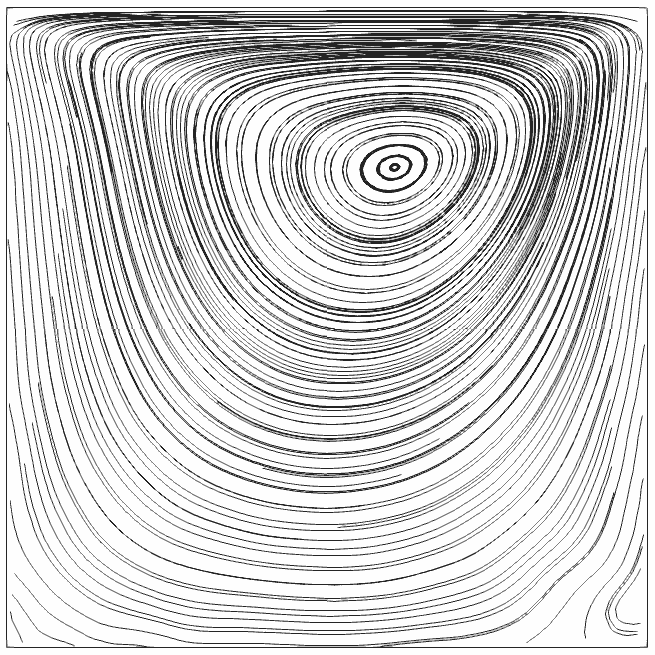}}\\
		\subfloat[{\tiny $100\times 100$ - Velocity (left) - Streamlines (right)}] {\includegraphics[width=0.12\textwidth]{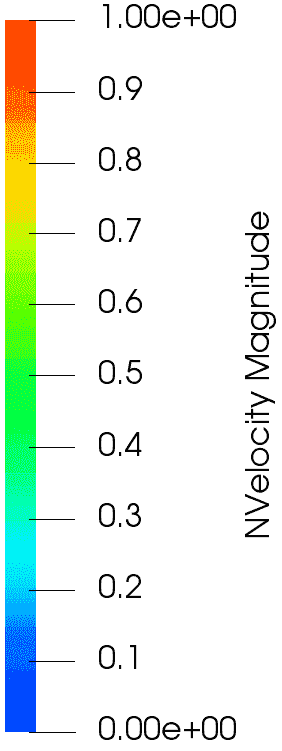}\includegraphics[width=0.4\textwidth]{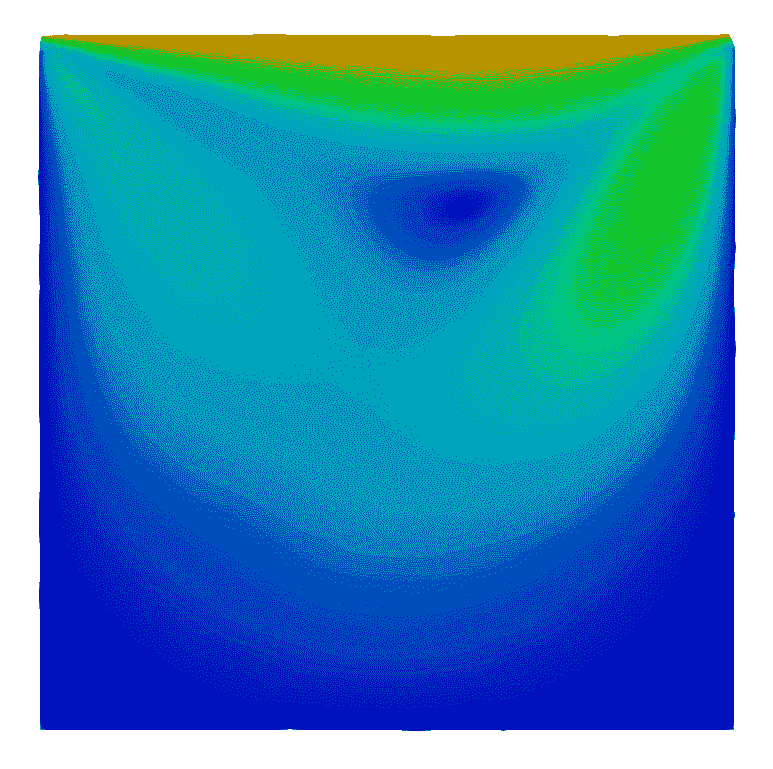}\includegraphics[width=0.4\textwidth]{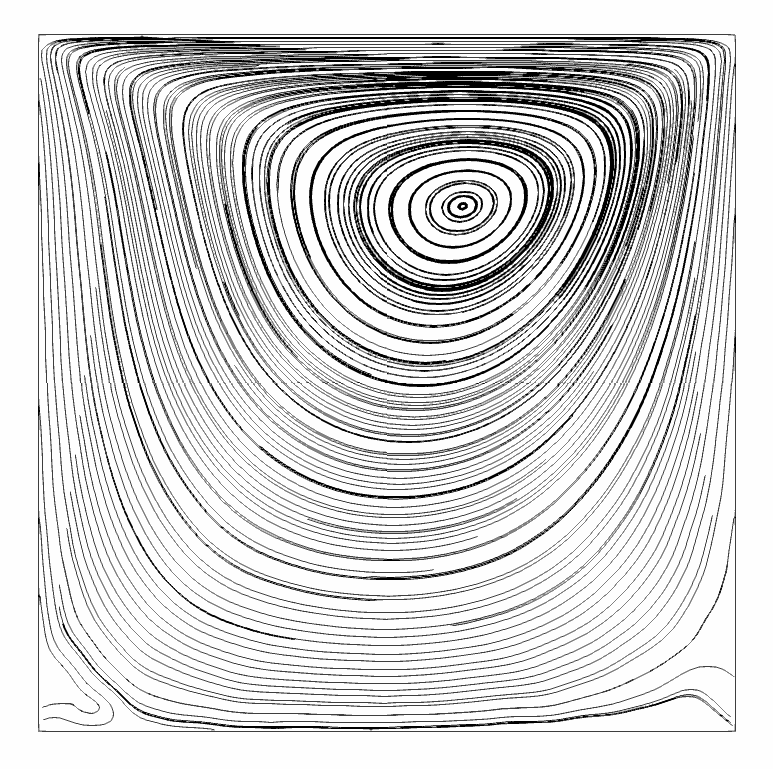}}\\	
		\subfloat[{\tiny $200\times 200$ - Velocity (left) - Streamlines (right)}] {\includegraphics[width=0.12\textwidth]{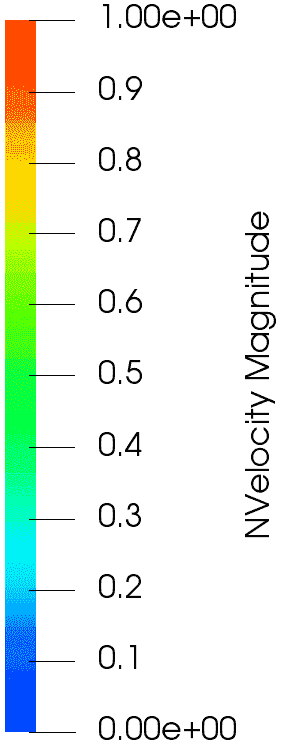}\includegraphics[width=0.4\textwidth]{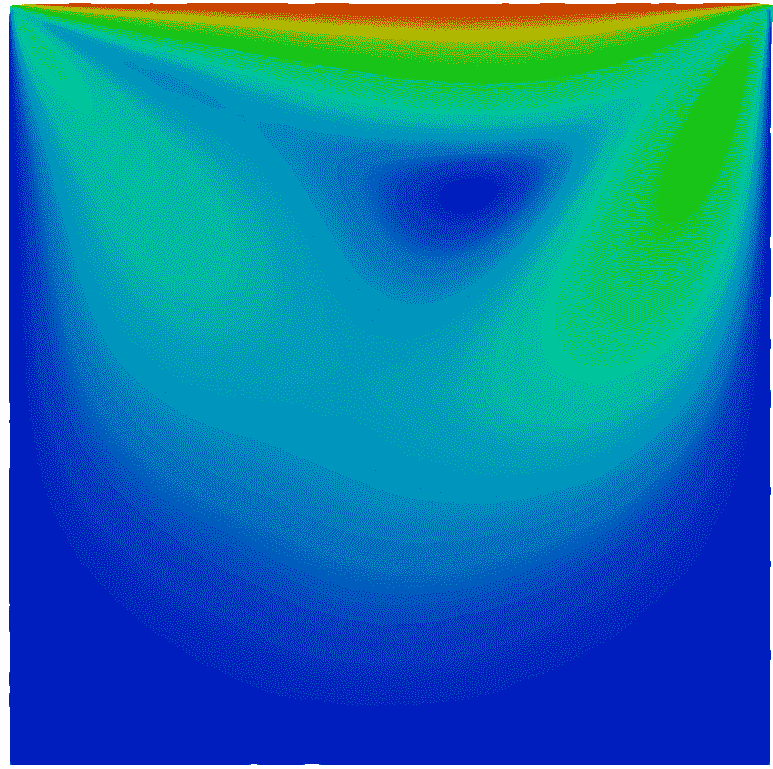}\includegraphics[width=0.4\textwidth]{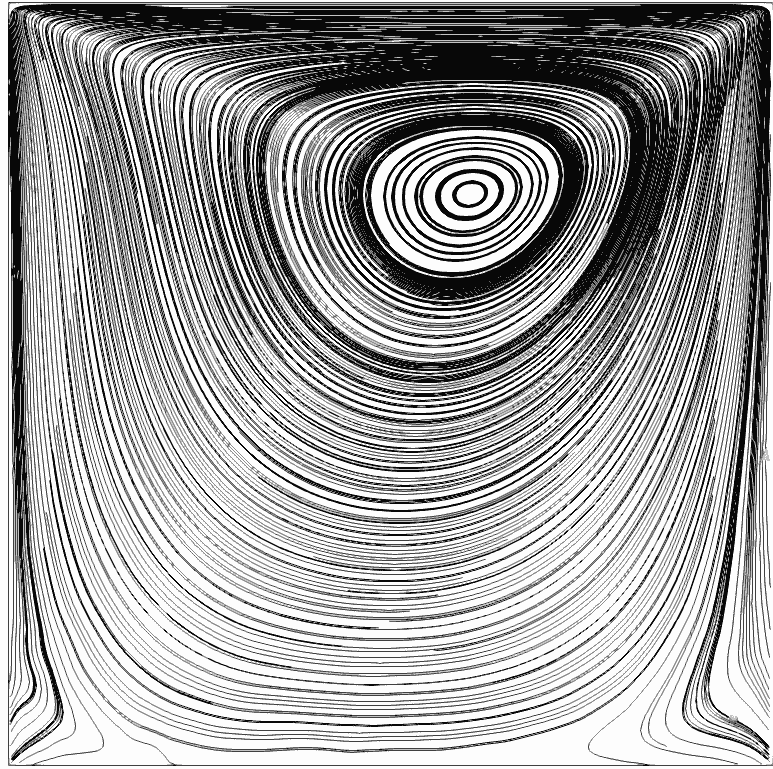}}\\
		\caption {SPH results for $Re=100$}
		\label{re100_sph}
	\end{minipage}\hfill
	\begin{minipage}{0.45\textwidth}	
		\captionsetup[subfigure]{labelformat=empty}
		\centering
		\subfloat[{\tiny $50\times 50$ - Velocity (left) - Streamlines (right)}] {\includegraphics[width=0.12\textwidth]{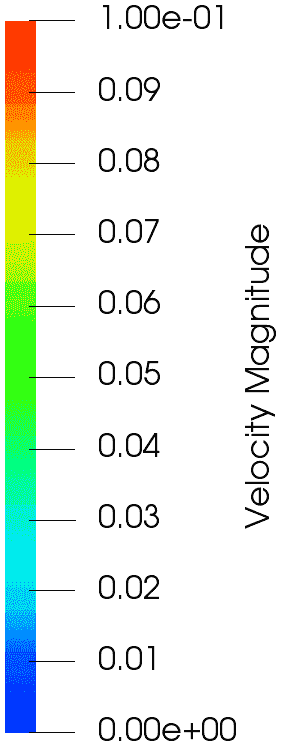}\includegraphics[width=0.4\textwidth]{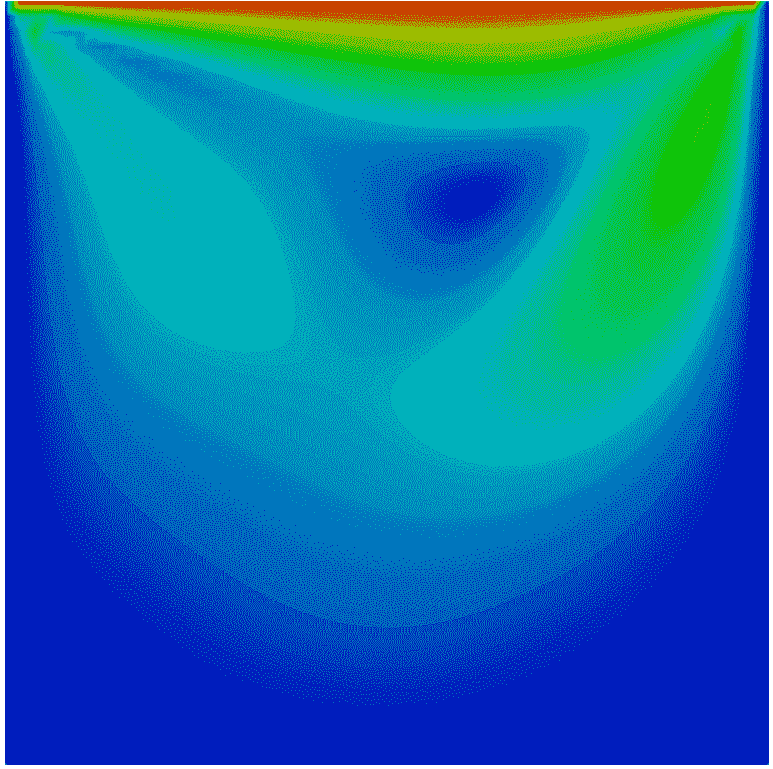}\includegraphics[width=0.4\textwidth]{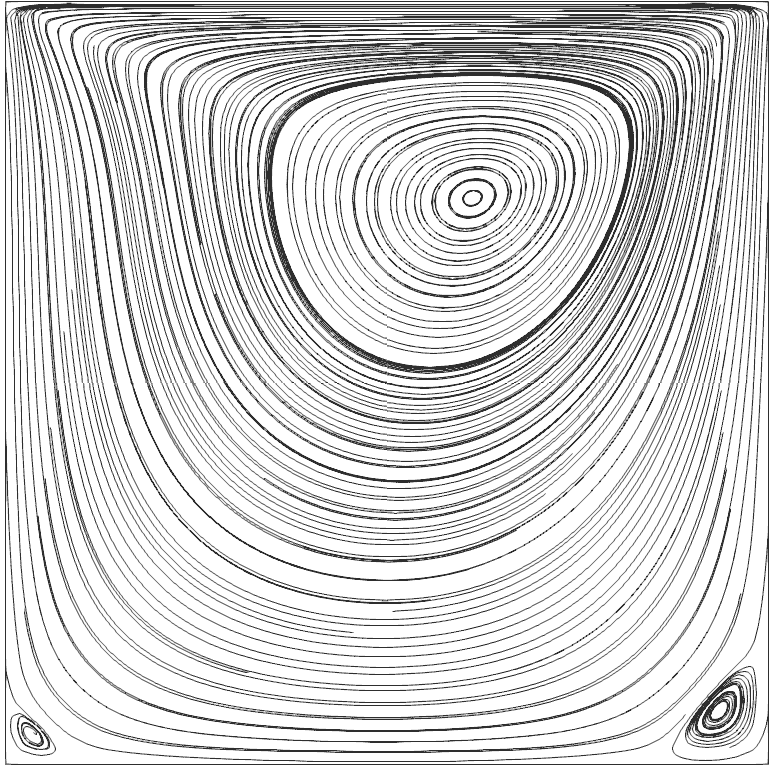}}\\
		\subfloat[{\tiny $100\times 100$ - Velocity (left) - Streamlines (right)}] {\includegraphics[width=0.12\textwidth]{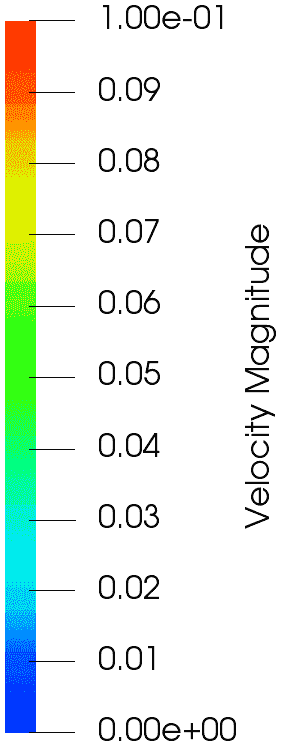}\includegraphics[width=0.4\textwidth]{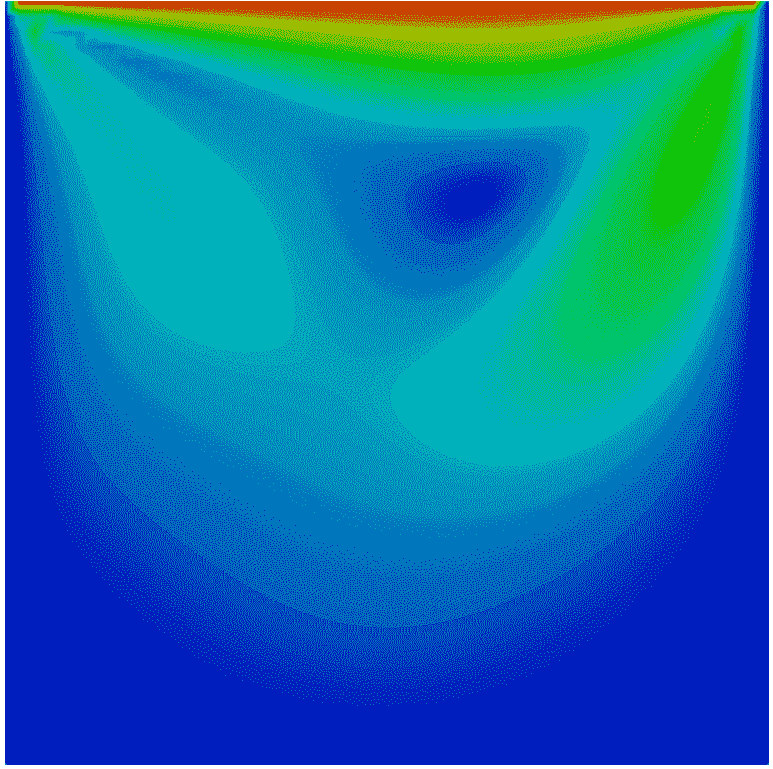}\includegraphics[width=0.4\textwidth]{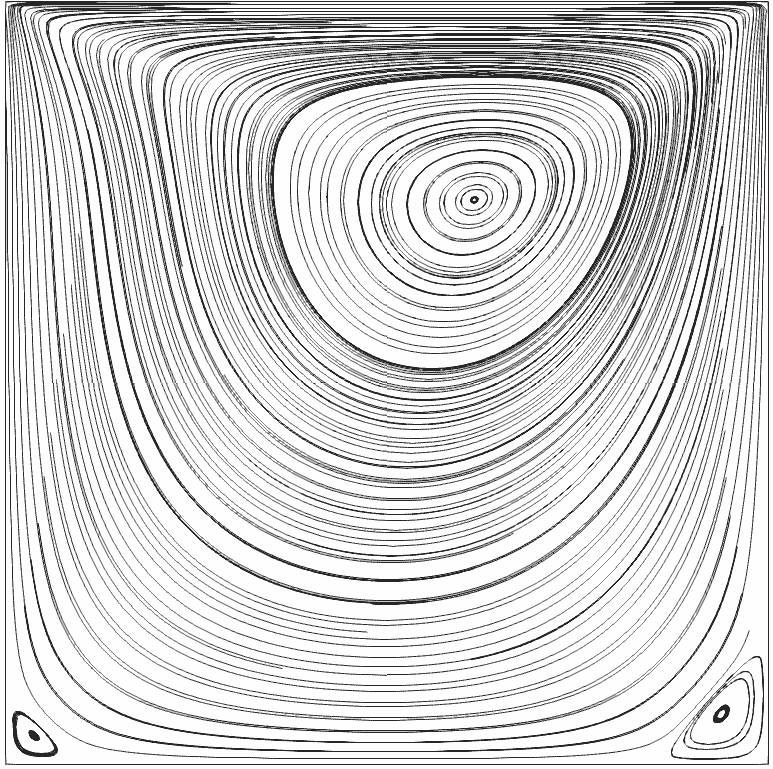}}\\	
		\subfloat[{\tiny $200\times 200$ - Velocity (left) - Streamlines (right)}] {\includegraphics[width=0.12\textwidth]{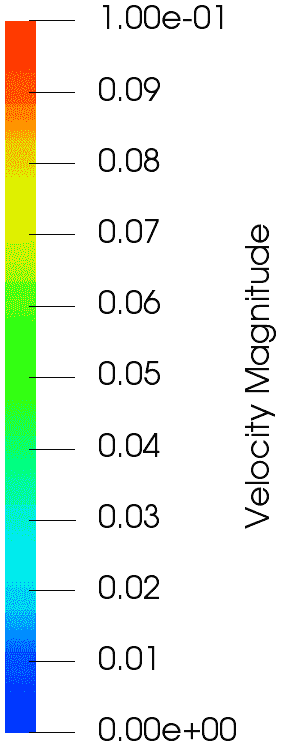}\includegraphics[width=0.4\textwidth]{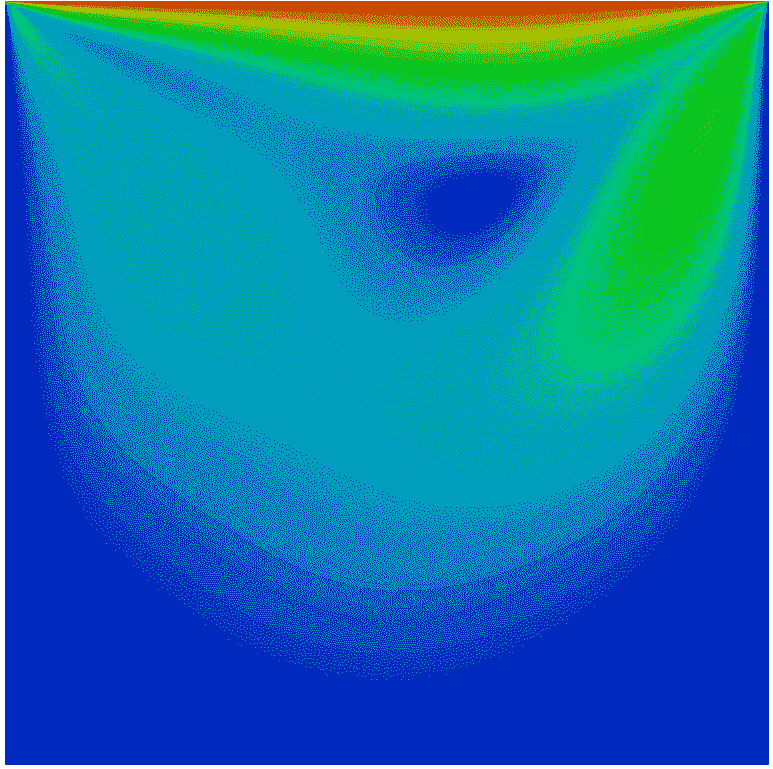}\includegraphics[width=0.4\textwidth]{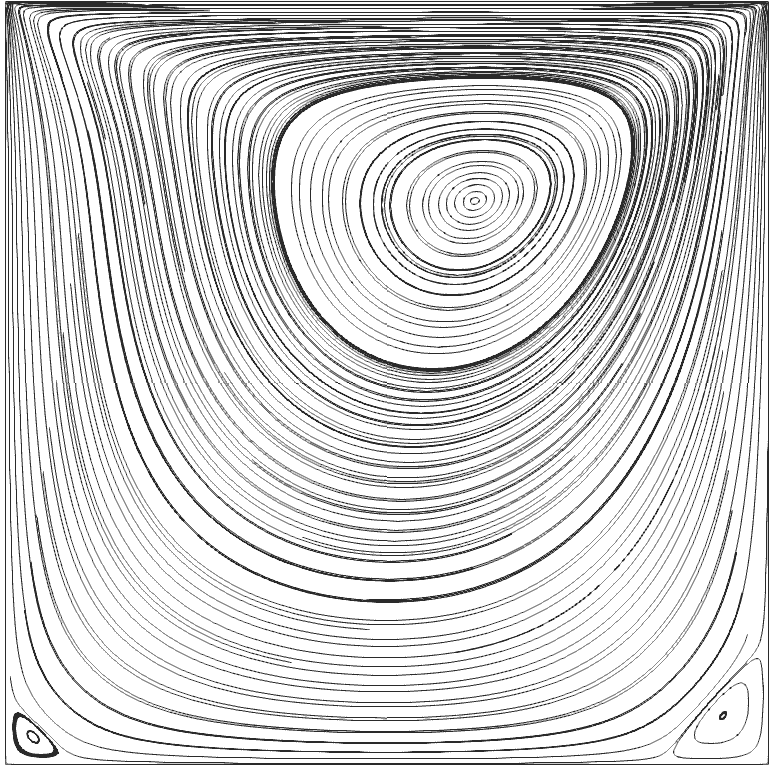}}\\
		\caption {LBM results for $Re=100$}
		\label{re100_lbm}
	\end{minipage}	
\end{figure}

When $Re=100$, the MRT operator for LBM with the standard relaxation times $\bm{S}$ is able to simulate the test case for all grid resolutions that were considered. As shown in Fig.~\ref{re_100}, both LBM and SPH are able to reproduce the velocity field more and more accurately as the lattice/particles resolution is increased. However, LBM always present a higher order of convergence ($\approx 2$ times faster). Moreover, LBM is the method that offers the best accuracy compared with Ghia et al.'s solution with an $L_2$ discrepancy of $\leq 0.025$ for the $200\times 200$ lattice resolution. On the other hand, the SPH method shows a higher $L_2$ discrepancy (at the boundaries in particular) with a maximum discrepancy of $\lesssim 0.06$ for the $200\times 200$ particles resolution.

Concerning the spatial distribution of the flow shown in Figs.~\ref{re100_lbm} and \ref{re100_sph}, LBM shows the appearance of two vortexes at the two bottom corners of the domain which is in accordance with the theory. On the contrary, SPH is not able to reproduce those two vertexes but instead has flow perturbations in the concerned areas.

In fact the two expected vertexes at the corners are  appearing during the SPH simulations but they are highly unstable. They keep forming (together or independently) and vanishing as the simulation progresses. It indicates that SPH captures an instability in the correct areas but fails to reach a steady state thus the formation of spurious perturbations. Those vertexes being of small intensity, their formation is probably affected by the boundary conditions.

\begin{figure}[bthp]
	\captionsetup[subfigure]{labelformat=empty}
	\centering
	\subfloat[{\tiny $50\times 50$ - $t=2.48\second$}] {\includegraphics[width=0.18\textwidth]{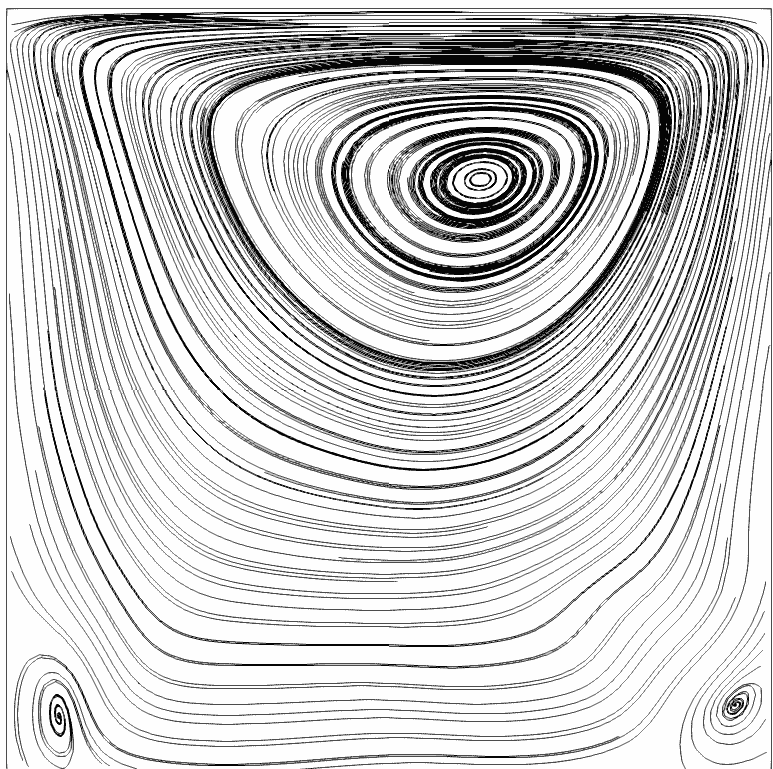}}\hspace{0.5cm}
	\subfloat[{\tiny $100\times 100$ - $t=46.66\second$}] {\includegraphics[width=0.18\textwidth]{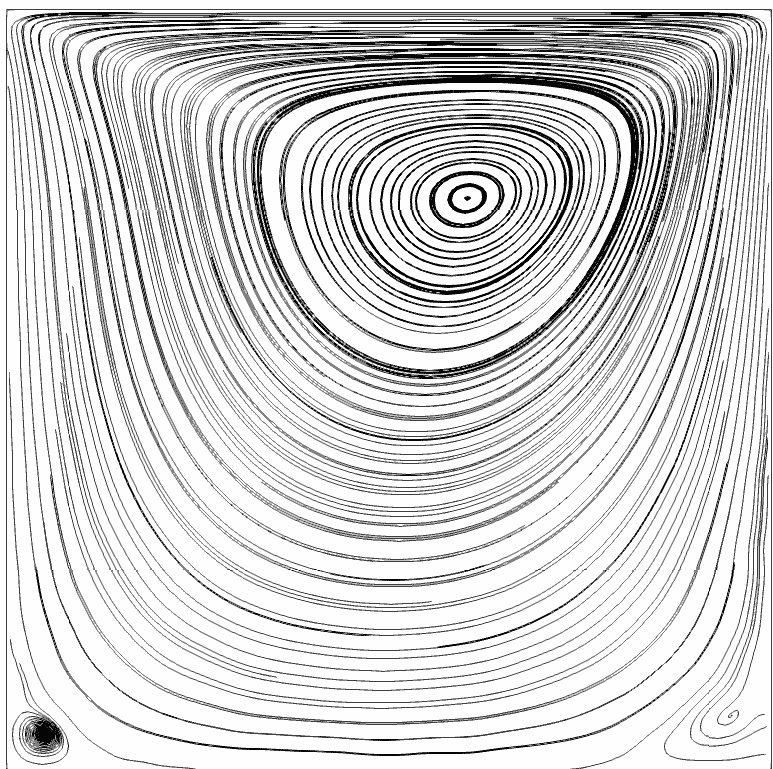}}\hspace{0.5cm}
	\subfloat[{\tiny $200\times 200$ - $t=54.31\second$}] {\includegraphics[width=0.18\textwidth]{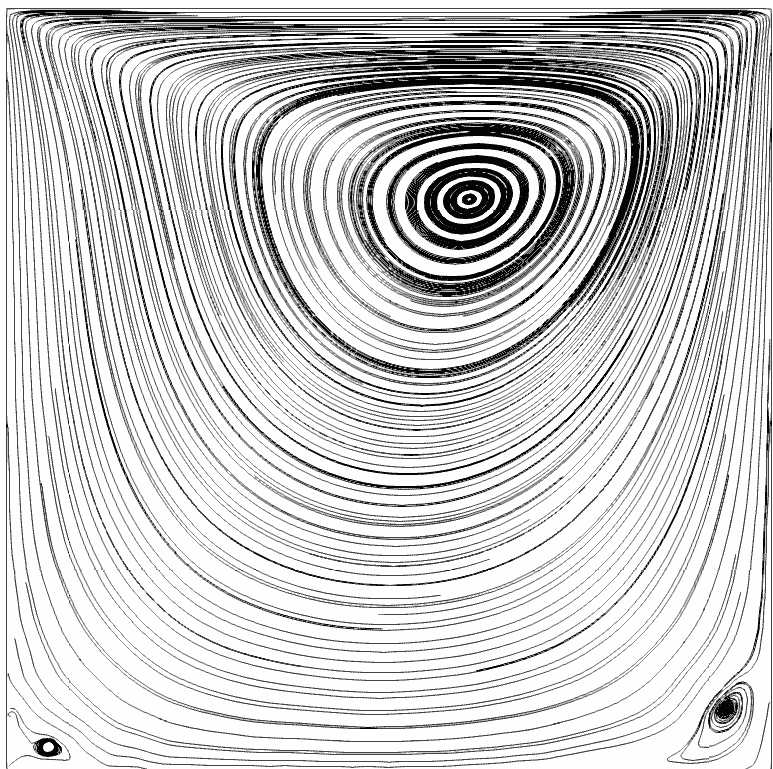}}
	\caption {SPH streamlines for $Re=100$ at selected timesteps}
	\label{re100_sl_sph}
\end{figure}

In Fig.~\ref{re100_sl_sph}, one can note that the two expected vertexes at the corners are in fact appearing during the SPH simulations but they are highly unstable. They keep forming (together or independently) and vanishing as the simulation progresses. It indicates that SPH captures an instability in the correct areas but fails to reach a steady state thus the formation of spurious perturbations. Those vertexes being of small intensity, their formation is probably affected by the boundary conditions.

\begin{figure}[bthp]
	\captionsetup[subfigure]{labelformat=empty}
	\centering
	\subfloat[] {\includegraphics[width=0.09\textwidth]{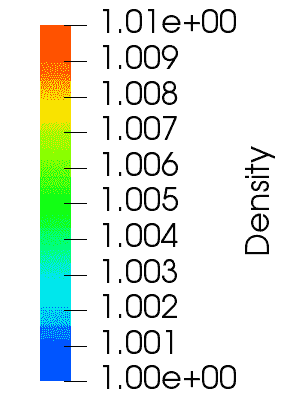}}
	\subfloat[{\tiny LBM - $200\times 200$}] {\includegraphics[width=0.2\textwidth]{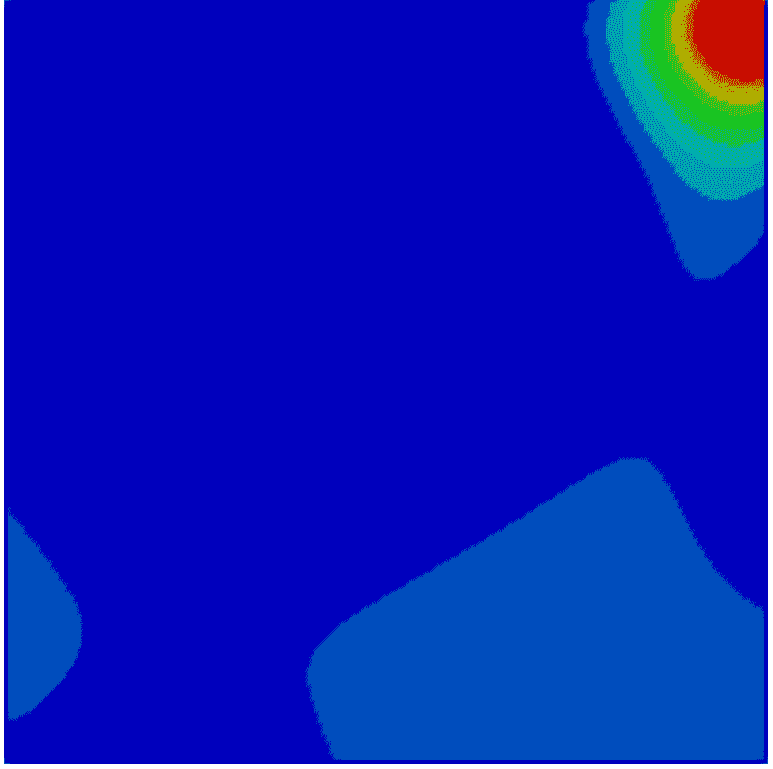}}\hspace{0.5cm}
	\subfloat[] {\includegraphics[width=0.09\textwidth]{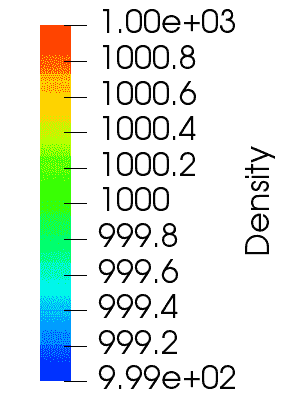}}
	\subfloat[{\tiny SPH - $200\times 200$}] {\includegraphics[width=0.2\textwidth]{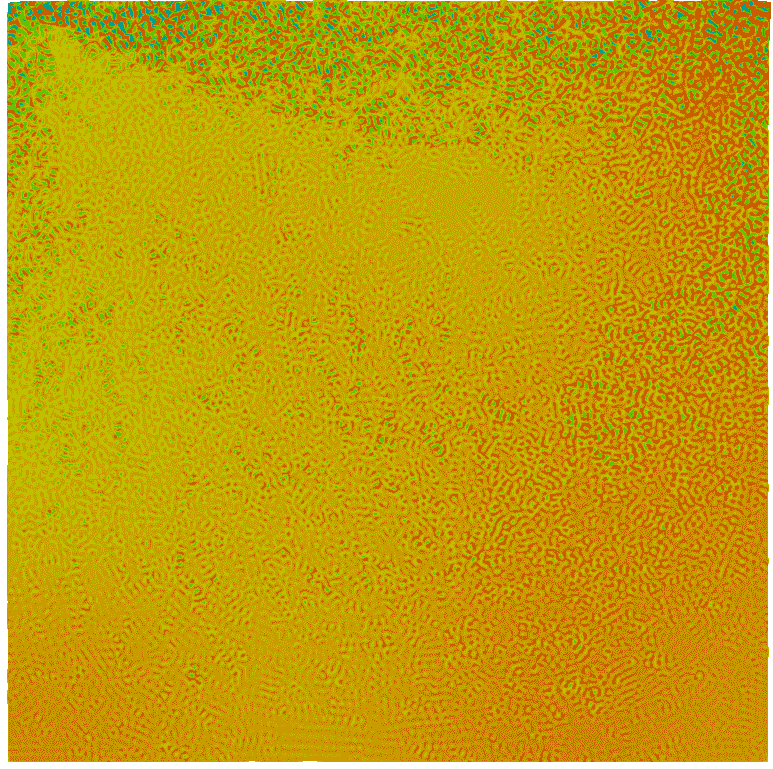}}
	\caption {Density fields for $Re=100$}
	\label{density_lbm_sph}
\end{figure}

The density fields of the two methods for $Re=100$ in Fig.~\ref{density_lbm_sph} show that LBM has a smoother density field compared with SPH. As expected, due to the choice to use the weakly compressible approach, SPH presents a noisy density field. It is expected that an incompressible approach (ISPH) with a Poisson solver would improve the quality of the density (and thus pressure) field. These observations are valid for all four Reynolds numbers studied in this section.

\begin{figure}[bthp]
	\begin{center}
		\captionsetup[subfigure]{labelformat=empty}
		\makebox[\textwidth][c]{
			\subfloat[]{\resizebox{0.5\textwidth}{!}{\input{Re100_bis}}}
			\subfloat[]{\resizebox{0.5\textwidth}{!}{%\pgfplotsset{label style={font=\tiny},
%	tick label style={font=\tiny} }
%
\definecolor{mycolor1}{rgb}{0.00,0.00,1.00}%
\definecolor{mycolor2}{rgb}{0.00,0.50,0.00}%
\definecolor{mycolor3}{rgb}{1.00,0.00,0.00}%
\definecolor{mycolor4}{rgb}{0.00,0.75,0.75}%
\definecolor{mycolor5}{rgb}{0.75,0.00,0.75}%
\definecolor{mycolor6}{rgb}{0.75,0.75,0.00}%
\definecolor{mycolor7}{rgb}{0.25,0.25,0.25}%
\definecolor{mycolor8}{rgb}{0.75,0.25,0.25}%
\definecolor{mycolor9}{rgb}{0.95,0.95,0.00}%
\definecolor{mycolor10}{rgb}{0.25,0.25,0.75}%
\definecolor{mycolor11}{rgb}{0.75,0.75,0.75}%
\definecolor{mycolor12}{rgb}{0.00,1.00,0.00}%
\definecolor{mycolor13}{rgb}{0.76,0.57,0.17}%
\definecolor{mycolor14}{rgb}{0.54,0.63,0.22}%
\definecolor{mycolor15}{rgb}{0.34,0.57,0.92}%
\definecolor{mycolor16}{rgb}{1.00,0.10,0.60}%
\definecolor{mycolor17}{rgb}{0.88,0.75,0.73}%
\definecolor{mycolor18}{rgb}{0.10,0.49,0.47}%
\definecolor{mycolor19}{rgb}{0.66,0.34,0.65}%
\definecolor{mycolor20}{rgb}{0.99,0.41,0.23}%
\begin{tikzpicture}

\begin{axis}[%
scaled ticks=false, 
tick label style={/pgf/number format/fixed},
xmajorgrids=false,
ymajorgrids=true,
grid style={dotted,gray},
width=3.0in,
height=1.5in,
at={(0in,0in)},
scale only axis,
xmode=log,
xmin=1000,
xmax=100000,
xminorticks=true,
yminorgrids=true,
ymode=log,
ymin=0.01,
ymax=0.15,
yminorticks=true,
ylabel near ticks,
xlabel near ticks,
xtick pos=left,
ytick pos=left,
xlabel={Number of nodes/particles},
ylabel={$L_2$ Discrepancy},
axis background/.style={fill=white},
legend style={legend style={nodes={scale=0.75, transform shape}},fill=white,align=left,draw=none,at={(0.65,0.25)}}
]
%\addplot [color=black,solid]
%  table[row sep=crcr]{%
%2500	0.115277146902742\\
%10000	0.074827381516425\\
%40000	0.06008943746863\\
%};
%\addlegendentry{SPH};
\addplot [color=mycolor4,only marks,mark=star,mark size=0.85, mark repeat=2, forget plot]
table[row sep=crcr]{%
	2500	0.115277146902742\\
};
\addplot [color=mycolor5,only marks,mark=square,mark size=0.85, mark repeat=10, forget plot]
table[row sep=crcr]{%
	10000	0.074827381516425\\
};
\addplot [color=mycolor6,only marks,mark=diamond,mark size=0.85, mark repeat=20, forget plot]
table[row sep=crcr]{%
	40000	0.06008943746863\\
};
%\addplot [color=black,densely dashed]
%table[row sep=crcr]{%
%	2500	0.102399879083338\\
%	10000	0.041930184458453\\
%	40000	0.0231505197968801\\
%};
%\addlegendentry{LBM};
\addplot [color=mycolor1,only marks,mark=x,mark options={solid},mark size=0.85, mark repeat=5, forget plot]
table[row sep=crcr]{%
	2500	0.102399879083338\\
};
\addplot [color=mycolor2,only marks,mark=o,mark options={solid},mark size=0.85, mark repeat=10, forget plot]
table[row sep=crcr]{%
	10000	0.041930184458453\\
};
\addplot [color=mycolor3,only marks,mark=triangle,mark options={solid,rotate=90},mark size=0.85, mark repeat=20, forget plot]
table[row sep=crcr]{%
	40000	0.0231505197968801\\
};
\addplot[color=black,solid, domain=1000:100000, samples=100, smooth] 
plot (\x, { (\x)^(-0.23498) *exp(-0.35736) } );
\addlegendentry{SPH ($S=-0.235$)};
\addplot[color=black,densely dashed, domain=1000:100000, samples=100, smooth] 
plot (\x, { (\x)^(-0.53628) *exp(1.86716) } );
\addlegendentry{LBM ($S=-0.536$)};
\end{axis}
\end{tikzpicture}%
}}}\\
		\makebox[\textwidth][c]{
			\subfloat[]{\resizebox{0.5\textwidth}{!}{\input{Re100_2_bis}}}
			\subfloat[]{\resizebox{0.5\textwidth}{!}{%\pgfplotsset{label style={font=\tiny},
%	tick label style={font=\tiny} }
%
\definecolor{mycolor1}{rgb}{0.00,0.00,1.00}%
\definecolor{mycolor2}{rgb}{0.00,0.50,0.00}%
\definecolor{mycolor3}{rgb}{1.00,0.00,0.00}%
\definecolor{mycolor4}{rgb}{0.00,0.75,0.75}%
\definecolor{mycolor5}{rgb}{0.75,0.00,0.75}%
\definecolor{mycolor6}{rgb}{0.75,0.75,0.00}%
\definecolor{mycolor7}{rgb}{0.25,0.25,0.25}%
\definecolor{mycolor8}{rgb}{0.75,0.25,0.25}%
\definecolor{mycolor9}{rgb}{0.95,0.95,0.00}%
\definecolor{mycolor10}{rgb}{0.25,0.25,0.75}%
\definecolor{mycolor11}{rgb}{0.75,0.75,0.75}%
\definecolor{mycolor12}{rgb}{0.00,1.00,0.00}%
\definecolor{mycolor13}{rgb}{0.76,0.57,0.17}%
\definecolor{mycolor14}{rgb}{0.54,0.63,0.22}%
\definecolor{mycolor15}{rgb}{0.34,0.57,0.92}%
\definecolor{mycolor16}{rgb}{1.00,0.10,0.60}%
\definecolor{mycolor17}{rgb}{0.88,0.75,0.73}%
\definecolor{mycolor18}{rgb}{0.10,0.49,0.47}%
\definecolor{mycolor19}{rgb}{0.66,0.34,0.65}%
\definecolor{mycolor20}{rgb}{0.99,0.41,0.23}%
\begin{tikzpicture}

\begin{axis}[%
scaled ticks=false, 
tick label style={/pgf/number format/fixed},
xmajorgrids=false,
ymajorgrids=true,
grid style={dotted,gray},
width=3.0in,
height=1.5in,
at={(0in,0in)},
scale only axis,
xmode=log,
xmin=1000,
xmax=100000,
xminorticks=true,
yminorgrids=true,
ymode=log,
ymin=0.01,
ymax=0.15,
yminorticks=true,
ylabel near ticks,
xlabel near ticks,
xtick pos=left,
ytick pos=left,
xlabel={Number of nodes/particles},
ylabel={$L_2$ Discrepancy},
axis background/.style={fill=white},
legend style={legend style={nodes={scale=0.75, transform shape}},fill=white,align=left,draw=none,at={(0.975,0.95)}}
]
%\addplot [color=black,solid]
%  table[row sep=crcr]{%
%2500	0.0756107874596331\\
%10000	0.0388494835217732\\
%40000	0.0252272833424413\\
%};
%\addlegendentry{SPH};
\addplot [color=mycolor4,only marks,mark=star,mark size=0.85, mark repeat=2, forget plot]
table[row sep=crcr]{%
	2500	0.0756107874596331\\
};
\addplot [color=mycolor5,only marks,mark=square,mark size=0.85, mark repeat=10, forget plot]
table[row sep=crcr]{%
	10000	0.0388494835217732\\
};
\addplot [color=mycolor6,only marks,mark=diamond,mark size=0.85, mark repeat=20, forget plot]
table[row sep=crcr]{%
	40000	0.0252272833424413\\
};
%\addplot [color=black,densely dashed]
%table[row sep=crcr]{%
%	2500	0.120381816940327\\
%	10000	0.051756144648029\\
%	40000	0.0209942646175514\\
%};
%\addlegendentry{LBM};
\addplot [color=mycolor1,only marks,mark=x,mark options={solid},mark size=0.85, mark repeat=5, forget plot]
table[row sep=crcr]{%
	2500	0.120381816940327\\
};
\addplot [color=mycolor2,only marks,mark=o,mark options={solid},mark size=0.85, mark repeat=10, forget plot]
table[row sep=crcr]{%
	10000	0.051756144648029\\
};
\addplot [color=mycolor3,only marks,mark=triangle,mark options={solid,rotate=90},mark size=0.85, mark repeat=20, forget plot]
table[row sep=crcr]{%
	40000	0.0209942646175514\\
};
\addplot[color=black,solid, domain=1000:100000, samples=100, smooth] 
plot (\x, { (\x)^(-0.3959) *exp(0.47638) } );
\addlegendentry{SPH ($S=-0.396$)};
\addplot[color=black,densely dashed, domain=1000:100000, samples=100, smooth] 
plot (\x, { (\x)^(-0.62989) *exp(2.82088) } );
\addlegendentry{LBM ($S=-0.630$)};
\end{axis}
\end{tikzpicture}%
}}}
		\caption {$Re=100$}
		\label{re_100}
	\end{center}
\end{figure}

\subsubsection{$Re=400$}

\begin{figure}[bthp]
	\begin{minipage}{0.45\textwidth}
		\captionsetup[subfigure]{labelformat=empty}
		\centering
		\subfloat[{\tiny $50\times 50$ - Velocity (left) - Streamlines (right)}] {\includegraphics[width=0.12\textwidth]{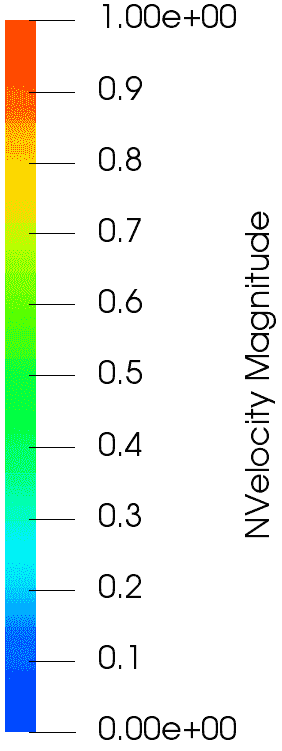}\includegraphics[width=0.4\textwidth]{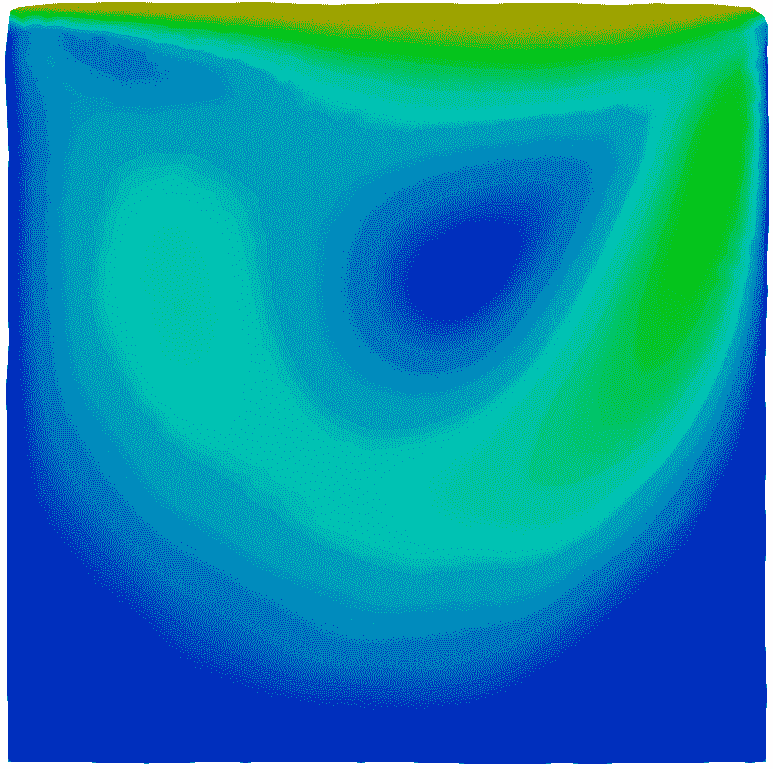}\includegraphics[width=0.4\textwidth]{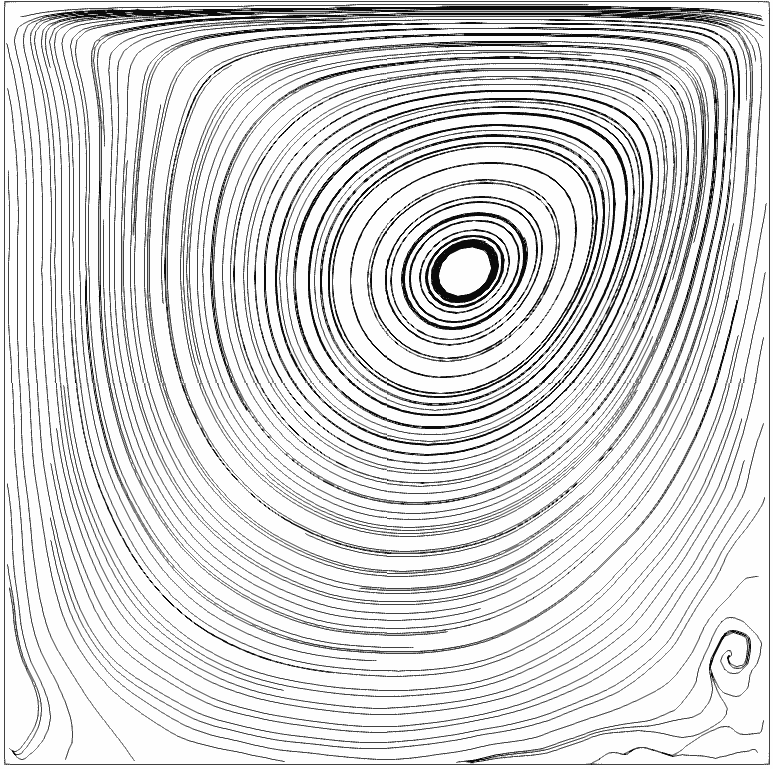}}\\
		\subfloat[{\tiny $100\times 100$ - Velocity (left) - Streamlines (right)}] {\includegraphics[width=0.12\textwidth]{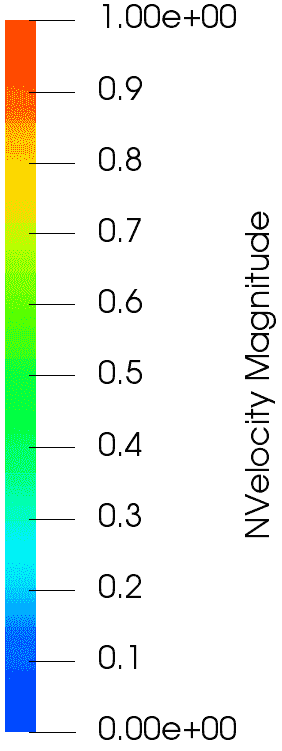}\includegraphics[width=0.4\textwidth]{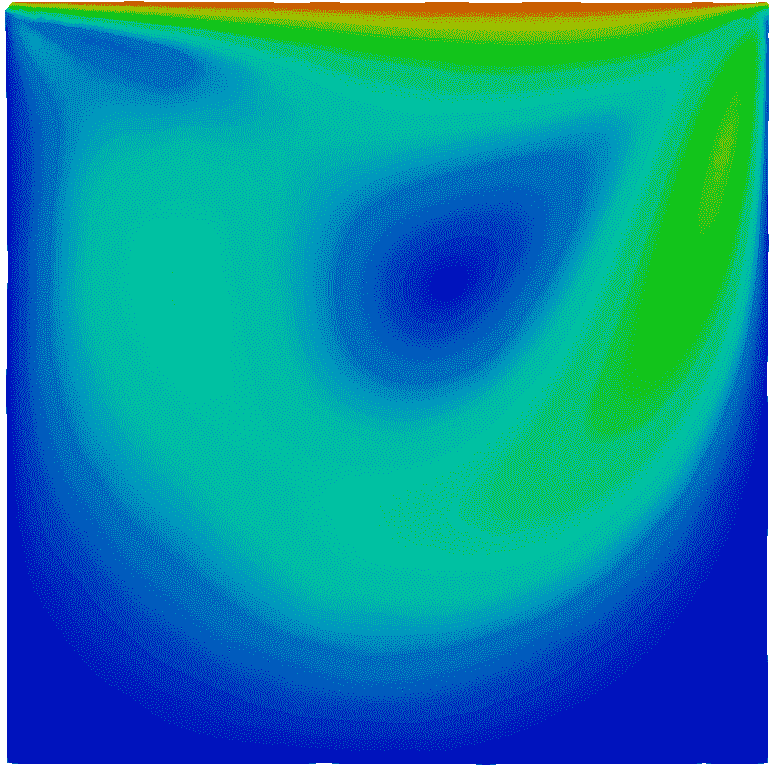}\includegraphics[width=0.4\textwidth]{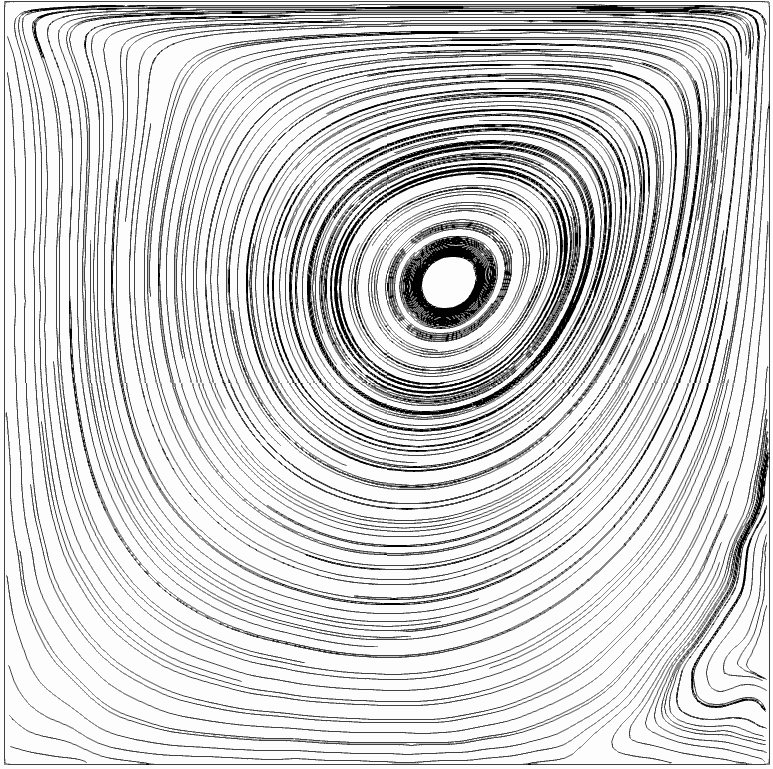}}\\	
		\subfloat[{\tiny $200\times 200$ - Velocity (left) - Streamlines (right)}] {\includegraphics[width=0.12\textwidth]{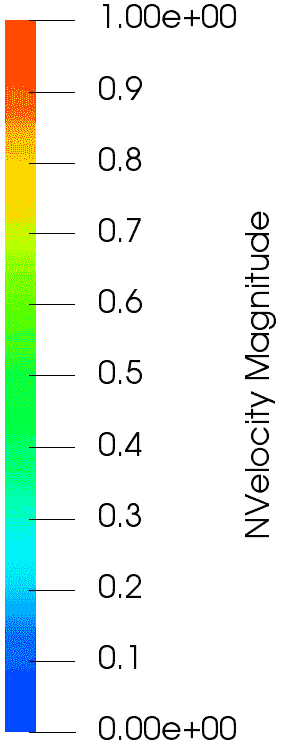}\includegraphics[width=0.4\textwidth]{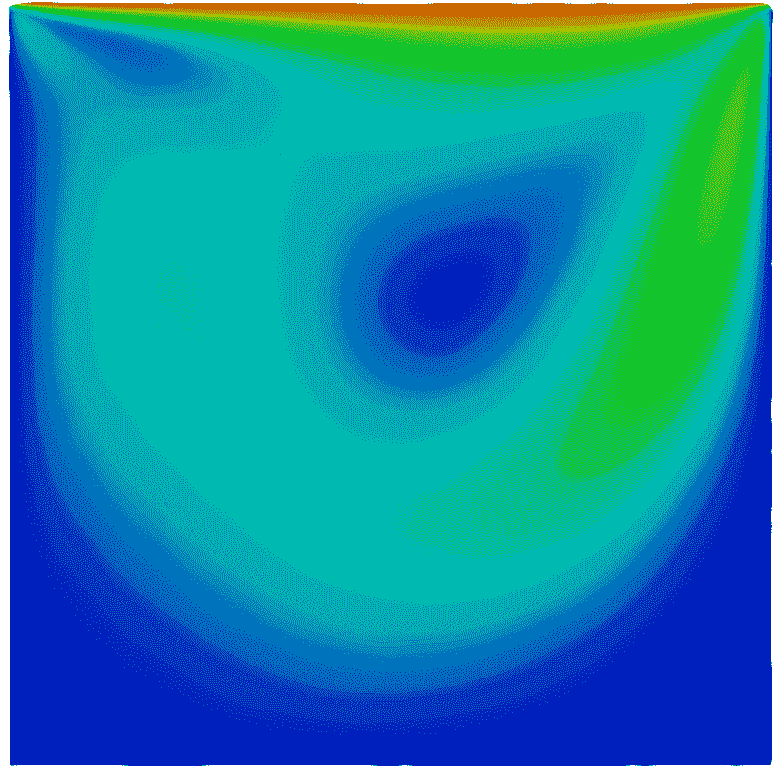}\includegraphics[width=0.4\textwidth]{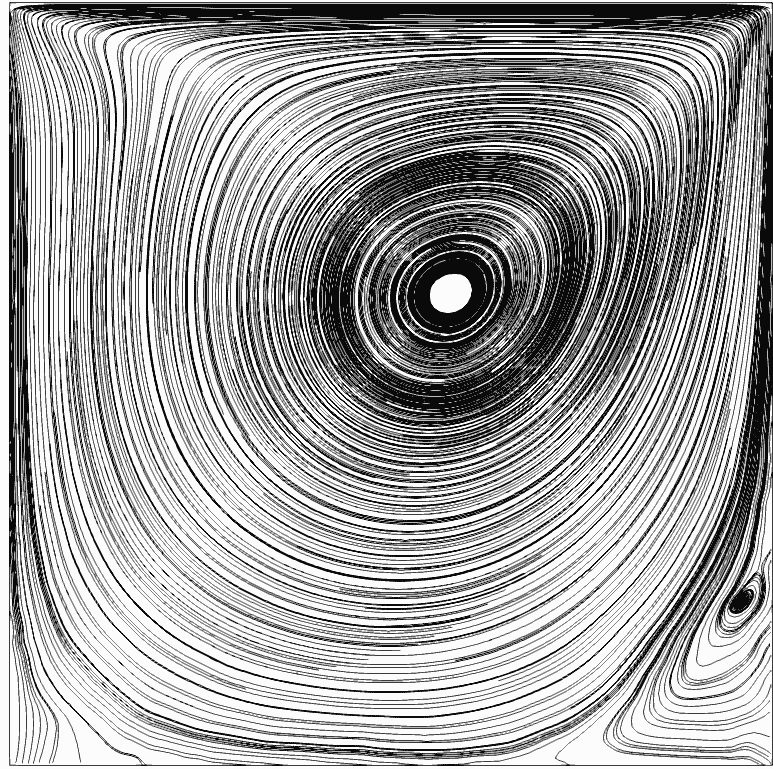}}\\
		\caption {SPH results for $Re=400$}
		\label{re400_sph}
	\end{minipage}\hfill
	\begin{minipage}{0.45\textwidth}	
		\captionsetup[subfigure]{labelformat=empty}
		\centering
		\subfloat[{\tiny $50\times 50$ - Velocity (left) - Streamlines (right)}] {\includegraphics[width=0.12\textwidth]{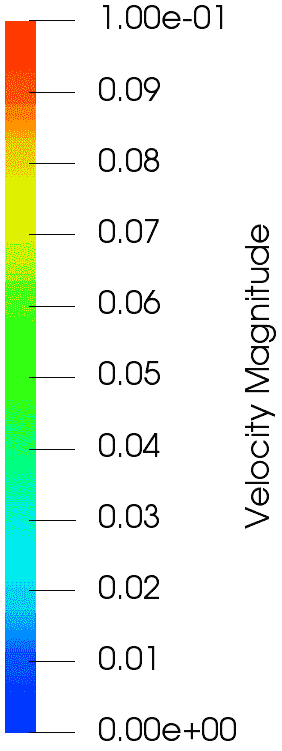}\includegraphics[width=0.4\textwidth]{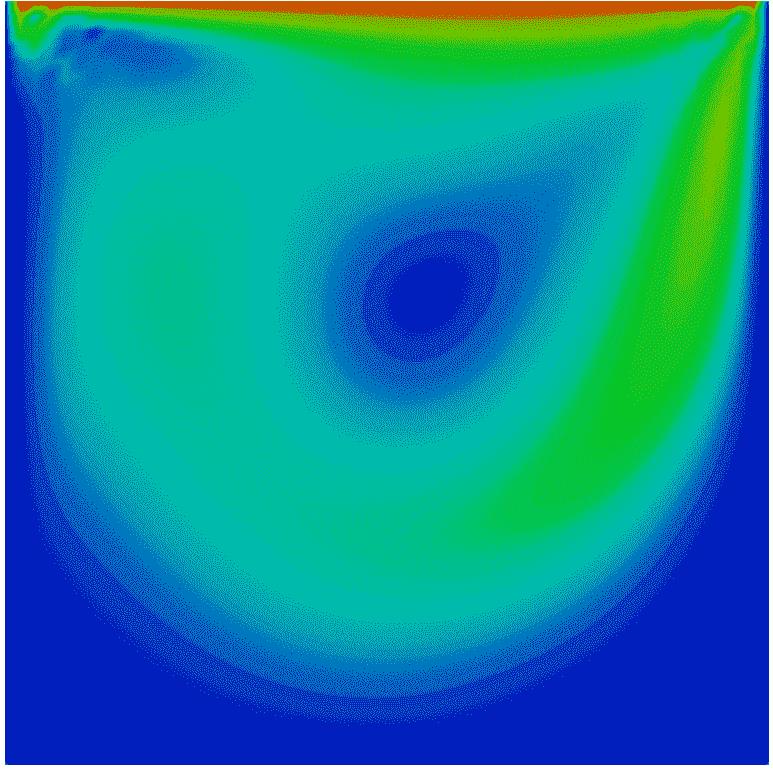}\includegraphics[width=0.4\textwidth]{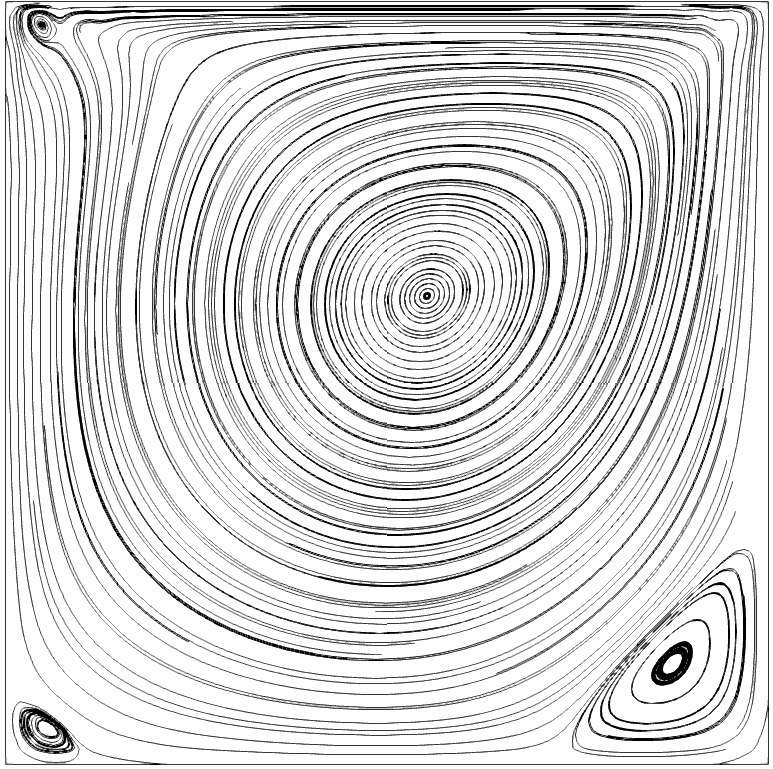}}\\
		\subfloat[{\tiny $100\times 100$ - Velocity (left) - Streamlines (right)}] {\includegraphics[width=0.12\textwidth]{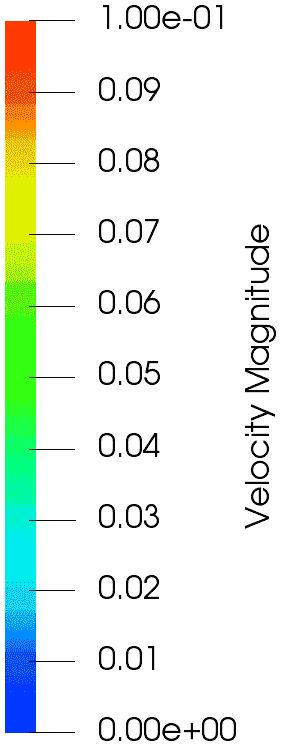}\includegraphics[width=0.4\textwidth]{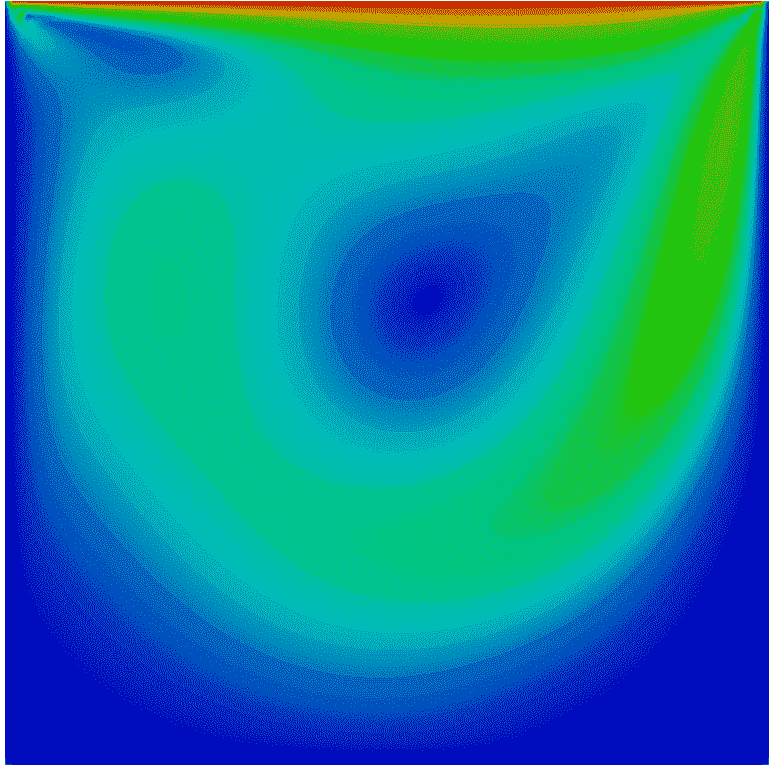}\includegraphics[width=0.4\textwidth]{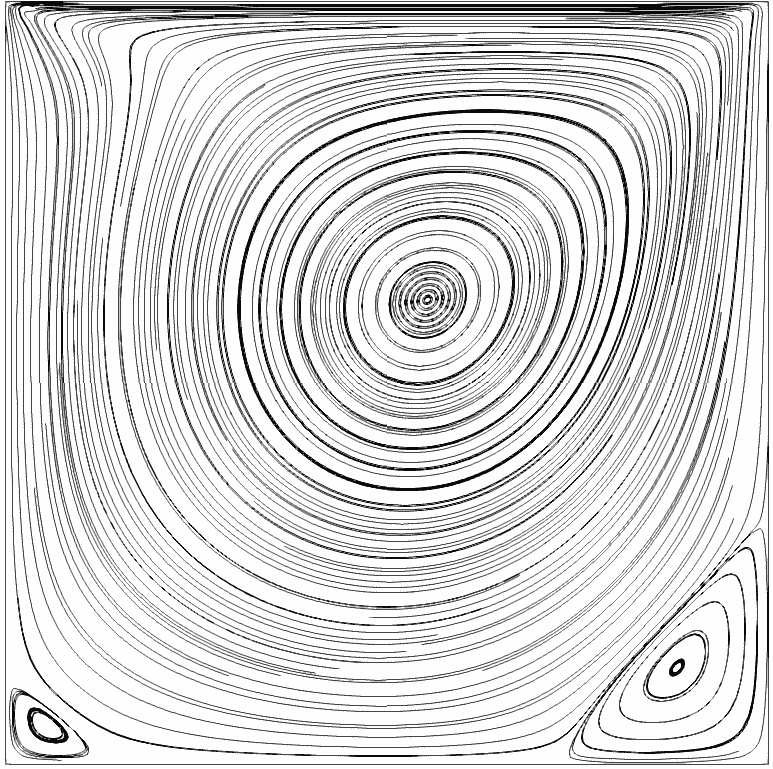}}\\	
		\subfloat[{\tiny $200\times 200$ - Velocity (left) - Streamlines (right)}] {\includegraphics[width=0.12\textwidth]{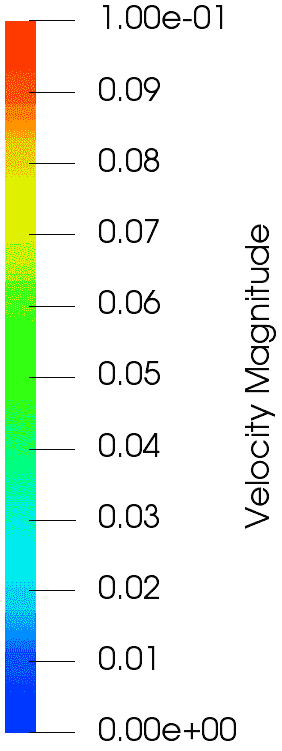}\includegraphics[width=0.4\textwidth]{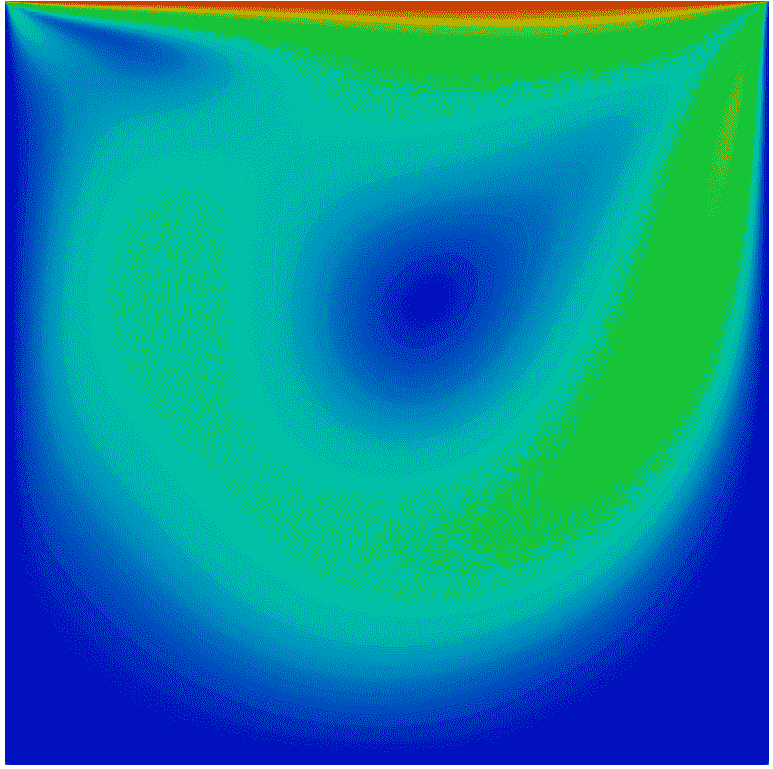}\includegraphics[width=0.4\textwidth]{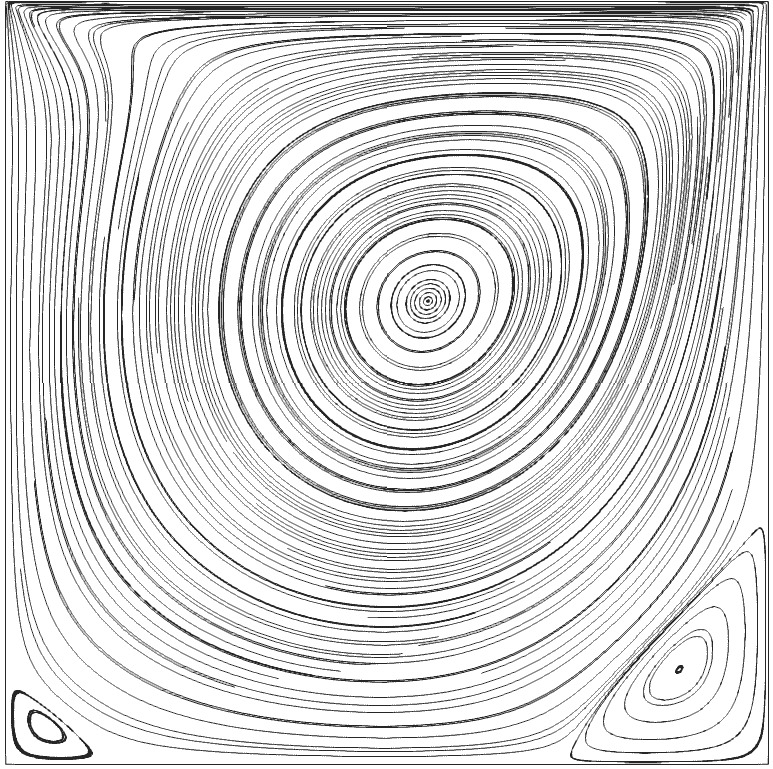}}\\
		\caption {LBM results for $Re=400$}
		\label{re400_lbm}
	\end{minipage}	
\end{figure}

For $Re=400$, the MRT operator for LBM with the standard relaxation times $\bm{S}$ is able to simulate the test case for all grid resolutions that were considered.

The superiority of LBM in terms of accuracy is magnified in that case. The LBM method shows more accurate results that SPH for all resolutions considered as shown in Fig.~\ref{re_400}. The maximum discrepancy is always located at the domain's boundaries. As an example, for the $200\times 200$ resolution, both LBM and SPH have a maximum difference with Ghia's reference $\geq 8 \%$ at the right boundary whereas it is $\leq 2 \%$ and $\leq 5 \%$ inside the domain for LBM and SPH respectively.

Once again, LBM shows a better global accuracy for the same resolution and a higher order of convergence than SPH as shown in Fig.~\ref{re_400}. In particular, for the $200\times 200$ resolution, the LBM $L_2$ discrepancy on $V_x$ along the vertical centerline is more than $3$ times lower than the SPH one. For $V_y$ along the horizontal centerline, both methods have a comparable discrepancy.

The streamlines plots of Figs.~\ref{re400_lbm} and \ref{re400_sph} are showing that LBM correctly reproduces the existence of two vertexes at the bottom corners of the domain. For the $50 \times 50$ case, a spurious vertex appears at the top left corner of the domain and is likely due to the combination of boundary conditions (bounceback + Zou-He) at this location as it is smoothed out when the resolution increases.

\begin{figure}[bthp]
	\captionsetup[subfigure]{labelformat=empty}
	\centering
	\subfloat[{\tiny $50\times 50$ - $t=31.6s$}] {\includegraphics[width=0.18\textwidth]{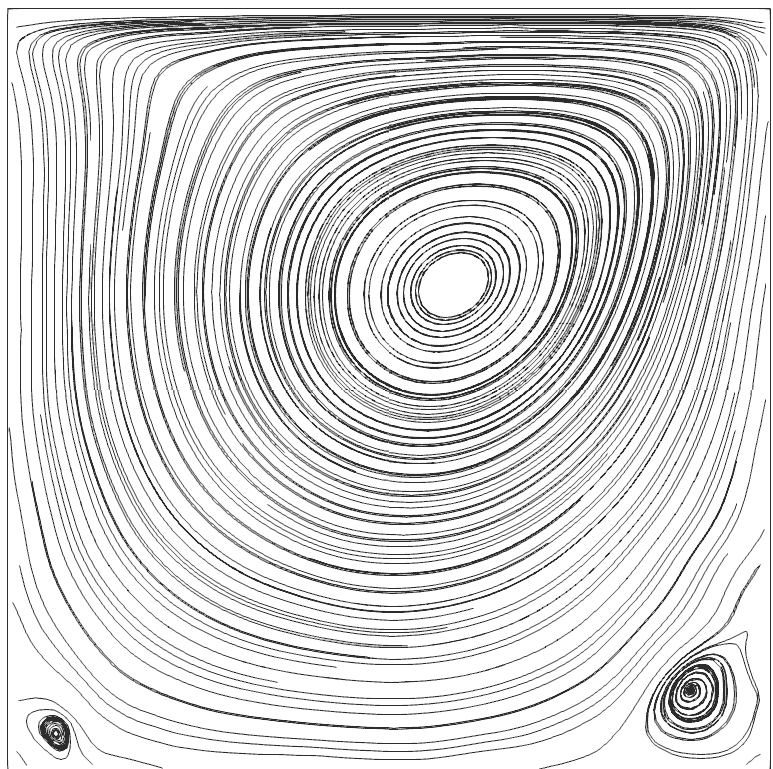}}\hspace{0.5cm}
	\subfloat[{\tiny $100\times 100$ - $t=58.17s$}] {\includegraphics[width=0.18\textwidth]{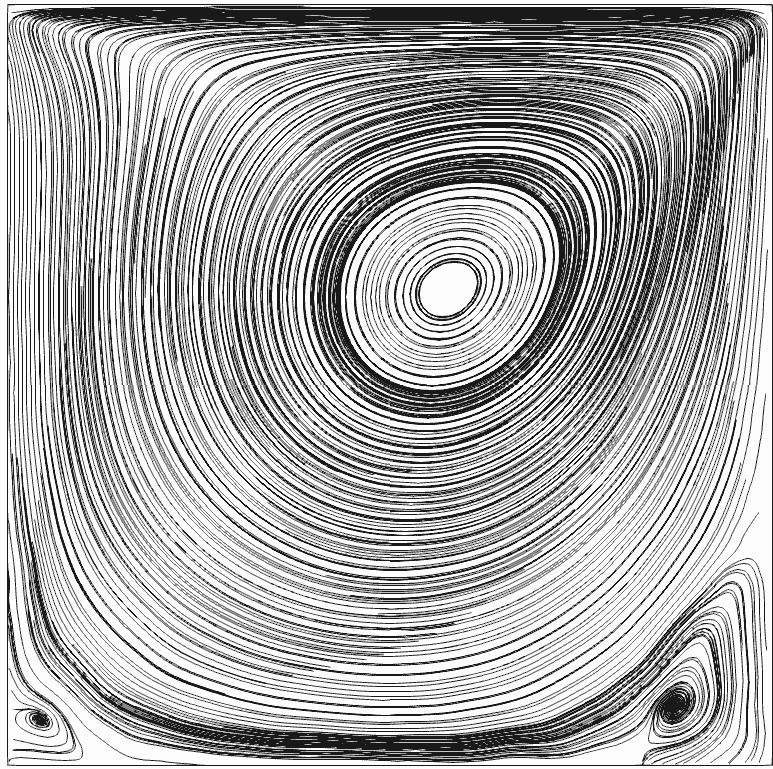}}\hspace{0.5cm}
	\subfloat[{\tiny $200\times 200$ - $t=52.6s$}] {\includegraphics[width=0.18\textwidth]{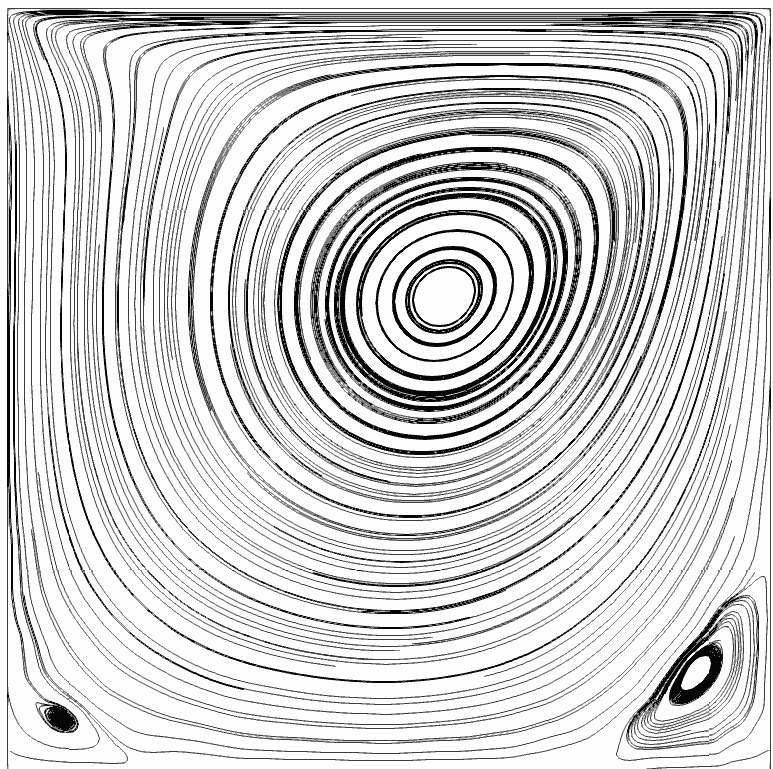}}
	\caption {SPH streamlines for $Re=400$ at selected timesteps}
	\label{re400_sl_sph}
\end{figure}

On the other hand, the SPH results are not able to simulate an established vertex pattern at the bottom corners. In the bottom right corner where the vertex is the strongest, for the $50\times 50$ and $200 \times 200$ cases, a vertex appears to be growing but is either not at the correct location or not with the correct amplitude. In fact, when looking at earlier streamlines plots as shown in Fig.~\ref{re400_sl_sph}, one can see that SPH does generate vortexes in the correct areas at selected instants during the simulation but they fail to stabilize and are continuously appearing and disappearing.

\begin{figure}[bthp]
	\captionsetup[subfigure]{labelformat=empty}
	\centering
	\makebox[\textwidth][c]{
		\subfloat[]{\resizebox{0.5\textwidth}{!}{\input{Re400}}}
		\subfloat[]{\resizebox{0.5\textwidth}{!}{%\pgfplotsset{label style={font=},
%	tick label style={font=\tiny} }
%
\definecolor{mycolor1}{rgb}{0.00,0.00,1.00}%
\definecolor{mycolor2}{rgb}{0.00,0.50,0.00}%
\definecolor{mycolor3}{rgb}{1.00,0.00,0.00}%
\definecolor{mycolor4}{rgb}{0.00,0.75,0.75}%
\definecolor{mycolor5}{rgb}{0.75,0.00,0.75}%
\definecolor{mycolor6}{rgb}{0.75,0.75,0.00}%
\definecolor{mycolor7}{rgb}{0.25,0.25,0.25}%
\definecolor{mycolor8}{rgb}{0.75,0.25,0.25}%
\definecolor{mycolor9}{rgb}{0.95,0.95,0.00}%
\definecolor{mycolor10}{rgb}{0.25,0.25,0.75}%
\definecolor{mycolor11}{rgb}{0.75,0.75,0.75}%
\definecolor{mycolor12}{rgb}{0.00,1.00,0.00}%
\definecolor{mycolor13}{rgb}{0.76,0.57,0.17}%
\definecolor{mycolor14}{rgb}{0.54,0.63,0.22}%
\definecolor{mycolor15}{rgb}{0.34,0.57,0.92}%
\definecolor{mycolor16}{rgb}{1.00,0.10,0.60}%
\definecolor{mycolor17}{rgb}{0.88,0.75,0.73}%
\definecolor{mycolor18}{rgb}{0.10,0.49,0.47}%
\definecolor{mycolor19}{rgb}{0.66,0.34,0.65}%
\definecolor{mycolor20}{rgb}{0.99,0.41,0.23}%
\begin{tikzpicture}

\begin{axis}[%
scaled ticks=false, 
tick label style={/pgf/number format/fixed},
xmajorgrids=false,
ymajorgrids=true,
grid style={dotted,gray},
width=3.0in,
height=1.5in,
at={(0in,0in)},
scale only axis,
xmode=log,
xmin=1000,
xmax=100000,
xminorticks=true,
yminorgrids=true,
ymode=log,
ymin=0.01,
ymax=1.0,
yminorticks=true,
ylabel near ticks,
xlabel near ticks,
xtick pos=left,
ytick pos=left,
xlabel={Number of nodes/particles},
ylabel={$L_2$ Discrepancy},
axis background/.style={fill=white},
legend style={legend style={nodes={scale=0.75, transform shape}},fill=white,align=left,draw=none,at={(0.95,0.95)}}
]
%\addplot [color=black,solid]
%  table[row sep=crcr]{%
%2500	0.487376956193963\\
%10000	0.131474016933709\\
%40000	0.115020743014738\\
%};
%\addlegendentry{SPH};
\addplot [color=mycolor4,only marks,mark=star,mark size=0.85, mark repeat=2, forget plot]
table[row sep=crcr]{%
	2500	0.487376956193963\\
};
\addplot [color=mycolor5,only marks,mark=square,mark size=0.85, mark repeat=10, forget plot]
table[row sep=crcr]{%
	10000	0.131474016933709\\
};
\addplot [color=mycolor6,only marks,mark=diamond,mark size=0.85, mark repeat=20, forget plot]
table[row sep=crcr]{%
	40000	0.115020743014738\\
};
%\addplot [color=black,densely dashed]
%table[row sep=crcr]{%
%	2500	0.152223629691821\\
%	10000	0.0709022706519475\\
%	40000	0.0337375911527078\\
%};
%\addlegendentry{LBM};
\addplot [color=mycolor1,only marks,mark=x,mark options={solid},mark size=0.85, mark repeat=5, forget plot]
table[row sep=crcr]{%
	2500	0.152223629691821\\
};
\addplot [color=mycolor2,only marks,mark=o,mark options={solid},mark size=0.85, mark repeat=10, forget plot]
table[row sep=crcr]{%
	10000	0.0709022706519475\\
};
\addplot [color=mycolor3,only marks,mark=triangle,mark options={solid,rotate=90},mark size=0.85, mark repeat=20, forget plot]
table[row sep=crcr]{%
	40000	0.0337375911527078\\
};
\addplot[color=black,solid, domain=1000:100000, samples=100, smooth] 
plot (\x, { (\x)^(-0.52079) *exp(3.15985) } );
\addlegendentry{SPH ($S=-0.521$)};
\addplot[color=black,densely dashed, domain=1000:100000, samples=100, smooth] 
plot (\x, { (\x)^(-0.54344) *exp(2.36596) } );
\addlegendentry{LBM ($S=-0.543$)};
\end{axis}
\end{tikzpicture}%
}}}\\
	\makebox[\textwidth][c]{
		\subfloat[]{\resizebox{0.5\textwidth}{!}{\input{Re400_2}}}
		\subfloat[]{\resizebox{0.5\textwidth}{!}{%\pgfplotsset{label style={font=},
%	tick label style={font=} }
%
\definecolor{mycolor1}{rgb}{0.00,0.00,1.00}%
\definecolor{mycolor2}{rgb}{0.00,0.50,0.00}%
\definecolor{mycolor3}{rgb}{1.00,0.00,0.00}%
\definecolor{mycolor4}{rgb}{0.00,0.75,0.75}%
\definecolor{mycolor5}{rgb}{0.75,0.00,0.75}%
\definecolor{mycolor6}{rgb}{0.75,0.75,0.00}%
\definecolor{mycolor7}{rgb}{0.25,0.25,0.25}%
\definecolor{mycolor8}{rgb}{0.75,0.25,0.25}%
\definecolor{mycolor9}{rgb}{0.95,0.95,0.00}%
\definecolor{mycolor10}{rgb}{0.25,0.25,0.75}%
\definecolor{mycolor11}{rgb}{0.75,0.75,0.75}%
\definecolor{mycolor12}{rgb}{0.00,1.00,0.00}%
\definecolor{mycolor13}{rgb}{0.76,0.57,0.17}%
\definecolor{mycolor14}{rgb}{0.54,0.63,0.22}%
\definecolor{mycolor15}{rgb}{0.34,0.57,0.92}%
\definecolor{mycolor16}{rgb}{1.00,0.10,0.60}%
\definecolor{mycolor17}{rgb}{0.88,0.75,0.73}%
\definecolor{mycolor18}{rgb}{0.10,0.49,0.47}%
\definecolor{mycolor19}{rgb}{0.66,0.34,0.65}%
\definecolor{mycolor20}{rgb}{0.99,0.41,0.23}%
\begin{tikzpicture}

\begin{axis}[%
scaled ticks=false, 
tick label style={/pgf/number format/fixed},
xmajorgrids=false,
ymajorgrids=true,
grid style={dotted,gray},
width=3.0in,
height=1.5in,
at={(0in,0in)},
scale only axis,
xmode=log,
xmin=1000,
xmax=100000,
xminorticks=true,
yminorgrids=true,
ymode=log,
ymin=0.01,
ymax=1.0,
yminorticks=true,
ylabel near ticks,
xlabel near ticks,
xtick pos=left,
ytick pos=left,
xlabel={Number of nodes/particles},
ylabel={$L_2$ Discrepancy},
axis background/.style={fill=white},
legend style={legend style={nodes={scale=0.75, transform shape}},fill=white,align=left,draw=none,at={(0.95,0.95)}}
]
%\addplot [color=black,solid]
%  table[row sep=crcr]{%
%2500	0.170296222978368\\
%10000	0.144642111999648\\
%40000	0.141072512810151\\
%};
%\addlegendentry{ SPH};
\addplot [color=mycolor4,only marks,mark=star,mark size=0.85, mark repeat=2, forget plot]
table[row sep=crcr]{%
	2500	0.170296222978368\\
};
\addplot [color=mycolor5,only marks,mark=square,mark size=0.85, mark repeat=10, forget plot]
table[row sep=crcr]{%
	10000	0.144642111999648\\
};
\addplot [color=mycolor6,only marks,mark=diamond,mark size=0.85, mark repeat=20, forget plot]
table[row sep=crcr]{%
	40000	0.141072512810151\\
};
%\addplot [color=black,densely dashed]
%table[row sep=crcr]{%
%	2500	0.216595348172944\\
%	10000	0.149112351577572\\
%	40000	0.13894866514398\\
%};
%\addlegendentry{ LBM};
\addplot [color=mycolor1,only marks,mark=x,mark options={solid},mark size=0.85, mark repeat=5, forget plot]
table[row sep=crcr]{%
	2500	0.216595348172944\\
};
\addplot [color=mycolor2,only marks,mark=o,mark options={solid},mark size=0.85, mark repeat=10, forget plot]
table[row sep=crcr]{%
	10000	0.149112351577572\\
};
\addplot [color=mycolor3,only marks,mark=triangle,mark options={solid,rotate=90},mark size=0.85, mark repeat=20, forget plot]
table[row sep=crcr]{%
	40000	0.13894866514398\\
};
\addplot[color=black,solid, domain=1000:100000, samples=100, smooth] 
plot (\x, { (\x)^(-0.0679) *exp(-1.26199) } );
\addlegendentry{ SPH ($S=-0.068$)};
\addplot[color=black,densely dashed, domain=1000:100000, samples=100, smooth] 
plot (\x, { (\x)^(-0.16011) *exp(-0.32745) } );
\addlegendentry{ LBM ($S=-0.160$)};
\end{axis}
\end{tikzpicture}%
}}}\\
	\caption {$Re=400$}
	\label{re_400}
\end{figure}

\subsubsection{$Re=1000$}

\begin{figure}[bthp]
	\begin{minipage}{0.45\textwidth}
		\captionsetup[subfigure]{labelformat=empty}
		\centering
		\subfloat[{\tiny $50\times 50$ - Velocity (left) - Streamlines (right)}] {\includegraphics[width=0.12\textwidth]{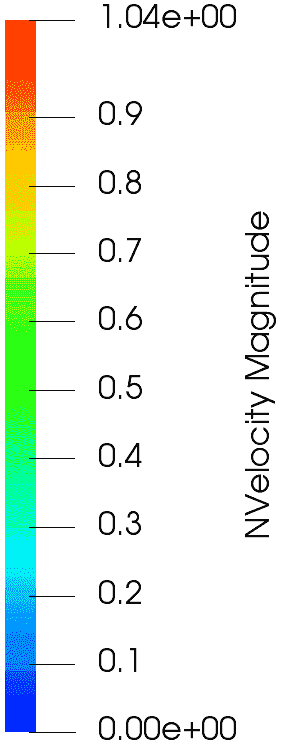}\includegraphics[width=0.4\textwidth]{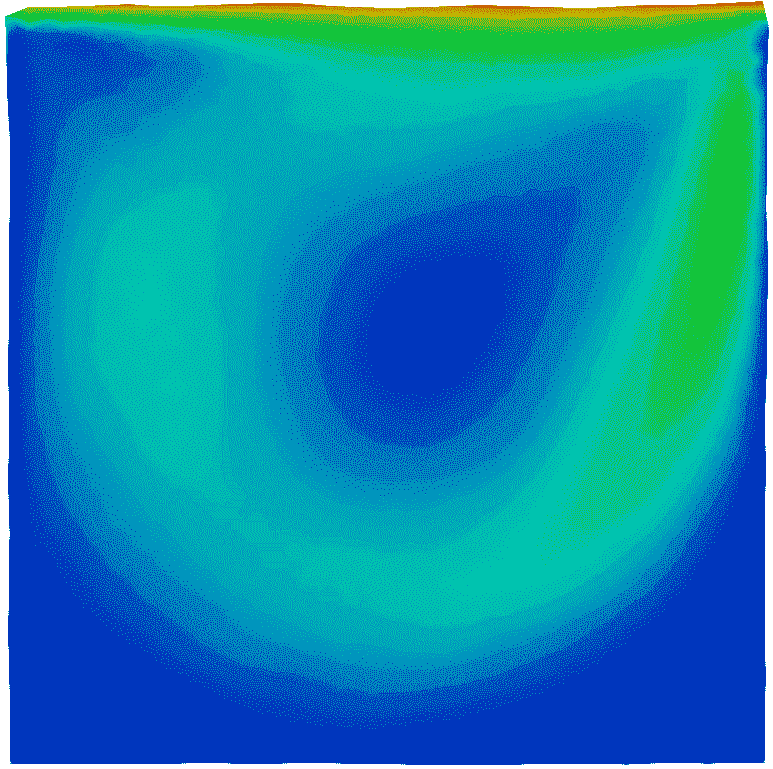}\includegraphics[width=0.4\textwidth]{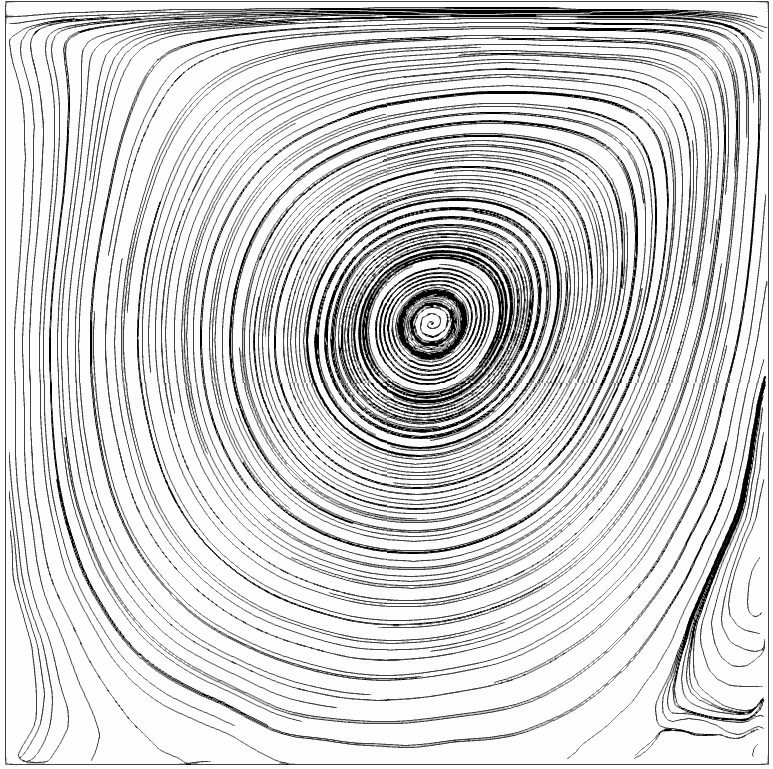}}\\
		\subfloat[{\tiny $100\times 100$ - Velocity (left) - Streamlines (right)}] {\includegraphics[width=0.12\textwidth]{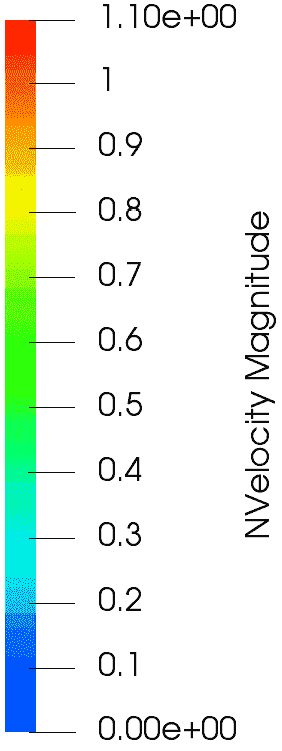}\includegraphics[width=0.4\textwidth]{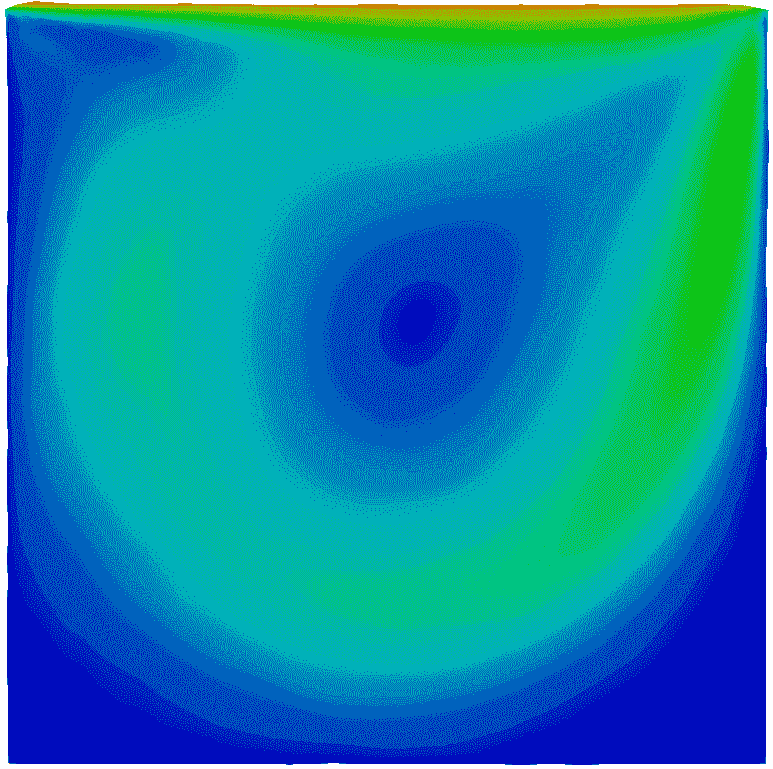}\includegraphics[width=0.4\textwidth]{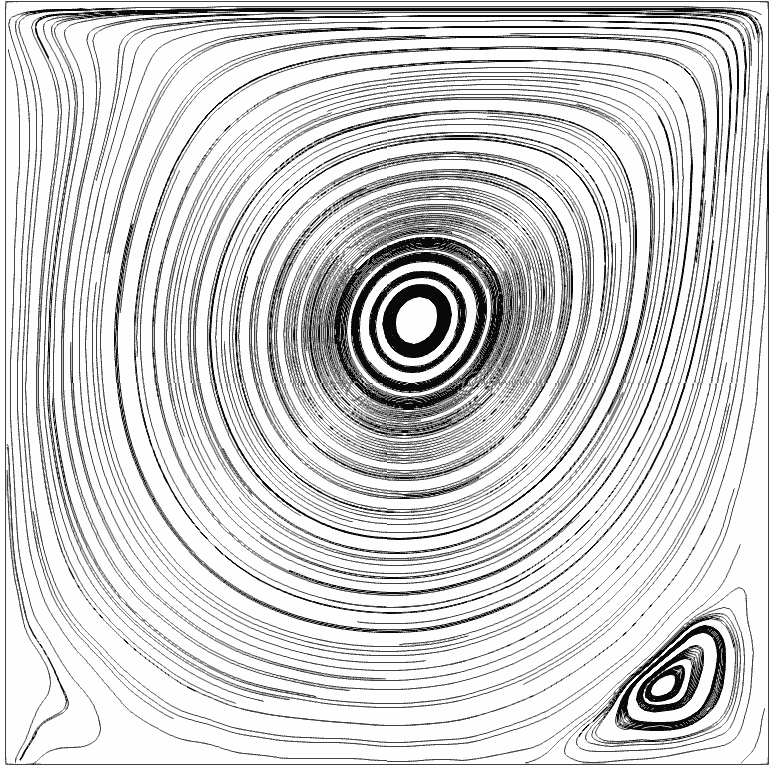}}\\	
		\subfloat[{\tiny $200\times 200$ - Velocity (left) - Streamlines (right)}] {\includegraphics[width=0.12\textwidth]{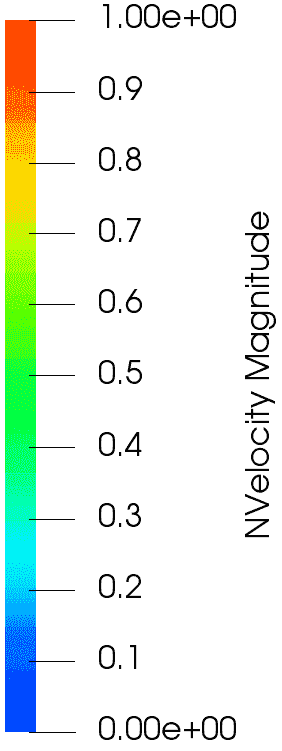}\includegraphics[width=0.4\textwidth]{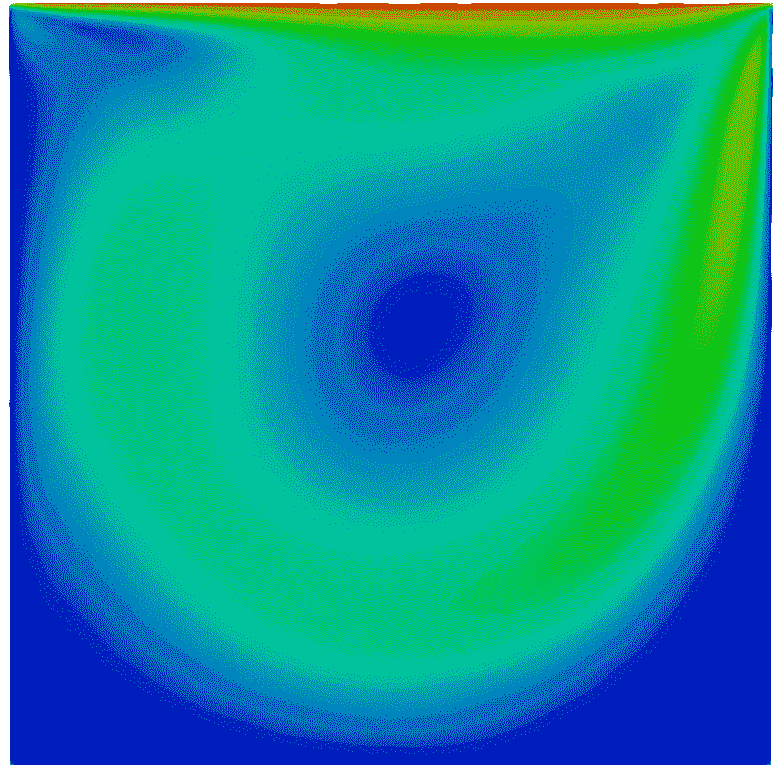}\includegraphics[width=0.4\textwidth]{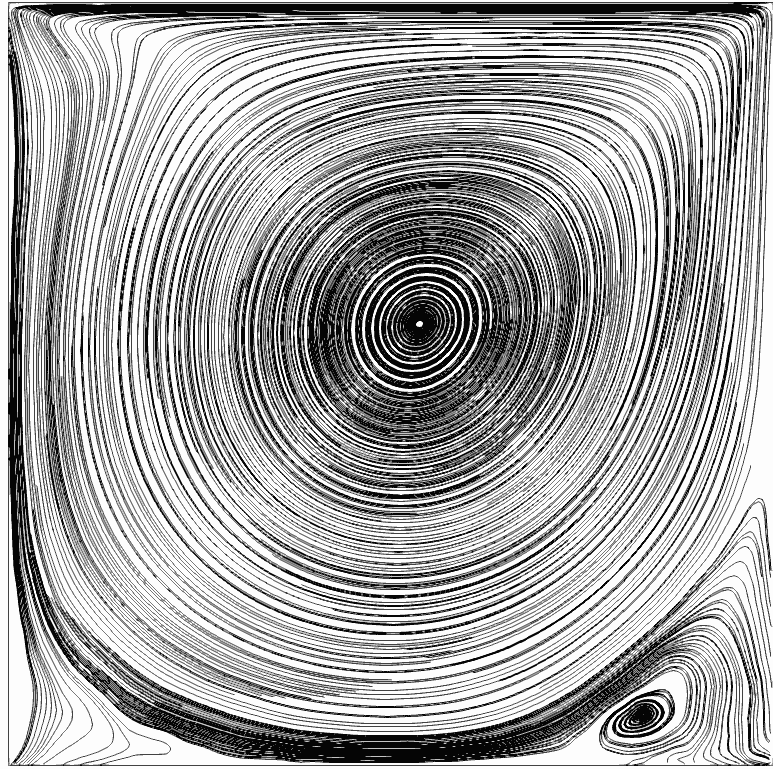}}\\
		\caption {SPH results for $Re=1000$}
		\label{re1000_sph}
	\end{minipage}\hfill
	\begin{minipage}{0.45\textwidth}	
		\captionsetup[subfigure]{labelformat=empty}
		\centering
		\subfloat[{\tiny $50\times 50$ - Velocity (left) - Streamlines (right)}] {\includegraphics[width=0.12\textwidth]{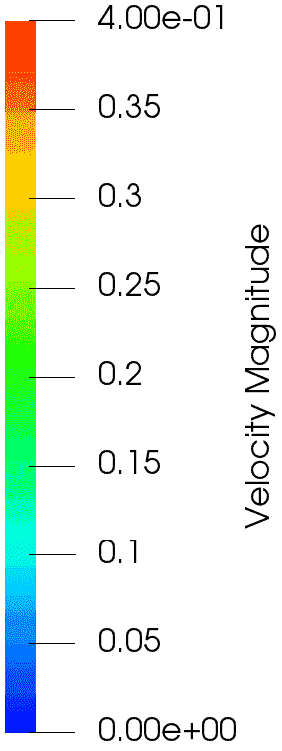}\includegraphics[width=0.4\textwidth]{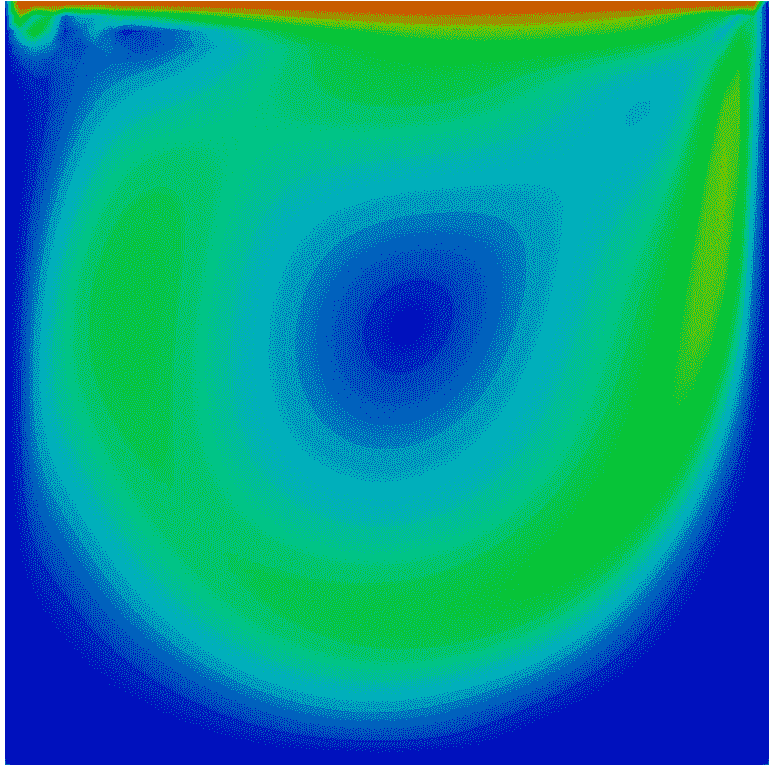}\includegraphics[width=0.4\textwidth]{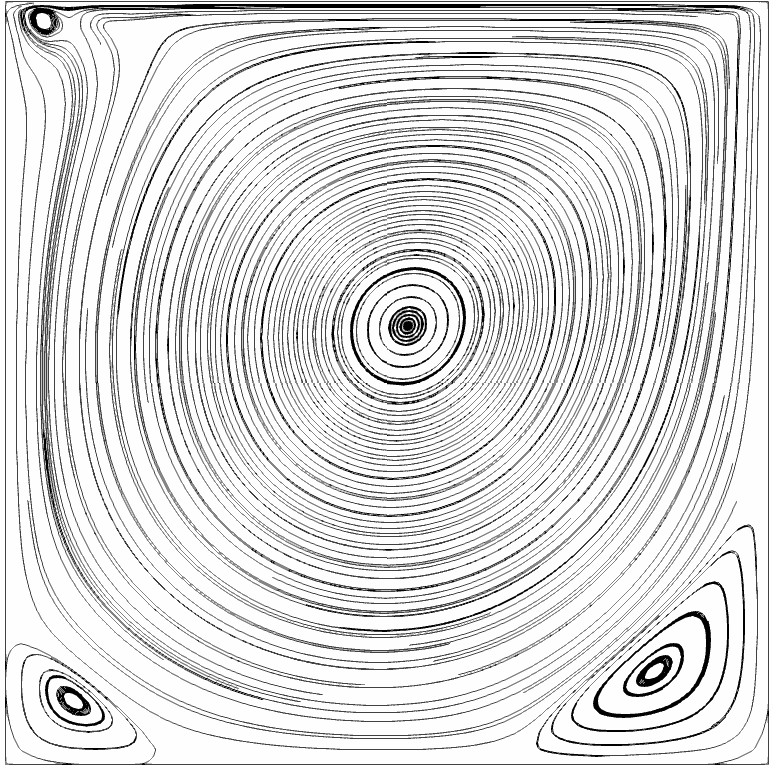}}\\
		\subfloat[{\tiny $100\times 100$ - Velocity (left) - Streamlines (right)}] {\includegraphics[width=0.12\textwidth]{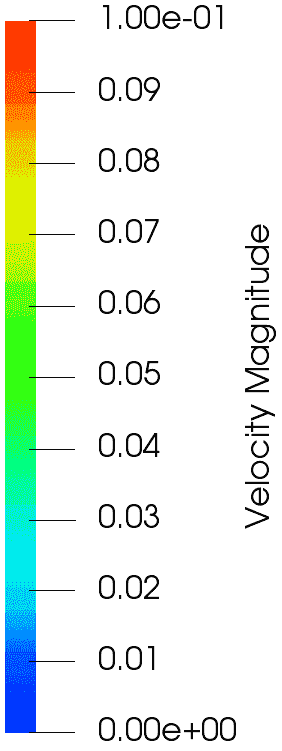}\includegraphics[width=0.4\textwidth]{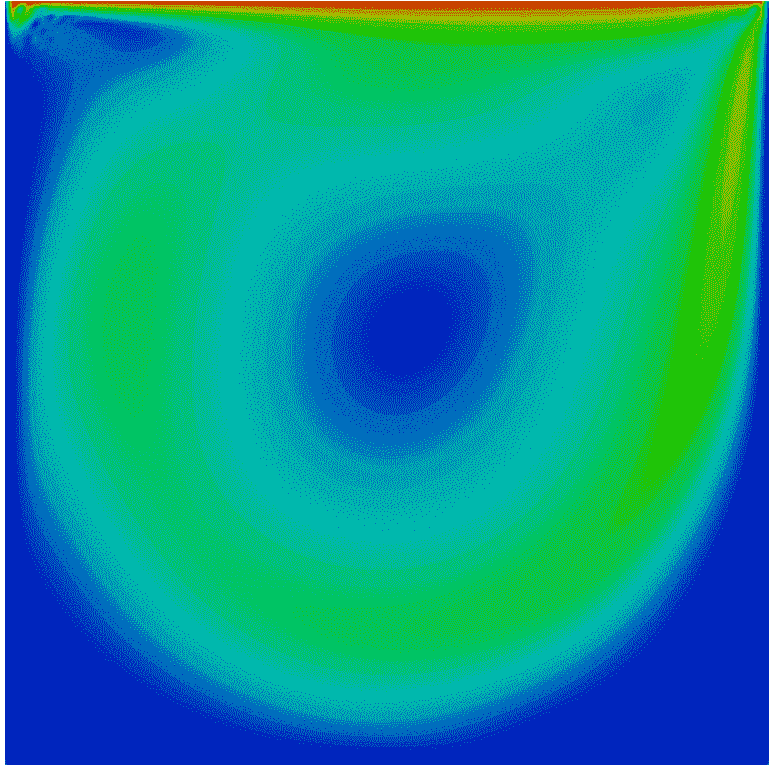}\includegraphics[width=0.4\textwidth]{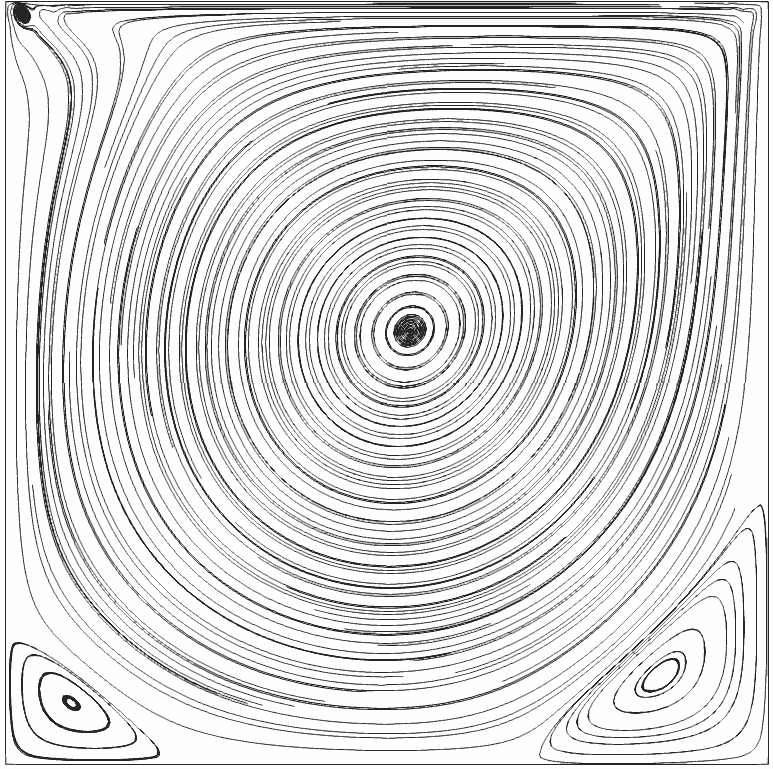}}\\	
		\subfloat[{\tiny $200\times 200$ - Velocity (left) - Streamlines (right)}] {\includegraphics[width=0.12\textwidth]{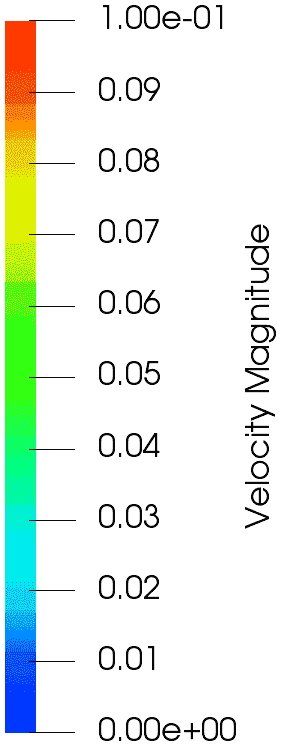}\includegraphics[width=0.4\textwidth]{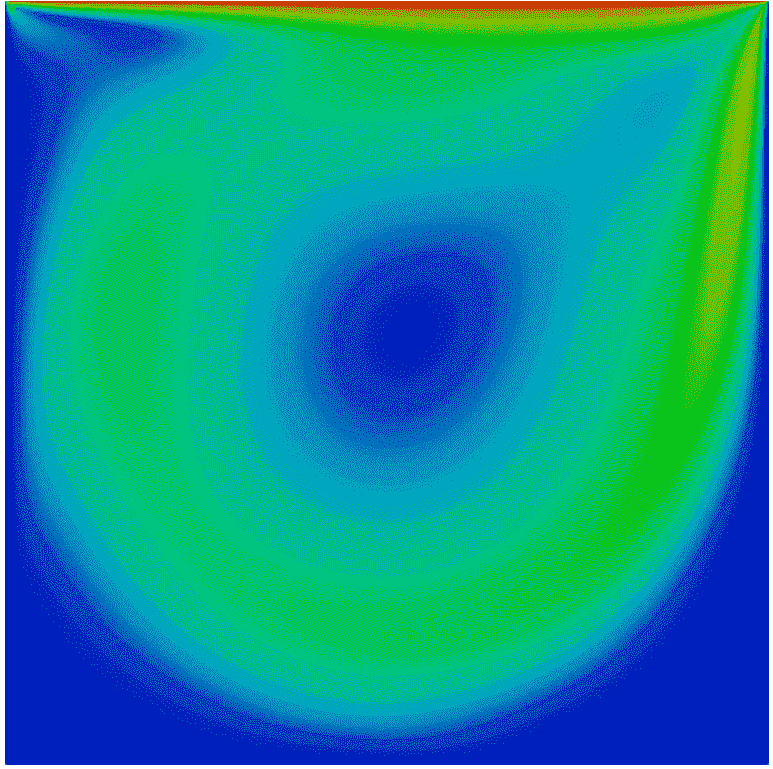}\includegraphics[width=0.4\textwidth]{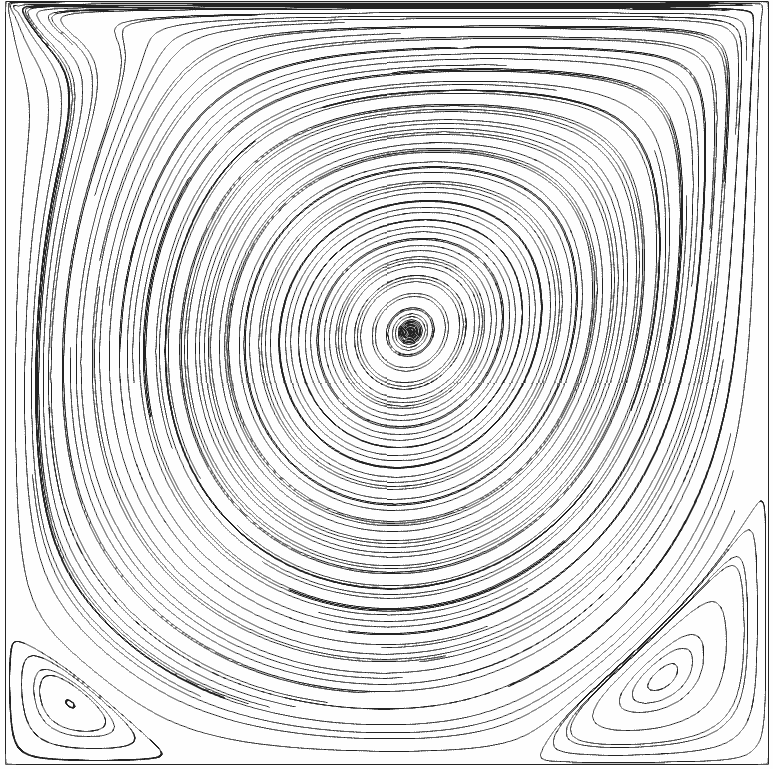}}\\
		\caption {LBM results for $Re=1000$}
		\label{re1000_lbm}
	\end{minipage}	
\end{figure}

For $Re=1000$, the MRT operator with the standard relaxation times $\bm{S}$ fails to give stable results for the $50 \times 50$ resolution. However, when using another set of relaxation times $\bm{S}^{*}$, one can obtain a stable solution. The impact of empirically adjusting the relaxation times to "make it work" remains to be investigated.

As in the previous cases, one can observe in Fig.~\ref{re_1000} that LBM exhibits a better accuracy than SPH for almost all resolutions. For the highest resolution, LBM has a maximum $L_2$ discrepancy of $\approx 0.07$. For the same resolution, SPH gives a $L_2$ discrepancy of $\approx 0.16$. Besides, the LBM order of convergence is still up to $2-3$ times higher than the SPH one.

For this Reynolds number, it can be seen in Figs.~\ref{re1000_lbm} and \ref{re1000_sph} that SPH is capable of generating a vertex pattern at the bottom right corner for the two highest resolutions but it is unstable for the smallest resolution. Moderate deviations of the flow indicating a potential growing vortex can be seen at the bottom left corner. When looking at the streamlines of the SPH simulations, we observe that all three resolutions are generating vertexes in the correct spots at selected instants but only the $200 \times 200$ case manage to stabilize one at the bottom right corner.

As previously said for the smaller Reynolds numbers, LBM is again showing the appearance of the two vertexes at the correct locations. An instability is growing at the top left corner but disappears at the highest resolution.

\begin{figure}[bthp]
	\captionsetup[subfigure]{labelformat=empty}
	\centering
	\subfloat[$50\times 50$ - $t=52.83s$] {\includegraphics[width=0.18\textwidth]{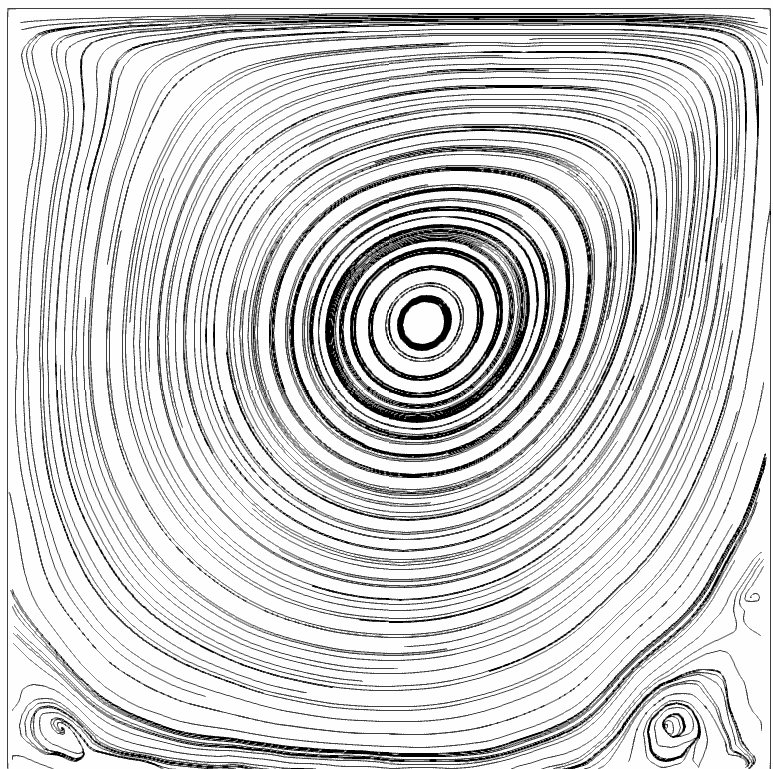}}\hspace{0.5cm}
	\subfloat[$100\times 100$ - $t=50.50s$] {\includegraphics[width=0.18\textwidth]{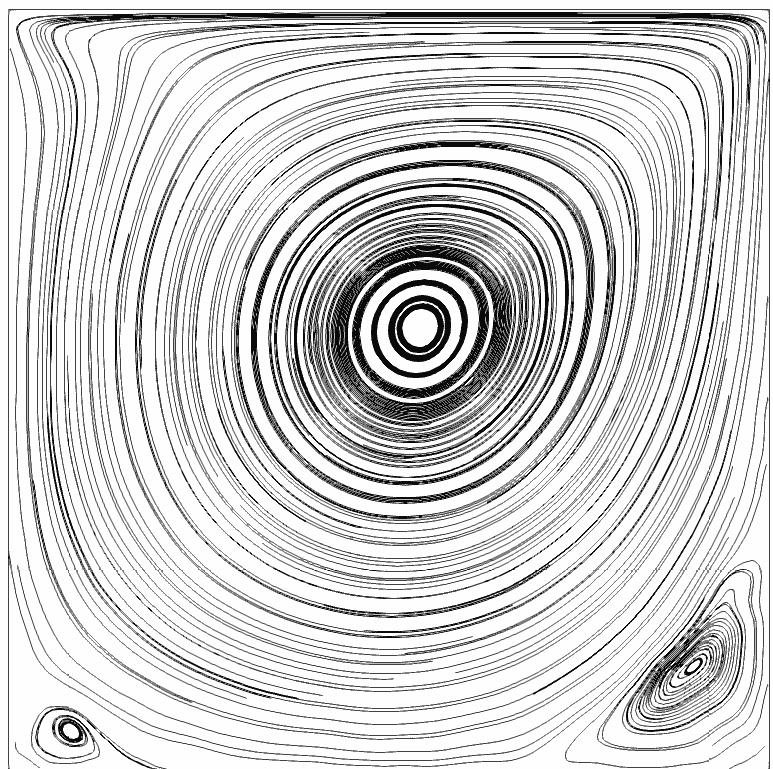}}\hspace{0.5cm}
	\subfloat[$200\times 200$ - $t=52.60s$] {\includegraphics[width=0.18\textwidth]{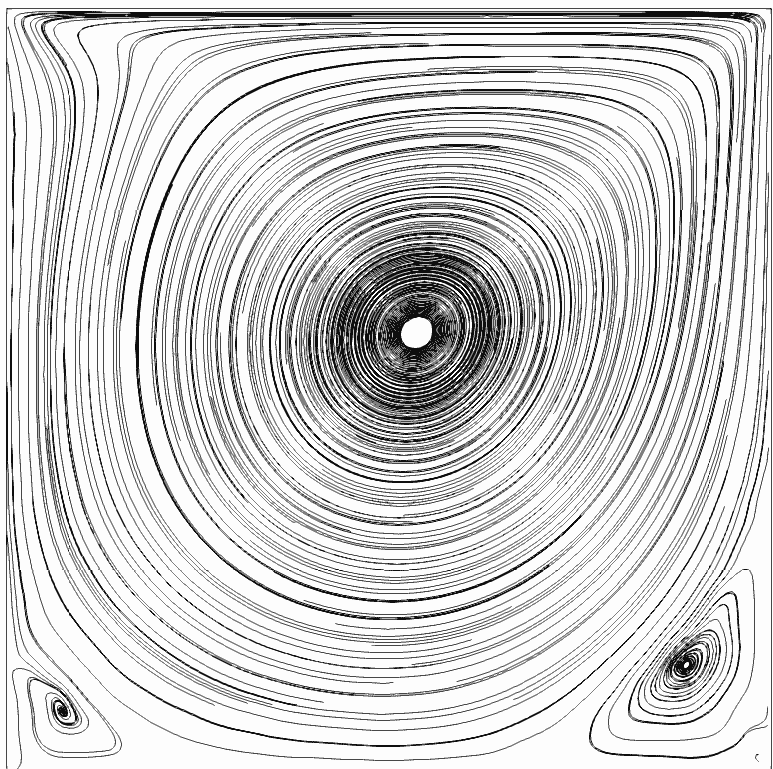}}
	\caption {SPH streamlines for $Re=1000$ at selected timesteps}
	\label{re1000_sl_sph}
\end{figure}

For this Reynolds number, it can be seen in Figs.~\ref{re1000_lbm} and \ref{re1000_sph} that SPH is capable of generating a vertex pattern at the bottom right corner for the two highest resolutions but it is unstable for the smallest resolution. Moderate deviations of the flow indicating a potential growing vortex can be seen at the bottom left corner. When computing the streamlines for selected timesteps of the SPH simulations as shown in Fig.~\ref{re1000_sl_sph}, it is seen that all three resolutions are generating vertexes in the correct spots but only the $200 \times 200$ case manage to stabilize one at the bottom right corner.

As previously said for the smaller Reynolds numbers, LBM is again showing the appearance of the two vertexes at the correct locations. An instability is growing at the top left corner but disappears at the highest resolution.

\begin{figure}[bthp]
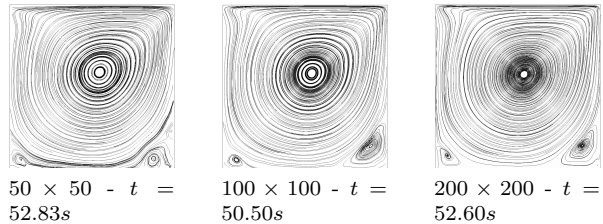

	\captionsetup[subfigure]{labelformat=empty}
	\centering
	\subfloat[$50\times 50$ - $t=52.83s$] {\includegraphics[width=0.18\textwidth]{SPH_Re1000_2500_SL_259.png}}\hspace{0.5cm}
	\subfloat[$100\times 100$ - $t=50.50s$] {\includegraphics[width=0.18\textwidth]{SPH_Re1000_10000_SL_500.png}}\hspace{0.5cm}
	\subfloat[$200\times 200$ - $t=52.60s$] {\includegraphics[width=0.18\textwidth]{SPH_Re1000_40000_SL_1048.png}}
	\caption {SPH streamlines for $Re=1000$ at selected timesteps}
	\label{re1000_sl_sph}
\end{figure}

\begin{figure}[bthp]
	\begin{center}
		\captionsetup[subfigure]{labelformat=empty}
		\makebox[\textwidth][c]{
			\subfloat[]{\resizebox{0.5\textwidth}{!}{\input{Re1000_bis}}}
			\subfloat[]{\resizebox{0.5\textwidth}{!}{%\pgfplotsset{label style={font=},
%	tick label style={font=\tiny} }
%
\definecolor{mycolor1}{rgb}{0.00,0.00,1.00}%
\definecolor{mycolor2}{rgb}{0.00,0.50,0.00}%
\definecolor{mycolor3}{rgb}{1.00,0.00,0.00}%
\definecolor{mycolor4}{rgb}{0.00,0.75,0.75}%
\definecolor{mycolor5}{rgb}{0.75,0.00,0.75}%
\definecolor{mycolor6}{rgb}{0.75,0.75,0.00}%
\definecolor{mycolor7}{rgb}{0.25,0.25,0.25}%
\definecolor{mycolor8}{rgb}{0.75,0.25,0.25}%
\definecolor{mycolor9}{rgb}{0.95,0.95,0.00}%
\definecolor{mycolor10}{rgb}{0.25,0.25,0.75}%
\definecolor{mycolor11}{rgb}{0.75,0.75,0.75}%
\definecolor{mycolor12}{rgb}{0.00,1.00,0.00}%
\definecolor{mycolor13}{rgb}{0.76,0.57,0.17}%
\definecolor{mycolor14}{rgb}{0.54,0.63,0.22}%
\definecolor{mycolor15}{rgb}{0.34,0.57,0.92}%
\definecolor{mycolor16}{rgb}{1.00,0.10,0.60}%
\definecolor{mycolor17}{rgb}{0.88,0.75,0.73}%
\definecolor{mycolor18}{rgb}{0.10,0.49,0.47}%
\definecolor{mycolor19}{rgb}{0.66,0.34,0.65}%
\definecolor{mycolor20}{rgb}{0.99,0.41,0.23}%
\begin{tikzpicture}

\begin{axis}[%
scaled ticks=false, 
tick label style={/pgf/number format/fixed},
xmajorgrids=false,
ymajorgrids=true,
grid style={dotted,gray},
width=3.0in,
height=1.5in,
at={(0in,0in)},
scale only axis,
xmode=log,
xmin=1000,
xmax=100000,
xminorticks=true,
yminorgrids=true,
ymode=log,
ymin=0.01,
ymax=1.0,
yminorticks=true,
ylabel near ticks,
xlabel near ticks,
xtick pos=left,
ytick pos=left,
xlabel={Number of nodes/particles},
ylabel={$L_2$ Discrepancy},
axis background/.style={fill=white},
legend style={legend style={nodes={scale=0.75, transform shape}},fill=white,align=left,draw=none,at={(0.95,0.95)}}
]
%\addplot [color=black,solid]
%  table[row sep=crcr]{%
%2500	0.307124180290415\\
%10000	0.192267596004043\\
%40000	0.129986830273123\\
%};
%\addlegendentry{SPH};
\addplot [color=mycolor4,only marks,mark=star,mark size=0.85, mark repeat=2, forget plot]
table[row sep=crcr]{%
	2500	0.307124180290415\\
};
\addplot [color=mycolor5,only marks,mark=square,mark size=0.85, mark repeat=10, forget plot]
table[row sep=crcr]{%
	10000	0.192267596004043\\
};
\addplot [color=mycolor6,only marks,mark=diamond,mark size=0.85, mark repeat=20, forget plot]
table[row sep=crcr]{%
	40000	0.129986830273123\\
};
%\addplot [color=black,densely dashed]
%table[row sep=crcr]{%
%	2500	0.24098117874095\\
%	10000	0.104977717748349\\
%	40000	0.046065153342566\\
%};
%\addlegendentry{LBM};
\addplot [color=mycolor1,only marks,mark=x,mark options={solid},mark size=0.85, mark repeat=5, forget plot]
table[row sep=crcr]{%
	2500	0.24098117874095\\
};
\addplot [color=mycolor2,only marks,mark=o,mark options={solid},mark size=0.85, mark repeat=10, forget plot]
table[row sep=crcr]{%
	10000	0.104977717748349\\
};
\addplot [color=mycolor3,only marks,mark=triangle,mark options={solid,rotate=90},mark size=0.85, mark repeat=20, forget plot]
table[row sep=crcr]{%
	40000	0.046065153342566\\
};
\addplot[color=black,solid, domain=1000:100000, samples=100, smooth] 
plot (\x, { (\x)^(-0.31011) *exp(1.23302) } );
\addlegendentry{SPH ($S=-0.310$)};
\addplot[color=black,densely dashed, domain=1000:100000, samples=100, smooth] 
plot (\x, { (\x)^(-0.59679) *exp(3.24507) } );
\addlegendentry{LBM ($S=-0.597$)};
\end{axis}
\end{tikzpicture}%
}}}\\
		\makebox[\textwidth][c]{	
			\subfloat[]{\resizebox{0.5\textwidth}{!}{\input{Re1000_2_bis}}}
			\subfloat[]{\resizebox{0.5\textwidth}{!}{%\pgfplotsset{label style={font=\tiny},
%	tick label style={font=\tiny} }
%
\definecolor{mycolor1}{rgb}{0.00,0.00,1.00}%
\definecolor{mycolor2}{rgb}{0.00,0.50,0.00}%
\definecolor{mycolor3}{rgb}{1.00,0.00,0.00}%
\definecolor{mycolor4}{rgb}{0.00,0.75,0.75}%
\definecolor{mycolor5}{rgb}{0.75,0.00,0.75}%
\definecolor{mycolor6}{rgb}{0.75,0.75,0.00}%
\definecolor{mycolor7}{rgb}{0.25,0.25,0.25}%
\definecolor{mycolor8}{rgb}{0.75,0.25,0.25}%
\definecolor{mycolor9}{rgb}{0.95,0.95,0.00}%
\definecolor{mycolor10}{rgb}{0.25,0.25,0.75}%
\definecolor{mycolor11}{rgb}{0.75,0.75,0.75}%
\definecolor{mycolor12}{rgb}{0.00,1.00,0.00}%
\definecolor{mycolor13}{rgb}{0.76,0.57,0.17}%
\definecolor{mycolor14}{rgb}{0.54,0.63,0.22}%
\definecolor{mycolor15}{rgb}{0.34,0.57,0.92}%
\definecolor{mycolor16}{rgb}{1.00,0.10,0.60}%
\definecolor{mycolor17}{rgb}{0.88,0.75,0.73}%
\definecolor{mycolor18}{rgb}{0.10,0.49,0.47}%
\definecolor{mycolor19}{rgb}{0.66,0.34,0.65}%
\definecolor{mycolor20}{rgb}{0.99,0.41,0.23}%
\begin{tikzpicture}

\begin{axis}[%
scaled ticks=false, 
tick label style={/pgf/number format/fixed},
xmajorgrids=false,
ymajorgrids=true,
grid style={dotted,gray},
width=3.0in,
height=1.5in,
at={(0in,0in)},
scale only axis,
xmode=log,
xmin=1000,
xmax=100000,
xminorticks=true,
yminorgrids=true,
ymode=log,
ymin=0.01,
ymax=1.0,
yminorticks=true,
ylabel near ticks,
xlabel near ticks,
xtick pos=left,
ytick pos=left,
xlabel={Number of nodes/particles},
ylabel={$L_2$ Discrepancy},
axis background/.style={fill=white},
legend style={legend style={nodes={scale=0.75, transform shape}},fill=white,align=left,draw=none,at={(0.95,0.95)}}
]
%\addplot [color=black,solid]
%  table[row sep=crcr]{%
%2500	0.257976953646187\\
%10000	0.198281295167554\\
%40000	0.15919111570858\\
%};
%\addlegendentry{SPH};
\addplot [color=mycolor4,only marks,mark=star,mark size=0.85, mark repeat=2, forget plot]
table[row sep=crcr]{%
	2500	0.257976953646187\\
};
\addplot [color=mycolor5,only marks,mark=square,mark size=0.85, mark repeat=10, forget plot]
table[row sep=crcr]{%
	10000	0.198281295167554\\
};
\addplot [color=mycolor6,only marks,mark=diamond,mark size=0.85, mark repeat=20, forget plot]
table[row sep=crcr]{%
	40000	0.15919111570858\\
};
%\addplot [color=black,densely dashed]
%table[row sep=crcr]{%
%	2500	0.302471843190694\\
%	10000	0.118983591721023\\
%	40000	0.0662529119957459\\
%};
%\addlegendentry{LBM};
\addplot [color=mycolor1,only marks,mark=x,mark options={solid},mark size=0.85, mark repeat=5, forget plot]
table[row sep=crcr]{%
	2500	0.302471843190694\\
};
\addplot [color=mycolor2,only marks,mark=o,mark options={solid},mark size=0.85, mark repeat=10, forget plot]
table[row sep=crcr]{%
	10000	0.118983591721023\\
};
\addplot [color=mycolor3,only marks,mark=triangle,mark options={solid,rotate=90},mark size=0.85, mark repeat=20, forget plot]
table[row sep=crcr]{%
	40000	0.0662529119957459\\
};
\addplot[color=black,solid, domain=1000:100000, samples=100, smooth] 
plot (\x, { (\x)^(-0.17412) *exp( 0.00017) } );
\addlegendentry{SPH ($S=-0.174$)};
\addplot[color=black,densely dashed, domain=1000:100000, samples=100, smooth] 
plot (\x, { (\x)^(-0.54769) *exp(3.03144) } );
\addlegendentry{LBM ($S=-0.548$)};
\end{axis}
\end{tikzpicture}%
}}}	
		\caption {$Re=1000$}
		\label{re_1000}
	\end{center}
\end{figure}

\subsubsection{$Re=10000$}

\begin{figure}[bthp]
	\begin{minipage}{0.45\textwidth}
		\captionsetup[subfigure]{labelformat=empty}
		\centering
		\subfloat[{\tiny $50\times 50$ - Velocity (left) - Streamlines (right)}] {\includegraphics[width=0.12\textwidth]{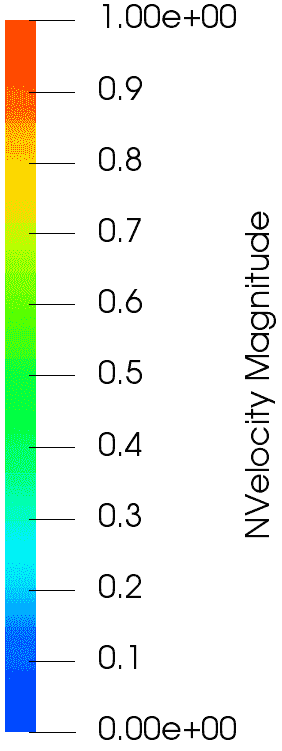}\includegraphics[width=0.4\textwidth]{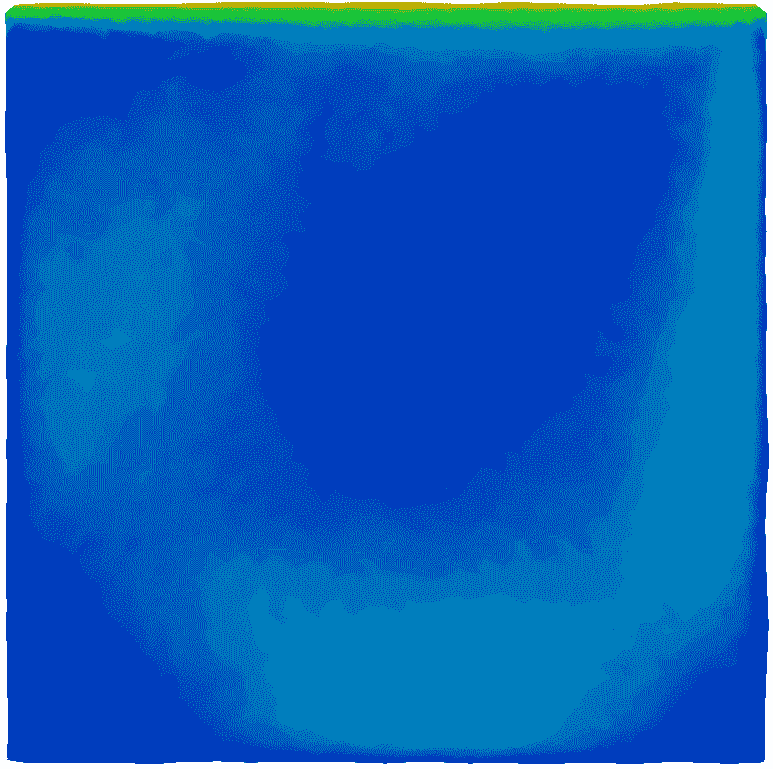}\includegraphics[width=0.4\textwidth]{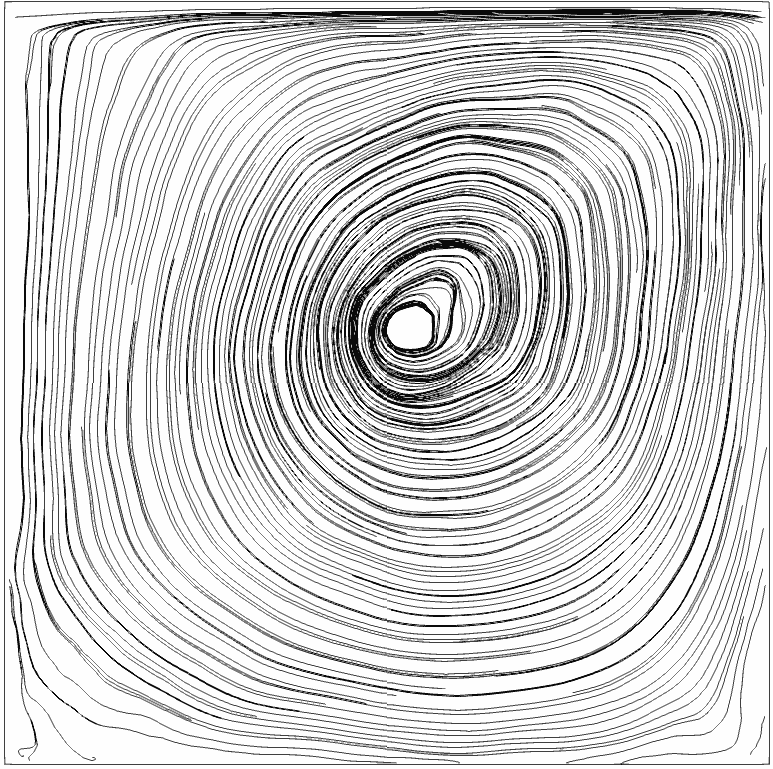}}\\
		\subfloat[{\tiny $100\times 100$ - Velocity (left) - Streamlines (right)}] {\includegraphics[width=0.12\textwidth]{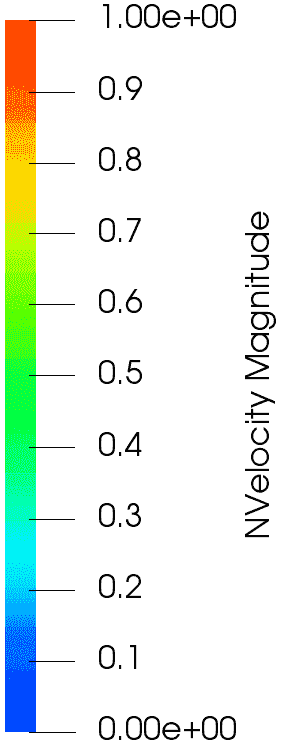}\includegraphics[width=0.4\textwidth]{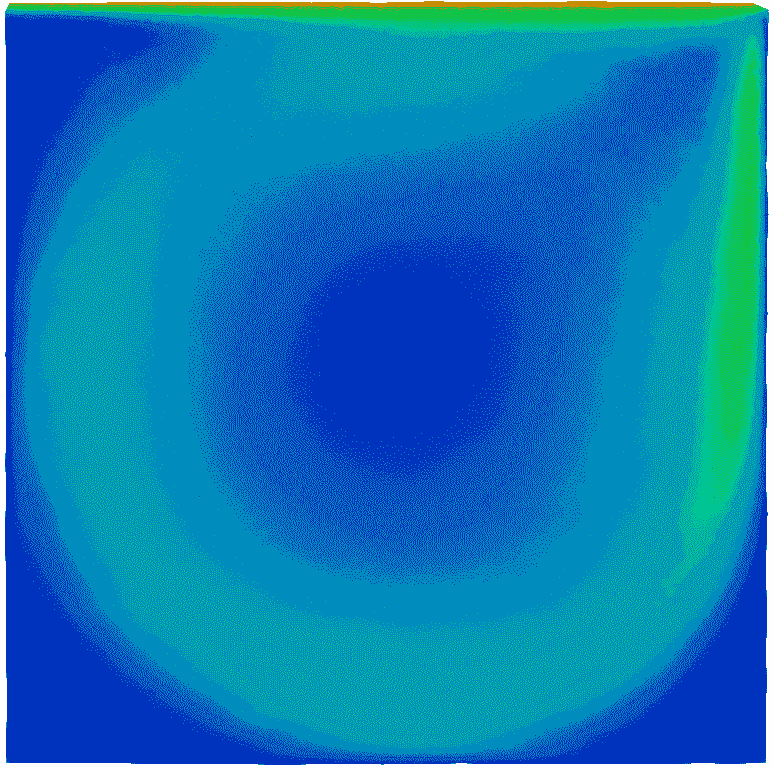}\includegraphics[width=0.4\textwidth]{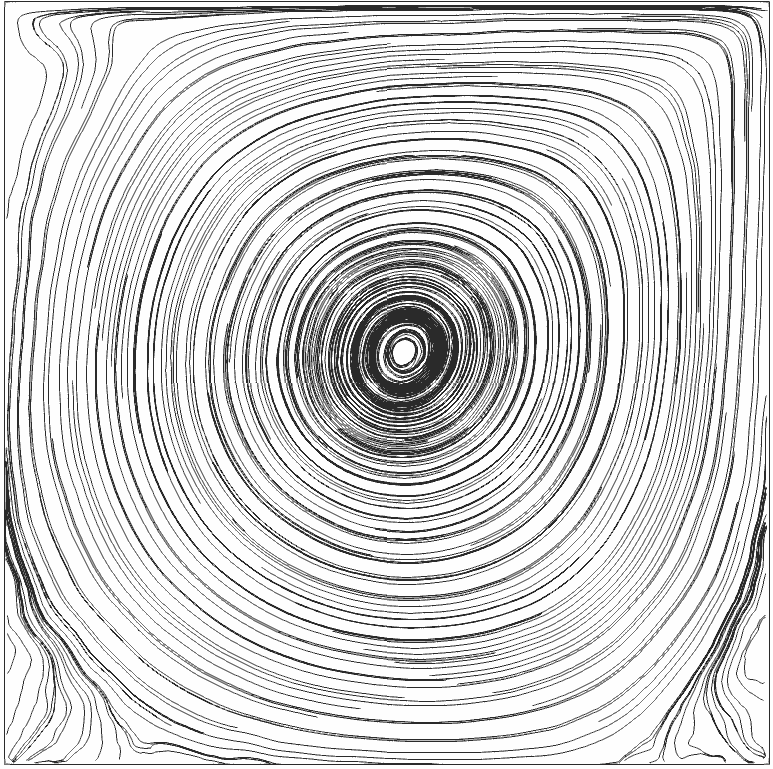}}\\	
		\subfloat[{\tiny $200\times 200$ - Velocity (left) - Streamlines (right)}] {\includegraphics[width=0.12\textwidth]{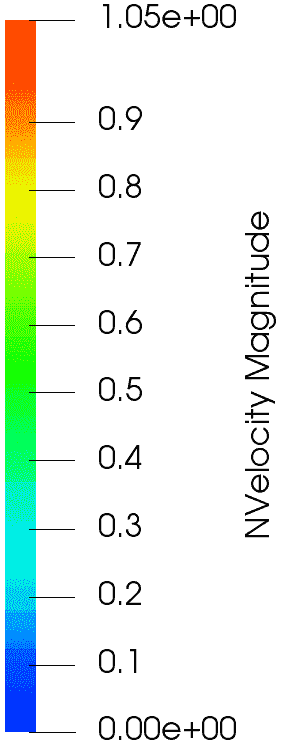}\includegraphics[width=0.4\textwidth]{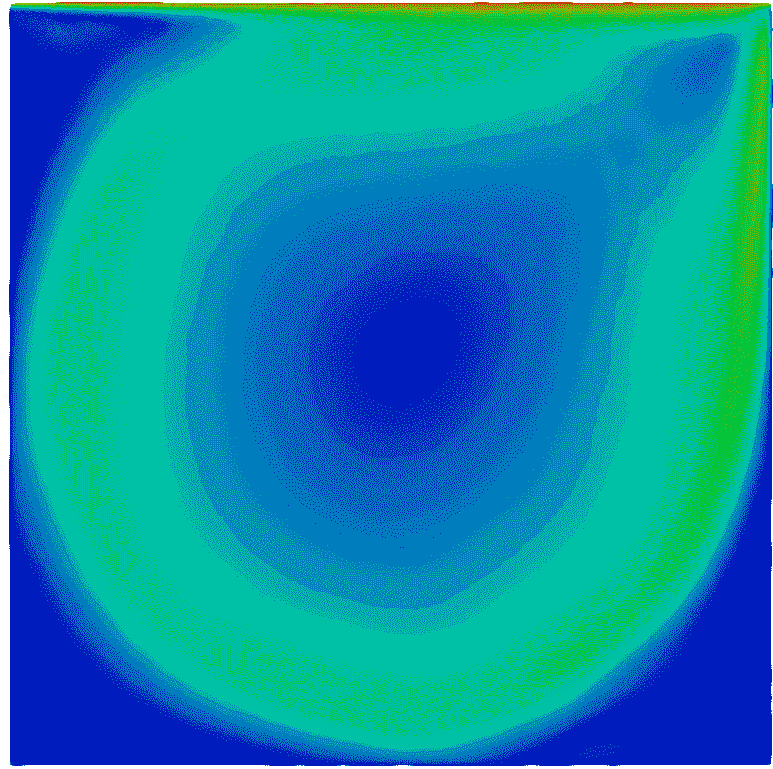}\includegraphics[width=0.4\textwidth]{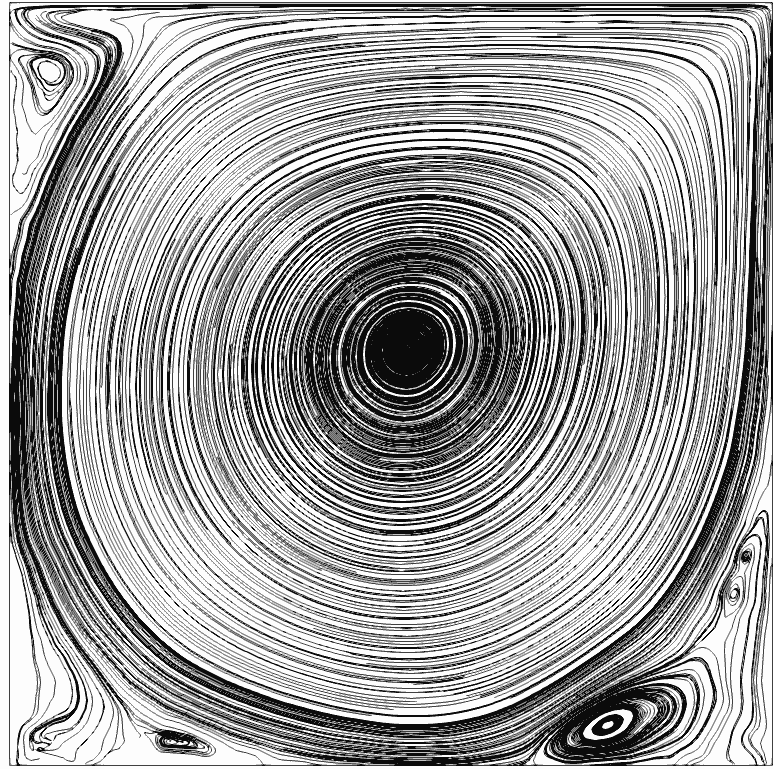}}\\
		\caption {SPH results for $Re=10000$}
		\label{re10000_sph}
	\end{minipage}\hfill
	\begin{minipage}{0.45\textwidth}	
		\captionsetup[subfigure]{labelformat=empty}
		\centering	
		\subfloat[{\tiny $200\times 200$ - Velocity (left) - Streamlines (right)}] {\includegraphics[width=0.12\textwidth]{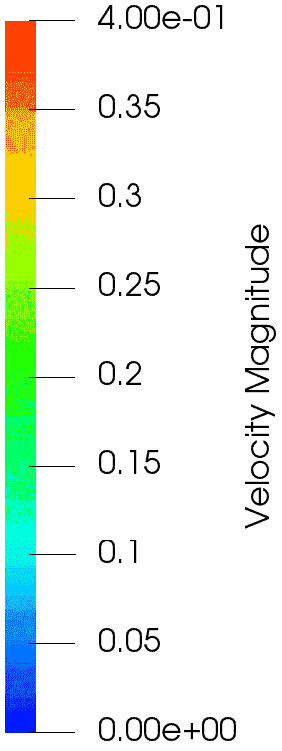}\includegraphics[width=0.4\textwidth]{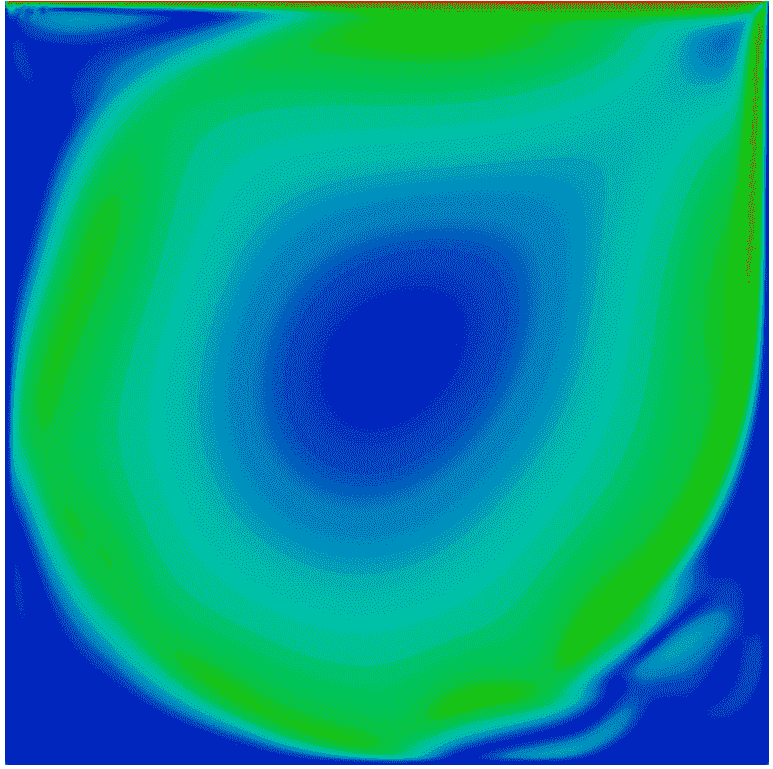}\includegraphics[width=0.4\textwidth]{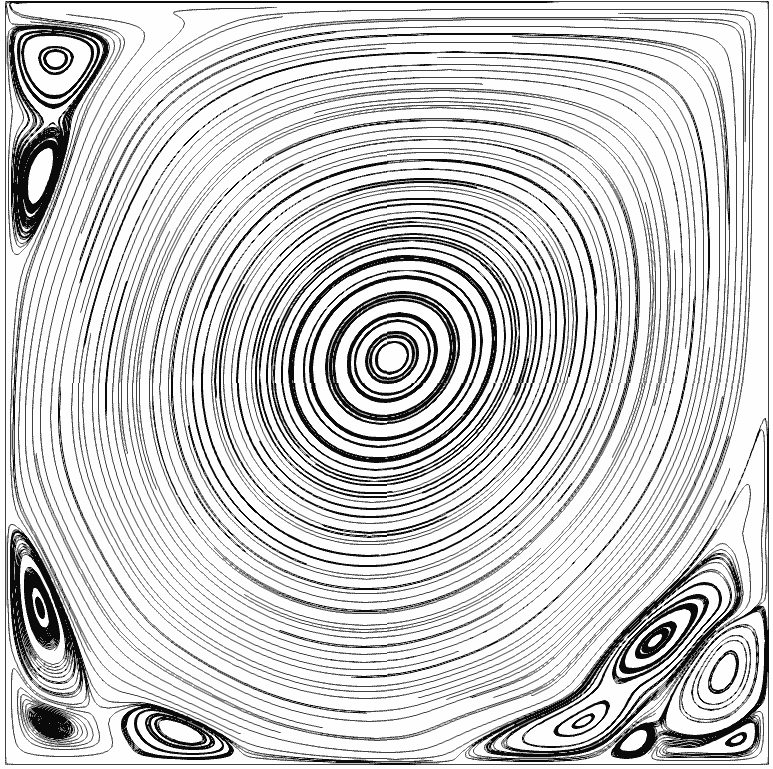}}\\
		\caption {LBM results for $Re=10000$}
		\label{re10000_lbm}
	\end{minipage}	
\end{figure}

For $Re=10000$, the MRT operator despite its superior stability properties compared to BGK is unable to give stable results for none of the considered lattice resolutions. Even using the set of relaxation times $\bm{S}^{*}$, only the highest lattice resolution $200 \times 200$ prevents the simulation to blow up. In fact, Zou-he boundary conditions are known to be unstable at high $Re$ and this is likely to be one the reasons the LBM simulations fail for the lowest resolutions considered. It is possible to enhance the stability of the velocity boundary conditions, see~\cite{Latt2008b} for example.

For high Reynolds numbers, the flow is typically considered turbulent. Since no LBM nor SPH models considered in this study include the effect of turbulence, results are to be taken with caution. In consequence, both methods are showing larger errors than in the previous cases where $Re$ was much smaller. Nevertheless, the pattern is the same. As observed in Fig.~\ref{re_10000}, LBM always offers a much better accuracy than SPH for the $200 \times 200$ resolution.

\begin{figure}[bthp]
	\begin{minipage}{0.45\textwidth}
		\captionsetup[subfigure]{labelformat=empty}
		\centering
		\subfloat[$50\times 50$ - $t=59.5s$] {\includegraphics[width=0.4\textwidth]{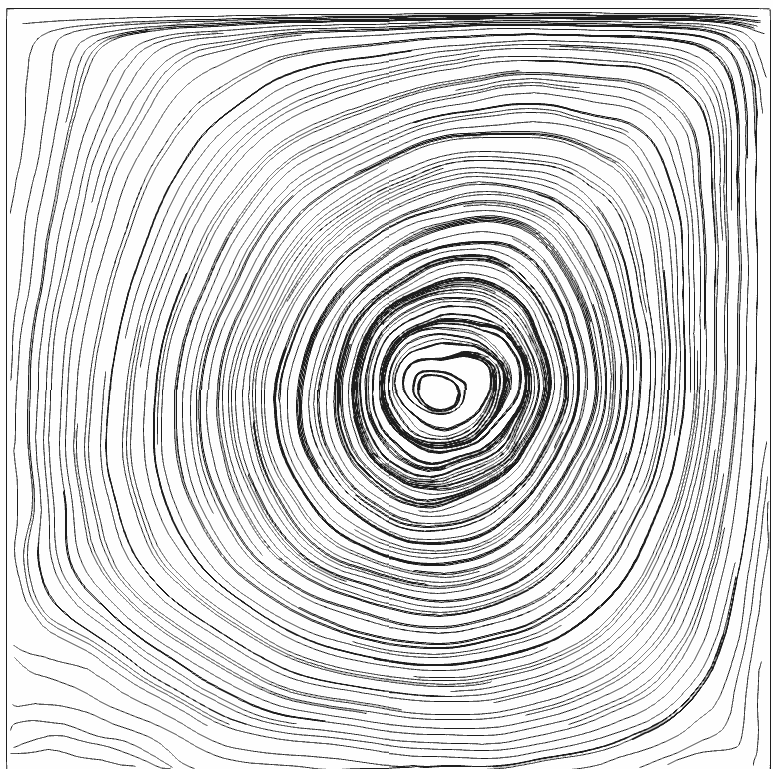}}\\
		\subfloat[$100\times 100$ - $t=46.56s$] {\includegraphics[width=0.4\textwidth]{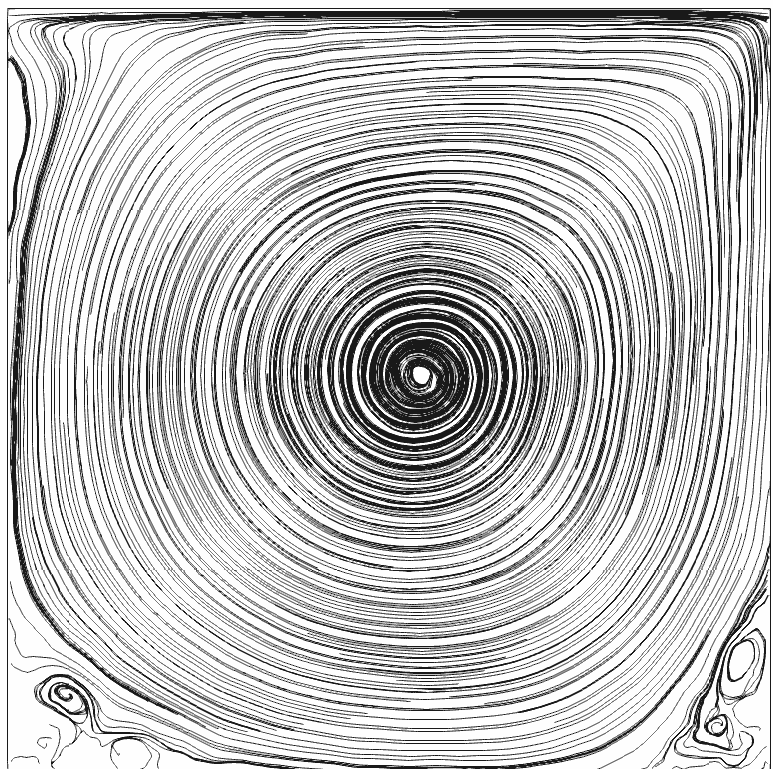}}\\
		\subfloat[$200\times 200$ - $t=51.05s$] {\includegraphics[width=0.4\textwidth]{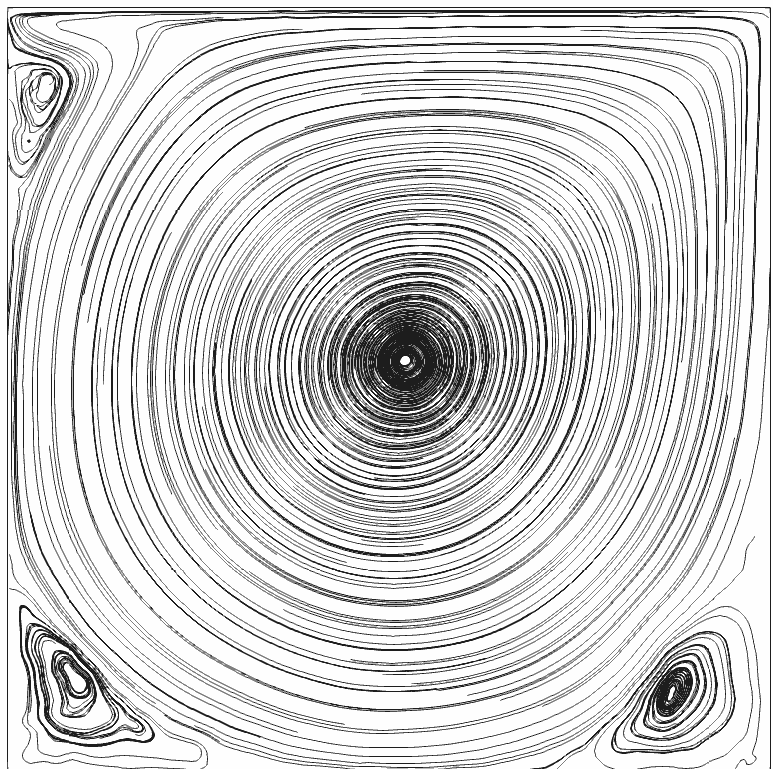}}
		\caption {SPH streamlines for $Re=10000$ at selected timesteps}
		\label{re10000_sl_sph}
	\end{minipage}\hfill
	\begin{minipage}{0.45\textwidth}	
		\captionsetup[subfigure]{labelformat=empty}
		\centering
		\subfloat[$200\times 200$] {\includegraphics[width=0.4\textwidth]{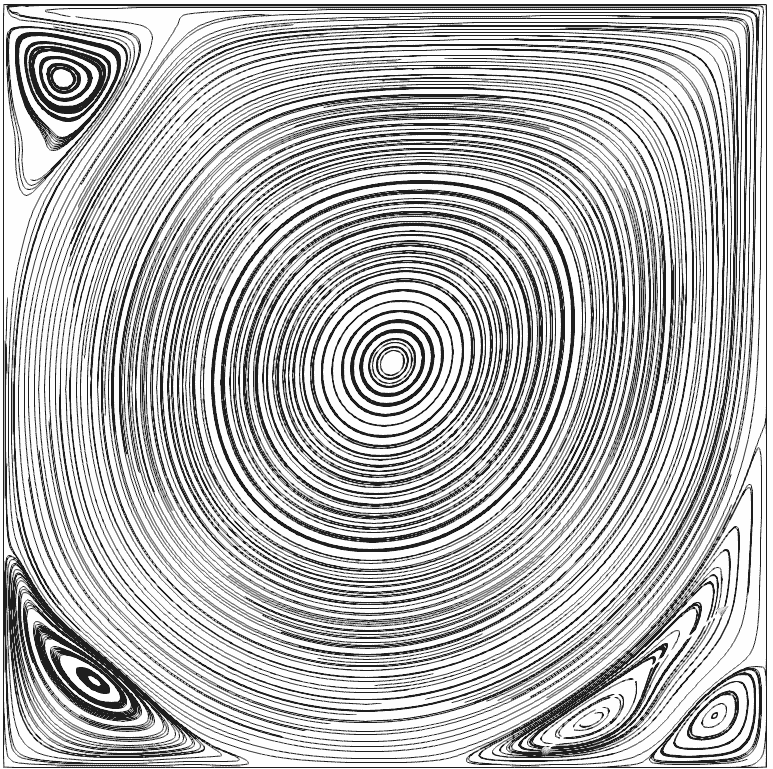}}
		\caption {LBM streamlines for $Re=10000$ at selected timestep}
		\label{re10000_sl_lbm}
	\end{minipage}	
\end{figure}

In Fig.~\ref{re10000_lbm}, the LBM results are showing a high number of vertexes at the bottom right corner (5 vertexes), the bottom left corner (3 vertexes) and the top left corner (2 vertexes). This is not agreeing with the theory where only 1 vertex is reported at the top left and bottom left corners and 2 vertexes at bottom right corner. These spurious vertexes could be due to the use of the MRT operator with relaxation times tuned based on a trial-and-error approach. The number of vortexes is variable during the simulation as shown in Fig.~\ref{re10000_sl_lbm} where the number of vortexes is correct. Extra vortexes keep appearing and disappearing throughout the simulation. No steady state is reached by the LBM in this case. The SPH streamlines plots of Fig.~\ref{re10000_sph} are not showing any vertex pattern until the highest resolution is reached. For this $200 \times 200$ case, one can note the appearance of a vertex at the top left corner, a small growing vertex at the bottom left corner and a growing vertex next to two very small vertexes at the bottom right corner. Those vertexes are stable through the simulation unlike the one at the bottom left corner as suggested by Figs.~\ref{re10000_sl_sph}. Those figures also show that at a smaller resolution, none of the expected vertexes are stable.

%\begin{figure}[bthp]
%	\captionsetup[subfigure]{labelformat=empty}
%	\centering
%	\subfloat[$200\times 200$] {\includegraphics[width=0.2\textwidth]{LBM_Re10000_40000_SL_1000.png}}
%	\caption {LBM streamlines for $Re=10000$ at selected timestep}
%	\label{re10000_sl_lbm}
%\end{figure}
%
%\begin{figure}[bthp]
%	\captionsetup[subfigure]{labelformat=empty}
%	\centering
%	\subfloat[$50\times 50$ - $t=59.5s$] {\includegraphics[width=0.2\textwidth]{SPH_Re10000_2500_SL_293.png}}\hspace{0.5cm}
%	\subfloat[$100\times 100$ - $t=46.56s$] {\includegraphics[width=0.2\textwidth]{SPH_Re10000_10000_SL_461.png}}\hspace{0.5cm}
%	\subfloat[$200\times 200$ - $t=51.05s$] {\includegraphics[width=0.2\textwidth]{SPH_Re10000_40000_SL_1017.png}}
%	\caption {SPH streamlines for $Re=10000$ at selected timesteps}
%	\label{re10000_sl_sph}
%\end{figure}

\begin{figure}[bthp]
	\captionsetup[subfigure]{labelformat=empty}
	\centering
	\makebox[\textwidth][c]{
		\subfloat[]{\resizebox{0.5\textwidth}{!}{\input{Re10000}}}
		\subfloat[]{\resizebox{0.5\textwidth}{!}{%\pgfplotsset{label style={font=\tiny},
%	tick label style={font=\tiny} }
%
\definecolor{mycolor1}{rgb}{0.00,0.00,1.00}%
\definecolor{mycolor2}{rgb}{0.00,0.50,0.00}%
\definecolor{mycolor3}{rgb}{1.00,0.00,0.00}%
\definecolor{mycolor4}{rgb}{0.00,0.75,0.75}%
\definecolor{mycolor5}{rgb}{0.75,0.00,0.75}%
\definecolor{mycolor6}{rgb}{0.75,0.75,0.00}%
\definecolor{mycolor7}{rgb}{0.25,0.25,0.25}%
\definecolor{mycolor8}{rgb}{0.75,0.25,0.25}%
\definecolor{mycolor9}{rgb}{0.95,0.95,0.00}%
\definecolor{mycolor10}{rgb}{0.25,0.25,0.75}%
\definecolor{mycolor11}{rgb}{0.75,0.75,0.75}%
\definecolor{mycolor12}{rgb}{0.00,1.00,0.00}%
\definecolor{mycolor13}{rgb}{0.76,0.57,0.17}%
\definecolor{mycolor14}{rgb}{0.54,0.63,0.22}%
\definecolor{mycolor15}{rgb}{0.34,0.57,0.92}%
\definecolor{mycolor16}{rgb}{1.00,0.10,0.60}%
\definecolor{mycolor17}{rgb}{0.88,0.75,0.73}%
\definecolor{mycolor18}{rgb}{0.10,0.49,0.47}%
\definecolor{mycolor19}{rgb}{0.66,0.34,0.65}%
\definecolor{mycolor20}{rgb}{0.99,0.41,0.23}%
\begin{tikzpicture}

\begin{axis}[%
scaled ticks=false, 
tick label style={/pgf/number format/fixed},
xmajorgrids=false,
ymajorgrids=true,
grid style={dotted,gray},
width=3.0in,
height=1.5in,
at={(0in,0in)},
scale only axis,
xmode=log,
xmin=1000,
xmax=100000,
xminorticks=true,
yminorgrids=true,
ymode=log,
ymin=0.1,
ymax=1.0,
yminorticks=true,
ylabel near ticks,
xlabel near ticks,
xtick pos=left,
ytick pos=left,
xlabel={Number of nodes/particles},
ylabel={$L_2$ Discrepancy},
axis background/.style={fill=white},
legend style={legend style={nodes={scale=0.75, transform shape}},fill=white,align=left,draw=none,at={(0.65,0.25)}}
]
%\addplot [color=black,solid]
%  table[row sep=crcr]{%
%2500	0.65856085090295\\
%10000	0.532249355465193\\
%40000	0.317231662674825\\
%};
%\addlegendentry{SPH};
\addplot [color=mycolor4,only marks,mark=star,mark size=0.85, mark repeat=2, forget plot]
table[row sep=crcr]{%
	2500	0.65856085090295\\
};
\addplot [color=mycolor5,only marks,mark=square,mark size=0.85, mark repeat=10, forget plot]
table[row sep=crcr]{%
	10000	0.532249355465193\\
};
\addplot [color=mycolor6,only marks,mark=diamond,mark size=0.85, mark repeat=20, forget plot]
table[row sep=crcr]{%
	40000	0.317231662674825\\
};
\addplot [color=black,densely dashed]
table[row sep=crcr]{%
	40000	0.113992921393078\\
};
\addlegendentry{LBM};
\addplot [color=mycolor3,only marks,mark=triangle,mark options={solid,rotate=90},mark size=0.85, mark repeat=20, forget plot]
table[row sep=crcr]{%
	40000	0.113992921393078\\
};
\addplot[color=black,solid, domain=1000:100000, samples=100, smooth] 
plot (\x, { (\x)^(-0.26344) *exp(1.69426) } );
\addlegendentry{SPH ($S=-0.263$)};
\end{axis}
\end{tikzpicture}%
}}}\\
	\makebox[\textwidth][c]{
		\subfloat[]{\resizebox{0.5\textwidth}{!}{\input{Re10000_2}}}
		\subfloat[]{\resizebox{0.5\textwidth}{!}{%\pgfplotsset{label style={font=\tiny},
%	tick label style={font=\tiny} }
%
\definecolor{mycolor1}{rgb}{0.00,0.00,1.00}%
\definecolor{mycolor2}{rgb}{0.00,0.50,0.00}%
\definecolor{mycolor3}{rgb}{1.00,0.00,0.00}%
\definecolor{mycolor4}{rgb}{0.00,0.75,0.75}%
\definecolor{mycolor5}{rgb}{0.75,0.00,0.75}%
\definecolor{mycolor6}{rgb}{0.75,0.75,0.00}%
\definecolor{mycolor7}{rgb}{0.25,0.25,0.25}%
\definecolor{mycolor8}{rgb}{0.75,0.25,0.25}%
\definecolor{mycolor9}{rgb}{0.95,0.95,0.00}%
\definecolor{mycolor10}{rgb}{0.25,0.25,0.75}%
\definecolor{mycolor11}{rgb}{0.75,0.75,0.75}%
\definecolor{mycolor12}{rgb}{0.00,1.00,0.00}%
\definecolor{mycolor13}{rgb}{0.76,0.57,0.17}%
\definecolor{mycolor14}{rgb}{0.54,0.63,0.22}%
\definecolor{mycolor15}{rgb}{0.34,0.57,0.92}%
\definecolor{mycolor16}{rgb}{1.00,0.10,0.60}%
\definecolor{mycolor17}{rgb}{0.88,0.75,0.73}%
\definecolor{mycolor18}{rgb}{0.10,0.49,0.47}%
\definecolor{mycolor19}{rgb}{0.66,0.34,0.65}%
\definecolor{mycolor20}{rgb}{0.99,0.41,0.23}%
\begin{tikzpicture}

\begin{axis}[%
scaled ticks=false, 
tick label style={/pgf/number format/fixed},
xmajorgrids=false,
ymajorgrids=true,
grid style={dotted,gray},
width=3.0in,
height=1.5in,
at={(0in,0in)},
scale only axis,
xmode=log,
xmin=1000,
xmax=100000,
xminorticks=true,
yminorgrids=true,
ymode=log,
ymin=0.1,
ymax=1.0,
yminorticks=true,
ylabel near ticks,
xlabel near ticks,
xtick pos=left,
ytick pos=left,
xlabel={Number of nodes/particles},
ylabel={$L_2$ Discrepancy},
axis background/.style={fill=white},
legend style={legend style={nodes={scale=0.75, transform shape}},fill=white,align=left,draw=none,at={(0.65,0.25)}}
]
%\addplot [color=black,solid]
%  table[row sep=crcr]{%
%2500	0.907238385220425\\
%10000	0.612989566627291\\
%40000	0.339147200443875\\
%};
%\addlegendentry{SPH};
\addplot [color=mycolor4,only marks,mark=star,mark size=0.85, mark repeat=2, forget plot]
table[row sep=crcr]{%
	2500	0.907238385220425\\
};
\addplot [color=mycolor5,only marks,mark=square,mark size=0.85, mark repeat=10, forget plot]
table[row sep=crcr]{%
	10000	0.612989566627291\\
};
\addplot [color=mycolor6,only marks,mark=diamond,mark size=0.85, mark repeat=20, forget plot]
table[row sep=crcr]{%
	40000	0.339147200443875\\
};
\addplot [color=black,densely dashed]
table[row sep=crcr]{%
	40000	0.111253227808534\\
};
\addlegendentry{LBM};
\addplot [color=mycolor3,only marks,mark=triangle,mark options={solid,rotate=90},mark size=0.85, mark repeat=20, forget plot]
table[row sep=crcr]{%
	40000	0.111253227808534\\
};
\addplot[color=black,solid, domain=1000:100000, samples=100, smooth] 
plot (\x, { (\x)^(-0.35489) *exp(2.71266) } );
\addlegendentry{SPH ($S=-0.355$)};
\end{axis}
\end{tikzpicture}%
}}}\\
	\caption {$Re=10000$}
	\label{re_10000}
\end{figure}

\subsection{Rayleigh-Taylor Instability}
\begin{figure}[bthp]
	\centering
	\makebox[\textwidth][c]{
		\subfloat[SPH] {\includegraphics[width=0.49\textwidth]{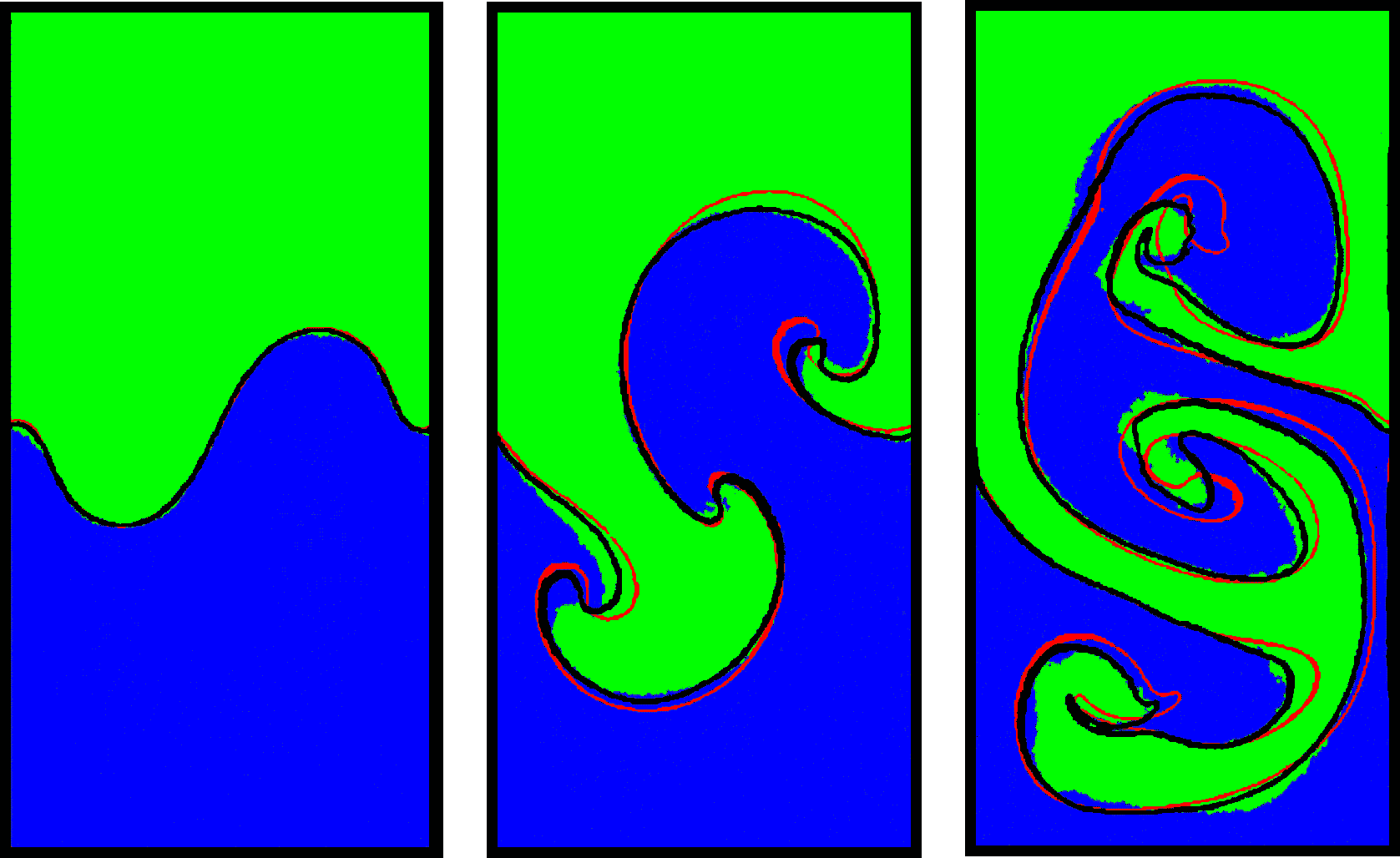}}\hfill
		\subfloat[LBM] {\includegraphics[width=0.49\textwidth]{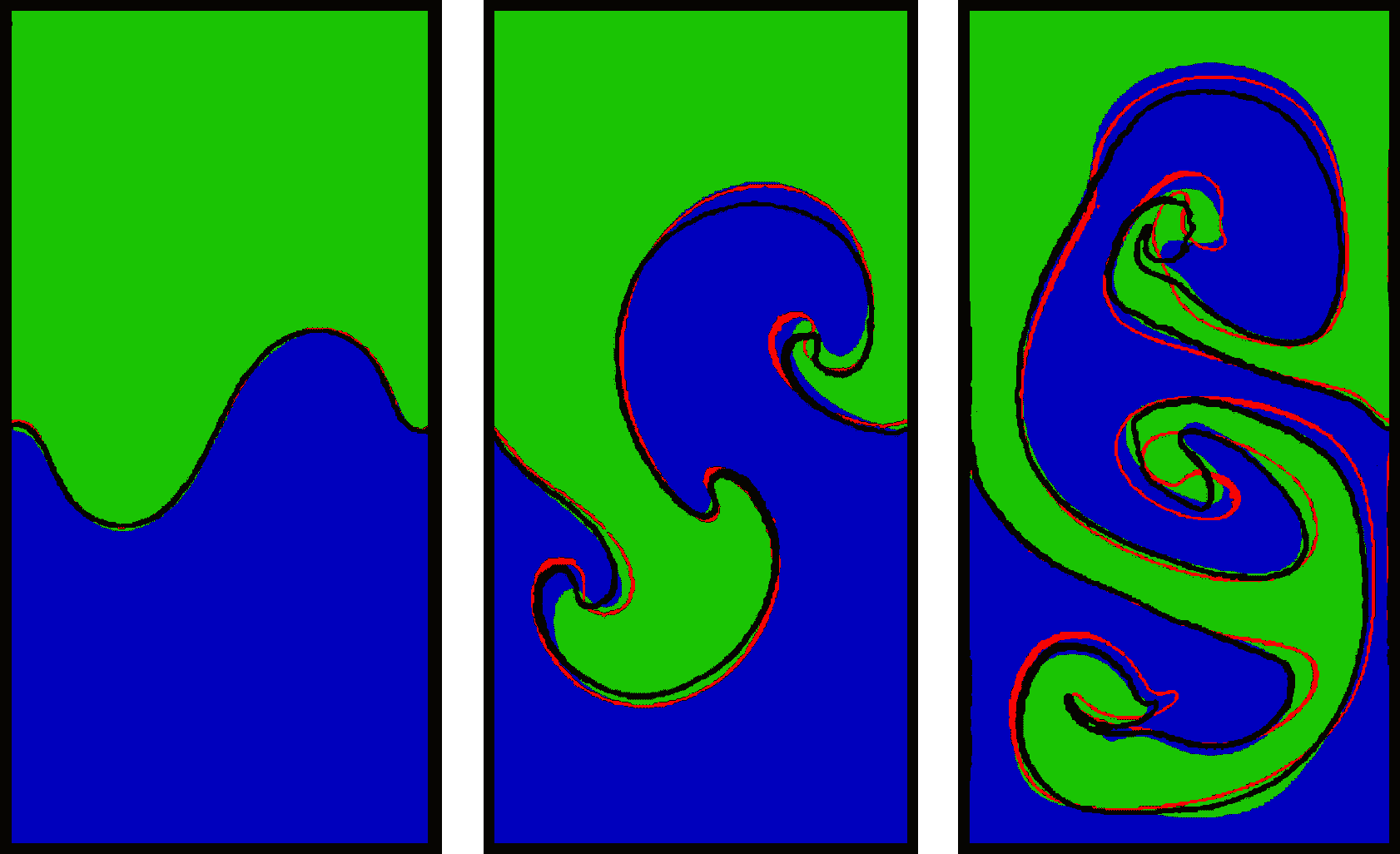}}}
	\caption{Phase distribution of Rayleigh-Taylor instability at selected timesteps : $t/t_g=1$, $3$ and $5$. Superposed with SPH interface (in black) and with Level-Set interface (in red) both extracted from~\cite{grenier2009}.}
	\label{rt}
\end{figure}

The Rayleigh-Taylor instability is a well-known two-phase problem in which a heavy fluid is placed on top of a light fluid with a given interface shape and submitted to gravity. Several previous works have reproduced this case with SPH or LBM, for example~\cite{grenier2009,szewc2013,Banari2012,Yan2016}. The test case and its parameters are borrowed from~\cite{grenier2009}. The computational domain is twice as high as long, $H\times L$ with $H=2L$ and populated with $40000$ nodes/particles. The density ratio is $1.8$ while the viscosity ratio is $1$. Gravity is set $g=9.81~\meter\per\second^{-2}$ for SPH and $g=1\times 10^{-4}~l.u.$ and oriented downwards. Therefore, the viscosity $\nu$ is adjusted to match the desired Reynolds number $Re = \sqrt{\frac{(H/2)^3 g}{\nu}} = 420$. No surface tension is used. No slip boundary conditions are applied to the walls. The interface is initialized as follows : $y= 1-\sin(2\pi x)$. Time $t$ is non-dimensionalized by $t_g=1/\sqrt{g/H}$. The distribution of the two phases is shown at selected timesteps in Fig.~\ref{rt}, superposed with results from~\cite{grenier2009}. Both methods are able to simulate the instability patterns as expected. Some differences are observable when $t/t_g \geq 3$ in particular when the interface is strongly distorted. LBM grows instabilities slightly faster than SPH and is closer to the behavior of the superposed Level-Set interface. On the other hand, our SPH results are naturally closer to the other SPH interface extracted from~\cite{grenier2009}. SPH appears to be more able than LBM (at the same resolution) to capture finer structures such the ones at $t/t_g = 5$ located on both ends of the mushroom-like shapes, but at an higher computational cost.

\section{Transformation matrices}\label{mrt_mat}
${\scriptstyle
	\bm{M}=\left(
	\begin{smallmatrix}
	\textcolor{white}{+}1 & \textcolor{white}{+}1 & \textcolor{white}{+}1 & \textcolor{white}{+}1 & \textcolor{white}{+}1 & \textcolor{white}{+}1 & \textcolor{white}{+}1 & \textcolor{white}{+}1 & \textcolor{white}{+}1\\
	-4 & -1 & -1 & -1 & -1 & \textcolor{white}{+}2 & \textcolor{white}{+}2 & \textcolor{white}{+}2 & \textcolor{white}{+}2\\
	\textcolor{white}{+}4 & -2 & -2 & -2 & -2 & \textcolor{white}{+}1 & \textcolor{white}{+}1 & \textcolor{white}{+}1 & \textcolor{white}{+}1\\
	\textcolor{white}{+}0 & \textcolor{white}{+}1 & \textcolor{white}{+}0 & -1 & \textcolor{white}{+}0 & \textcolor{white}{+}1 & -1 & -1 & \textcolor{white}{+}1\\
	\textcolor{white}{+}0 & -2 & \textcolor{white}{+}0 & \textcolor{white}{+}2 & \textcolor{white}{+}0 & \textcolor{white}{+}1 & -1 & -1 & \textcolor{white}{+}1\\
	\textcolor{white}{+}0 & \textcolor{white}{+}0 & \textcolor{white}{+}1 & \textcolor{white}{+}0 & -1 & \textcolor{white}{+}1 & \textcolor{white}{+}1 & -1 & -1\\
	\textcolor{white}{+}0 & \textcolor{white}{+}0 & -2 & \textcolor{white}{+}0 & \textcolor{white}{+}2 & \textcolor{white}{+}1 & \textcolor{white}{+}1 & -1 & -1\\
	\textcolor{white}{+}0 & \textcolor{white}{+}1 & -1 & \textcolor{white}{+}1 & -1 & \textcolor{white}{+}0 & \textcolor{white}{+}0 & \textcolor{white}{+}0 & \textcolor{white}{+}0\\
	\textcolor{white}{+}0 & \textcolor{white}{+}0 & \textcolor{white}{+}0 & \textcolor{white}{+}0 & \textcolor{white}{+}0 & \textcolor{white}{+}1 & -1 & \textcolor{white}{+}1 & -1
	\end{smallmatrix}\right),\quad
	\bm{M}^{-1}=\frac{1}{36}\left(
	\begin{smallmatrix}
	\textcolor{white}{+}4 & -4 & \textcolor{white}{+}4 & \textcolor{white}{+}0 & \textcolor{white}{+}0 & \textcolor{white}{+}0 & \textcolor{white}{+}0 & \textcolor{white}{+}0 & \textcolor{white}{+}0\\
	\textcolor{white}{+}4 & -1 & -2 & \textcolor{white}{+}6 & -6 & \textcolor{white}{+}0 & \textcolor{white}{+}0 & \textcolor{white}{+}9 & \textcolor{white}{+}0\\
	\textcolor{white}{+}4 & -1 & -2 & \textcolor{white}{+}0 & \textcolor{white}{+}0 & \textcolor{white}{+}6 & -6 & -9 & \textcolor{white}{+}0\\
	\textcolor{white}{+}4 & -1 & -2 & -6 & \textcolor{white}{+}6 & \textcolor{white}{+}0 & \textcolor{white}{+}0 & \textcolor{white}{+}9 & \textcolor{white}{+}0\\
	\textcolor{white}{+}4 & -1 & -2 & \textcolor{white}{+}0 & \textcolor{white}{+}0 & -6 & \textcolor{white}{+}6 & -9 & \textcolor{white}{+}0\\
	\textcolor{white}{+}4 & \textcolor{white}{+}2 & \textcolor{white}{+}1 & \textcolor{white}{+}6 & \textcolor{white}{+}3 & \textcolor{white}{+}6 & \textcolor{white}{+}3 & \textcolor{white}{+}0 & \textcolor{white}{+}9\\
	\textcolor{white}{+}4 & \textcolor{white}{+}2 & \textcolor{white}{+}1 & -6 & -3 & \textcolor{white}{+}6 & \textcolor{white}{+}3 & \textcolor{white}{+}0 & -9\\
	\textcolor{white}{+}4 & \textcolor{white}{+}2 & \textcolor{white}{+}1 & -6 & -3 & -6 & -3 & \textcolor{white}{+}0 & \textcolor{white}{+}9\\
	\textcolor{white}{+}4 & \textcolor{white}{+}2 & \textcolor{white}{+}1 & \textcolor{white}{+}6 & \textcolor{white}{+}3 & -6 & -3 & \textcolor{white}{+}0 & -9
	\end{smallmatrix}\right)}$

\end{document}